\documentclass[oneside,11pt]{book}
\usepackage{amssymb,amsthm,amsmath,graphicx,hyperref,color}
\usepackage{abntcite}
\usepackage{esint}
\usepackage[top=1.25in,bottom=1.25in,left=1.25in,right=1.25in,bindingoffset=0.25in,heightrounded,]{geometry}
\usepackage{breakcites}
\usepackage{fancyhdr}

\begin{document}
	\clearpage 
	\thispagestyle{empty}
	\begin{center}
		{\Large On the Linear Regime of the Characteristic formulation of General relativity in the Minkowski and Schwarzschild's Backgrounds}\\[4cm]
		{\large Carlos Eduardo Cede\~no Monta\~na} \\[2cm] 
			\href{http://link.aps.org/doi/10.1103/PhysRevD.92.124015}{PRD,{\bf 92}, 124015}
			\\
			\href{http://dx.doi.org/10.1007/s10714-016-2038-1}{Gen. Rel. Grav. ,{\bf 48}, 1}\\
			\href{http://stacks.iop.org/0264-9381/33/i=10/a=105010}{Class. Quantum. Grav. ,{\bf 33}, 105010}		
	\end{center}
\vfill
	\begin{center}
	\begin{tabular}{ll}
	\hspace*{5cm}& \begin{minipage}{5cm}
		Doctorate thesis of the Post Graduation Course in Astrophysics, advised by Dr. Jos\'e Carlos N. de Araujo, aproved in February, 17 2016.
	\end{minipage}
	\end{tabular}
	\end{center}
\vfill
\begin{center}
	{\Large 
	Instituto Nacional de Pesquisas Espaciais - INPE\\
	S\~ao Jos\'e dos Campos\\
	2016}
\end{center}
\newpage
\thispagestyle{empty}
\vspace*{6cm}
\begin{center}
\begin{minipage}{14cm}
\begin{center}{\bf Abstract}\end{center}	
We present here the linear regime of the Einstein's field equations in the characteristic formulation. Through a simple decomposition of the metric variables in spin-weighted spherical harmonics, the field equations are expressed as a system of coupled ordinary differential equations. The process for decoupling them leads to a simple equation for $J$ - one of the Bondi-Sachs metric variables - known in the literature as the master equation. Then, this last equation is solved in terms of Bessel's functions of the first kind for the Minkowski's background, and in terms of the Heun's function in the Schwarzschild's case. In addition, when a matter source is considered, the boundary conditions across the time-like world tubes bounding the source are taken into account. These boundary conditions are computed for all multipole modes. Some examples as the point particle binaries in circular and eccentric orbits, in the Minkowski's background  are shown as particular cases of this formalism.
\end{minipage}
\end{center}
\newpage
\bibliographystyle{abnt-alf}
\thispagestyle{empty}
\vspace*{19cm} 
\begin{minipage}{14cm}
{\itshape\Large This text is dedicated specially to my father Ricardo, in Memoriam, my mother Elicenia, my brother Ricardo and my wife, Sandra. Thanks for always being with me.}
\end{minipage}
\thispagestyle{empty}
\vspace*{6cm} 
\begin{minipage}{14cm}
\begin{center}{\bf Acknowledgements}\end{center}
I feel grateful to my parents by their continuous support during all instants of my life. I appreciate very much their guidance and patience. I would like to express my gratitude for furnishing me a real model to follow. I don't know how to express my deep grateful to my brother, who with his criticisms and conscientious reading, help me to improve my text. Also, I owe a special mention to Sandra, my wife, who helps me every day with her support, happiness and for encouraging me to improve in all aspects. Thanks also go to her for all her suggestions and critical readings of my manuscript. I would like to express my deep and sincere gratitude to my advisor Dr. Jos\'e Carlos N. de Araujo. His continuous support, his patient guidance and enthusiastic encouragement during my PhD study and related research have been valuable. For give me hope in the most difficult circumstances. I think that this project would not have been possible without his advices and hope.
I would also like to thank the Brazilian agencies CAPES, FAPESP (2013/11990-1) and CNPq (308983/2013-0) for the financial support.
\end{minipage}
\tableofcontents
\chapter{INTRODUCTION}
The high complexity of the Einstein's field equations, given their non-linearity, makes impossible to find analytical solutions valid for all gravitational systems. However, in addition to the exact solutions, which are valid for some restricted geometries and situations, the perturbative methods and the numerical relativity are two of the most promising ways to solve the Einstein's field equations in presence of strong gravitational fields in a wide variety of matter configurations. 

The holy grail of numerical relativity is to obtain the gravitational radiation patterns produced by black hole - black hole (BH-BH), neutron star - neutron star (NS-NS) or neutron star - black hole (NS-BH) binary systems, because of their relevance in astrophysics. Actually, there are highly accurate and strongly convergent numerical codes, capable of performing simulations of binaries taking into account the mass and momentum transfer \cite{FBSTR06,LF14,KKHMK15}, the hydrodynamic evolution \cite{TBFS05,BMCM13,MBM14}, the magneto-hydrodynamic evolution \cite{F08}, the  electromagnetic and gravitational signatures produced by binaries \cite{PLPLANM13,PLLPANM15,KKHMK15}; and recently,  the  spin-spin and the spin-orbit interactions in binary systems have been also studied \cite{DLZ08,I12,ZL15}.

All these advances were possible thanks to the Lichnerowicz, Choquet-Bruhat and Geroch works \cite{L44,F52,CG69}, which opened the possibility to evolve a space-time from a set of initial data; putting the principles of the Initial Value Problem (IVP) \cite{Gourgoulhon07,A08_book,BS10_Book} and checking that this is a local and a global well-posed problem, that are necessary conditions to guarantee stable numerical evolutions. 

A different point of view to carry out the evolution of a given space-time was proposed by Bondi et. al. in the 1960s decade \cite{B_VII,B_VIII}. They studied the problem of evolving a given metric, from the specification of it and its first derivatives, by using the radiation coordinates, assuming that the initial data is given on a null initial hypersurface and on a prescribed time-like world tube. This is known as the Characteristic Initial Value Problem (CIVP) \cite{SF82} and was effectively proved as a well-posed problem when the field equations are written in terms only of first-order derivatives \cite{F05}.

In the literature, there are essentially three possible ways to evolve space-times and sources from a specific initial data, see  e. g. \cite{C00,L01,MM03,GM07,W12,CGHS15} for detailed descriptions and status of the formalisms available in numerical relativity. The first one is the Regge calculus, in which the space-time is decomposed in a network of 4-dimensional flat simplices.\footnote{Simplices (Simplexes) are the generalisation of triangles for bi-dimensional and tetrahedron for three-dimensional spaces to four or more dimensional spaces. In the Regge calculus these simplices are supposed flat and the curvature is given just at the vertices of the structure, just like when a sphere is covered using flat triangles.} The Riemann tensor and consequently the field equations are expressed in a discrete version of such atomic structures. It extends the calculus to the most general spaces than differentiable manifolds \cite{R61}. The second are the Arnowitt-Deser-Misner (ADM) based formulations in which the space-time is foliated into space-like hypersurfaces which are locally orthogonal to the tangent vectors of a central time-like geodesic \cite{ADM59,ADM60_1,ADM60_2,Y71,Y79}. The third are the characteristic formalisms, which are based on the Bondi et. al. works in which the space-time is foliated into null cones emanated from a central time-like geodesic or a world tube, and hypersurfaces that are related to the unit sphere through diffeomorphisms \cite{B_VII,B_VIII,W83,W84,W12}. 

Most of the recent work have been constructed using the ADM formalisms,\footnote{These formalisms are known also as 3+1 because of the form in which the field equations are decomposed.} whereas the null cone formalisms are less known. One of the biggest problems in these last formulations is their mathematical complexity. However, these formalisms result particularly useful for constructing waveform extraction schemes, because they are based on radiation coordinates. Impressive advances in the characteristic formulation have been carried out recently, in particular in the development of matching algorithms, which evolving from the Cauchy-Characteristic-Extraction (CCE) to the Cauchy-Characteristic-Matching (CCM) \cite{BGLW96,BGLMW05,RBLTS07,BBSW09,BSWZ11,ROSS11}. 

A cumbersome aspect of the null-cone formulation is the formation of caustics in the non-linear regime, because at these points the coordinates are meaningless. The caustics are formed when the congruences of light beams bend, focusing and forming points where the coordinate system is not well defined. This problem is not present in the CCM algorithms because the characteristic evolution is performed for the vacuum, where the light beams  not bend outside of the time-like world tube \cite{W12}. Therefore, the characteristic evolutions have been usually performed only for the vacuum, considering the sources as bounded by such time-like hypersurface. Inside of the time-like world tube, the matter is evolved from the conservation laws. However, there are some works in which the gravitational collapse of scalar fields, massive or not, are performed using only characteristic schemes, but obeying restrictive geometries and taking into account the no-development of caustics \cite{GBF07,B14a,B14b}. At this point it is worth mentioning that the finite difference schemes are not the unique methods to solve efficiently the Einstein's field equations. There are significative advances in the spectral methods applied to the characteristic formulation using the Galerkin method, see e.g. \cite{L08,LO07,OR08,OR11}

One way to calibrate these complex and accurate codes is to make tests of validity in much simpler systems and geometries than those used in such kind of simulations. In order to do so, toy models for these codes can be obtained with the linear version of the field equations. Depending on the background, the linearised equations can lead to several regimes of validity. One example of this is that the linear regime of the field equations on a Minkowski or on a Schwarzschild's background leads to waveforms and behaviours of the gravitational fields completely different. There is a great quantity of possibilities to perform approximations to the field equations. Among them, there are different orders of the Post-Newtonian approximations, the post-Minkowskian approximations, the approximations using spectral decompositions, and so on. 

Despite lack of real physical meaning near to the sources, the linear approximations of the characteristic formulation of general relativity exhibit an interesting point of view even from the theoretical perspective. It is possible to construct exact solutions to the Einstein's field equations for these space-times in a easy way. It allows us to reproduce at first approximation some interesting features of simple radiative systems. In the weak field limit, it is possible to write the field equations as a system of coupled ordinary differential equations, that can be easily solved analytically. 

Here we present exact solutions for space-times resulting from small perturbations to the Minkowski and Schwarzschild's space-times. Also, we construct three simple toy models, a thin shell, a circular point particle binary system of unequal masses, and a generalisation to this last model including eccentricity. These gravitational radiating systems were treated and solved from the formalism developed from the perturbations for the metrics mentioned above. In order to present that, perturbations to a generic space-time at first and higher order are shown in chapter 2. Gravitational wave equations for these orders are obtained as well as their respective eikonal equations. Additionally, chapter 2 introduces the Green functions and the multipolar expansion. In chapter 3, the eth formalism is explained in detail, separately from the characteristic formulation. It is an efficient method to regularise angular derivative operators. The spin-weighted spherical harmonics are introduced from the usual harmonics through successive applications of the eth derivatives. In chapter 4 the initial value problem, the ADM formulation and the outgoing characteristic formalism with and without eth expressions are shown. In chapter 5 the linear regime in the outgoing characteristic formulation is obtained, the field equations are simplified and solved analytically. In order to do that a differential equation (the {\it master equation}) for $J$, a Bondi-Sachs variable, is found. This equation is solved for the Minkowski (Schwarzschild) background in terms of Hypergeometric (Heun) functions. Finally in chapter 6, two examples are presented, the point particle binaries without and with eccentricity. At the end, the conclusions and some final considerations are discussed. 
\chapter[Linear Regime]{LINEAR REGIME OF THE EINSTEIN'S FIELD EQUATIONS AND GRAVITATIONAL WAVES}
This chapter explores the linear regime of Einstein's field equations and the gravitational waves. In general, the linearisation of the Einstein's field equations is performed assuming a flat background  \cite{B07,C101,C102,C103}. However, this approximation turns inapplicable to the cases in which strong fields are involved. Eisenhart in 1926 and Komar in 1957 made perturbations to the metric tensor at the first order showing how the gravitational waves are propagated away from the sources \cite{RW57}. \citeonline{RW57} considered small perturbations in a spherical symmetric space-time to explore the stability of the Schwarzschild's solution,  obtaining a radial wave equation in presence of an effective gravitational potential, namely the {\it Reege-Wheeler equation}, which appears for odd-parity perturbations. On the other hand, \citeonline{Z70} made even-parity perturbations obtaining a different radial wave equation, namely the {\it Zerilli equation} obeying a different effective potential. After that, by using the vector and tensor harmonics, Moncrief extended the Zerilli's works to the Reissner-Nordstr\"om exterior space-time and to stellar models by using a perfect fluid stress-energy tensor \cite{M74_I,M74_II,M74_III,M74_IV}. \citeonline{BH64} explored the stability of the {\it Geons}, which are objects composed of electromagnetic fields held together by gravitational attraction in the linear regime of the field equations, off the flat space-time, but considering spherical symmetry and asymptotically flat space-times.  Isaacson found a generalisation to the gravitational wave equation when an arbitrary background is considered. He proved that the gravitational waves for high and low frequencies are found by performing perturbations to distinct orders in the metric tensor \cite{I68_1,I68_2}. 

Here some of the aspects of the linearisation approximation to first and higher orders are examined. By using the Wentzel-Kramers-Brillouin approximation (WKB) the eikonal equation is found, relating the tensor of amplitudes to the metric perturbations with its propagation vector. After that, in the Minkowski's background, the gravitational waves are expressed in terms of the Green's functions. In addition, a multipolar expansion is made as usual. Finally, following \cite{PM63}, the quadrupole radiation formula is used to find the energy lost by emission of gravitational waves by a binary system of unequal masses. 

The convention used here with respect to the indices is: $x^\mu$ represent coordinates, $\mu=1$ for temporal components, $\mu=i,j,k,\cdots$ for spatial coordinates. The adopted signature is $+2$. 

Finally, it is worth mentioning that the linearisation process of the Einstein's field equation presented in this section and the general results shown here are important because we linearise the characteristic equations in the same way, by just perturbing the Bondi-Sachs metric. The equations obtained for these perturbations (See Chapters \ref{LRME} and \ref{APP}) are equivalent to those obtained in this section. For this reason it is not surprise that in the characteristic formulation we obtain radiative solutions and that they are characterised by the Bondi's News function. 
\section{First Order Perturbations}
In this section, we will explore in some detail the linear regime of the Einstein's field equations when an arbitrary background is considered. Despite the derivation of the wave equation at first order does not differ from that in which a Minkowski's background is taken into account, additional terms related to the background Riemann tensor and the correct interpretation of the D'Alembertian is shown. We follow the same convention and procedures exposed by \citeonline{I68_1} and subsequently used in \cite{MTW73}

As a starting point, perturbations to an arbitrary background $\overset{\hspace{-0.3cm}(0)}{g_{\mu\nu}}$ at first order are chosen, i.e.,
\begin{equation}
g_{\mu\nu}=\overset{\hspace{-0.3cm}(0)}{g_{\mu\nu}}+\epsilon \overset{\hspace{-0.3cm}(1)}{g_{\mu\nu}},
\label{eq2.1}
\end{equation}
where $\epsilon$ is a parameter that measures the perturbation, satisfying $\epsilon\ll 1$. It is worth stressing that it guaranties that the second term is smaller than the first, because the characteristic length of such perturbations, $\lambda$, must be very small compared to the characteristic length of the radius of curvature of the background, $L$. This limit is known as the high frequency approximation \cite{I68_1}. 

Considering that the inverse metric $g^{\mu\nu}$ is given as a background term plus a first order perturbation with respect to the background, i.e.,
\begin{equation}
g^{\mu\nu}=\overset{\hspace{-0.3cm}(0)}{g^{\mu\nu}}+\epsilon\, \overset{\hspace{-0.3cm}(1)}{g^{\mu\nu}},
\label{eq2.2}
\end{equation}
and that $g_{\mu\nu}g^{\eta\nu}= \overset{\hspace{-0.3cm}(0)}{g_{\mu\nu}} \overset{\hspace{-0.3cm}(0)}{g^{\eta\nu}} = \delta_\mu^{~\eta}$, then
\begin{align}
g_{\mu\sigma}g^{\sigma\nu}&= 
\overset{\hspace{-0.3cm}(0)}{g_{\mu\sigma}} \overset{\hspace{-0.3cm}(0)}{g^{\sigma\nu}}+ \epsilon\left(\, \overset{\hspace{-0.3cm}(1)}{g^{\sigma\nu}}\, \overset{\hspace{-0.3cm}(0)}{g_{\mu\sigma}} + \overset{\hspace{-0.3cm}(1)}{g_{\mu\sigma}} \overset{\hspace{-0.3cm}(0)}{g^{\sigma\nu}} \right)+O(\epsilon^2). 
\label{eq2.3}
\end{align}
Therefore, the perturbation of the inverse metric is given by
\begin{equation}
\overset{\hspace{-0.3cm}(1)}{g^{\eta\nu}}=-\overset{\hspace{-0.3cm}(1)}{g_{\mu\sigma}} \overset{\hspace{-0.3cm}(0)}{g^{\mu\eta}}\overset{\hspace{-0.3cm}(0)}{g^{\sigma\nu}}.
\label{eq2.5}
\end{equation}
As a result, the Christoffel's symbols of the first kind, reads
\begin{equation}
\Gamma_{\mu\nu\gamma}=\frac{1}{2}\left(g_{\mu\nu,\gamma}+g_{\gamma\mu,\nu}-g_{\nu\gamma,\mu}\right),
\label{eq2.7}
\end{equation}
where the comma indicates partial derivative. These symbols can be separated as a term referred to the background plus a perturbation, namely,
\begin{equation}
\Gamma_{\mu\nu\gamma}=\overset{\hspace{-0.6cm}(0)}{\Gamma_{\mu\nu\gamma}}+\epsilon \overset{\hspace{-0.6cm}(1)}{\Gamma_{\mu\nu\gamma}},
\label{eq2.8}
\end{equation}
where
\begin{equation}
\overset{\hspace{-0.6cm}(i)}{\Gamma_{\mu\nu\gamma}}=\frac{1}{2}\left(\,\overset{\hspace{-0.5cm}(i)}{g_{\mu\nu,\gamma}}+\overset{\hspace{-0.5cm}(i)}{g_{\gamma\mu,\nu}}
-
\overset{\hspace{-0.5cm}(i)}{g_{\nu\gamma,\mu}}\right),\hspace{0.5cm}i=0,1.
\label{eq2.9}
\end{equation}

Thus, the Christoffel's symbols of the second kind,
\begin{align}
\Gamma^\mu_{~\nu\gamma}&=g^{\mu\sigma}\Gamma_{\sigma\nu\gamma}, 
\label{eq2.10}
\end{align}
can also be separated \cite{I68_1} as,
\begin{align}
\Gamma^\mu_{~\nu\gamma}&=
\overset{\hspace{-0.6cm}(0)}{\Gamma^\mu_{~\nu\gamma}} + \epsilon\overset{\hspace{-0.6cm}(1)}{\Gamma^\mu_{~\nu\gamma}}+O(\epsilon^2), 
\label{eq2.11}
\end{align}
where,
\begin{align}
\overset{\hspace{-0.6cm}(0)}{\Gamma^\mu_{~\nu\gamma}}=\overset{\hspace{-0.3cm}(0)}{g^{\mu\sigma}}\overset{\hspace{-0.6cm}(0)}{\Gamma_{\sigma\nu\gamma}} 
\hspace{0.5cm} \text{and}\hspace{0.5cm}
\overset{\hspace{-0.6cm}(1)}{\Gamma^\mu_{~\nu\gamma}}= \overset{\hspace{-0.3cm}(0)}{g^{\mu\sigma}}\overset{\hspace{-0.6cm}(1)}{\Gamma_{\sigma\nu\gamma}}
+
\overset{\hspace{-0.3cm}(1)}{g^{\mu\sigma}}\overset{\hspace{-0.6cm}(0)}{\Gamma_{\sigma\nu\gamma}}.
\label{eq2.12}
\end{align}

Consequently the Riemann's tensor is written as a term associated to the background plus a term corresponding to a perturbation, i.e.,
\begin{align}
R^\mu_{~~\nu\gamma\delta}&= \overset{\hspace{-0.7cm}(0)}{R^\mu_{~~\nu\gamma\delta}}+\epsilon \overset{\hspace{-0.7cm}(1)}{R^\mu_{~~\nu\gamma\delta}},
\label{eq2.13}
\end{align}
where the background Riemann tensor is given by
\begin{equation}
\overset{\hspace{-0.7cm}(0)}{R^\mu_{~~\nu\gamma\delta}}=
2\overset{\hspace{-0.8cm}(0)}{\Gamma^\mu_{~\nu[\delta,\gamma]}}
+2\overset{\hspace{-0.6cm}(0)}{\Gamma^\sigma_{~\nu[\delta}}\overset{\hspace{-0.6cm}(0)}{\Gamma^\mu_{~\gamma]\sigma}},
\label{eq2.14}
\end{equation}
and the term associated with the perturbation reads
\begin{align}
\overset{\hspace{-0.7cm}(1)}{R^\mu_{~~\nu\gamma\delta}}
&=
2\,\overset{\hspace{-0.8cm}(1)}{\Gamma^\mu_{~\nu[\delta,\gamma]}}
+ 2\, \overset{\hspace{-0.5cm}(1)}{\Gamma^\mu_{~\sigma[\gamma}}\overset{\hspace{-0.5cm}(0)}{\Gamma^\sigma_{~\delta]\nu}}   
+2\,\overset{\hspace{-0.5cm}(1)}{\Gamma^\sigma_{~\nu[\delta}}\overset{\hspace{-0.5cm}(0)}{\Gamma^\mu_{~\gamma]\sigma}}.
\label{eq2.15}
\end{align}

As usual, the square brackets indicate anti-symmetrisation, i.e.,
\begin{equation}
A_{[\alpha_1\cdots\alpha_n]}=\frac{1}{n!}\epsilon_{\alpha_1\cdots\alpha_n}^{\hspace{0.9cm}\beta_1\cdots\beta_n}A_{\beta_1\cdots\beta_n},
\end{equation}
where $\epsilon_{\alpha_1\cdots\alpha_n}^{\hspace{0.9cm}\beta_1\cdots\beta_n}$ is the generalised Levi-Civita permutation symbol \cite{MTW73}.

From $\overset{\hspace{-0.8cm}(1)}{\Gamma^\mu_{~\nu\delta:\gamma}}$, where the colon indicates covariant derivative  associated with the background metric $\overset{(0)}{g}_{\mu\nu}$, one obtains
\begin{equation}
\overset{\hspace{-0.7cm}(1)}{\Gamma^\mu_{~\nu[\delta,\gamma]}}= \overset{\hspace{-0.7cm}(1)}{\Gamma^\mu_{~\nu[\delta:\gamma]}} 
- \overset{\hspace{-0.5cm}(0)}{\Gamma^\mu_{~\sigma[\gamma}} \overset{\hspace{-0.5cm}(1)}{\Gamma^\sigma_{~\delta]\nu}}
+\overset{\hspace{-0.5cm}(0)}{\Gamma^\sigma_{~\nu[\gamma}}
\overset{\hspace{-0.5cm}(1)}{\Gamma^\mu_{~\delta]\sigma}}
+\overset{\hspace{-0.5cm}(0)}{\Gamma^\sigma_{~[\gamma\delta]}} \overset{\hspace{-0.5cm}(1)}{\Gamma^\mu_{~\nu\sigma}}.
\label{eq2.16}
\end{equation}

Thus, substituting \eqref{eq2.16} into the Riemann's tensor \eqref{eq2.15}, one immediately obtains
\begin{align}
\overset{\hspace{-0.7cm}(1)}{R^\mu_{~~\nu\gamma\delta}}&=
2\overset{\hspace{-0.9cm}(1)}{\Gamma^\mu_{~\nu[\delta:\gamma]}}.
\label{eq2.17}
\end{align}

From \eqref{eq2.10} it follows that
\begin{align}
\overset{\hspace{-0.9cm}(1)}{\Gamma^\mu_{~\nu[\delta:\gamma]}}&= \overset{\hspace{-0.6cm}(1)}{\Gamma_{\sigma\nu[\delta}}\overset{\hspace{-0.8cm}(0)}{g^{\mu\sigma}_{~~~:\gamma]}} +\overset{\hspace{-0.3cm}(0)}{g^{\mu\sigma}}\overset{\hspace{-0.8cm}(1)}{\Gamma_{\sigma\nu[\delta:\gamma]}}
-\overset{\hspace{-0.6cm}(0)}{\Gamma_{\sigma\nu[\delta}}\overset{\hspace{-0.6cm}(1)}{g^{\mu\sigma}_{~~~:\gamma]}}
-\overset{\hspace{-0.3cm}(1)}{g^{\mu\sigma}}\overset{\hspace{-0.8cm}(0)}{\Gamma_{\sigma\nu[\delta:\gamma]}}\nonumber\\
&=\overset{\hspace{-0.3cm}(0)}{g^{\mu\sigma}}\overset{\hspace{-0.9cm}(1)}{\Gamma_{\sigma\nu[\delta:\gamma]}}
-\overset{\hspace{-0.6cm}(0)}{\Gamma_{\sigma\nu[\delta}}\overset{\hspace{-0.7cm}(1)}{g^{\mu\sigma}_{~~~:\gamma]}}
-\overset{\hspace{-0.3cm}(1)}{g^{\mu\sigma}}\overset{\hspace{-0.8cm}(0)}{\Gamma_{\sigma\nu[\delta:\gamma]}}.
\label{eq2.18}
\end{align}

Then, substituting \eqref{eq2.18} into \eqref{eq2.17}
\begin{align}
\overset{\hspace{-0.7cm}(1)}{R^\mu_{~~\nu\gamma\delta}} 
=&\overset{\hspace{-0.3cm}(0)}{g^{\mu\sigma}}\overset{\hspace{-0.8cm}(1)}{\Gamma_{\sigma\nu\delta:\gamma}} - \overset{\hspace{-0.3cm}(0)}{g^{\mu\sigma}}\overset{\hspace{-0.8cm}(1)}{\Gamma_{\sigma\nu\gamma:\delta}}
-\overset{\hspace{-0.6cm}(0)}{\Gamma_{\sigma\nu\delta}}\overset{\hspace{-0.6cm}(1)}{g^{\mu\sigma}_{~~~:\gamma}}
+\overset{\hspace{-0.6cm}(0)}{\Gamma_{\sigma\nu\gamma}}\overset{\hspace{-0.6cm}(1)}{g^{\mu\sigma}_{~~~:\delta}}\nonumber\\
&-\overset{\hspace{-0.3cm}(1)}{g^{\mu\sigma}}\overset{\hspace{-0.7cm}(0)}{\Gamma_{\sigma\nu\delta:\gamma}}
+\overset{\hspace{-0.3cm}(1)}{g^{\mu\sigma}}\overset{\hspace{-0.7cm}(0)}{\Gamma_{\sigma\nu\gamma:\delta}}.
\label{eq2.19}
\end{align}

In order to compute the Riemann's tensor for the perturbation \eqref{eq2.19}, it is necessary to calculate
\begin{align}
\overset{\hspace{-0.8cm}(1)}{\Gamma_{\sigma\nu\delta:\gamma}}&=
\frac{1}{2}\left(\,\overset{\hspace{-0.7cm}(1)}{g_{\sigma\nu,\delta:\gamma}}+\overset{\hspace{-0.7cm}(1)}{g_{\delta\sigma,\nu:\gamma}}-\overset{\hspace{-0.7cm}(1)}{g_{\nu\delta,\sigma:\gamma}}\right),
\label{eq2.20}
\end{align}
where
\begin{align}
\overset{\hspace{-0.7cm}(1)}{g_{\sigma\nu,\delta:\gamma}}&=\left(\overset{\hspace{-0.5cm}(1)}{g_{\sigma\nu:\delta}}+\overset{\hspace{-0.5cm}(0)}{\Gamma^{\lambda}_{~\sigma\delta}} \overset{\hspace{-0.3cm}(1)}{g_{\lambda\nu}}+\overset{\hspace{-0.7cm}(0)}{\Gamma^{\lambda}_{~~\delta\nu}}\overset{\hspace{-0.3cm}(1)}{g_{\sigma\lambda}}\right)_{:\gamma}\nonumber\\
&=
\overset{\hspace{-0.7cm}(1)}{g_{\sigma\nu:\delta\gamma}}
+\overset{\hspace{-0.7cm}(0)}{\Gamma^{\lambda}_{~\sigma\delta:\gamma}} \overset{\hspace{-0.3cm}(1)}{g_{\lambda\nu}}
+\overset{\hspace{-0.5cm}(0)}{\Gamma^{\lambda}_{~\sigma\delta}} \overset{\hspace{-0.5cm}(1)}{g_{\lambda\nu:\gamma}}
+\overset{\hspace{-0.8cm}(0)}{\Gamma^{\lambda}_{~~\delta\nu:\gamma}}\overset{\hspace{-0.3cm}(1)}{g_{\sigma\lambda}}
+\overset{\hspace{-0.5cm}(0)}{\Gamma^{\lambda}_{~~\delta\nu}}\overset{\hspace{-0.5cm}(1)}{g_{\sigma\lambda:\gamma}}.
\label{eq2.21}
\end{align}

Substituting \eqref{eq2.21} into \eqref{eq2.20} it is found that
\begin{align}
\overset{\hspace{-0.7cm}(1)}{\Gamma_{\sigma\nu\delta:\gamma}}=\frac{1}{2}\left(\,\overset{\hspace{-0.7cm}(1)}{g_{\sigma\nu:\delta\gamma}} + \overset{\hspace{-0.7cm}(1)}{g_{\delta\sigma:\nu\gamma}} - \overset{\hspace{-0.7cm}(1)}{g_{\nu\delta:\sigma\gamma}} \right)+\overset{\hspace{-0.8cm}(0)}{\Gamma^\lambda_{~\nu\delta:\gamma}}\overset{\hspace{-0.3cm}(1)}{g_{\sigma\lambda}}+\overset{\hspace{-0.5cm}(0)}{\Gamma^\lambda_{~\nu\delta}}\overset{\hspace{-0.6cm}(1)}{g_{\sigma\lambda:\gamma}}.
\label{eq2.22}
\end{align}

Therefore, substituting \eqref{eq2.22} into \eqref{eq2.19} one obtains that the Riemann tensor corresponding to the perturbations is given by
\begin{align}
\overset{\hspace{-0.7cm}(1)}{R^\mu_{~~\nu\gamma\delta}}
=&\frac{1}{2}\left(\,
\overset{\hspace{-0.6cm}(1)}{g^\mu_{~\nu:\delta\gamma}}
+\overset{\hspace{-0.6cm}(1)}{g^{~\mu}_{\delta~:\nu\gamma}}
+\overset{\hspace{-0.6cm}(1)}{g^{~~~\mu}_{\nu\gamma:~\delta}}
-\overset{\hspace{-0.6cm}(1)}{g^{~~~\mu}_{\nu\delta:~\gamma}}
-\overset{\hspace{-0.6cm}(1)}{g^{\mu}_{~\nu:\gamma\delta}}
-\overset{\hspace{-0.6cm}(1)}{g^{~\mu}_{\gamma~:\nu\delta}}
\right).
\label{eq2.23}
\end{align}

Now, writing the field equations as
\begin{equation}
R_{\mu\nu}=8\pi \left(T_{\mu\nu}-\frac{1}{2}g_{\mu\nu}T\right),
\label{eq2.24}
\end{equation}
where, $T_{\mu\nu}$ and $T$ are the energy-stress tensor and its trace respectively, and using \eqref{eq2.11} then
\begin{equation}
\overset{\hspace{-0.3cm}(0)}{R_{\mu\nu}}+\epsilon\overset{\hspace{-0.3cm}(1)}{R_{\mu\nu}}=8\pi \left(T_{\mu\nu}-\frac{1}{2}g_{\mu\nu}T\right).
\label{eq2.25}
\end{equation}

Assuming that the background satisfies the Einstein's field equations
\begin{equation}
\overset{\hspace{-0.3cm}(0)}{R_{\mu\nu}}=8\pi \left(T_{\mu\nu}-\frac{1}{2}g_{\mu\nu}T\right),
\label{eq2.26}
\end{equation}
i.e., disregarding perturbations on the stress-energy tensor, we found that the perturbation to the Ricci's tensor satisfies
\begin{equation}
\overset{\hspace{-0.3cm}(1)}{R_{\mu\nu}}=0.
\label{eq2.27}
\end{equation}

Contracting \eqref{eq2.23}, and substituting in \eqref{eq2.27}
\begin{equation}
\frac{1}{2}\left(
\overset{(1)}{g^\mu_{~\nu:\delta\mu}}
+\overset{(1)}{g^{~\mu}_{\delta~:\nu\mu}}
-\overset{(1)}{g^{~~~\mu}_{\nu\delta:~\mu}}
-\overset{(1)}{g^{~\mu}_{\mu~:\nu\delta}}
\right)=0
\label{eq2.28}
\end{equation}
which corresponds to a first order wave equation for the metric perturbations.

It is worth stressing that \eqref{eq2.28} can be re-written as
\begin{equation}
\overset{\hspace{-0.3cm}(0)}{g^{\mu\sigma}}\left(
2\,\overset{\hspace{-0.8cm}(1)}{g_{\sigma\nu:[\delta\mu]}} +\overset{\hspace{-0.7cm}(1)}{g_{\sigma\nu:\mu\delta}}
+2\,\overset{\hspace{-0.8cm}(1)}{g_{\delta\sigma:[\nu\mu]}}
+\overset{\hspace{-0.7cm}(1)}{g_{\delta\sigma:\mu\nu}}
-\overset{\hspace{-0.7cm}(1)}{g_{\nu\delta:\sigma\mu}}
-\overset{\hspace{-0.7cm}(1)}{g_{\mu\sigma:\nu\delta}}\right)
=0,
\label{eq2.29}
\end{equation}
where
\begin{equation}
\overset{\hspace{-0.9cm}(1)}{g_{\sigma\nu:[\delta\mu]}}=
\overset{\hspace{-0.9cm}(1)}{g_{\sigma\nu,[\delta:\mu]}}
-\overset{\hspace{-0.9cm}(0)}{\Gamma^{\lambda}_{~\sigma[\delta:\mu]}}\overset{\hspace{-0.3cm}(1)}{g_{\nu\lambda}}
-\overset{\hspace{-0.5cm}(1)}{g_{\nu\lambda:[\mu}}\overset{\hspace{-0.6cm}(0)}{\Gamma^{\lambda}_{~\delta]\sigma}}
-\overset{\hspace{-0.9cm}(0)}{\Gamma^{\lambda}_{~\nu[\delta:\mu]}}\overset{\hspace{-0.3cm}(1)}{g_{\sigma\lambda}}
-\overset{\hspace{-0.5cm}(1)}{g_{\sigma\delta:[\mu}}\overset{\hspace{-0.6cm}(0)}{\Gamma^{\lambda}_{~\delta]\nu}}.
\label{eq2.30}
\end{equation}

Explicitly, \eqref{eq2.30} is
\begin{align*}
\overset{\hspace{-0.8cm}(1)}{g_{\sigma\nu:[\delta\mu]}}=&
\overset{\hspace{-0.8cm}(1)}{g_{\sigma\nu,[\delta\mu]}}-\overset{\hspace{-0.5cm}(1)}{g_{\lambda\nu,[\delta}}\overset{\hspace{-0.7cm}(0)}{\Gamma^{\lambda}_{~\mu]\sigma}}-\overset{\hspace{-0.5cm}(1)}{g_{\sigma\lambda,[\delta}}\overset{\hspace{-0.7cm}(0)}{\Gamma^\lambda_{~\mu]\nu}}-\overset{\hspace{-0.7cm}(0)}{\Gamma^\lambda_{~[\delta\mu]}}\overset{\hspace{-0.5cm}(0)}{g_{\sigma\nu,\lambda}}
\nonumber\\
&-\left(\,\overset{\hspace{-0.9cm}(0)}{\Gamma^\lambda_{~\sigma[\delta,\mu]}}+\overset{\hspace{-0.5cm}(0)}{\Gamma^\lambda_{~\epsilon[\mu}}\overset{\hspace{-0.5cm}(0)}{\Gamma^\epsilon_{~\delta]\sigma}}-\overset{\hspace{-0.5cm}(0)}{\Gamma^\epsilon_{~\sigma[\mu}}\overset{\hspace{-0.5cm}(0)}{\Gamma^\lambda_{~\delta]\epsilon}}-\overset{\hspace{-0.5cm}(0)}{\Gamma^\epsilon_{~[\delta\mu]}}\overset{\hspace{-0.5cm}(0)}{\Gamma^\lambda_{~\sigma\epsilon}}\right)\overset{\hspace{-0.3cm}(1)}{g_{\nu\lambda}}
\nonumber\\
&-\overset{\hspace{-0.6cm}(1)}{g_{\nu\lambda,[\mu}}\overset{\hspace{-0.6cm}(0)}{\Gamma^\lambda_{~\delta]\sigma}}+\overset{\hspace{-0.5cm}(0)}{\Gamma^\lambda_{~\sigma[\delta}}\overset{\hspace{-0.5cm}(0)}{\Gamma^\epsilon_{~\mu]\nu}}\overset{\hspace{-0.3cm}(1)}{g_{\epsilon\lambda}}+\overset{\hspace{-0.5cm}(0)}{\Gamma^\lambda_{~\sigma[\delta}}\overset{\hspace{-0.5cm}(0)}{\Gamma^\epsilon_{~\mu]\lambda}}\overset{\hspace{-0.3cm}(1)}{g_{\nu\epsilon}}
\nonumber\\
&-\left(\,\overset{\hspace{-0.9cm}(0)}{\Gamma^\lambda_{~\nu[\delta,\mu]}}+\overset{\hspace{-0.5cm}(0)}{\Gamma^\lambda_{~\epsilon[\mu}}\overset{\hspace{-0.5cm}(0)}{\Gamma^\epsilon_{~\delta]\nu}}-\overset{\hspace{-0.5cm}(0)}{\Gamma^\epsilon_{~\nu[\mu}}\overset{\hspace{-0.5cm}(0)}{\Gamma^\lambda_{~\delta]\epsilon}}-\overset{\hspace{-0.5cm}(0)}{\Gamma^\epsilon_{~[\delta\mu]}}\overset{\hspace{-0.5cm}(0)}{\Gamma^\lambda_{~\nu\epsilon}}\right)\overset{\hspace{-0.3cm}(1)}{g_{\sigma\lambda}}
\nonumber\\
&-\overset{\hspace{-0.6cm}(1)}{g_{\sigma\lambda,[\mu}}\overset{\hspace{-0.5cm}(0)}{\Gamma^\lambda_{~\delta]\nu}}+\overset{\hspace{-0.5cm}(0)}{\Gamma^\lambda_{~\nu[\delta}}\overset{\hspace{-0.5cm}(0)}{\Gamma^\epsilon_{~\mu]\sigma}}\overset{\hspace{-0.3cm}(1)}{g_{\epsilon\lambda}}+\overset{\hspace{-0.5cm}(0)}{\Gamma^\lambda_{~\nu[\delta}}\overset{\hspace{-0.5cm}(0)}{\Gamma^\epsilon_{~\mu]\lambda}}\overset{\hspace{-0.3cm}(1)}{g_{\sigma\epsilon}},
\end{align*}
or 
\begin{align}
\overset{\hspace{-0.8cm}(1)}{g_{\sigma\nu:[\delta\mu]}}
&=-\left(\,\overset{\hspace{-0.9cm}(0)}{\Gamma^\lambda_{~\sigma[\delta,\mu]}}+\overset{\hspace{-0.5cm}(0)}{\Gamma^\lambda_{~\epsilon[\mu}}\overset{\hspace{-0.5cm}(0)}{\Gamma^\epsilon_{~\delta]\sigma}}\right)\overset{\hspace{-0.3cm}(1)}{g_{\nu\lambda}}
-\left(\,\overset{\hspace{-0.9cm}(0)}{\Gamma^\lambda_{~\nu[\delta,\mu]}}+\overset{\hspace{-0.5cm}(0)}{\Gamma^\lambda_{~\epsilon[\mu}}\overset{\hspace{-0.5cm}(0)}{\Gamma^\epsilon_{~\delta]\nu}}\right)\overset{\hspace{-0.3cm}(1)}{g_{\sigma\lambda}},\nonumber\\
&=-\frac{1}{2}\left(
\overset{\hspace{-0.6cm}(0)}{R^\lambda_{~\sigma\mu\delta}}\overset{\hspace{-0.3cm}(1)}{g_{\nu\lambda}}
+\overset{\hspace{-0.6cm}(0)}{R^\lambda_{~\nu\mu\delta}}\overset{\hspace{-0.3cm}(1)}{g_{\sigma\lambda}}
\right).
\label{eq2.31}
\end{align}

Substituting \eqref{eq2.31} into \eqref{eq2.29}
\begin{align}
& 
+\overset{\hspace{-0.7cm}(1)}{g^\mu_{~\nu:\mu\delta}}
+\overset{\hspace{-0.7cm}(1)}{g^\mu_{~\delta:\mu\nu}}
-\overset{\hspace{-0.7cm}(1)}{g^{~~~\mu}_{\nu\delta:~\mu}}
-\overset{\hspace{-0.7cm}(1)}{g^\mu_{~\mu:\nu\delta}}\nonumber\\
&
-\overset{\hspace{-0.6cm}(0)}{R^\lambda_{~\sigma\mu\delta}}\overset{\hspace{-0.3cm}(1)}{g_{\nu\lambda}}\overset{\hspace{-0.3cm}(0)}{g^{\mu\sigma}}
-\overset{\hspace{-0.6cm}(0)}{R^\lambda_{~\nu\mu\delta}}\overset{\hspace{-0.3cm}(1)}{g^\mu_{~\lambda}}
-\overset{\hspace{-0.6cm}(0)}{R^\lambda_{~\delta\mu\nu}}\overset{\hspace{-0.3cm}(1)}{g^\mu_{~\lambda}}-\overset{\hspace{-0.6cm}(0)}{R^\lambda_{~\sigma\mu\nu}}\overset{\hspace{-0.3cm}(1)}{g_{\delta\lambda}}\overset{\hspace{-0.3cm}(0)}{g^{\mu\sigma}}
=0,
\label{eq2.32}
\end{align}
or
\begin{align}
&
\overset{\hspace{-0.7cm}(1)}{g^\mu_{~\nu:\mu\delta}}
+\overset{\hspace{-0.7cm}(1)}{g^\mu_{~\delta:\mu\nu}}
-\overset{\hspace{-0.7cm}(1)}{g^{~~~\mu}_{\nu\delta:~\mu}}
-\overset{\hspace{-0.7cm}(1)}{g^\mu_{~\mu:\nu\delta}}
+2\overset{\hspace{-0.6cm}(0)}{R^{}_{\lambda\nu\delta\mu}}\overset{\hspace{-0.3cm}(1)}{g^{\mu\lambda}}
+\overset{\hspace{-0.3cm}(0)}{R^{}_{\lambda\nu}}\overset{\hspace{-0.3cm}(1)}{g^{~\lambda}_{\delta~}}
+\overset{\hspace{-0.3cm}(0)}{R^{}_{\lambda\delta}}\overset{\hspace{-0.3cm}(1)}{g^{~\lambda}_{\nu~}}
=0.
\label{eq2.33}
\end{align}
Defining now a reverse trace tensor $h_{\mu\nu}$ as
\begin{equation}
h_{\mu\nu}=\overset{\hspace{-0.3cm}(1)}{g_{\mu\nu}}-\frac{1}{2}\overset{(1)}{g}\,\overset{\hspace{-0.3cm}(0)}{g_{\mu\nu}},
\label{eq2.34}
\end{equation}
and contracting \eqref{eq2.34} one obtains $h=-\overset{(1)}{g}$. Therefore,
\begin{equation}
\overset{\hspace{-0.3cm}(1)}{g_{\mu\nu}}=h_{\mu\nu}-\frac{1}{2}h\,\overset{\hspace{-0.3cm}(0)}{g_{\mu\nu}}.
\label{eq2.35}
\end{equation}

Substituting \eqref{eq2.35} into \eqref{eq2.33} one obtains 
\begin{align}
&
h^\mu_{~\nu:\mu\delta}+h^\mu_{~\delta:\mu\nu}
-h_{\nu\delta:~\mu}^{~~~\mu}-\frac{1}{2}h_{:~\mu}^{\,\,\mu}\,\overset{\hspace{-0.3cm}(0)}{g_{\nu\delta}}
+
2\overset{\hspace{-0.6cm}(0)}{R^{}_{\lambda\nu\delta\mu}}h^{\mu\lambda}
+\overset{\hspace{-0.3cm}(0)}{R^{}_{\lambda\nu}}h^{~\lambda}_{\delta~}
+\overset{\hspace{-0.3cm}(0)}{R^{}_{\lambda\delta}} h^{~\lambda}_{\nu~}
=0.
\label{eq2.36}
\end{align}

Under the transformation of coordinates
\begin{align}
\overline{x}^{\alpha}:=\overline{x}^{\alpha} (x^\beta),
\label{eq2.37}
\end{align}
the metric transforms as
\begin{align}
g^{\overline{\mu}\,\overline{\nu}}=g^{\mu\nu}\Delta^{\overline{\mu}}_{~\mu} \Delta^{\overline{\nu}}_{~\nu},
\label{eq2.38}
\end{align}
where $g^{\overline{\mu}\,\overline{\nu}}$ and $g^{\mu\nu}$ are referred to the $\overline{x}^{\alpha}$ and $x^{\alpha}$ coordinates respectively and the transformation matrix  $\Delta^{\overline{\mu}}_{~\mu}$ is given in terms of partial derivatives, i.e.,
\begin{equation}
\Delta^{\overline{\mu}}_{~\mu}=x^{\overline{\mu}}_{~,\mu}.
\label{eq2.39}
\end{equation}

Additionally, from the transformation \eqref{eq2.38} and the perturbation \eqref{eq2.1}, it follows that
\begin{equation*}
\overset{(0)}{g_{\overline{\mu}\,\overline{\nu}}}+\epsilon\overset{(1)}{g_{\overline{\mu}\,\overline{\nu}}}=\Delta^\mu_{~\overline{\mu}}\Delta^\nu_{~\overline{\nu}}\left(\overset{(0)}{g_{\mu\nu}}+\epsilon\overset{(1)}{g_{\mu\nu}}\right), 
\end{equation*}
which implies that the perturbation obeys the transformation rules for tensor under Lorentz transformations, namely
\begin{equation}
\overset{(1)}{g_{\overline{\mu}\,\overline{\nu}}}= \Delta^\mu_{~\overline{\mu}}\Delta^\nu_{~\overline{\nu}} \overset{(1)}{g_{\mu\nu}}.
\label{eq2.40}
\end{equation}

In particular, considering an infinitesimal boost, i.e.,
\begin{equation}
\overline{x}^a=x^a+\epsilon\, \zeta^a,
\label{eq2.41}
\end{equation}
where $|\epsilon \zeta^a|\ll |x^a|$ are infinitesimal displacements, then the matrices \eqref{eq2.31} become
\begin{equation}
\overline{x}^{\alpha}_{~,\beta}=\delta^\alpha_{~,\beta}+\epsilon\,\zeta^\alpha_{~,\beta}.
\label{eq2.42}
\end{equation}

Thus, substituting \eqref{eq2.42} into \eqref{eq2.40},
\begin{align}
g^{\overline{\mu}\,\overline{\nu}}(\overline{x}^\beta)
&=g^{\overline{\mu}\,\overline{\nu}}(x^\alpha)+\epsilon\left(\overset{(1)}{g^{\overline{\mu}\,\nu}}\zeta^{\overline{\nu}}_{~,\nu} + \overset{(1)}{g^{\mu\,\overline{\nu}}}\zeta^{\overline{\mu}}_{~,\mu}\right)+O(\zeta^2),
\label{eq2.43}
\end{align}
expanding the metric around $\zeta$,
\begin{equation}
g^{\overline{\mu}\,\overline{\nu}}(x^\alpha)\simeq g^{\overline{\mu}\,\overline{\nu}}-\epsilon\,\zeta^\sigma g^{\overline{\mu}\,\overline{\nu}}_{~~~,\sigma},
\label{eq2.44}
\end{equation}
and substituting it into \eqref{eq2.43}, one obtains
\begin{align}
g^{\overline{\mu}\,\overline{\nu}}(\overline{x}^\beta)&\simeq g^{\overline{\mu}\,\overline{\nu}}-\epsilon\left(\zeta^\sigma g^{\overline{\mu}\,\overline{\nu}}_{~~~,\sigma} - g^{\overline{\mu}\,\nu}\zeta^{\overline{\nu}}_{~,\nu} - g^{\mu\,\overline{\nu}}\zeta^{\overline{\mu}}_{~,\mu}\right).
\label{eq2.45}
\end{align}

Now, from the covariant derivative of the inverse metric one has
\begin{equation}
g^{\mu\nu}_{~~~,\delta}=-g^{\sigma\nu}\Gamma^\mu_{~\sigma\delta}-g^{\mu\sigma}\Gamma^\nu_{~\sigma\delta}.
\label{eq2.46}
\end{equation}

Substituting \eqref{eq2.46} into \eqref{eq2.45}, one obtains
\begin{align}
g^{\overline{\mu}\,\overline{\nu}}(\overline{x}^\beta)&\simeq g^{\overline{\mu}\,\overline{\nu}}(x^\beta)
+\epsilon\left(\zeta^\sigma g^{\overline{\mu}\,\eta}\Gamma^{\overline{\nu}}_{~\sigma\eta}+\zeta^\sigma g^{\eta\,\overline{\nu}}\Gamma^{\overline{\mu}}_{~\sigma\eta} 
+g^{\overline{\mu}\,\nu}\zeta^{\overline{\nu}}_{~,\nu} 
+ g^{\mu\,\overline{\nu}}\zeta^{\overline{\mu}}_{~,\mu}\right),\nonumber\\
&\simeq g^{\overline{\mu}\,\overline{\nu}}(x^\beta)
+\epsilon\left(g^{\overline{\mu}\,\nu}\left(\zeta^{\overline{\nu}}_{~,\nu}
+\zeta^\sigma \Gamma^{\overline{\nu}}_{~\sigma\nu}\right)
+ g^{\mu\,\overline{\nu}}\left(\zeta^{\overline{\mu}}_{~,\mu}
+\zeta^\sigma \Gamma^{\overline{\mu}}_{~\sigma\mu}\right)\right),\nonumber\\
&\simeq g^{\overline{\mu}\,\overline{\nu}}(x^\beta)
+2\,\epsilon\,\zeta^{(\overline{\nu}:\overline{\mu})},
\label{eq2.47}
\end{align}
where, as usual the round brackets indicates symmetrisation. The symmetrisation is defined as
\begin{equation}
A_{(\alpha_1\cdot\alpha_n)}=\frac{1}{n!}\sum_{n} A_{\alpha_{\sigma_1}\cdot\alpha_{\sigma_n}},
\end{equation}
where the sum is performed over all index permutations.

Thus, the metric is invariant under such transformation whenever
\begin{equation}
\zeta_{(\nu:\mu)}=0,
\label{eq2.48}
\end{equation}
in which $\zeta^\alpha$ are just the Killing vectors associated with the background space-time \cite{landau75}.

Lowering the indices of \eqref{eq2.47} with the metric, and using \eqref{eq2.1} one immediately obtains a gauge condition for the perturbations, i.e.,
\begin{equation}
\overset{\hspace{-0.3cm}(1)}{g_{\mu\nu}}(\overline{x}^\beta)=\overset{\hspace{-0.3cm}(1)}{g_{\mu\nu}}(x^\beta)+2\zeta_{(\nu:\mu)}.
\label{eq2.49}
\end{equation}

From this last equation, one immediately reads 
\begin{equation}
\overset{\hspace{-0.7cm}(1)}{\overline{g}_{\mu\nu}^{~~~:\mu}}=\overset{\hspace{-0.7cm}(1)}{g_{\mu\nu}^{~~~:\mu}}+2\zeta_{(\nu:\mu)}^{~~~~~\mu},
\label{eq2.50}
\end{equation}
where the overline indicates the metric in the new coordinate system, i.e., $\overset{\hspace{-0.3cm}(1)}{\overline{g}_{\mu\nu}}=\overset{\hspace{-0.3cm}(1)}{g_{\mu\nu}}(\overline{x}^\alpha)$ which allows to impose 
\begin{equation}
\overset{\hspace{-0.7cm}(1)}{\overline{g}_{\mu\nu}^{~~~:\mu}}=0.
\label{eq2.51}
\end{equation}
This gauge is known as De Donder or Hilbert gauge.

The form of the gauge for $h_{\mu\nu}$ is found when \eqref{eq2.34} is substituted into \eqref{eq2.49}, it results in

\begin{align}
\overline{h}_{\mu\nu}
&=\overset{\hspace{-0.3cm}(1)}{g_{\mu\nu}}-\frac{1}{2}\overset{(1)}{g}\,\overset{\hspace{-0.3cm}(0)}{g_{\mu\nu}}+2\zeta_{(\nu:\mu)}-\zeta^\sigma_{~:\sigma}\,\overset{\hspace{-0.3cm}(0)}{g_{\mu\nu}},\nonumber\\
&=h_{\mu\nu}+2\zeta_{(\nu:\mu)}-\zeta^\sigma_{~:\sigma}\,\overset{\hspace{-0.3cm}(0)}{g_{\mu\nu}},
\label{eq2.52}
\end{align}
which implies that its trace is given by
\begin{align}
\overline{h}&=h+2\zeta^\mu_{~:\mu}-\zeta^\sigma_{~:\sigma}\,\delta^\mu_{~\mu},\nonumber\\
&=h-2\zeta^\mu_{~:\mu}.
\label{eq2.53}
\end{align}

Therefore, computing the covariant derivative of \eqref{eq2.52}, one has
\begin{align}
\overline{h}_{\mu\nu:}^{~~~\nu}&=h_{\mu\nu:}^{~~~\nu}+2\,\zeta_{(\nu:\mu)}^{~~~~\,\nu}-\zeta^{\sigma~\nu}_{~:\sigma}\,\overset{\hspace{-0.3cm}(0)}{g_{\mu\nu}},\nonumber\\
&=h_{\mu\nu:}^{~~~\nu}+2\,\zeta_{(\nu:\mu)}^{~~~~\,\nu}-\zeta^{\sigma}_{~:\sigma\mu},\nonumber\\
&=h_{\mu\nu:}^{~~~\nu}+\zeta_{\mu:~\nu}^{~\,\nu}+2\,\zeta^{\sigma}_{~:[\sigma\mu]}.
\label{eq2.54}
\end{align}

Considering that
\begin{align}
2\,\zeta^{\sigma}_{~:[\sigma\mu]}&=\overset{\hspace{-0.5cm}(0)}{R^\sigma_{\lambda\sigma\mu}}\zeta^\lambda,\nonumber\\
&=\overset{\hspace{-0.3cm}(0)}{R_{\lambda\mu}}\zeta^\lambda,
\label{eq2.55}
\end{align}
then,
\begin{align}
\overline{h}_{\mu\nu:}^{~~~\nu}
&=h_{\mu\nu:}^{~~~\nu}+\zeta_{\mu:~\nu}^{~\,\nu}+\overset{\hspace{-0.3cm}(0)}{R_{\lambda\mu}}\zeta^\lambda.
\label{eq2.56}
\end{align}

Thus, \eqref{eq2.53} and \eqref{eq2.56} can be re-written as
\begin{equation}
h_{\mu\nu:}^{~~~\nu}=\overline{h}_{\mu\nu:}^{~~~\nu}-\zeta_{\mu:~\nu}^{~\,\nu}-\overset{\hspace{-0.3cm}(0)}{R_{\lambda\mu}}\zeta^\lambda,\hspace{0.5cm}h=\overline{h}+2\zeta^\mu_{~:\mu},
\label{eq2.57}
\end{equation}
then the tensor field $h_{\mu\nu}$ can be recalibrated making the selection
\begin{align}
h=0, \hspace{0.5cm}h_{\mu\nu:}^{~~~\nu}=0,
\label{eq2.58}
\end{align}
only if the following conditions are met,
\begin{align}
\overline{h}_{\mu\nu:}^{~~~\nu}=\zeta_{\mu:~\nu}^{~\,\nu}+\overset{\hspace{-0.3cm}(0)}{R_{\lambda\mu}}\zeta^\lambda, \hspace{0.5cm}\overline{h}=-2\zeta^\mu_{~:\mu}.
\label{eq2.59}
\end{align}

Substituting \eqref{eq2.58} into \eqref{eq2.36}, one obtains
\begin{align}
&
h_{\nu\delta:~\mu}^{~~~\mu}
-2\overset{\hspace{-0.6cm}(0)}{R^{}_{\lambda\nu\delta\mu}}h^{\mu\lambda}
-\overset{\hspace{-0.3cm}(0)}{R^{}_{\lambda\nu}}h^{~\lambda}_{\delta~}
-\overset{\hspace{-0.3cm}(0)}{R^{}_{\lambda\delta}} h^{~\lambda}_{\nu~}
=0,
\label{eq2.60}
\end{align}
which is just a wave equation for $h_{\nu\delta}$ \cite{I68_1}. This equation includes the terms related to the background's curvature.
\section{Higher Order Perturbations}
At this point, there appears the question how are the forms of the higher order perturbations to the Ricci's tensor.  Different approximations can be made considering different expansions for the metric $g_{\mu\nu}$ or for the inverse metric $g^{\mu\nu}$. The perturbation method can vary depending on which quantity is expanded and how it is done. In particular \citeonline{I68_1} shows the Ricci's tensor for higher order perturbation, expanding only the inverse metric $g^{\mu\nu}$; however, other perturbation schemes were explored with interesting results, for example \citeonline{B69} expands the metric and its inverse supposing {\it ab initio} that both quantities depends on two parameters, a frequency and a phase, which leaves to different versions of the perturbed Ricci tensor. 

As a starting point, the procedure exposed by \citeonline{I68_1} is followed. Thus, the metric is expanded as 
\begin{equation}
g_{\mu\nu}=\overset{\hspace{-0.3cm}(0)}{g_{\mu\nu}}
+\epsilon\, \overset{\hspace{-0.3cm}(1)}{g_{\mu\nu}},
\label{eq2.2.1}
\end{equation}
whereas its inverse metric, $g^{\mu\nu}$, is expanded as
\begin{equation}
g^{\mu\nu}=\overset{\hspace{-0.3cm}(0)}{g^{\mu\nu}}+\sum_{i=1}^{n}\epsilon^i\,\overset{\hspace{-0.3cm}(i)}{g^{\mu\nu}}+O(\epsilon^{n+1}).
\label{eq2.2.2}
\end{equation}

Thus, from \eqref{eq2.2.1} and \eqref{eq2.2.2} 
\begin{equation}
g_{\mu\nu}g^{\nu\delta}=\overset{\hspace{-0.3cm}(0)}{g_{\mu\nu}}\overset{\hspace{-0.3cm}(0)}{g^{\nu\delta}}+\sum_{i=1}^{n}\epsilon^i\left(\,\overset{\hspace{-0.3cm}(0)}{g_{\mu\nu}}\overset{\hspace{-0.3cm}(i)}{g^{\nu\delta}}+\overset{\hspace{-0.3cm}(1)}{g_{\mu\nu}} \ \overset{\hspace{-0.1cm}(i-1)}{g^{\nu\delta}}\right)+O(\epsilon^{n+1}),
\label{eq2.2.3}
\end{equation}
which implies 
\begin{equation}
\overset{\hspace{-0.3cm}(i)}{g^{\zeta\delta}}=-\overset{\hspace{-0.3cm}(1)}{g_{\mu\nu}}\ \overset{\hspace{-0.3cm}(0)}{g^{\mu\zeta}}\ \overset{\hspace{-0.1cm}(i-1)}{g^{\nu\delta}},\hspace{1cm}i=1,2,\cdots
\label{eq2.2.4}
\end{equation}
then, 
\begin{equation}
\overset{\hspace{-0.3cm}(1)}{g^{\zeta\delta}}=-\overset{\hspace{-0.3cm}(1)}{g_{\mu\nu}}\ \overset{\hspace{-0.3cm}(0)}{g^{\mu\zeta}}\ \overset{\hspace{-0.3cm}(0)}{g^{\nu\delta}},\hspace{0.5cm}
\overset{\hspace{-0.3cm}(2)}{g^{\zeta\delta}}=-\overset{\hspace{-0.3cm}(1)}{g_{\mu\nu}}\ \overset{\hspace{-0.3cm}(0)}{g^{\mu\zeta}}\ \overset{\hspace{-0.3cm}(1)}{g^{\nu\delta}},\hspace{0.5cm}\overset{\hspace{-0.3cm}(3)}{g^{\zeta\delta}}=-\overset{\hspace{-0.3cm}(1)}{g_{\mu\nu}}\ \overset{\hspace{-0.3cm}(0)}{g^{\mu\zeta}}\ \overset{\hspace{-0.3cm}(2)}{g^{\nu\delta}}, \hspace{0.5cm}\cdots
\label{eq2.2.5}
\end{equation}
Substituting recursively the last equations, one finds
\begin{align}
&\overset{\hspace{-0.3cm}(1)}{g^{\zeta\delta}}=-\overset{\hspace{-0.3cm}(1)}{g_{\mu\nu}}\ \overset{\hspace{-0.3cm}(0)}{g^{\mu\zeta}}\ \overset{\hspace{-0.3cm}(0)}{g^{\nu\delta}},
\hspace{1cm}
\overset{\hspace{-0.3cm}(2)}{g^{\zeta\delta}}=\overset{\hspace{-0.3cm}(1)}{g_{\mu\nu}} \ \overset{\hspace{-0.3cm}(1)}{g_{\alpha\beta}}\ \overset{\hspace{-0.3cm}(0)}{g^{\mu\zeta}}\ \overset{\hspace{-0.3cm}(0)}{g^{\alpha\nu}} \overset{\hspace{-0.3cm}(0)}{g^{\beta\delta}},
\hspace{1cm}\nonumber
\\
&\overset{\hspace{-0.3cm}(3)}{g^{\zeta\delta}}=-\overset{\hspace{-0.3cm}(1)}{g_{\mu\nu}}\ \overset{\hspace{-0.3cm}(1)}{g_{\alpha\beta}}\ \overset{\hspace{-0.3cm}(1)}{g_{\gamma\eta}}\ \overset{\hspace{-0.3cm}(0)}{g^{\mu\zeta}}\ \overset{\hspace{-0.3cm}(0)}{g^{\alpha\nu}}\ \overset{\hspace{-0.3cm}(0)}{g^{\gamma\beta}}\ \overset{\hspace{-0.3cm}(0)}{g^{\eta\delta}},
\hspace{1cm}\cdots
\label{eq2.2.6}
\end{align}

In this approximation, the Christoffel symbols of the first kind can be separated just as in \eqref{eq2.8} where each addend is given by \eqref{eq2.9}. Using \eqref{eq2.2.2} to raise the first index in \eqref{eq2.8}, it is found
\begin{equation}
\Gamma^{\alpha}_{~\beta\gamma}=\overset{\hspace{-0.6cm}(0)}{\Gamma^{\alpha}_{~\beta\gamma}}+\sum_{i=1}^{n}\epsilon^i\,\overset{\hspace{-0.6cm}(i)}{\Gamma^{\alpha}_{~\beta\gamma}}+O(\epsilon^{n+1}),
\label{eq2.2.8}
\end{equation}
where
\begin{equation}
\overset{\hspace{-0.6cm}(0)}{\Gamma^{\alpha}_{~\beta\gamma}}=\overset{\hspace{-0.3cm}(0)}{g^{\alpha\eta}}\ \overset{\hspace{-0.6cm}(0)}{\Gamma^{}_{\eta\beta\gamma}},
\hspace{1cm}\overset{\hspace{-0.6cm}(k)}{\Gamma^{\alpha}_{~\beta\gamma}}=\ 
 \overset{\hspace{-0.3cm}(k-1)}{g^{\alpha\eta}}\ \overset{\hspace{-0.6cm}(1)}{\Gamma^{}_{\eta\beta\gamma}}+\overset{\hspace{-0.3cm}(k)}{g^{\alpha\eta}}\overset{\hspace{-0.6cm}(0)}{\Gamma^{}_{\eta\beta\gamma}}, \hspace{1cm}k=1,2,\cdots,
\label{eq2.2.9}
\end{equation}
which is just one of the possibilities to generalise \eqref{eq2.12}. The separation of the Christoffel's symbols of the second kind allows to write the Riemann's tensor as
\begin{equation}
R^\mu_{~\nu\gamma\delta}=\overset{\hspace{-0.6cm}(0)}{R^\mu_{~\nu\gamma\delta}}+\sum_{i=1}^{n}\epsilon^i\,\overset{\hspace{-0.6cm}(i)}{R^\mu_{~\nu\gamma\delta}}+O(\epsilon^{n+1}),
\label{eq2.2.10}
\end{equation}
where $\overset{\hspace{-0.6cm}(0)}{R^{\mu}_{~\nu\gamma\delta}}$ is given in \eqref{eq2.14} and $\overset{\hspace{-0.6cm}(i)}{R^{\mu}_{~\nu\gamma\delta}}$ corresponds to
\begin{equation}
\overset{\hspace{-0.6cm}(k)}{R^{\mu}_{~\nu\gamma\delta}}=2\,\overset{\hspace{-0.9cm}(k)}{\Gamma^{\mu}_{~\nu[\delta:\gamma]}}+2\,\sum_{i=1}^{k}\ \ \overset{\hspace{-0.7cm}(k-i)}{\Gamma^{\mu}_{~\sigma[\gamma}} \ \overset{\hspace{-0.6cm}(i)}{\Gamma^{\sigma}_{~\delta]\nu}}.
\label{eq2.2.11}
\end{equation}
Computing the derivative of \eqref{eq2.2.9} and anti-symmetrising it, one obtains
\begin{equation}
\overset{\hspace{-0.9cm}(k)}{\Gamma^{\alpha}_{~\beta[\gamma:\delta]}}=\ 
 \overset{\hspace{-0.7cm}(1)}{\Gamma^{}_{\eta\beta[\gamma}}\,\overset{\hspace{-0.5cm}(k-1)}{g^{\alpha\eta}_{~~~:\delta]}} 
 +\ \overset{\hspace{-0.3cm}(k-1)}{g^{\alpha\eta}}\ \overset{\hspace{-0.9cm}(1)}{\Gamma^{}_{\eta\beta[\gamma:\delta]}}
 +\overset{\hspace{-0.7cm}(0)}{\Gamma^{}_{\eta\beta[\gamma}}\,\overset{\hspace{-0.6cm}(k)}{g^{\alpha\eta}_{~~~:\delta]}}+\overset{\hspace{-0.3cm}(k)}{g^{\alpha\eta}}\,\,\overset{\hspace{-1cm}(0)}{\Gamma^{}_{\eta\beta[\gamma:\delta]}}. 
\label{eq2.2.12}
\end{equation}
Noting that
\begin{align}
\overset{\hspace{-0.6cm}(1)}{\Gamma^{}_{\eta\beta\gamma}}&=
\overset{\hspace{-0.6cm}(1)}{g_{\eta\beta:\gamma}}
+\overset{\hspace{-0.6cm}(1)}{g_{\gamma\eta:\beta}}
-\overset{\hspace{-0.6cm}(1)}{g_{\beta\gamma:\eta}}
+2\,\overset{\hspace{-0.6cm}(0)}{\Gamma^\sigma_{~\gamma\beta}}\,\overset{\hspace{-0.3cm}(1)}{g_{\sigma\eta}},
\label{eq2.2.13}
\end{align}
then,
\begin{align}
\,\overset{\hspace{-0.9cm}(1)}{\Gamma^{}_{\eta\beta[\gamma:\delta]}}=&
\,\overset{\hspace{-0.7cm}(1)}{g_{\eta\beta:[\gamma\delta]}}
+\overset{\hspace{-0.9cm}(1)}{g_{[\gamma|\eta:\beta|\delta]}}
-\overset{\hspace{-1cm}(1)}{g_{\beta[\gamma|:\eta|\delta]}} 
+2\,\overset{\hspace{-0.6cm}(0)}{\Gamma^\sigma_{~\beta[\gamma:\delta]}}\,\overset{\hspace{-0.3cm}(1)}{g_{\sigma\eta}}
+2\,\overset{\hspace{-0.5cm}(1)}{g_{\sigma\eta:[\delta}}\,\overset{\hspace{-0.6cm}(0)}{\Gamma^\sigma_{~\gamma]\beta}}.
\label{eq2.2.14}
\end{align}
The first term in \eqref{eq2.2.12} is
\begin{align}
\overset{\hspace{-0.7cm}(1)}{\Gamma^{}_{\eta\beta[\gamma}}\,\overset{\hspace{-0.5cm}(k-1)}{g^{\alpha\eta}_{~~~:\delta]}}
&=
\overset{\hspace{-0.6cm}(1)}{g_{\eta\beta:[\gamma}}\,\overset{\hspace{-0.5cm}(k-1)}{g^{\alpha\eta}_{~~~:\delta]}}
+\overset{\hspace{-0.5cm}(k-1)}{g^{\alpha\eta}_{~~~:[\delta}}\,\overset{\hspace{-0.6cm}(1)}{g_{\gamma]\eta:\beta}}
-\overset{\hspace{-0.5cm}(k-1)}{g^{\alpha\eta}_{~~~:[\delta}}\,\overset{\hspace{-0.6cm}(1)}{g_{\gamma]\beta:\eta}}
+2\,\overset{\hspace{-0.5cm}(k-1)}{g^{\alpha\eta}_{~~~:[\delta}}\,\overset{\hspace{-0.6cm}(0)}{\Gamma^\sigma_{~\gamma]\beta}}\,\overset{\hspace{-0.3cm}(1)}{g_{\sigma\eta}},
\label{eq2.2.15}
\end{align}
and the second term in \eqref{eq2.2.12} is
\begin{align}
\overset{\hspace{-0.3cm}(k-1)}{g^{\alpha\eta}}\ \overset{\hspace{-0.9cm}(1)}{\Gamma^{}_{\eta\beta[\gamma:\delta]}}=& 
\ \ \overset{\hspace{-0.3cm}(k-1)}{g^{\alpha\eta}}\
\,\overset{\hspace{-0.7cm}(1)}{g_{\eta\beta:[\gamma\delta]}}
+\ \ \overset{\hspace{-0.3cm}(k-1)}{g^{\alpha\eta}}\
\,\overset{\hspace{-1cm}(1)}{g_{[\gamma|\eta:\beta|\delta]}}
-\ \ \overset{\hspace{-0.3cm}(k-1)}{g^{\alpha\eta}}\
\,\overset{\hspace{-1cm}(1)}{g_{\beta[\gamma|:\eta|\delta]}}
\nonumber\\
&\ +2\ \ \overset{\hspace{-0.3cm}(k-1)}{g^{\alpha\eta}}\
\overset{\hspace{-0.9cm}(0)}{\Gamma^\sigma_{~\beta[\gamma:\delta]}}\ \overset{\hspace{-0.3cm}(1)}{g_{\sigma\eta}}
\ +2\ \ \overset{\hspace{-0.3cm}(k-1)}{g^{\alpha\eta}}\
\overset{\hspace{-0.6cm}(1)}{g_{\sigma\eta:[\delta}}
\overset{\hspace{-0.7cm}(0)}{\Gamma^\sigma_{~\gamma]\beta}}. 
\label{eq2.2.16}
\end{align}
Therefore the first term in \eqref{eq2.2.11} is
\begin{align}
2\,\overset{\hspace{-0.9cm}(k)}{\Gamma^{\mu}_{~\nu[\delta:\gamma]}}=&
2\overset{\hspace{-0.6cm}(1)}{g_{\eta\beta:[\gamma}}\,\overset{\hspace{-0.5cm}(k-1)}{g^{\alpha\eta}_{~~~:\delta]}}
+2\overset{\hspace{-0.5cm}(k-1)}{g^{\alpha\eta}_{~~~:[\delta}}\,\overset{\hspace{-0.6cm}(1)}{g_{\gamma]\eta:\beta}}
-2\overset{\hspace{-0.5cm}(k-1)}{g^{\alpha\eta}_{~~~:[\delta}}\,\overset{\hspace{-0.6cm}(1)}{g_{\gamma]\beta:\eta}}
+4\,\overset{\hspace{-0.5cm}(k-1)}{g^{\alpha\eta}_{~~~:[\delta}}\,\overset{\hspace{-0.6cm}(0)}{\Gamma^\sigma_{~\gamma]\beta}}\,\overset{\hspace{-0.3cm}(1)}{g_{\sigma\eta}}\nonumber\\
&+ 2\ \ \overset{\hspace{-0.3cm}(k-1)}{g^{\alpha\eta}}\
\,\overset{\hspace{-0.7cm}(1)}{g_{\eta\beta:[\gamma\delta]}}
+2\ \ \overset{\hspace{-0.3cm}(k-1)}{g^{\alpha\eta}}\
\,\overset{\hspace{-1cm}(1)}{g_{[\gamma|\eta:\beta|\delta]}}
-2\ \ \overset{\hspace{-0.3cm}(k-1)}{g^{\alpha\eta}}\
\,\overset{\hspace{-1cm}(1)}{g_{\beta[\gamma|:\eta|\delta]}}
\nonumber\\
&+4\ \ \overset{\hspace{-0.3cm}(k-1)}{g^{\alpha\eta}}\
\overset{\hspace{-0.9cm}(0)}{\Gamma^\sigma_{~\beta[\gamma:\delta]}}\ \overset{\hspace{-0.3cm}(1)}{g_{\sigma\eta}}
\ +4\ \ \overset{\hspace{-0.3cm}(k-1)}{g^{\alpha\eta}}\
\overset{\hspace{-0.6cm}(1)}{g_{\sigma\eta:[\delta}}
\overset{\hspace{-0.7cm}(0)}{\Gamma^\sigma_{~\gamma]\beta}} +2\overset{\hspace{-0.7cm}(0)}{\Gamma^{}_{\eta\beta[\gamma}}\,\overset{\hspace{-0.6cm}(k)}{g^{\alpha\eta}_{~~~:\delta]}} \nonumber\\
&+2\overset{\hspace{-0.3cm}(k)}{g^{\alpha\eta}}\,\,\overset{\hspace{-1cm}(0)}{\Gamma^{}_{\eta\beta[\gamma:\delta]}}.
\label{eq2.2.17}
\end{align}
Using \eqref{eq2.2.13} the second term in \eqref{eq2.2.11} is given by
\begin{align}
2\,\sum_{i=1}^{k}\ \ \overset{\hspace{-0.7cm}(k-i)}{\Gamma^{\mu}_{~\sigma[\gamma}} \ \overset{\hspace{-0.6cm}(i)}{\Gamma^{\sigma}_{~\delta]\nu}}=&2\,\sum_{i=1}^{k} \left(
\ \ \ \overset{\hspace{-0.4cm}(k-i-1)}{g^{\mu\eta}}\ \ \overset{\hspace{-0.3cm}(i-1)}{g^{\sigma\zeta}}\ \overset{\hspace{-0.6cm}(1)}{\Gamma_{\eta\sigma[\gamma}}\ \overset{\hspace{-0.7cm}(1)}{\Gamma_{|\zeta|\delta]\nu}}+ \ \ \ \overset{\hspace{-0.4cm}(k-i-1)}{g^{\mu\eta}}\ \ \overset{\hspace{-0.3cm}(i)}{g^{\sigma\zeta}}\ \overset{\hspace{-0.6cm}(0)}{\Gamma_{\zeta\nu[\delta}}\ \overset{\hspace{-0.7cm}(1)}{\Gamma_{|\eta\sigma|\gamma]}} \right.\nonumber\\
&\left.\ \ \ \ \ \ \ +\ \  \overset{\hspace{-0.4cm}(k-i)}{g^{\mu\eta}}\ \ \overset{\hspace{-0.3cm}(i-1)}{g^{\sigma\zeta}}\ \overset{\hspace{-0.6cm}(0)}{\Gamma_{\eta\sigma[\gamma}}\ \overset{\hspace{-0.7cm}(1)}{\Gamma_{|\zeta|\delta]\nu}}+\ \  \overset{\hspace{-0.4cm}(k-i)}{g^{\mu\eta}}\ \ \overset{\hspace{-0.3cm}(i)}{g^{\sigma\zeta}}\overset{\hspace{-0.6cm}(0)}{\Gamma_{\eta\sigma[\gamma}}\ \overset{\hspace{-0.7cm}(0)}{\Gamma_{|\zeta|\delta]\nu}}\right).
\label{eq2.2.18}
\end{align}

One wave equation for the vacuum for each perturbation order is obtained contracting \eqref{eq2.2.11}, i.e.,
\begin{equation}
\overset{\hspace{-0.3cm}(k)}{R_{~\nu\delta}}=2\,\overset{\hspace{-0.9cm}(k)}{\Gamma^{\mu}_{~\nu[\delta:\mu]}}+2\,\sum_{i=1}^{k}\ \ \overset{\hspace{-0.7cm}(k-i)}{\Gamma^{\mu}_{~\sigma[\mu}} \ \overset{\hspace{-0.6cm}(i)}{\Gamma^{\sigma}_{~\delta]\nu}}=0.
\label{eq2.2.19}
\end{equation}
where, the first term in \eqref{eq2.2.19} is obtained from the contraction of \eqref{eq2.2.17}
\begin{align}
2\,\overset{\hspace{-0.9cm}(k)}{\Gamma^{\mu}_{~\nu[\delta:\mu]}}=&
2\overset{\hspace{-0.6cm}(1)}{g_{\eta\beta:[\mu}}\,\overset{\hspace{-0.5cm}(k-1)}{g^{\alpha\eta}_{~~~:\delta]}}
+2\overset{\hspace{-0.5cm}(k-1)}{g^{\alpha\eta}_{~~~:[\delta}}\,\overset{\hspace{-0.6cm}(1)}{g_{\mu]\eta:\beta}}
-2\overset{\hspace{-0.5cm}(k-1)}{g^{\alpha\eta}_{~~~:[\delta}}\,\overset{\hspace{-0.6cm}(1)}{g_{\mu]\beta:\eta}}
+4\,\overset{\hspace{-0.5cm}(k-1)}{g^{\alpha\eta}_{~~~:[\delta}}\,\overset{\hspace{-0.6cm}(0)}{\Gamma^\sigma_{~\mu]\beta}}\,\overset{\hspace{-0.3cm}(1)}{g_{\sigma\eta}}\nonumber\\
&+ 2\ \ \overset{\hspace{-0.3cm}(k-1)}{g^{\alpha\eta}}\
\,\overset{\hspace{-0.7cm}(1)}{g_{\eta\beta:[\mu\delta]}}
+2\ \ \overset{\hspace{-0.3cm}(k-1)}{g^{\alpha\eta}}\
\,\overset{\hspace{-1cm}(1)}{g_{[\mu|\eta:\beta|\delta]}}
-2\ \ \overset{\hspace{-0.3cm}(k-1)}{g^{\alpha\eta}}\
\,\overset{\hspace{-1cm}(1)}{g_{\beta[\mu|:\eta|\delta]}}
\nonumber\\
&+4\ \ \overset{\hspace{-0.3cm}(k-1)}{g^{\alpha\eta}}\
\overset{\hspace{-0.9cm}(0)}{\Gamma^\sigma_{~\beta[\mu:\delta]}}\ \overset{\hspace{-0.3cm}(1)}{g_{\sigma\eta}}
\ +4\ \ \overset{\hspace{-0.3cm}(k-1)}{g^{\alpha\eta}}\
\overset{\hspace{-0.6cm}(1)}{g_{\sigma\eta:[\delta}}
\overset{\hspace{-0.7cm}(0)}{\Gamma^\sigma_{~\mu]\beta}} +2\overset{\hspace{-0.7cm}(0)}{\Gamma^{}_{\eta\beta[\mu}}\,\overset{\hspace{-0.6cm}(k)}{g^{\alpha\eta}_{~~~:\delta]}} \nonumber\\
&+2\overset{\hspace{-0.3cm}(k)}{g^{\alpha\eta}}\,\,\overset{\hspace{-1cm}(0)}{\Gamma^{}_{\eta\beta[\mu:\delta]}}.
\label{eq2.2.20}
\end{align}
and the second term results from the contraction of \eqref{eq2.2.18}
\begin{align}
2\,\sum_{i=1}^{k}\ \ \overset{\hspace{-0.7cm}(k-i)}{\Gamma^{\mu}_{~\sigma[\mu}} \ \overset{\hspace{-0.6cm}(i)}{\Gamma^{\sigma}_{~\delta]\nu}}=&2\,\sum_{i=1}^{k} \left(
\ \ \ \overset{\hspace{-0.4cm}(k-i-1)}{g^{\mu\eta}}\ \ \overset{\hspace{-0.3cm}(i-1)}{g^{\sigma\zeta}}\ \overset{\hspace{-0.6cm}(1)}{\Gamma_{\eta\sigma[\mu}}\ \overset{\hspace{-0.7cm}(1)}{\Gamma_{|\zeta|\delta]\nu}}+ \ \ \ \overset{\hspace{-0.4cm}(k-i-1)}{g^{\mu\eta}}\ \ \overset{\hspace{-0.3cm}(i)}{g^{\sigma\zeta}}\ \overset{\hspace{-0.6cm}(0)}{\Gamma_{\zeta\nu[\delta}}\ \overset{\hspace{-0.7cm}(1)}{\Gamma_{|\eta\sigma|\mu]}} \right.\nonumber\\
&\left.\ \ \ \ \ \ \ +\ \  \overset{\hspace{-0.4cm}(k-i)}{g^{\mu\eta}}\ \ \overset{\hspace{-0.3cm}(i-1)}{g^{\sigma\zeta}}\ \overset{\hspace{-0.6cm}(0)}{\Gamma_{\eta\sigma[\mu}}\ \overset{\hspace{-0.7cm}(1)}{\Gamma_{|\zeta|\delta]\nu}}+\ \  \overset{\hspace{-0.4cm}(k-i)}{g^{\mu\eta}}\ \ \overset{\hspace{-0.3cm}(i)}{g^{\sigma\zeta}}\overset{\hspace{-0.6cm}(0)}{\Gamma_{\eta\sigma[\mu}}\ \overset{\hspace{-0.7cm}(0)}{\Gamma_{|\zeta|\delta]\nu}}\right).
\label{eq2.2.21}
\end{align}

It is worth nothing here some of the most important aspects of this last results. First, observe that \eqref{eq2.2.6} expresses the perturbations $\overset{\hspace{-0.3cm}(k)}{g^{\mu\nu}}$ in terms of power of the perturbations $\overset{\hspace{-0.3cm}(1)}{g_{\mu\nu}}$. Thus, each order in \eqref{eq2.2.19} corresponds to a wave equation related to such powers. Second, \eqref{eq2.2.19} can be read as inhomogeneous wave equations because the second derivatives for the metric becomes from the first term, thus the second term, formed from products of Christoffel symbols, contributes like a barrier that affects the frequency of the waves. Third, different eikonal equations are obtained from the substitution of solutions like $g_{\mu\nu}:=A_{\mu\nu}e^{i\phi}$ with $\phi_{,\alpha}=k_\alpha$, namely WKB solutions. Note that high order non-linear terms will appear given the factors  $\overset{\hspace{-0.3cm}(k)} {g^{\mu\nu}}$.  

As an example, substituting the WKB solutions into \eqref{eq2.60} one obtains
\begin{equation}
-k_{\mu}k^{\mu} A_{\nu\delta} + A_{\nu\delta:~\mu}^{~~~\mu} +2ik_{\mu} A_{\nu\delta:}^{~~~\mu}+ik^{\mu}_{~:\mu}A_{\nu\delta} 
-2\overset{\hspace{-0.6cm}(0)}{R^{}_{\lambda\nu\delta\mu}}A^{\mu\lambda}
-\overset{\hspace{-0.3cm}(0)}{R^{}_{\lambda\nu}}A^{~\lambda}_{\delta~}
-\overset{\hspace{-0.3cm}(0)}{R^{}_{\lambda\delta}} A^{~\lambda}_{\nu~}
=0
\end{equation}
and from Equations \eqref{eq2.58} one has
\begin{equation}
A^\mu_{~\mu}=\overset{\hspace{-0.3cm}(0)}{g^{\mu\nu}}A_{\nu\mu}=0 \hspace{0.5cm} \text{and} \hspace{0.5cm} A_{\mu\nu:}^{~~~\nu}=-ik^{\nu}A_{\mu\nu}=0
\end{equation}
The last equation implies that these waves are transversal. Assuming that the gravitational waves are propagated in geodesics, i.e., that the wave vector is null, 
\begin{equation}
k^\mu k_\mu=0,
\end{equation}
one finds immediately 
\begin{equation}
 A_{\nu\delta:~\mu}^{~~~\mu} +2ik_{\mu} A_{\nu\delta:}^{~~~\mu}
-2\overset{\hspace{-0.6cm}(0)}{R^{}_{\lambda\nu\delta\mu}}A^{\mu\lambda}
-\overset{\hspace{-0.3cm}(0)}{R^{}_{\lambda\nu}}A^{~\lambda}_{\delta~}
-\overset{\hspace{-0.3cm}(0)}{R^{}_{\lambda\delta}} A^{~\lambda}_{\nu~}
=0
\label{eikon}
\end{equation}
that corresponds to the Eikonal equation, which relates the tensor of amplitudes and the wave vector for space-times perturbed to first order (see \cite{I68_1,I68_2}). Space-times corresponding to higher order perturbations include the terms appearing in \eqref{eikon}.

Finally, given that the higher order perturbations are linked with the first order perturbation for the metric, then the TT gauge can be imposed only from a simple coordinate gauge, as shown for the first order perturbation in the precedent section. It implies that these infinitesimal coordinate transformation leads to gauge conditions which simplify the uncalibrated wave equation \eqref{eq2.2.19}. Other approximations, in which higher order perturbation in the metric and in its inverse, without considering averages on the stress-energy tensor have been carry out \cite{CG69}.
\section{Green's Functions for the Flat Background and Perturbations of First Order}
In this section, the Green's functions are introduced with the aim to perform a multipolar expansion. Also, the decomposition of the wave functions in terms of advance and retarded potentials is needed to explain the back reaction effects, which appears in the presence of curvature in the non-linear as well as in the linear case. However, here it is considered only the flat case, where only the retarded Green function is not null.

From \eqref{eq2.60}, the inhomogeneous gravitational wave equation in the TT gauge for a Minkowski's background reads
\begin{equation}
\label{green1}
\square h_{\mu\nu}+16\pi T_{\mu\nu}=0,
\end{equation}
where the d'Alembertian is given by
\begin{equation}
\label{green2}
\square =-\partial^2_t + \nabla^2.
\end{equation}
Therefore, the wave equation for the flat background takes the form
\begin{equation}
\label{green3}
(-\partial^2_t + \nabla^2) h_{\mu\nu}+16\pi T_{\mu\nu}=0,
\end{equation}
where, the perturbations and the source term are functions of the coordinates, i.e.,
\begin{equation}
\label{green4}
h_{\mu\nu}:=h_{\mu\nu}(t,\mathbf x),\hspace{1cm} T_{\mu\nu}:=T_{\mu\nu}(t,\mathbf x).
\end{equation}

In particular, the perturbations and the source can be described in terms of the Fourier transforms 
\begin{align}
\label{green5}
h_{\mu\nu}(t,\mathbf x)&=\frac{1}{\sqrt{2\pi}}\int_{-\infty}^\infty d\omega \ \tilde{h}_{\mu\nu}(\omega,\mathbf x)e^{-i\omega t},
\nonumber 
\\
T_{\mu\nu}(t,\mathbf x)&=\frac{1}{\sqrt{2\pi}}\int_{-\infty}^\infty d\omega \ \tilde{T}_{\mu\nu}(\omega,\mathbf x)e^{-i\omega t},
\end{align}
where, it is assumed that the inverse transformation exists. Consequently, it is possible to return again to the original variables. Thus, the inverse transform is given by 
\begin{align}
\label{green6}
\tilde h_{\mu\nu}(\omega,\mathbf x)=&\frac{1}{\sqrt{2\pi}}\int_{-\infty}^\infty dt \ h_{\mu\nu}(t,\mathbf x)e^{i\omega t},
\nonumber \\
\tilde T_{\mu\nu}(\omega,\mathbf x)=&\frac{1}{\sqrt{2\pi}}\int_{-\infty}^\infty dt \ T_{\mu\nu}(t,\mathbf x)e^{i\omega t}.
\end{align}
Substituting \eqref{green5} into \eqref{green3} one obtains
\begin{equation}
\label{green7}
\frac{1}{\sqrt{2\pi}}\int_{-\infty}^\infty d\omega \ \left[(\omega^2 + \nabla^2) \tilde h_{\mu\nu}(\omega,\mathbf x)+16\pi \tilde T_{\mu\nu}(\omega,\mathbf x)\right] e^{-i\omega t}=0,
\end{equation}
which will be satisfied only if the integrand is null, i.e., 
\begin{equation}
\label{green8}
(\omega^2 + \nabla^2) \tilde h_{\mu\nu}(\omega,\mathbf x)+16\pi \tilde T_{\mu\nu}(\omega,\mathbf x)=0.
\end{equation}
This equation is known as a {\it Helmholtz equation} \cite{J75}. Redefining the second term in the last equation, as  $4\tilde T_{\mu\nu}(\omega,\mathbf x)=\tilde F_{\mu\nu}$, then,
\begin{equation}
\label{green9}
(\omega^2 + \nabla^2) \tilde h_{\mu\nu}(\omega,\mathbf x)+4\pi \tilde F_{\mu\nu}(\omega,\mathbf x)=0,
\end{equation}
and from the fact that the wave vector is null, one has
\begin{equation}
\label{green10}
(k_ik^i + \nabla^2) \tilde h_{\mu\nu}(\omega,\mathbf x)+4\pi \tilde F_{\mu\nu}(\omega,\mathbf x)=0.
\end{equation}
The Green's function used to construct the solution must satisfy
\begin{equation}
\label{green11}
(k_ik^i + \nabla^2_x) G_k(\mathbf x;\mathbf x')+4\pi \delta (\mathbf x-\mathbf x')=0,
\end{equation}
where $\mathbf{x}$ indicates the observer position and $\mathbf{x'}$ indicates each point in the source.

The Laplacian can be decomposed as a Legendrian plus a radial operator, namely
\begin{equation}
\label{green12}
\nabla^2_x = \frac{1}{r^2}\frac{\partial }{\partial r}\left( r^2\frac{\partial }{\partial r}\right) + \mathcal{L}^2,
\end{equation}
where, the Legendrian is explicitly defined as
\begin{equation}
\label{121}
\mathcal{L}^2=\frac{1}{\sin\theta}\frac{\partial}{\partial \theta}\left( \sin\theta \frac{\partial}{\partial \theta}\right) + \frac{1}{\sin^2\theta}\frac{\partial^2}{\partial \phi^2},
\end{equation}
and $r=|\mathbf x-\mathbf x'|$. Assuming that  $\mathbf x'=0$, i.e., the source is at the coordinate origin, then
\begin{equation}
\label{green13}
\mathcal{L}^2 G_k(\mathbf x;\mathbf x')=0.
\end{equation}

Since far enough from the source the gravitational waves must be spherical, then \eqref{green11} is reduced to
\begin{equation}
\label{green14}
\frac{1}{r}\frac{d^2 }{d r^2}\left( rG_k( r)\right) +k_ik^i G_k(r) + 4\pi \delta (r)=0.
\end{equation}
Then, for all points in the 3-space except for the origin, we have that the homogeneous version of \eqref{green14} is given by
\begin{equation}
\label{green15}
\frac{d^2 }{d r^2}\left( r G_k(r)\right) +k_ik^i (r G_k(r)) =0,
\end{equation}
whose family of solutions is
\begin{equation}
\label{green16}
G_k(r)=\frac{C_{+}}{r}e^{ikr}+\frac{C_{-}}{r}e^{-ikr},
\end{equation}
where, $k=|\mathbf k|=\sqrt{k_ik^i}$ and $C_{\pm}$ are arbitrary constants. The physically acceptable solutions must satisfy 
\begin{equation}
\label{green17}
\lim_{kr\rightarrow0} G_k(r)=\frac{1}{r}.
\end{equation}
Therefore, the solutions take the form
\begin{equation}
\label{green18}
G_k(r)=\frac{C}{r}e^{ikr}+\frac{1-C}{r}e^{-ikr}.
\end{equation}
Now, we notice that the inverse Fourier transform of \eqref{green11} leads immediately to the wave equation for the Green's function 
\begin{equation}
\label{green19}
\square G(t,\mathbf x;t',\mathbf x')+4\pi\delta(\mathbf x-\mathbf x')\delta(t-t')=0,
\end{equation}
with
\begin{eqnarray}
\label{green20}
G(t,\mathbf x;t',\mathbf x')&=&\frac{1}{2\pi}\int_{-\infty}^\infty d\omega\ G_k(\mathbf x,\mathbf x')e^{-i\omega (t-t')},\nonumber\\
&=&\frac{C}{|\mathbf x-\mathbf x'|}\delta\left(\tau -|\mathbf x-\mathbf x'|\right)+\frac{1-C}{|\mathbf x-\mathbf x'|}\delta\left(\tau +|\mathbf x-\mathbf x'|\right),\nonumber\\
&=&CG^{(+)}(t,\mathbf x;t',\mathbf x')+(1-C)G^{(-)}(t,\mathbf x;t',\mathbf x'),
\end{eqnarray}
where, $\tau=t-t'$ and
\begin{equation}
\label{green21}
G^{(\pm)}(t,\mathbf x;t',\mathbf x')=\frac{1}{|\mathbf x-\mathbf x'|}\delta\left(\tau \mp|\mathbf x-\mathbf x'|\right).
\end{equation}
Note that the solution for the Green's function \eqref{green18} is written in terms of two functions, one for the advance time $G^{(+)}$ and the other for the retarded time $G^{(-)}$. 

If the second term in the wave equation \eqref{green3} is written as
\begin{eqnarray}
\label{green22}
16\pi T_{\mu\nu}(t,\mathbf x)&=&4\pi F_{\mu\nu}(t,\mathbf x),\nonumber\\
&=&4\pi\int_{-\infty}^\infty\int_{V} dt' d^3x' \ F_{\mu\nu}(t',\mathbf x') \delta(\mathbf x-\mathbf x')\delta(t-t'),
\end{eqnarray}
where $V$ is the source volume, then $\overline h_{\mu\nu}$ must be
\begin{eqnarray}
\label{green23}
\overline h_{\mu\nu}(x^\alpha)&=&4\int d^4x'\ T_{\mu\nu}(x^{\alpha'}) G(x^\alpha;x^{\alpha'}),
\end{eqnarray}
where the integral is defined for all times and for the volume occupied by the source. Substituting \eqref{green20} into \eqref{green23}, one has
\begin{equation}
\label{green24}
\overline h_{\mu\nu}(x^\alpha)=4\int d^4x'\ T_{\mu\nu}(x^{\alpha'}) \left( CG^{(+)}(x^\alpha;x^{\alpha'})+(1-C)G^{(-)}(x^\alpha;x^{\alpha'})\right).
\end{equation}
Now, observing the structure of the Green's function \eqref{green21}, the delta distribution argument is
\begin{equation}
\label{green25}
t_\pm=t'\mp|\mathbf x-\mathbf x'|.
\end{equation}
This means that the Green's function is describing two travelling waves, one {\it outgoing} and other {\it ingoing}. However, the advanced Green's function is physically unacceptable in the flat background because of the causality principle. Thus, the solution is restricted only to the retarded Green's function, which indicates that one wave will be detected at the point $\mathbf x$ in a time $t$ after generated at a point  $\mathbf x'$ in a time $t'$. This wave propagates from $\mathbf x'$ to $\mathbf x$ with velocity $c$. Therefore \eqref{green24} takes the form
\begin{equation}
\label{green26}
\overline h_{\mu\nu}(x^\alpha)=4\int d^4x'\ T_{\mu\nu}(x^{\alpha'}) G^{(-)}(x^\alpha;x^{\alpha'}).
\end{equation}
Substituting explicitly the Green's function \eqref{green21} one obtains the expression for the wave function in term of the sources
\begin{eqnarray}
\label{green27}
\overline h_{\mu\nu}(t,\mathbf x)&=&4\int_{-\infty}^\infty \int_{V} dt' d^3x'\ \frac{T_{\mu\nu}(t',\mathbf x') \delta\left(t-t' + |\mathbf x-\mathbf x'|\right)}{|\mathbf x-\mathbf x'|}\nonumber\\
&=&4 \int_{V} d^3x'\  \frac{T_{\mu\nu}(t-|\mathbf x-\mathbf x'|,\mathbf x')}{|\mathbf x-\mathbf x'|}.
\end{eqnarray}

It is worth mentioning that the advance Green's function in the presence of curved space-time must be taken into account, because both terms, advance and retarded, appears in back reaction phenomena. As a consequence of the effective potential in the radial equations, for example, when the Schwarzschild's space-time is axially perturbed, two radial waves will travel between the source and the spatial infinity. 
\section{Multipolar Expansion}
\graphicspath{{./Figuras/Multipolar/}}
A series expansion is a way to compute the contribution of the sources to the gravitational radiation in  Equation \eqref{green27}. This kind of procedure is known in the literature as multipolar expansion.

\begin{figure}[h!]
\begin{center}
\vspace{1.5cm}
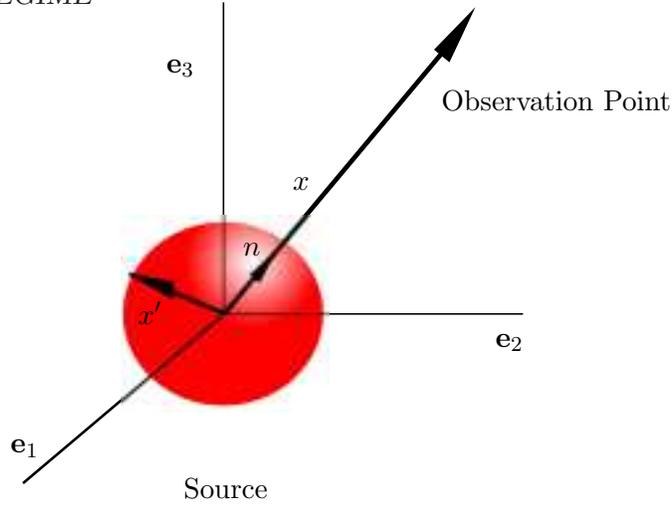
\end{center}
\caption{Source and observer's position.}	
\label{multipolar}
\end{figure}

Note that 
\begin{equation}
\label{mul1}
\frac{1}{|\mathbf x - \mathbf x'|}=\frac{1}{\left(r^2-2x_ix^{i'}+x'_ix^{i'}\right)^{1/2}},
\end{equation}
where, $r^2=x_i x^i$. The observer, at $\mathbf x$, is far from the source, then $\|\mathbf{x}\|\gg \|\mathbf{x}'\|$, as sketched in Figure \ref{multipolar}. 
\\Thus, $r^2\gg x'_ix^{i'}$, then,
\begin{equation}
\label{mul2}
\frac{1}{|\mathbf x - \mathbf x'|}=\frac{1}{\left(r^2-2x_ix^{i'}\right)^{1/2}}.
\end{equation}
Expanding in McLaurin series for $x^k$,
\begin{equation}
\label{mul3}
\frac{1}{|\mathbf x - \mathbf x'|}=\frac{1}{r}+\frac{x'_k x^k}{r^3}+\frac{1}{2}\frac{3x'_k x'_m x^k x^m}{r^5}+\cdots
\end{equation}
Then, whenever $r\rightarrow\infty$, i.e., whenever the observer is far from the source,
\begin{equation}
\label{mul4}
\frac{1}{|\mathbf x - \mathbf x'|}\approx\frac{1}{r}
\end{equation}
and therefore \eqref{green27} can be written \cite{C101} as follows
\begin{eqnarray}
\label{mul5}
\overline h_{\mu\nu}(t,\mathbf x)=\frac{4}{r}\int_{V}  d^3x'\ T_{\mu\nu}(t-|\mathbf x-\mathbf x'|,\mathbf x').
\end{eqnarray}
On the other hand, 
\begin{eqnarray}
\label{mul6}
|\mathbf x - \mathbf x'|=r\left(1-2\frac{x^ix'_i}{r^2}\right)^{1/2},
\end{eqnarray}
which, can be expanded in McLaurin series for $x^i$ 
\begin{eqnarray}
\label{mul7}
|\mathbf x - \mathbf x'|=r\left(1-\frac{x'_k}{r}n^k-\frac{1}{2}\frac{x'_k x'_l}{r^2}n^k n^l+\cdots\right),
\end{eqnarray}
where, 
\begin{eqnarray}
\label{mul8}
n^k=\frac{x^k}{r},\hspace{1cm}n^kn_k=1.
\end{eqnarray}
Therefore, far from the gravitational wave sources
\begin{eqnarray}
\label{mul9}
t-|\mathbf x - \mathbf x'|=t-r+x'_kn^k,
\end{eqnarray}
which, implies that
\begin{eqnarray}
\label{mul10}
T_{\mu\nu}(t-|\mathbf x - \mathbf x'|)=T_{\mu\nu}(t-r+x'_kn^k).
\end{eqnarray}

Defining $t'=t-r$, \eqref{mul8} can be written in the form
\begin{eqnarray}
\label{mul11}
T_{\mu\nu}(t-|\mathbf x - \mathbf x'|)=T_{\mu\nu}(t'+x'_kn^k).
\end{eqnarray}
Thus, the stress-energy tensor can be expanded as
\begin{align}
\label{mul12}
T_{\mu\nu}(t'+x'_kn^k)&=T_{\mu\nu}(t')+T_{\mu\nu,1}(t')x'_kn^k + \frac{1}{2!}T_{\mu\nu,11}(t')x'_kx'_jn^kn^j\nonumber \\ 
 &+\frac{1}{3!}T_{\mu\nu,111}(t')x'_ix'_jx'_kn^in^jn^k+\cdots
\end{align}
where is assumed that the source is moving slowly with respect to the speed of light $c$, or in other words, $r\gg \lambda/2\pi$, with $\lambda$ the gravitational wave length. Substituting the last equation in the expression for the wave function \eqref{mul5}, one obtains
\begin{align}
\label{mul13}
&\overline h_{\mu\nu}(t,\mathbf x)=\nonumber\\
&\frac{4}{r}\int_{V} d^3x'\  \bigg( T_{\mu\nu}(t')+T_{\mu\nu,1}(t')x'_kn^k 
 + \frac{1}{2!}T_{\mu\nu,11}(t')x'_kx'_jn^kn^j  \nonumber \\ 
& \hspace{2.2cm}+\frac{1}{3!}T_{\mu\nu,111}(t')x'_ix'_jx'_kn^in^jn^k+\cdots \bigg) 
\end{align}
Thus, it is possible to define the following momenta of the stress-energy tensor \cite{P64,SR10}
\begin{align}
\label{mul14}
M(t)&=\int_{V} d^3x\ T^{11}(t,x^m),\hspace{2.1cm}
M^i(t)=\int_{V} d^3x\ T^{11}(t,x^m)x^i,\nonumber\\
M^{ij}(t)&=\int_{V} d^3x\ T^{11}(t,x^m)x^ix^j,\hspace{1cm}
M^{ijk}(t)=\int_{V} d^3x\ T^{11}(t,x^m)x^ix^jx^k,\nonumber\\
P^i(t)&=\int_{V} d^3x\ T^{1i}(t,x^m),\hspace{2.0cm}
P^{ij}(t)=\int_{V} d^3x\ T^{1i}(t,x^m)x^j,\nonumber\\
P^{ijk}(t)&=\int_{V} d^3x\ T^{1i}(t,x^m)x^jx^k,\hspace{1.2cm}
S^{ij}(t)=\int_{V} d^3x\ T^{ij}(t,x^m),\nonumber\\
S^{ijk}(t)&=\int_{V} d^3x\ T^{ij}(t,x^m)x^k,\hspace{1.4cm}
S^{ijkl}(t)=\int_{V} d^3x\ T^{ij}(t,x^m)x^kx^l.
\end{align}

Using the conservation equation
\begin{equation}
\label{mul15}
T^{\mu\nu}_{~ ~ ~;\mu}=0,
\end{equation}
that in the case of the linear theory reads
\begin{equation}
\label{mul16}
T^{1\nu}_{~ ~ ~,1}=-T^{i\nu}_{~ ~,i}.
\end{equation}
Since $T^{11}_{~ ~ ~,1}=-T^{i1}_{~ ~,i}$, it is possible to re-express 
\begin{align}
\label{mul17}
\dot M(t)&=\int_{V} d^3x\ T^{11}_{~ ~ ~,1}(t,x^m),\nonumber\\
&=-\int_{V} d^3x\ T^{1i}_{~ ~ ~,i}(t,x^m),\nonumber\\
&=-\oint_{\partial V} d^2x\ T^{1i}(t,x^m)n_i, \nonumber\\
&=0,
\end{align}
where, $\partial V$ is the source surface and $n^i$ is its normal vector. Thus, one has
\begin{eqnarray}
\label{mul18}
\dot M^j(t)&=&\left(\int_{V} d^3x\ T^{11}(t,x^m) x^{j}\right)_{,1},\nonumber\\
&=&\int_{V} d^3x\ T^{11}_{~ ~ ~,1}(t,x^m)x^{j}+\int_{V} d^3x\ T^{11}(t,x^m) \dot x^{j},\nonumber\\
&=&-\int_{V} d^3x\ T^{1i}_{~ ~ ~,i}(t,x^m) x^{j}+\int_{V} d^3x\ T^{11}(t,x^m) \dot x^{j},\nonumber \\
&=&\int_{V} d^3x\ T^{1i}(t,x^k)\delta_i^{~ j},\nonumber\\
&=&P^j(t).
\end{eqnarray}

Also, the following relation between the momenta for the stress-energy tensor $T_{\mu\nu}$ are established
\begin{subequations}
\begin{align}
\label{mul19}
\dot M^{ij}&=P^{ij}+P^{ji},\hspace{0.5cm}
\dot M^{ijk}=P^{ijk}+P^{jki}+P^{kij},\\
\dot P^{j}&=0,\hspace{0.5cm}
\dot P^{ij}=S^{ij},\hspace{0.5cm}
\dot P^{ijk}=S^{ijk}+S^{ikj},\\
&\ddot M^{jk}=2S^{jk},\hspace{0.5cm}
\dddot M^{ijk}=3!S^{(ijk)}.
\end{align}
\end{subequations}

Thus, from \eqref{mul13} it is obtained that
\begin{eqnarray}
\label{mul22}
\overline h^{11}&=&\frac{4}{r}M+\frac{4}{r}P^in_i+\frac{4}{r}S^{ij}n_in_j +\frac{4}{r}\dot S^{ijk}n_in_jn_k+\cdots,\nonumber\\
\overline h^{1i}&=&\frac{4}{r}P^i+\frac{4}{r}S^{ij}n_j+\frac{4}{r}\dot S^{ijk}n_jn_k+\cdots,\nonumber\\
\overline h^{ij}&=&\frac{4}{r}S^{ij}+\frac{4}{r}\dot S^{ijk}n_k+\cdots,
\end{eqnarray}
which is known as the multipolar expansion \cite{SR10}. From the gauge condition \eqref{eq2.49}, one obtains
\begin{equation}
\label{mul23}
\overset{\text{New}}{\overline h^{\mu\nu}}=\overset{\text{Old}}{\overline h^{\mu\nu}}+2\xi^{(\mu,\nu)}-\eta^{\mu\nu}\xi^\gamma_{~ ~,\gamma},
\end{equation}
from which, the changes in the different components of the metric result as
\begin{eqnarray}
\label{mul24}
\delta \overline h^{11}&=&\xi^{1,1}+\xi^i_{~ ~,i}\nonumber,\\
\delta \overline h^{1j}&=&\xi^{1,j}+\xi^{j,1}\nonumber,\\
\delta \overline h^{jl}&=&\xi^{j,l}+\xi^{l,j}-\delta^{jl}\zeta^\mu_{~ \mu}.
\end{eqnarray}

One can select the gauge functions
\begin{eqnarray}
\label{mul25}
\xi^1&=&\frac{1}{r}P^i_{~i}+\frac{1}{r}P^{jl}n_jn_l+\frac{1}{r}S^i_{~ij}n^j+\frac{1}{r}S{ijk}n_in_jn_k,\nonumber\\
\xi^i&=&\frac{4}{r}M^i+\frac{4}{r}P^{ij}n_j-\frac{1}{r}P^j_{~ j}n^i-\frac{1}{r}P^{jk}n_jn_kn^i+\frac{4}{r}S^{ijk}n_jn_k\nonumber \\ &&-\frac{1}{r}S^l_{~lk}n^kn^i-\frac{1}{r}S^{jlk}n_jn_ln_kn^i,
\end{eqnarray}
such that in the TT gauge, the components $\overline{h}^{TT\mu\nu}$ take the form
\begin{eqnarray}
\label{mul26}
\overline h^{TT11}&=&\frac{4M}{r},\nonumber\\
\overline h^{TT1i}&=&0,\nonumber\\
\overline h^{TTij}&=&\frac{4}{r}\left[ \perp^{ik} \perp^{jl} S_{lk} +\frac{1}{2}\perp^{ij}\left( S_{kl}n^kn^l-S^k_{~k}\right)\right].
\end{eqnarray}
Observe that $\overline h^{TT11}$ is not a radiative term, it corresponds to the Newtonian potential which falls as $\sim1/r$. From $\overline h^{TTij}$ we see that the radiative terms have quadrupolar nature or higher. The projection tensor $\perp^{ij}$ is defined \cite{W73,B07} as
\begin{eqnarray}
\label{mul27}
\perp^{ij}=\delta^{ij}-n^in^j
\end{eqnarray}
so that
\begin{eqnarray}
\label{mul28}
\perp^{ij}n_j=0,\hspace{1cm}\perp^{ij}\perp_{j}^{~k}=\perp^{ik}.
\end{eqnarray}
Since, $\overline h^{TTij}$ does not depend on the trace of $S$, then one can define
\begin{eqnarray}
\label{mul29}
\overline S^{ij}=S^{ij}-\frac{1}{3}\delta{ij}S^k_{~k},
\end{eqnarray}
which is the trace-free part of $S^{ij}$. In the same manner, the trace-free part of $M^{ij}$ is defined as
\begin{eqnarray}
\label{mul30}
\overline M^{ij}=M^{ij}-\frac{1}{3}\delta{ij}M^k_{~k},\hspace{1cm}\overline S^{ij}=\frac{1}{2}\ddot{\overline{M}}^{ij}.
\end{eqnarray}
Therefore
\begin{eqnarray}
\label{mul31}
\overline h^{TTij}=\frac{2}{r}\left(\perp^{ik}\perp^{jl}\ddot{\overline M}_{kl}+\frac{1}{2}\perp^{ij}\ddot{\overline M}_{lk}n^ln^k \right),
\end{eqnarray}
which depends strictly on the quadrupolar contribution of the source. 

Returning to the weak field approximation, it can be shown that another form to expand the left side of \eqref{mul1} is in spherical harmonics \cite{J75}, i.e.,
\begin{equation}
\label{esfe1}
\frac{1}{\mathbf x -\mathbf x'}=4\pi \sum_{l=0}^\infty \sum_{m=-l}^l \frac{1}{2l+1}\frac{r_<^l}{r_>^{l+1}}\bar Y_{lm}(\theta',\phi')Y_{lm}(\theta,\phi),
\end{equation}
where, $r_>=\max(|\mathbf x|,|\mathbf x'|)$ and $r_<=\min(|\mathbf x|,|\mathbf x'|)$. 
Hence, \eqref{green27} can be written as
\begin{equation}
\label{esfe2}
\overline h_{\mu\nu}(x^\alpha)=16\pi\int_{V} d^3x'\ T_{\mu\nu}(t-|\mathbf x-\mathbf x'|,x^{j'})\sum_{l,m}\frac{1}{2l+1}\frac{r_<^l}{r_>^{l+1}}\bar Y_lm(\theta',\phi')Y_lm(\theta,\phi),
\end{equation}
where, the volume element is
\begin{equation}
\label{esfe3}
d^3x'=r^{'2}dr'd\Omega',
\end{equation}
and $d\Omega'=\sin^2\theta' d\theta' d\phi'$, and the symbol $\sum_{l,m}$ represents the double sum that appears in \eqref{esfe1}.

It is worth noting that \eqref{esfe2} can be written as
\begin{equation}
\label{esfe4}
\overline h_{\mu\nu}(x^\alpha)=16\pi\sum_{l=0}^\infty \sum_{m=-l}^l \frac{1}{2l+1} \frac{Y_{lm}(\theta,\phi)}{r^l+1}q^{~ ~ ~ lm}_{\mu\nu},
\end{equation}
where the multipolar moments are defined as
\begin{equation}
\label{esfe5}
q^{~ ~ ~ lm}_{\mu\nu}=\int_{V} d^3x'\ r^{l'} T_{\mu\nu}(t-|\mathbf x-\mathbf x'|,x^{j'})\bar Y_{lm}(\theta',\phi'),
\end{equation}
which are equivalent to those multipolar moments defined in \eqref{mul14}.
\section{Gravitational Radiation from Point Particle Binary System}
It is worth recalling the main steps given by Peters and Mathews \cite{PM63} to obtain the well-known and widespread equation used for the power radiated by two point masses in a Keplerian orbit.

\begin{figure}[h!]
	\begin{center}
		\includegraphics[height=5cm]{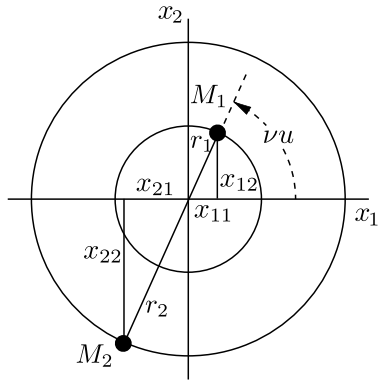}
	\end{center}
	\caption{Binary system as viewed from the top. The coordinates $x_{ai}$  of the particles are indicated, as well as the angle $\nu u$ with respect to the $x$ axis.}
	\label{fig4}
\end{figure}

As it is well known in the literature, in the weak field limit of the Einstein's field equations, i.e., when the metric can be written as a perturbation $h_{\mu\nu}$ of the Minkowski metric $\eta_{\mu\nu}$, namely 
\begin{equation}
g_{\mu\nu}=\eta_{\mu\nu}+h_{\mu\nu}, \hspace{0.5cm}|h_{\mu\nu}|\ll |\eta_{\mu\nu}|,
\end{equation}
the power emitted by any discrete mass distribution in the limit of low velocities, as shown, e.g., in Ref. \cite{PM63}, is given by
\begin{equation}
P=\frac{1}{5}\left(\dddot{Q}_{ij}\dddot{Q}_{ij}-\frac{1}{3}\dddot{Q}_{ii}\dddot{Q}_{jj}\right),
\label{PM_pot}
\end{equation}
where the dots indicate derivative with respect to the retarded time $u$ and
\begin{equation}
Q_{ij}=\sum_a m_a x_{ai}x_{aj},
\label{PM_Q}
\end{equation}
in which $a$ labels each particle of the system and $x_{ai}$ is the projection of the position vector of each mass along the $x$ and $y$ axes. Particularly, for a point particle binary system of different masses in circular orbits, when the Lorentz factor is considered to be $\gamma=1$,  one can write
\begin{equation}
x_{a1}=r_a \cos(\nu u-\pi \delta_{a2}), \hspace{0.5cm}x_{a2}=r_a \sin(\nu u-\pi \delta_{a2}),\hspace{0.5cm} a=1,2
\label{PM_x}
\end{equation}
as shown in Figure \ref{fig4}

Here, $r_a$ is given by \eqref{rdef}, $\nu$ by \eqref{freqnu},  and the Kronecker delta discriminates each  particle. Then, the components of $Q_{ij}$ read
\begin{eqnarray}
Q_{ij}=
\begin{pmatrix}
\mu d_0^2 \cos^2(\nu u) & \mu d_0^2 \sin(\nu u)\cos(\nu u) \\ 
\mu d_0^2 \sin(\nu u)\cos(\nu u) & \mu d_0^2 \sin^2(\nu u)
\end{pmatrix},
\label{PM_Q1}
\end{eqnarray}
thus,
\begin{eqnarray}
\dddot Q_{ij}=
\begin{pmatrix} 
4 \nu^3 \mu d_0^2\sin (2 \nu  u) & -4 \nu^3 \mu d_0^2\cos (2 \nu  u) \\ 
-4 \nu^3 \mu d_0^2\cos (2 \nu  u) & -4 \nu^3 \mu d_0^2 \sin (2 \nu  u)
	\end{pmatrix}.
\label{PM_Q2}
\end{eqnarray}
Finally, substituting the above equation in \eqref{PM_pot}, one obtains
\begin{equation}
P=\frac{32}{5} \mu ^2 \nu ^6 d_0^4 = \frac{32 {m_1}^2 {m_2}^2 (m_1+m_2)}{5 d_0^5},
\end{equation}
where Kepler's third law is used in the last equality.

If the eccentricity of the orbits are taken into account, the expression for the power lost by emission in gravitational waves \cite{PM63} become
\begin{equation}
P=\frac{8}{15} \frac{{m_1}^2 {m_2}^2 (m_1+m_2)}{a^5(1-\epsilon^2)^5}\left(1+\epsilon\cos\tilde{\phi}\right)^4\left(12\left(1+\epsilon\cos\tilde{\phi}\right)^2+\epsilon^2\sin^2\tilde{\phi}\right),
\label{PM_pot_ecc}
\end{equation}
where
\begin{equation}
\dot{\tilde{\phi}}=\frac{((m_1+m_2)a(1-\epsilon^2))^{1/2}}{d^2}\hspace{0.5cm}\text{and}\hspace{0.5cm}d=\frac{a(1-\epsilon^2)}{1+\epsilon\cos\phi},
\label{PM_ang_vel_and_rad}
\end{equation}
where $d$ is the separation of the particles, $a$ is the semi-major axis of the ellipse described by the particles and $\tilde{\phi}$ is the angle between the line that connects both particles and the $x$ axis.
\chapter[THE {\it Eth} FORMALISM]{THE {\it Eth} FORMALISM AND THE SPIN-WEIGHTED SPHERICAL HARMONICS }
Before introducing the outgoing characteristic formulation of the general relativity, it is convenient to consider a standard tool to regularise the angular differential operators, namely the {\it eth formalism}, which is based on a non-conformal mapping of the regular coordinate charts to make a finite coverage of the unit sphere. This kind of mapping was originally used in global weather studies \cite{W70,S72,RIP96}, and is based on the stereographic and gnomonic projections. It is worth mentioning that these projections that make the finite coverage of the unit sphere, remove the singular points related to the fact that the sphere can not be covered by only one coordinate chart. 

The {\it eth formalism} \cite{NP66,GMNRS66,GLPW97,stewart93,T07} is a variant of the Newman-Penrose formalism. As such in this last formalism, scalars and associated functions, and operators related to the projections onto the null tangent vectors to the unit sphere are present. The projection onto the tangent vectors to a topological sphere (a diffeomorphism to the unit sphere) can also be generalised. 

In order to present the eth formalism, the non-conformal mapping using stereographic coordinates is given. After that, a decomposition to the unit sphere and the transformation of vectors and one-forms are shown. These transformation rules are extended to the dyads and their spin-weights are found. It is worth mentioning that the spin-weight induced into the scalar functions comes from the transformation rules associated with the stereographic dyads. However, this property is not exclusive of this kind of coordinates, and appears as a transformation associated with the coordinate maps needed to make the finite coverage to the unit sphere. Then the spin-weighted scalars are constructed from the irreducible representation for tensors of type $(0,2)$ and then, the general form for a spin-weighted scalar of spin-weight $s$ is shown. The rising and lowering operators are presented from the projection of the covariant derivative associated with the unit sphere metric and the Legendrian operator is then expressed in terms of these rising and lowering operators. 

Subsequently, some properties of spin-weighted scalars are shown and the orthonormality of such functions is defined. It is shown that the spin-weighted spherical harmonics $_sY_{lm}$ constitute a base of functions in which any spin-weighted function on the sphere can be decomposed. The spin-weighted spherical harmonics $_sY_{lm}$ and the action of the rising and lowering operators in them are constructed. Finally, another base of functions to decompose spin-weighted functions on the sphere, composed by the spin-weighted spherical harmonics $_sZ_{lm}$, is defined as linear combinations of the $_sY_{lm}$. 
\section{Non-conformal Mappings in the Sphere}
There are infinite forms to make up finite coverage of the sphere. The principal aim here is to show an atlas, with at least two coordinate charts, in which all points in $S^2$ are mapped. In the context of the global weather studies diverse  useful schemes were proposed, from the numerical point of view, to make finite coverages to the sphere \cite{SAM68,W70,S72,Thacker80, BF85,CK15}. Only two of these schemes become important in numerical relativity. The first one is the stereographic projection in two maps and the second one is the gnomonic projection in six maps, also known as cubed sphere. Both offer great numerical advantages, as the simplification of all angular derivatives, in the case of the stereographic coordinates and simplification in the numerical computation as in the case of the cubed sphere projection. It is worth stressing that the {\it eth} formalism is totally independent on the selection of the coordinates, as we will show in the next sections. However, given the simplification in some of the calculations and its use in those numerical computations, we present in  details the connection between the stereographic coordinates and the spin-weighted scalars.
\section{Stereographic Coordinates}
\begin{figure}[h!]
\begin{center}		
	\vspace{-1.5cm}
\begin{tabular}{cc}
\includegraphics[height=10cm]{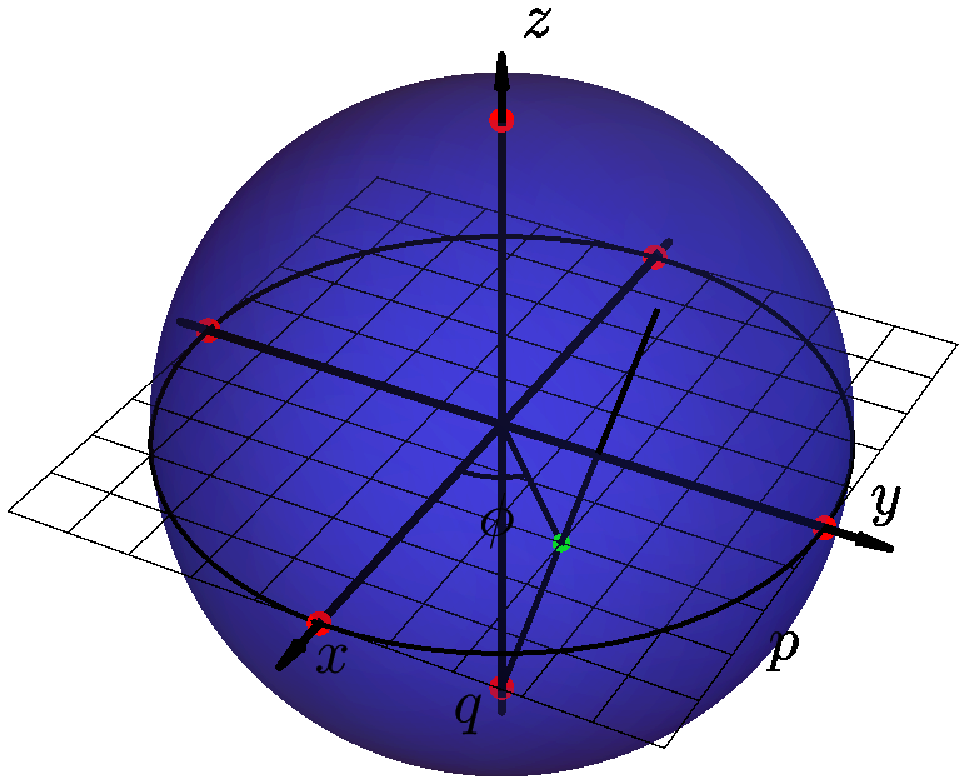}&\begin{minipage}{10cm}
\vspace{-11cm}\hspace{-1cm}\includegraphics[height=6cm]{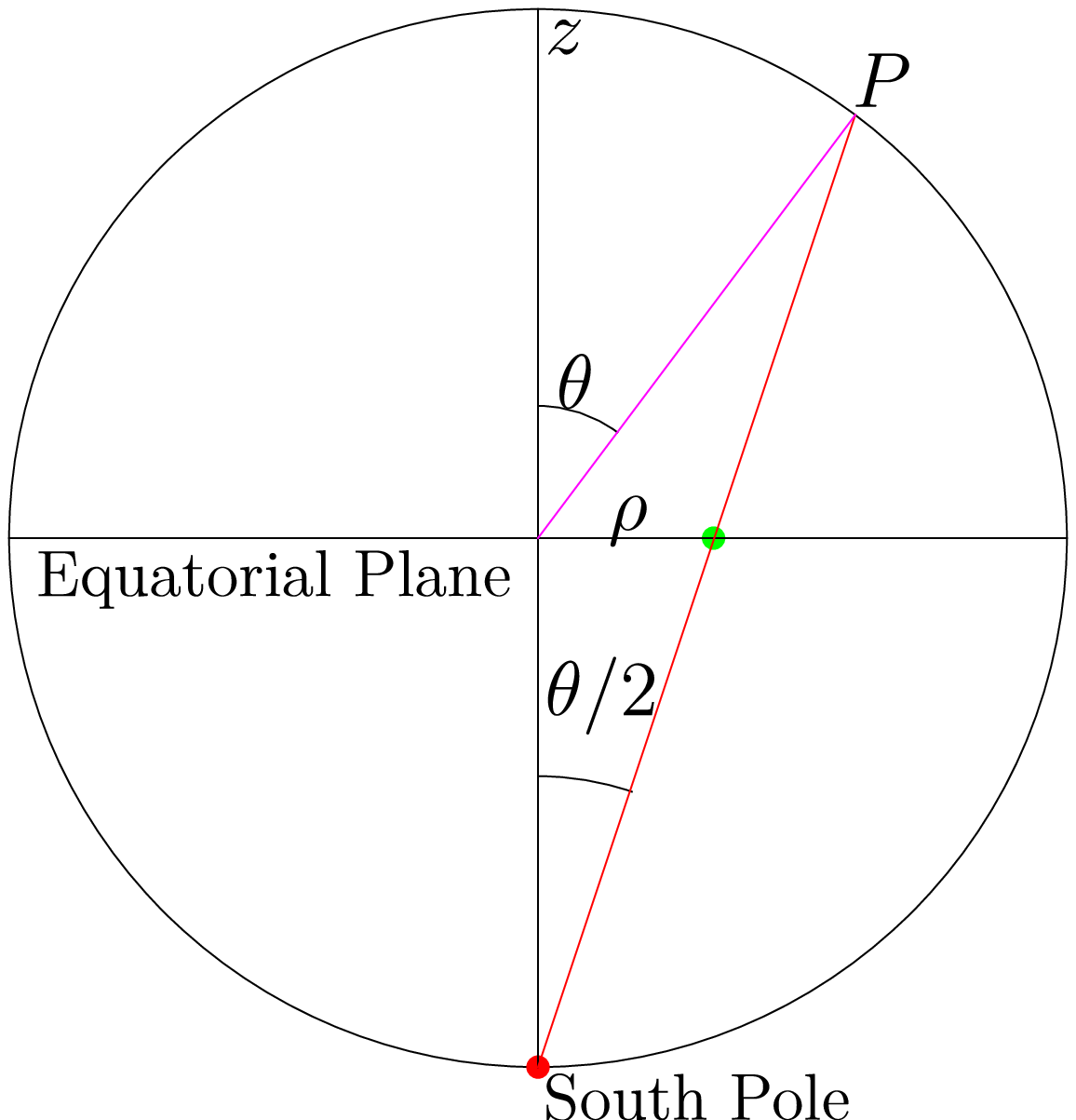}
\end{minipage}	
\end{tabular}
\end{center}
\vspace{-1.5cm}
\caption{Stereographic coordinates construction: the equatorial plane is projected from the south pole to the surface of the unit sphere. The interior points to the equator are projected to the north hemisphere, whereas the exterior points are projected to the south.}
\label{fstereo1}
\end{figure}

This section starts with the description of the construction of the stereographic atlas which covers the entire sphere. As an example, a point (in green) in the equatorial plane is projected into the north hemisphere from the south pole as sketched in Figure \ref{fstereo1}

The coordinates on the equatorial plane (the green point) are represented as the ordered pair $(q,p)$ and the point to be represented in the sphere $P$ as the ordered triad $(x,y,z)$. From Figure \ref{fstereo1}, one has
\begin{equation}
\rho=\tan\left(\frac{\theta}{2}\right), \hspace{1cm} q=\rho\cos\phi,\hspace{1cm} p=\rho\sin\phi.
\label{stereo1}
\end{equation}
\newpage
Then, it is possible to represent the coordinates through a complex quantity $\zeta$ \cite{NP66}, in the form
\begin{equation}
\zeta=\tan\left(\dfrac{\theta}{2}\right)e^{i\phi};
\label{stereo2}
\end{equation}
thus, $\Re(\zeta)=q$ and $\Im(\zeta)=p$. It is worth stressing that it is not possible to map all points in the spherical surface into the equatorial plane, even if the plane is extended to the infinity. Thus, it is necessary to appeal to at least two coordinate charts. One possible way to do this is by selecting one for each hemisphere north $(N)$ and south $(S)$ \cite{GMNRS66}, namely
\begin{equation}
\zeta_N=\tan\left(\dfrac{\theta}{2}\right)e^{i\phi}, \hspace{1cm}\zeta_S=\cot\left(\dfrac{\theta}{2}\right)e^{-i\phi}, \hspace{1cm} \zeta_{N\atop S}=q_{N\atop S}+ip_{N\atop S};
\label{stereo3}
\end{equation}
such that
\begin{equation}
|q_{N}|\le 1, \hspace{1cm} |p_{N}|\le 1, 
\label{stereo4}
\end{equation}
which defines a rectangular domain in the plane to be mapped  into the sphere.

From the definition \eqref{stereo1}, one immediately has
\begin{equation}
q=\tan\left(\frac{\theta}{2}\right)\cos\phi,\hspace{1cm} p=\tan\left(\frac{\theta}{2}\right)\sin\phi.
\label{stereo5}
\end{equation}
Taken into account that
\begin{align}
\tan\left(\frac{\theta}{2}\right)&=\frac{\sin\theta}{2}\left(1+\tan^2\left(\frac{\theta}{2}\right)\right),
\label{stereo6}
\end{align}
then the relationship between the rectangular and the $q,p$ reads
\begin{equation}
x=\frac{2q}{1+q^2+p^2},\hspace{1cm}y=\frac{2p}{1+q^2+p^2}.
\label{stereo7}
\end{equation}
This allows to write the $z$ coordinate as
\begin{align}
z&=\cos\theta,\nonumber\\
&=\frac{1-q^2-p^2}{1+q^2+p^2}.
\label{stereo8}
\end{align} 

\begin{figure}[h!]
\begin{center}
\vspace{-1.5cm}
\includegraphics[height=10cm]{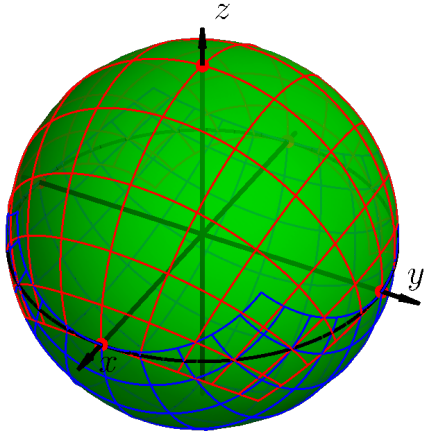}
\vspace{-1.5cm}
\end{center}	
\caption{Coordinate atlas in the sphere. Coordinate lines as result of the mapping of the plane maps contructed from the equator of the sphere.}
\label{fstereo2}
\end{figure}

With equations \eqref{stereo7} and \eqref{stereo8} the coordinate lines $(q,p)$ on the surface of the sphere are constructed, as shown in Figure \ref{fstereo2}, which shows how the atlas $\{\{q_N,p_N\},\{q_S,p_S\}\}$ for the unit sphere is constructed. 

\begin{figure}[h!]
	\begin{center}
			\includegraphics[height=8cm]{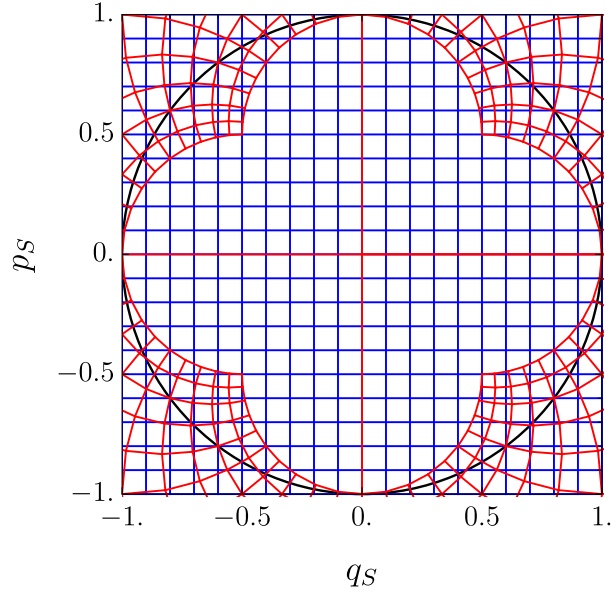}
	\end{center}
	\caption{Coordinate lines of north hemisphere into the south region. The equatorial line is indicated as a circle in black.
		}
	\label{fig:stereo1}
\end{figure}

From \eqref{stereo3}, for all points except the poles,
\begin{align}
\zeta_N 
&=\dfrac{1}{\zeta_S}
\label{stereo9}
\end{align}
or
\begin{align}
\zeta_N&=\dfrac{\bar{\zeta_S}}{\zeta_S\bar{\zeta_S}}.
\label{stereo10}
\end{align}
In terms of the $q$ and $p$ coordinates, \eqref{stereo10} reads
\begin{align}
q_N=\dfrac{q_S}{q_S^2+p_S^2}, \hspace{1cm} p_N=\dfrac{-p_S}{q_S^2+p_S^2},
\label{stereo11}
\end{align}
which define the relationship between the north and south coordinates, and therefore it defines the transformation between the corresponding charts. Thus, the form of the coordinate lines $(q_N,p_N)$, corresponding to the north map when $p_S$ or $q_S$ are considered as constant, can be traced (see Figure \ref{fig:stereo1}). It is particularly useful when a discretisation scheme of the angular operators in the sphere is implemented, because it shows clearly that a bi-dimensional interpolation is needed to pass information from one to another coordinate map. 
\section{Decomposition of the Metric of the Unit Sphere}
The square of the line element that describes the $S^2$ manifold (the unit sphere) in spherical coordinates is given by
\begin{equation}
ds^2=d\theta^2+\sin^2\theta d\phi^2.
\label{angular1}
\end{equation}
Now, from \eqref{stereo2} the total differential of $\zeta$ and $\overline{\zeta}$ are computed, namely
	\begin{align}
	d\zeta=\zeta_{,\theta}d\theta+\zeta_{,\phi}d\phi\hspace{0.5cm}\text{and}\hspace{0.5cm} 
	d\overline{\zeta}=\overline{\zeta}_{,\theta}d\theta+\overline{\zeta}_{,\phi}d\phi. 
	\label{angular2}%
	\end{align}
Here the absence of the indices $N$ or $S$ means that the results are equal for both hemispheres. Thus, from \eqref{angular2} one obtains that
\begin{align*}
d\zeta d\overline{\zeta} &= \zeta_{,\theta}\overline{\zeta}_{,\theta}d\theta^2 +  \left(\zeta_{,\theta} \overline{\zeta}_{,\phi} + \zeta_{,\phi} \overline{\zeta}_{,\theta}\right) d\phi d\theta + \zeta_{,\phi} \overline{\zeta}_{,\phi}d\phi^2 \\
&= \dfrac{1}{4}\left(1+\zeta\overline{\zeta}\right)^2 \left(d\theta^2 + \sin^2\theta d\phi^2\right).
\end{align*}
Therefore, the unit sphere metric in terms of $\zeta,\overline{\zeta}$ takes the non-diagonal form \cite{stewart93},
\begin{equation}
d\theta^2 + \sin^2\theta d\phi^2=\dfrac{4}{\left(1+\zeta\overline{\zeta}\right)^2}d\zeta d\overline{\zeta}.
\label{angular3}
\end{equation}

Expressing the total derivatives $d\zeta$ and $d\overline{\zeta}$ as
\begin{equation}
d\zeta=dq+i dp, \hspace{1cm}d\overline{\zeta}=dq-i dp,
\label{angular4}
\end{equation}
then
\begin{equation}
d\zeta d\overline{\zeta}=dq^2+dp^2.
\label{angular5}
\end{equation}
For this reason, the element of line \eqref{angular3} can be written as \cite{GLPW97},
\begin{equation}
d\theta^2 + \sin^2\theta d\phi^2=\dfrac{4}{\left(1+\zeta\overline{\zeta}\right)^2}\left(dq^2+dp^2\right).
\label{angular6}
\end{equation}

Now, it is considered that the metric \eqref{angular6} can be decomposed in terms of a new complex vector field $q_A$ \cite{NP66,GMNRS66} as follows
\begin{equation}
q_{AB}=q_{(A}\overline{q}_{B)}.
\label{decomposition1}
\end{equation}
These vectors are related to the tangent vectors to the unit sphere along the coordinate lines. These two vector fields, $q_A$ and $\overline{q}_A$, that allow to decompose the unit sphere metric, are known as dyads and it is said that the metric is written in terms of dyadic products. The metric and its inverse are related as
\begin{equation}
q_{AB}q^{BC}=\delta_A^{~~C},
\label{decomposition2}
\end{equation}
then,  in terms of these dyads one obtains
\begin{align}
\delta_A^{~~C} &=q_{(A}\overline{q}_{B)}q^{(B}\overline{q}^{C)} .
\label{decomposition3}
\end{align}
Imposing that
\begin{equation}
\overline{q}_{B}q^{B}=2, \hspace{1cm}\overline{q}_{B}\overline{q}^{B}=0,
\label{decomposition4}
\end{equation}
the expression \eqref{decomposition3} is reduced to
\begin{align}
\delta_A^{~~C}&=q_{(A}\overline{q}^{C)}. 
\label{decomposition5}
\end{align}
From \eqref{angular6} and \eqref{decomposition1}
\begin{equation}
\dfrac{4}{\left(1+\zeta\overline{\zeta}\right)^2}\delta_{AB}=q_{(A}\overline{q}_{B)},
\label{decomposition7}
\end{equation}
one obtains
\begin{equation*}
|q_{3}|^2=\dfrac{4}{\left(1+\zeta\overline{\zeta}\right)^2}\hspace{0.5cm}\text{and} \hspace{0.5cm}|q_{4}|^2=\dfrac{4}{\left(1+\zeta\overline{\zeta}\right)^2}.
\end{equation*}
Thus, it is possible to make the choice
\begin{equation*}
q_{3}=\dfrac{2}{\left(1+\zeta\overline{\zeta}\right)}\hspace{0.5cm}\text{and} \hspace{0.5cm}q_{4}=\dfrac{2i}{\left(1+\zeta\overline{\zeta}\right)}.
\end{equation*}
For this reason, the complex vectors $q_A$ can be written \cite{NP66,GMNRS66} as,
\begin{equation}
q_{A}=\dfrac{2}{\left(1+\zeta\overline{\zeta}\right)}\left(\delta^3_{~A}+i\delta^4_{~A}\right)\hspace{0.5cm}\text{and} \hspace{0.5cm} \overline{q}_{A}=\dfrac{2}{\left(1+\zeta\overline{\zeta}\right)}\left(\delta^3_{~A}-i\delta^4_{~A}\right).
\label{decomposition8}
\end{equation}
Raising the index of $q_A$ with the metric $q^{AB}$ one obtains
\begin{equation}
q^{A}=\dfrac{\left(1+\zeta\overline{\zeta}\right)}{2}\left(\delta^A_{~~3}+i\delta^A_{~~4}\right)\hspace{0.5cm}\text{and}\hspace{0.5cm}
\overline{q}^{A}=\dfrac{\left(1+\zeta\overline{\zeta}\right)}{2}\left(\delta^A_{~~3}-i\delta^A_{~~4}\right).
\label{decomposition10}
\end{equation}
If the spherical coordinates are used, then \eqref{decomposition1} can be written as
\begin{equation}
\begin{pmatrix}
1 & 0 \\
0 & \sin^2\theta
\end{pmatrix}=\begin{pmatrix}
q_3\overline{q}_3 & q_3\overline{q}_4 + \overline{q}_3q_4 \\
q_3\overline{q}_4 + \overline{q}_3q_4 & q_4\overline{q_4}
\end{pmatrix},
\label{eth_spherical_1}
\end{equation}
which implies that the spherical dyads \cite{T07} take the form
\begin{subequations}
\begin{align}
q_A=\delta^{~~3}_A+i\sin\theta\delta^{~~4}_A, \hspace{1cm} \overline{q}_A=\delta^{~~3}_A-i\sin\theta\delta^{~~4}_A,\\
q^A=\delta^A_{~~3}+i\csc\theta\delta^A_{~~4}, \hspace{1cm} \overline{q}^A=\delta^A_{~~3}-i\csc\theta\delta^A_{~~4}. 
\label{eth_spherical_2}
\end{align}
\end{subequations} 
\section{Transformation Rules for Vectors and One-forms}
In order to establish the transformation rules for the dyads, it is necessary to understand how the differential operators transform between one map to another. Thus, as $q_N:=q_N(q_S,p_S)$ and $p_N:=p_N(q_S,p_S)$ as shown explicitly in \eqref{stereo11}, then the one-forms $\partial_{q_N}$ and $\partial_{q_S}$ transform as
\begin{subequations}
\begin{align}
\partial_{q_N}&=(\partial_{q_N}q_S)\partial_{q_S}+(\partial_{q_N}p_S)\partial_{p_S},\label{diff3.1}\\
\partial_{p_N}&=(\partial_{p_N}q_S)\partial_{q_S}+(\partial_{p_N}p_S)\partial_{p_S}.\label{diff3.2}
\end{align}
\label{diff3}%
\end{subequations}
Computing each coefficient in Equations \eqref{diff3}, one obtains
\begin{align*}
\partial_{q_N} q_S=\dfrac{p_N^2-q_N^2}{(q_N^2+p_N^2)^2},\hspace{1cm} \partial_{q_N} p_S=\dfrac{2q_Np_N}{(q_N^2+p_N^2)^2},\\
\partial_{p_N} q_S=-\dfrac{2q_Np_N}{(q_N^2+p_N^2)^2},\hspace{1cm} \partial_{p_N} p_S=\dfrac{p_N^2-q_N^2}{(q_N^2+p_N^2)^2}.
\end{align*}
It means that the differential operators \eqref{diff3} become
\begin{subequations}
\begin{align}
\partial_{q_N}&=\dfrac{1}{\left(q_N^2+p_N^2\right)^2}\left(\left(p_N^2-q_N^2\right)\partial_{q_S}+2q_Np_N\partial_{p_S}\right), \label{diff4.1}\\
\partial_{p_N}&=\dfrac{1}{\left(q_N^2+p_N^2\right)^2}\left(-2q_Np_N\partial_{q_S} +\left(p_N^2-q_N^2\right)\partial_{p_S}\right).
\label{diff4.2}
\end{align}
\label{diff4}%
\end{subequations}
Now, the transformation rule for the vectors will be examined 
\begin{equation}
dx^A_N=\partial_{x^B_S} x^A_Ndx^B_S.
\end{equation}
Specifically, the transformation rules for the vectors $dq$ and $dp$ are given by
\begin{subequations}
\begin{align}
dq_N&=\partial_{q_{S}} q_{N} dq_S+ \partial_{p_{S}} q_{N} dp_S,\label{diff5.1}\\
dp_N&=\partial_{q_{S}} p_{N} dq_S+ \partial_{p_{S}} p_{N} dp_S.\label{diff5.2}
\end{align}
\label{diff5}%
\end{subequations}
Here, it is important to point out that the equations \eqref{stereo11} are symmetrical  with respect to the interchange of indices $N$ and $S$, i.e., the same expressions are obtained if $q_S$ and $p_S$ are considered as functions of $q_N$ and $p_S$, therefore
\begin{align*}
\partial_{q_S} q_N=\dfrac{p_S^2-q_S^2}{(q_S^2+p_S^2)^2},\hspace{1cm} \partial_{q_S} p_N=\dfrac{2q_Sp_S}{(q_S^2+p_S^2)^2},\\
\partial_{p_S} q_N=-\dfrac{2q_Sp_S}{(q_S^2+p_S^2)^2},\hspace{1cm} \partial_{p_S} p_N=\dfrac{p_S^2-q_S^2}{(q_S^2+p_S^2)^2}.
\end{align*}
Then, the vectors \eqref{diff5} transform as
\begin{subequations}
\begin{align}
dq_N&=\dfrac{1}{\left(q_S^2+p_S^2\right)^2}\left(\left(p_S^2-q_S^2\right) dq_S - 2q_Sp_S dp_S\right),
\label{diff6.1}\\
dp_N&=\dfrac{1}{\left(q_S^2+p_S^2\right)^2}\left(2q_Sp_S dq_S+ \left(p_S^2-q_S^2\right) dp_S\right).
\label{diff6.2}
\end{align}
\label{diff6}%
\end{subequations}
Notice that, by virtue of the interchangeability of the indices in \eqref{stereo11}, the relations \eqref{diff4} and \eqref{diff6} for one-forms and vectors are symmetrical with respect to the interchange of the indices $N$ and $S$. Therefore the same rules are applied to construct the inverse transformation from north to south. 
\section{Transformation Rules for the Dyads and Spin-weight}
Any vector field $\boldsymbol{v}$ can be expanded in terms of a basis of one-forms $\boldsymbol{e}_A$, namely $\boldsymbol{v}=v^A\boldsymbol{e}_A$. Thus, for each hemisphere
\begin{equation}
\boldsymbol{v}_N=v^A_N\boldsymbol{e}_{A_N}\hspace{0.5cm}\text{and} \hspace{0.5cm}\boldsymbol{v}_S=v^A_S\boldsymbol{e}_{A_S}.
\label{spin1}
\end{equation}
In particular for a local coordinate basis $\{\partial_{A_N}\}$ and $\{\partial_{A_S}\}$, the complex vectors $\boldsymbol{q}_N$ and $\boldsymbol{q}_S$ can be expressed as the linear combinations, i.e.,
\begin{equation}
\boldsymbol{q}_N=q^A_N\partial_{A_N}\hspace{0.5cm}\text{and} \hspace{0.5cm} \boldsymbol{q}_S=q^A_S\partial_{A_S}.
\label{spin2}
\end{equation}
Using the explicit expression for the dyads components $q_N^A$ given in \eqref{decomposition10}, \eqref{spin2} take the explicit form
\begin{align}
\boldsymbol{q}_N&=\dfrac{\left(1+\zeta_N\overline{\zeta}_N\right)}{2}\left(\partial_{q_N}+i\partial_{p_N}\right).
\label{spin3}
\end{align}
Then, transforming the basis in \eqref{spin3}, using for this \eqref{diff4}, one obtains
\begin{align}
\boldsymbol{q}_N 
=&-\dfrac{\overline{\zeta}_S}{\zeta_S}\boldsymbol{q}_S,
\label{spin4}
\end{align}
which is the transformation rule for the dyads. It is worth stressing that, apparently, this transformation appears as induced by the stereographic mapping used to make the finite coverage to the unit sphere. However, it is a vector property that appears by the fact that the atlas is constructed from two local charts, whose centres are diametrically opposed. This result can be written in terms of components, as
\begin{equation}
q^A_N=e^{i \alpha}q^A_S,
\label{spin5}
\end{equation}
where the complex factor
\begin{equation}
e^{i \alpha}=-\dfrac{\overline{\zeta}_S}{\zeta_S},
\label{spin6}
\end{equation}
is the spin-weight associated with the transformation of coordinates \cite{NP66,GMNRS66,stewart93}.

From \eqref{spin5} it is obtained immediately the rule for the complex conjugate dyads components, namely
\begin{equation}
\overline{q}^A_N=e^{-i \alpha}\overline{q}^A_S.
\label{spin7}
\end{equation}
In order to complete this description, it is necessary to examine the transformation rules of the covariant components of the dyads. Thus, expressing the dyads as linear combinations of the vectors
 \begin{align}
\boldsymbol{q}_N=q_{A_N}\boldsymbol{e}^{A_N}\hspace{0.5cm}\text{and} \hspace{0.5cm}\boldsymbol{q}_S=q_{A_S}\boldsymbol{e}^{A_S}.
\label{spin8}
 \end{align}
Then, using a local coordinate basis, one has
 \begin{align}
\boldsymbol{q}_N &= \dfrac{2}{\left(1+\zeta_N\overline{\zeta}_N\right)}\left(dq_N+idp_N\right).
\label{spin9}
 \end{align}
Using the transformation rules \eqref{diff6}, one obtains
 \begin{align*}
\boldsymbol{q}_N&=-\dfrac{\overline{\zeta}_S}{\zeta_S}\boldsymbol{q}_S.
 \end{align*} 
This shows that these transformation rules are completely consistent. Thus, it allows one to lower the index with the unit sphere metric, i.e. from \eqref{spin5} and \eqref{spin7}, one has
\begin{equation}
q_{A_N}=e^{i\alpha}q_{A_S}\hspace{0.5cm}\text{and} \hspace{0.5cm} \overline{q}_{A_N}=e^{-i\alpha}\overline{q}_{A_S}.
\label{spin10}
\end{equation} 
It is worth mentioning that the unit sphere metric \eqref{decomposition1} has spin-weight zero, namely
\begin{align*}
{q_N}_{(A}{\overline{q}_N}_{B)}&=e^{i\alpha}{q_S}_{(A}e^{-i\alpha}{\overline{q}_S}_{B)}\\
&={q_S}_{(A}{\overline{q}_S}_{B)}.
\end{align*}
The spin-weight of a finite product of these tangent vectors depends on the number of $q_A$, $\overline{q}_A$, $q^A$ and $\overline{q}^A$ considered. For example, if the product $\prod_{i=1}^n {q}_{Ai}$ of tangent vectors is considered, then its transformation from north to south hemisphere is given by
\begin{align}
\prod_{i=1}^n {q_N}_{Ai} 
&=\left(e^{i\alpha}\right)^n\prod_{i=1}^n {q_S}_{Ai},
\label{spin11}
\end{align}
which implies that this product has a spin-weight of $s=n$. As another example, if the product $\prod_{i=1}^n {q}_{Ai} \prod_{j=n+1}^m {\overline{q}}_{Aj}$ is considered, then it transforms as
\begin{align}
\prod_{i=1}^n {q_N}_{Ai} \prod_{j=n+1}^m {\overline{q}_N}_{Aj}
&=\left(e^{i \alpha}\right)^{(2n-m)}\prod_{i=1}^n {q_S}_{Ai} \prod_{j=n+1}^m {\overline{q}_S}_{Aj}.
\label{spin12}
\end{align}
which means that its spin-weight is $s=2n-m$. Therefore, if scalar quantities involving products like those given above are considered, then these scalars must have spin-weight induced by these products. Thus, the scalar functions constructed through the projection of the tensors onto these dyads, inherits the spin-weight carried by these dyads. This crucial point will be clarified in the next section.
\section{Spin-weighted Scalars and Spin-weight}
Here, we will show that any tensor field of rank $2$ of type $(0,2)$, namely $\omega_{AB}$, in the tangent space of the unit sphere admits a irreducible decomposition in spin-weighted functions \cite{stewart93,GLPW97}. In order to show that, it is first considered that $\omega_{AB}$ can be decomposed into its symmetric and anti-symmetric part, i.e.,
\begin{equation}
\omega_{AB}=\omega_{(AB)}+ \omega_{[AB]}.
\label{spinscalars1}
\end{equation}
The symmetric part can be separated in two parts, one trace-free and other corresponding to its trace 
\begin{equation}
\omega_{(AB)}=t_{AB}+\dfrac{q_{AB}}{2}\omega,
\label{spinscalars2}
\end{equation}
where $t=q^{AB}t_{AB}=0$, i.e., $t_{AB}$ is the trace-free symmetric part of $\omega_{AB}$, and the second term is its trace, i.e., 
\begin{align}
\omega&=\omega_{AB}q^{AB}\nonumber\\
&=2\omega_{AB}q^{(A}\overline{q}^{B)}. 
\label{spinscalars3}
\end{align}
Thus, $\omega_{AB}$ can be written as
\begin{equation}
\omega_{AB}=t_{AB}+\dfrac{q_{AB}}{2}\omega+ \omega_{[AB]}.
\label{spinscalars4}
\end{equation}
The anti-symmetric part can be expressed as
\begin{align*}
\omega_{[AB]}&=\omega_{CD}\delta^C_{~~[A}\delta^D_{~~B]}
\end{align*}
where using \eqref{decomposition5}
\begin{align*}
\omega_{[AB]}
&=\dfrac{\omega_{CD}}{2}\left(\overline{q}^{(C}q_{A)}\overline{q}^{(D}q_{B)}-\overline{q}^{(C}q_{B)}\overline{q}^{(D}q_{A)}\right)\nonumber\\
&=\dfrac{1}{4}\dfrac{\omega_{CD}}{2}\Bigg(\left(\overline{q}_{A}q_{B}-q_{A}\overline{q}_{B}\right)\left(q^{C}\overline{q}^{D}-\overline{q}^{C}q^{D}\right)\Bigg),\nonumber
\label{spinscalars5}
\end{align*}
i.e.,
\begin{equation}
\omega_{[AB]}=\dfrac{1}{2}\overline{q}_{[A}q_{B]}u,
\label{spinscalars6}
\end{equation}
where
\begin{equation}
u=\omega_{CD}q^{[C}\overline{q}^{D]}.
\label{spinscalars7}
\end{equation}
For this reason, \eqref{spinscalars4} can be written as
\begin{equation}
\omega_{AB}=t_{AB}+\dfrac{\omega}{2}q_{AB}+ \dfrac{1}{2}\overline{q}_{[A}q_{B]}u.
\label{spinscalars8}
\end{equation}
Here, it is important to notice that $\omega$ and $u$ are scalar functions with spin weight zero, as given in \eqref{spinscalars3} and \eqref{spinscalars7} respectively. The symmetric traceless part, $t_{AB}$ admits a irreducible decomposition in scalar spin-weighted functions as follow
\begin{equation}
I=t_{AB}q^Aq^B, \hspace{1cm} L=t_{AB}\overline{q}^A\overline{q}^B, \hspace{1cm} S=t_{AB}\overline{q}^Aq^B,
\label{spinscalars9}
\end{equation}
where, if it is considered that $t_{AB} \in \mathbb{R}$, then $L=\overline{I}$, and $S=\overline{S}$.
\\Consequently, any tensor field of type $(0,2)$, say $\omega_{AB}$, is completely determined by a linear combination of spin-weighted scalar fields of weight $0$, $2$ and $-2$ \cite{GMNRS66,stewart93}. In general, it is possible to construct spin-weighted scalars from tensor fields into the tangent space to the unitary sphere, in the form
\begin{equation}
_s\Psi=\prod_{i=1}^{n}q_{A_i} \prod_{j=n+1}^{r}\overline{q}_{A_{j}}\prod_{k=1}^{m}q^{B_k}\prod_{l=m+1}^{\tilde{s}}\overline{q}^{B_l} \Psi^{A_1\cdots A_n A_{n+1}\cdots A_r}_{\hspace{2.3cm}B_1\cdots B_m B_{m+1}\cdots B_{\tilde{s}}}.
\label{spinscalars10}
\end{equation}
Then, it is possible to compute the spin-weight of $_s\Psi$ taking advantage of \eqref{spin12}. Considering the expression \eqref{spinscalars10} for the north or south hemisphere, and making the transformation from one region to another, one obtains that the spin-weight for the $_s\Psi$ function is,
\begin{equation}
s=2(n+m)-r-\tilde{s}.
\label{spinscalars11}
\end{equation}
\section{Raising and Lowering Operators}
Here it will be shown the action of the differential operators induced by the projection of the covariant derivative of the tensor field defined in \eqref{spinscalars10}. In order to do this, it is useful to compact the notation in the form
\begin{equation}
\tilde{\Lambda}_{\tilde{A}_{ab}}=\prod\limits_{i=a}^{b}\Lambda_{A_i},\hspace{1cm}\tilde{\Lambda}^{\tilde{A}_{ab}}=\prod\limits_{i=a}^{b}\Lambda^{A_i},
\label{ladder1}
\end{equation}
and for the tensor field
\begin{equation}
\Psi^{\tilde{B}_{1m}}_{\hspace{0.6cm}\tilde{A}_{1n}}=\Psi^{B_{1}\cdots B_{m}}_{\hspace{1.1cm}A_{1}\cdots A_{n}}.
\label{ladder2}
\end{equation}
Thus, \eqref{spinscalars10} will be written as
\begin{equation}
_s\Psi=\tilde{\Lambda}_{\tilde{B}_{1m}} \tilde{\Lambda}^{\tilde{A}_{1n}}\Psi^{\tilde{B}_{1m}}_{\hspace{0.6cm}\tilde{A}_{1n}}.
\label{ladder3}
\end{equation}
The {\it eth} operator $\eth$ is defined through the projection of the covariant derivative of $\Psi^{B_1\cdots B_m}_{\hspace{1.1cm}A_1\cdots A_n}$ associated with $q_{AB}$ noted by
\begin{equation}
\triangle_A\Psi^{B_1\cdots B_m}_{\hspace{1.1cm}A_1\cdots A_n}=\Psi^{B_1\cdots B_m}_{\hspace{1.1cm}A_1\cdots A_n|A},
\end{equation} 
onto the dyads $\boldsymbol{q}$, i.e.,
\begin{align}
\eth \ _s\Psi&=q^D\ _s\Psi_{|D}\nonumber\\
&=q^D\tilde{\Lambda}_{\tilde{B}_{1m}}\tilde{\Lambda}^{\tilde{A}_{1n}}  \Psi^{\tilde{B}_{1m}}_{\hspace{0.65cm}\tilde{A}_{1n}|D},
\label{ladder4}
\end{align}
where the symbols $\Lambda_{B_i}$ and the $\Lambda^{A_j}$ are defined as
\begin{subequations}
	\begin{equation}
	\Lambda_{B_i}=\begin{cases}
	q_{B_i} & \text{if} \hspace{0.5cm} i\le x\\
	\overline{q}_{B_i} & \text{if} \hspace{0.5cm} i> x
	\end{cases},
	\label{ethgen2.1}
	\end{equation}
	and
	\begin{equation}
	\Lambda^{A_j}=\begin{cases}
	q^{A_j} & \text{if}\hspace{0.5cm} j\le y\\
	\overline{q}^{A_j} & \text{if} \hspace{0.5cm} j> y\\
	\end{cases},
	\label{ethgen2.2}
	\end{equation}
	\label{ethgen2}%
\end{subequations}
for $1\le x\le m$ and $1\le y \le n$. In this case, the spin-weight of this function, in agreement with \eqref{spinscalars11}, will be
\begin{equation}
s=2(x+y)-(m+n).
\end{equation}
On the other hand, the {\it eth bar} operator is defined as
\begin{equation}
\overline{\eth} \ _s\Psi=\overline{q}^D \tilde{\Lambda}_{\tilde{B}_{1m}} \tilde{\Lambda}^{\tilde{A}_{1n}}  \Psi^{\tilde{B}_{1m}}_{\hspace{0.6cm}\tilde{A}_{1n}|D}.
\label{ethgen18_v2}
\end{equation}
After some algebra, it is shown that the $\eth$ and $\overline{\eth}$ operators acting on a spin-weighted function $_s\Psi$ can be expressed as
\begin{equation}
{\eth} \ _s\Psi=q^D\partial_D \ _s\Psi +s\Omega\ _s\Psi\hspace{0.5cm} \text{and} \hspace{0.5cm} 
\overline{\eth} \ _s\Psi=\overline q^D\partial_D \ _s\Psi -s\overline \Omega\ _s\Psi
\label{ethgen19}
\end{equation}
(see \href{apendiceA}{Appendix A} for further details).

It is worth stressing that from \eqref{ethgen19} the $\eth$ and $\overline{\eth}$ operators can be written in general \cite{NP66,GMNRS66,GLPW97} as
\begin{align}
\eth = q^D\partial_D + s\, \Omega ,\hspace{0.5cm} \overline{\eth} = \overline{q}^D\partial_D - s\, \overline{\Omega}.
\label{ethgen251}
\end{align}
where $\Omega$ is defined from \eqref{ethgen11}, i.e.
\begin{equation}
\Omega=\frac{1}{2}q^A\overline{q}^Bq_{AB}.
\label{ethgen26}
\end{equation}
Note that, \eqref{ethgen19} allows to operate directly on the spin-weighted functions. Furthermore, they put in evidence their character to raise and lower the spin-weight of the function $_s\Psi$. Under a transformation of coordinates between north and south hemispheres, one has
\begin{equation}
\left(\eth \ _s\Psi\right)_N=e^{i\alpha(s+1)}\left(\eth \ _s\Psi\right)_S,\hspace{0.5cm}\text{and}\hspace{0.5cm}\left(\overline{\eth} \ _s\Psi\right)_N=e^{i\alpha(s-1)}\left(\overline{\eth} \ _s\Psi\right)_S.
\end{equation}
Despite using the stereographic coordinates in each chart, this property does not depend on the coordinates chosen to be used in each coordinate map. The last equations show that $\eth\ _{s}\Psi$ and $\overline{\eth}\ _{s}\Psi$ are functions with $s+1$ and $s-1$ spin-weight, then
\begin{equation}
\eth\ _s\Psi=A_{s+1}\ _{s+1}\Psi\hspace{0.5cm}\text{and} \hspace{0.5cm} \overline{\eth}\ _s\Psi=A_{s-1}\ _{s-1}\Psi,
\label{ethgen27}
\end{equation}
where $A_{s+1}$ and $A_{s-1}$ are multiplicative constants. 

The explicit forms of the $\eth$ and $\bar{\eth}$ operators in spherical coordinates \cite{T07} read
\begin{equation}
\eth= \partial_\theta + i\csc\theta\partial_\phi - s\cot\theta \hspace{0.5cm}\text{and} \hspace{0.5cm} \overline{\eth}= \partial_\theta - i\csc\theta\partial_\phi + s\cot\theta,
\label{eth_spherical_7}
\end{equation}
where \eqref{eth_spherical_1} and \eqref{eth_spherical_2} were used. Using these last equations we found that \eqref{ethgen26} results in
\begin{equation}
\Omega=-\cot\theta.
\label{ethgen28}
\end{equation}
\section{Transforming the Coordinate Basis}
Here we will show the explicit form of the $\partial_{q}$, $\partial_{p}$ and $\partial_{qp}$ operators in terms of the $\eth$, $\overline{\eth}$ operators and its commutator $[\eth,\overline{\eth}]$. Also, we will show that the commutator $[\eth,\overline{\eth}]$ satisfies an eigenvalue equation, fixing the algebra for the eth operators. 

Developing explicitly \eqref{ethgen19} and substituting the tangent vector components \eqref{decomposition10}, one has
\begin{align}
\eth\ _s\Psi &=\dfrac{1+\zeta\overline{\zeta}}{2}\left(\ _s\Psi_{,q}+i\ _s\Psi_{,p}\right)+s\zeta\ _s\Psi.
\label{abase1}
\end{align}
and
\begin{align}
\overline{\eth}\ _s\Psi&= \dfrac{1+\zeta\overline{\zeta}}{2} \left(\ _s\Psi_{,q}-i \ _s\Psi_{,p}\right)-s\overline{\zeta}\ _s\Psi.
\label{abase2}
\end{align}
Then, from \eqref{abase1} and \eqref{abase2}, one obtains 
\begin{subequations}
\begin{align}
_s\Psi_{,q}=\dfrac{\eth\ _s\Psi+\overline{\eth}\ _s\Psi -s(\zeta-\overline{\zeta}) \ _s\Psi}{1+\zeta\overline{\zeta}}
\label{abase3.1},\\
_s\Psi_{,p}=i\dfrac{\overline{\eth}\ _s\Psi - \eth\ _s\Psi+ s(\zeta+\overline{\zeta}) \ _s\Psi}{1+\zeta\overline{\zeta}}
\label{abase3.2},
\end{align}
\label{abase3}%
\end{subequations}
which written in terms of $q$ and $p$ result in
\begin{subequations}
\begin{align}
_s\Psi_{,q}=\dfrac{\eth\ _s\Psi+\overline{\eth}\ _s\Psi -2isp \ _s\Psi}{1+q^2+p^2}
\label{abase4.1},\\
_s\Psi_{,p}=i\dfrac{\overline{\eth}\ _s\Psi - \eth\ _s\Psi+ 2sq \ _s\Psi}{1+q^2+p^2}
\label{abase4.2}.
\end{align}
\label{abase4}%
\end{subequations}
Thus, the base vectors (or conversely the differential operators) $\partial_q$ and $\partial_p$ can be written as
\begin{subequations}
\begin{align}
\partial_{q}&=\dfrac{1}{1+\zeta\overline{\zeta}}\left(\eth +\overline{\eth} -s(\zeta-\overline{\zeta}) \right), \label{abase5.1}\\
\partial_{p}&=\dfrac{i}{1+\zeta\overline{\zeta}}\left(\overline{\eth}  - \eth + s(\zeta+\overline{\zeta})  \right).\label{abase5.2}
\end{align}
\label{abase5}%
\end{subequations}
It is worth stressing that, in these expressions appear the spin-weight $s$ associated with the functions. Consequently, these operators must be applied carefully in future computations, in order to avoid errors.  

From \eqref{abase1}, \eqref{abase2} and \eqref{abase3} it is possible to obtain immediately the expressions for $\partial_{qq}$, $\partial_{qp}$ and $\partial_{pp}$. Here, we will start with $\partial_{qq}$. There are at least two forms to do it. Here, we follow two ways with the aim to check the resulting expressions. First, the action of the derivative with respect to $q$ on $ _s\Psi_{,q}$ will be considered. Thus, using \eqref{abase3.1} one obtains
\begin{align}
	\partial_{qq}
		=&\left(\dfrac{1}{1+\zeta\overline{\zeta}}\right)_{,q}\left(\eth +\overline{\eth}  -s(\zeta-\overline{\zeta}) \right) 
	+\left(\dfrac{1}{1+\zeta\overline{\zeta}}\right)\partial_q\left(\eth +\overline{\eth}  -s(\zeta-\overline{\zeta}) \right),
	\label{abase6}
\end{align}
where 
\begin{align}
\left(\dfrac{1}{1+\zeta\overline{\zeta}}\right)_{,q}=-\dfrac{\zeta+\overline{\zeta}}{\left(1+\zeta\overline{\zeta}\right)^2},
\label{abase7}
\end{align}
and
\begin{align}
	\partial_q\left(\eth +\overline{\eth}  -s(\zeta-\overline{\zeta}) \right)
	&=\partial_q\eth+\partial_q\overline{\eth} -s (\zeta-\overline{\zeta})\partial_{q},
	\label{abase8}
\end{align}
because $(\zeta-\overline{\zeta})_{,q}=0$. The first term in \eqref{abase6} is given by
\begin{align}
	\left(\dfrac{1}{1+\zeta\overline{\zeta}}\right)_{,q}\left(\eth +\overline{\eth} -s(\zeta-\overline{\zeta}) \right) 	=-\dfrac{\left(\zeta+\overline{\zeta}\right)\eth+\left(\zeta+\overline{\zeta}\right)\overline{\eth}
		-s\left(\zeta^2-\overline{\zeta}^2\right) }{\left(1+\zeta\overline{\zeta}\right)^2}.
	\label{abase9}
\end{align}
Each derivative in \eqref{abase8} is computed considering \eqref{ethgen27} and \eqref{abase3.1}. Then
\begin{subequations}
\begin{align}
\partial_q\eth
	&=\dfrac{\eth^2 +\overline{\eth}\eth  -(s+1)(\zeta-\overline{\zeta}) \eth }{1+\zeta\overline{\zeta}},
	\label{abase10.1}
\end{align}
and
\begin{align}
	\partial_q \overline{\eth}
		&=\dfrac{\eth\overline{\eth} +\overline{\eth}^2  -(s-1)(\zeta-\overline{\zeta}) \overline{\eth}  }{1+\zeta\overline{\zeta}}.
	\label{abase10.2}
\end{align}
\label{abase10}%
\end{subequations}
Thus, substituting the relations \eqref{abase10} into \eqref{abase8} one obtains
\begin{align}
\partial_q\left(\eth +\overline{\eth}  -s(\zeta-\overline{\zeta}) \right) 
= &
\dfrac{1}{1+\zeta\overline{\zeta}}\left(\eth^2  + \overline{\eth}^2+ (\overline{\eth},\eth)  -(2s+1)(\zeta-\overline{\zeta}) \eth  \right. \nonumber\\
&
\left.-(2s-1)(\zeta-\overline{\zeta}) \overline{\eth}  + s^2(\zeta-\overline{\zeta})^2 \right),
\label{abase11}
\end{align}
where we used the anti-commutator
\begin{equation}
\left(\overline{\eth},\eth\right)\ _s\Psi=\overline{\eth} \eth\ _s\Psi+ \eth\overline{\eth} \ _s\Psi.
\label{abase12}
\end{equation}
Then, the second order differential operator $\partial_{qq}$ can be written as
\begin{align}
\partial_{qq}
&=\dfrac{1}{\left(1+\zeta\overline{\zeta}\right)^2}\Bigg(\eth^2\  + \overline{\eth}^2 + (\overline{\eth},\eth) +2\left(s\overline{\zeta}-(s+1)\zeta\right) \eth    \nonumber\\
&-2\left(s\zeta-(s-1)\overline{\zeta}\right)\overline{\eth}+ s\left(s(\zeta-\overline{\zeta})^2   +\left(\zeta^2-\overline{\zeta}^2\right) \right)\Bigg).
\label{abase13}
\end{align}
After that,  $\partial_{pp}$ is computed using \eqref{abase3.2}, thus
\begin{align}
	\partial_{pp}
	= 
	i\left(\left(\dfrac{1}{1+\zeta\overline{\zeta}}\right)_{,p}(\overline{\eth} -\eth +s(\zeta+\overline{\zeta}) )\right. 
	\left.+\left(\dfrac{1}{1+\zeta\overline{\zeta}}\right)\partial_p(\overline{\eth}-\eth +s(\zeta+\overline{\zeta}))\right),
	\label{abase14}
\end{align}
where
\begin{equation}
\left(\dfrac{1}{1+\zeta\overline{\zeta}}\right)_{,p}=\dfrac{i(\zeta-\overline{\zeta})}{\left(1+\zeta\overline{\zeta}\right)^2}.
\label{abase15}
\end{equation}
Then the first term in \eqref{abase14} is given by
\begin{align}
&\left(\dfrac{1}{1+\zeta\overline{\zeta}}\right)_{,p}(\overline{\eth} -\eth  +s(\zeta+\overline{\zeta}) )
= 
i\dfrac{(\zeta-\overline{\zeta})\overline{\eth} -(\zeta-\overline{\zeta})\eth\   +s(\zeta^2-\overline{\zeta}^2) }{\left(1+\zeta\overline{\zeta}\right)^2}.
\label{abase16}
\end{align}
The second term can be spanned as
\begin{align}
	\partial_p(\overline{\eth} -\eth  +s(\zeta+\overline{\zeta}) )
	&=\partial_p\overline{\eth}-\partial_p\eth +s(\zeta+\overline{\zeta})\partial_p,
	\label{abase17}
\end{align}
where it is considered that $(\zeta+\overline{\zeta})_{,p}=0$.

Each term in the last equation can be computed by using \eqref{ethgen27} and \eqref{abase3.2}, thus 
\begin{subequations}
\begin{align}
	\partial_p\overline{\eth}
	&=i\dfrac{\overline{\eth}^2 -\eth\overline{\eth}  +(s-1)(\zeta+\overline{\zeta})\overline{\eth} }{1+\zeta\overline{\zeta}},
	\label{abase18.1}
\end{align}
and
\begin{align}
	\partial_p \eth
	&=i\dfrac{\overline{\eth}\eth -\eth^2  +(s+1)(\zeta+\overline{\zeta})\eth }{1+\zeta\overline{\zeta}}.
	\label{abase18.2}
\end{align}
\label{abase18}%
\end{subequations}
The substitution of \eqref{abase18} into \eqref{abase17} yields
\begin{align}
	\partial_p(\overline{\eth} -\eth  +s(\zeta+\overline{\zeta}) )
	= &
	\dfrac{i}{1+\zeta\overline{\zeta}}\Bigg(\eth^2  + \overline{\eth}^2 -(\eth,\overline{\eth}) +(s-1)(\zeta+\overline{\zeta})\overline{\eth}    \nonumber\\
	&-(s+1)(\zeta+\overline{\zeta})\eth + s(\zeta+\overline{\zeta})\overline{\eth} -s(\zeta+\overline{\zeta}){\eth} +s^2(\zeta+\overline{\zeta})^2 \Bigg).
	\label{abase19}
\end{align}
Then, substituting \eqref{abase16} and \eqref{abase19} into \eqref{abase14} one obtains
 the second order operator $\partial_{pp}$, which is given by
\begin{align}
\partial_{pp}
&=-\dfrac{1}{\left(1+\zeta\overline{\zeta}\right)^2}\Bigg( \eth^2 + \overline{\eth}^2 - (\eth,\overline{\eth}) +2(s\zeta +(s-1)\overline{\zeta})\overline{\eth}   
\nonumber\\
& -2(s\overline{\zeta}+(s+1)\zeta)\eth  +s\left(s\left(\zeta+\overline{\zeta}\right)^2  +\left(\zeta^2-\overline{\zeta}^2\right)\right)\Bigg).
\label{abase20}
\end{align}
Now, we compute the mixed operator $\partial_{qp}$ by means of \eqref{ethgen27} and \eqref{abase3}, i.e.,
\begin{align}
	\partial_{qp} 
	=&\left(\dfrac{1}{1+\zeta\overline{\zeta}}\right)_{,p}\left(\eth +\overline{\eth}  -s(\zeta-\overline{\zeta})  \right) 
	+\dfrac{1}{1+\zeta\overline{\zeta}}\partial_p\left(\eth +\overline{\eth}  -s(\zeta-\overline{\zeta}) \right).
	\label{abase21}
\end{align}
The first term in the last equation is given by
\begin{align}
&\left(\dfrac{1}{1+\zeta\overline{\zeta}}\right)_{,p}\left(\eth +\overline{\eth}  -s(\zeta-\overline{\zeta}) \right) 
= 
\dfrac{i}{\left(1+\zeta\overline{\zeta}\right)^2}\left((\zeta-\overline{\zeta})\eth +(\zeta-\overline{\zeta})\overline{\eth}  -s(\zeta-\overline{\zeta})^2  \right),
\label{abase22}
\end{align}
where \eqref{abase15} has been used. The second term is computed making use of equations \eqref{abase3} and \eqref{abase18}, thus
\begin{align}	
	\dfrac{1}{1+\zeta\overline{\zeta}}\partial_p\left(\eth +\overline{\eth}  -s(\zeta-\overline{\zeta}) \right) 
	&=\dfrac{i}{\left(1+\zeta\overline{\zeta}\right)^2}\Bigg( \overline{\eth}^2 - \eth^2  + [\overline{\eth},\eth] + \left((2s+1)\zeta + \overline{\zeta}\right)\eth  \nonumber\\
	& +\left((2s-1)\overline{\zeta} -\zeta\right)\overline{\eth}  - s\left(2(1+\zeta\overline{\zeta}) + s (\zeta^2-\overline{\zeta}^2) \right)  \Bigg),
	\label{abase23}
\end{align}
where we use the commutator
\begin{equation}
\left[\overline{\eth},\eth\right]=\overline{\eth}\eth - \eth\overline{\eth}.
\label{abase24}
\end{equation}
Consequently, it is possible to write the operator $\partial_{qp}$ as
\begin{align}
\partial_{qp} 
&=\dfrac{i}{\left(1+\zeta\overline{\zeta}\right)^2}\Bigg( \overline{\eth}^2 - \eth^2 + [\overline{\eth},\eth]  + 2(s+1)\zeta \ \eth + 2(s-1)\overline{\zeta} \ \overline{\eth} \nonumber\\
&   - s\left(2 + \zeta^2 +\overline{\zeta}^2 + s (\zeta^2-\overline{\zeta}^2)   \right)  
\Bigg).
\label{abase25}
\end{align}
In order to test the consistency of this formalism, and with the goal to confirm \eqref{abase25}, we will compute the mixed operator $\partial_{pq}$, i.e., 
\begin{align}
	\partial_{pq}
	=&i\left(\left(\dfrac{1}{1+\zeta\overline{\zeta}}\right)_{,q} \left(\overline{\eth} - \eth + s(\zeta+\overline{\zeta})  \right)\right. 
	\left. + 
	\dfrac{1}{1+\zeta\overline{\zeta}} \partial_q\left(\overline{\eth} - \eth + s(\zeta+\overline{\zeta}) \right)\right).
	\label{abase26}
\end{align}
The first term in the last equation is given by
\begin{align}
	&\left(\dfrac{1}{1+\zeta\overline{\zeta}}\right)_{,q} \left(\overline{\eth}  - \eth + s(\zeta+\overline{\zeta})  \right)
	=-\dfrac{1}{\left(1+\zeta\overline{\zeta}\right)^2} \left(\left(\zeta+\overline{\zeta}\right)\overline{\eth} - \left(\zeta+\overline{\zeta}\right)\eth + s\left(\zeta+\overline{\zeta}\right)^2 \right),
	\label{abase27}
\end{align}
where \eqref{abase7}  was used. The second term in \eqref{abase26} is computed taking into account equations \eqref{abase3.1} and \eqref{abase10} 
\begin{align}
	\dfrac{1}{1+\zeta\overline{\zeta}} \partial_q\left(\overline{\eth} - \eth + s(\zeta+\overline{\zeta}) \right) 
	=&\dfrac{1}{\left(1+\zeta\overline{\zeta}\right)^2}\Bigg(\overline{\eth}^2 - \eth^2 + [\eth,\overline{\eth}]  +((2s-1)\overline{\zeta}+\zeta) \overline{\eth}   \nonumber\\
	& +((2s+1)\zeta-\overline{\zeta}) \eth +s\left(2(1+\zeta\overline{\zeta}) -s\left(\zeta^2-\overline{\zeta}^2\right)\right)  \Bigg).
	\label{abase28}
\end{align}
Then, substituting \eqref{abase27} and \eqref{abase28} into \eqref{abase26} one obtains
\begin{align}
\partial_{pq}
&= \dfrac{i}{\left(1+\zeta\overline{\zeta}\right)^2}\Bigg( \overline{\eth}^2 - \eth^2  + \left[\eth,\overline{\eth}\right]  +2(s-1)\overline{\zeta}\ \overline{\eth}  +2(s+1)\zeta \eth   \nonumber\\
& +s\left(2- \zeta^2 - \overline{\zeta}^2 -s\left(\zeta^2-\overline{\zeta}^2\right) \right) \Bigg).
\label{abase29}%
\end{align}
Now, noting that
\begin{equation}
\left[\partial_q,\partial_p\right]\ _s\Psi=0,
\label{abase30}
\end{equation}
because $_s\Psi$ is supposed to be a complex function with at least continuous second derivatives. Then,  using \eqref{abase25} and \eqref{abase29}, one has 
\begin{align*}
_s\Psi_{,qp}-\ _s\Psi_{,pq} &=\dfrac{i}{\left(1+\zeta\overline{\zeta}\right)^2}\left(\left[\overline{\eth},\eth\right]-\left[\eth,\overline{\eth}\right]-4s\right)\ _s\Psi;
\end{align*}
which implies that
\begin{align*}
\left(\left[\overline{\eth},\eth\right]-\left[\eth,\overline{\eth}\right]-4s\right)\ _s\Psi
&=0,
\end{align*}
i.e., the commutator of the $\eth$ and $\overline{\eth}$ satisfy an eigenvalue equation,
\begin{equation}
\left[\overline{\eth},\eth\right]\ _s\Psi=2s\ _s\Psi.
\label{abase31}
\end{equation}
It is worth stressing that by using \eqref{ethgen27} one obtains
\begin{align}
\left[\overline{\eth},\eth\right]\ _s\Psi&= \overline{\eth}\eth \ _s\Psi - \eth\overline{\eth} \ _s\Psi \nonumber\\
&= \overline{\eth}\left(A_{s+1} \ _{s+1}\Psi\right) - \eth \left(A_{s-1} \ _{s-1}\Psi\right) \nonumber\\
&= A_{s+1}\overline{\eth}\left(  _{s+1}\Psi\right) - A_{s-1} \eth \left(  _{s-1}\Psi\right) \nonumber\\
&= A_{s}\left(A_{s+1}    - A_{s-1} \right) \ _{s}\Psi,
\label{abase32}
\end{align}
which defines the constant of structure for the group of functions that satisfy \eqref{abase31}, i.e., 
\begin{equation}
A_{s}\left(A_{s+1} - A_{s-1} \right)=2s.
\label{abase33}
\end{equation}
Thus, the explicit form for the partial derivatives $\partial_q\ _s\Psi$ and $\partial_p \ _s\Psi$ as expressed in equations \eqref{abase3} was obtained. 
With these expressions, the explicit form for the second order operators $\partial_{qq}$, $\partial_{pp}$, $\partial_{qp}$ and $\partial_{pq}$ were expressed as in \eqref{abase13}, \eqref{abase20}, \eqref{abase25} and \eqref{abase29} respectively. However, it is important to highlight that $\partial_q$ and $\partial_p$ are commutable. With this last fact the commutation rule for $\eth$ and $\overline{\eth}$ was derived, which is given in \eqref{abase31}. The last relation is particularly important because from it, the eigenfunctions for this eigenvalue equation are constructed. 
\section{Legendrian Operator}
This section is dedicated to the treatment of the Legendrian operator and its relationship with the spherical harmonics $_0Y_{lm}$. Here this operator is expressed in terms of the raising and lowering spin-weighted operators $\eth$ and $\overline{\eth}$. 

As it is well known, the Laplace equation  
\begin{equation}
\nabla^2\Psi=0
\label{legend1}
\end{equation}
can be written as
\begin{equation}
\dfrac{1}{r}\partial_{rr}\left(r\Psi\right)+\dfrac{1}{r^2}\mathcal{L}^2\Psi=0,
\label{legend2}
\end{equation}
where the Legendrian operator $\mathcal L^2$ is given by
\begin{equation}
\mathcal L^2=\dfrac{1}{\sin\theta}\partial_\theta\left(\sin\theta\partial_\theta\right)+\dfrac{1}{\sin^2\theta}\partial_{\phi\phi}.
\label{legend3}
\end{equation}
The partial differential equation \eqref{legend1} is hyperbolic and hence their solutions can be written as
\begin{equation}
\Psi(\theta,\phi)=\dfrac{R(r)}{r}P(\theta)Q(\phi),
\label{legend4}
\end{equation}
which yields a set of ordinary differential equations for the functions $R(r)$, $P(\theta)$ and $Q(\phi)$, namely
\begin{subequations}
	\begin{align}
	&\dfrac{d^2R(r)}{dr^2}+\dfrac{l(l+1)R(r)}{r^2}=0, \label{legend5.1}\\
	&\dfrac{d^2Q(\phi)}{d\phi^2}+m^2Q(\phi)=0, \label{legend5.2}\\
	&\dfrac{1}{\sin\theta}\dfrac{d}{d\theta}\left(\sin\theta\dfrac{dP(\theta)}{d\theta}\right)+\left(l(l+1)-\dfrac{m^2}{\sin^2\theta}\right)P(\theta)=0. \label{legend5.3}
	\end{align}
	\label{legend5}%
\end{subequations}
The solutions for \eqref{legend5.3}, for any $l\in \mathbb{Z^+}$ and $m\in \mathbb{Z}$ in which   $-\left(l+1\right)\le m\le l+1$, are the associated Legendre polynomials, $P^m_{l}(x)$, which satisfy the orthogonality relation
\begin{equation}
\int_{-1}^{1}dxP^m_{l'}(x)P^m_{l}(x)=\dfrac{2}{2l+1}\dfrac{\left(l+m\right)!}{\left(l-m\right)!}\delta_{ll'}.
\label{legend6}
\end{equation}
With these polynomials and with the solution of \eqref{legend5.2}, i.e., 
\begin{equation}
Q(\phi)=e^{im\phi},
\end{equation}
a base for all angular functions are constructed. Such base is called spherical harmonics \cite{J75}, which read
\begin{equation}
Y_{lm}(\theta,\phi)=\sqrt{\dfrac{2l+1}{4\pi}\dfrac{(l-m)!}{(l+m)!}}P^m_l(\cos\theta)e^{im\phi}.
\label{legend7}
\end{equation}
Thus, particular solutions for the Laplace equation can be constructed in the following form
\begin{equation*}
\Psi_{lm}=\dfrac{R_{l}(r)}{r}Y_{lm}(\theta,\phi).
\end{equation*}
Substituting the last equation into \eqref{legend2} and using \eqref{legend5.1}, one obtains that the spherical harmonics are eigenfunctions of the Legendrian operator, corresponding to the eigenvalues $-l(l+1)$, i.e.,
\begin{equation}
\mathcal{L}^2Y_{lm}=-l(l+1)Y_{lm}.
\label{legend8}
\end{equation}
Now, it is possible to write \eqref{legend3} in the following form
\begin{align}
\mathcal L^2 
&=\dfrac{1-\tan^2\left(\theta/2\right)}{2\tan\left(\theta/2\right)}\partial_\theta +\partial_{\theta\theta}+\left(\dfrac{1}{2\tan\left(\theta/2\right)\cos^2\left(\theta/2\right)}\right)^2\partial_{\phi\phi},
\label{legend9}
\end{align}
where,
\begin{equation}
\dfrac{1-\tan^2\left(\theta/2\right)}{2\tan\left(\theta/2\right)}=\dfrac{1-\zeta\overline{\zeta}}{2\left(\zeta\overline{\zeta}\right)^{1/2}},
\label{legend10}
\end{equation}
and
\begin{equation}
\left(\dfrac{1}{2\tan\left(\theta/2\right)\cos^2\left(\theta/2\right)}\right)^2=\dfrac{\left(1+\zeta\overline{\zeta}\right)^2}{4\zeta\overline{\zeta}}.
\label{legend11}
\end{equation}
The operator $\partial_\theta$ can be written as
\begin{align}
\partial_\theta&=q_{,\theta}\partial_q+p_{,\theta}\partial_p, 
\label{legend12}
\end{align}
where the factors $q_{,\theta}$ and $p_{,\theta}$ are computed using \eqref{stereo1}, namely
\begin{subequations}
\begin{align}
q_{,\theta}=\dfrac{\left(\zeta+\overline{\zeta}\right)\left(1+\zeta\overline{\zeta}\right)}{4\left(\zeta\overline{\zeta}\right)^{1/2}},
\hspace{0.5cm}
\text{and}
\hspace{0.5cm}
p_{,\theta}=\dfrac{i\left(\overline{\zeta}-\zeta\right)\left(1+\zeta\overline{\zeta}\right)}{4\left(\zeta\overline{\zeta}\right)^{1/2}}.
\end{align}
\label{legend16s}%
\end{subequations}
Using Equations \eqref{legend16s}, \eqref{legend12} takes the form
\begin{align}
\partial_\theta&=\dfrac{\left(1+\zeta\overline{\zeta}\right)}{4\left(\zeta\overline{\zeta}\right)^{1/2}}\left(\left(\zeta+\overline{\zeta}\right)\partial_q + i\left(\overline{\zeta}-\zeta\right)\partial_p\right). 
\label{legend17}
\end{align}
Since the spherical harmonics defined in \eqref{legend7} have spin-weight $s$ zero, then the operators $\partial_q$ and $\partial_p$, given in \eqref{abase5}, are reduced to
\begin{subequations}
	\begin{align}
	\partial_{q}=\dfrac{1}{1+\zeta\overline{\zeta}}\left(\eth +\overline{\eth}  \right), 
	\hspace{0.5cm}
	\text{and}
	\hspace{0.5cm}
	\partial_{p}=\dfrac{i}{1+\zeta\overline{\zeta}}\left(\overline{\eth}  - \eth \right).
	\end{align}
	\label{legend18}%
\end{subequations}
Then, the operators given in Equations \eqref{legend18} allow to re-express \eqref{legend17} as
\begin{align}
\partial_\theta
&=\dfrac{ \overline{\zeta}\ \eth+ \zeta\ \overline{\eth} }{2\left(\zeta\overline{\zeta}\right)^{1/2}}.
\label{legend19}
\end{align}
In order to compute the second order derivative $\partial_{\theta\theta}$, it is necessary to make the calculation of the quantities $q_{,\theta\theta}$ and $p_{,\theta\theta}$. Thus, from 
\eqref{legend16s} one has
\begin{subequations}
\begin{align}
q_{,\theta\theta}
=\dfrac{\left(\zeta+\overline{\zeta}\right)\left(1+\zeta\overline{\zeta}\right)}{4},
\hspace{0.5cm}
\text{and}
\hspace{0.5cm}
p_{,\theta\theta}
=-i\dfrac{\left(\zeta-\overline{\zeta}\right)\left(1+\zeta\overline{\zeta}\right)}{4}.
\end{align}
\label{legend21}%
\end{subequations}
The second order operator $\partial_{\theta\theta}$ is directly computed using \eqref{legend12}, thus
\begin{align*}
\partial_{\theta\theta} 
&= q_{,\theta\theta}\partial_q + p_{,\theta\theta}\partial_p + q_{,\theta}\partial_{\theta q}  +p_{,\theta}\partial_{\theta p},
\end{align*}
where
\begin{align*}
q_{,\theta}\partial_\theta\partial_q 
&= q_{,\theta}^2 \partial_{qq} + q_{,\theta}p_{,\theta} \partial_{pq}, 
\end{align*}
and
\begin{align*}
p_{,\theta}\partial_\theta\partial_p
&=p_{,\theta}q_{,\theta}\partial_{qp}+p_{,\theta}^2\partial_{pp}.
\end{align*}
Consequently, the second order operator $\partial_{\theta\theta}$ reads
\begin{align}
\partial_{\theta\theta}&= q_{,\theta\theta}\partial_q + p_{,\theta\theta}\partial_p + q_{,\theta}^2 \partial_{qq} + 2q_{,\theta}p_{,\theta} \partial_{qp}   +p_{,\theta}^2\partial_{pp}.
\label{legend22}
\end{align}
The commutator for the $\overline{\eth}$ and $\eth$ given in \eqref{abase31} for zero spin-weighted functions becomes 
\begin{equation}
\left[\overline{\eth},\eth\right]\ _0\Psi=0,
\label{legend23}
\end{equation}
then the anti-commutator for these functions takes the form
\begin{equation}
\left(\overline{\eth},\eth\right)\ _0\Psi=2\overline{\eth}\eth\ _0\Psi.
\label{legend24}
\end{equation}
For functions of this type, the second order differential operators $\partial_{qq}$, $\partial_{pp}$ and $\partial_{qp}$, given in \eqref{abase13}, \eqref{abase20} and \eqref{abase25} respectively, are strongly simplified to
\begin{subequations}
	\begin{align}
	\partial_{qq}
	&=\dfrac{1}{\left(1+\zeta\overline{\zeta}\right)^2}\left(\eth^2\  + \overline{\eth}^2 + 2\overline{\eth}\eth -2\zeta \ \eth  -2\overline{\zeta}\ \overline{\eth}  \right), \label{legend25.1} \\
	\partial_{pp}
	&=-\dfrac{1}{\left(1+\zeta\overline{\zeta}\right)^2}\left( \eth^2 + \overline{\eth}^2 - 2\overline{\eth}\eth -2\overline{\zeta}\ \overline{\eth} -2\zeta\ \eth   \right), \label{legend25.2} \\
	\partial_{qp} 
	&=\dfrac{i}{\left(1+\zeta\overline{\zeta}\right)^2}\left( \overline{\eth}^2 - \eth^2 + 2\zeta \ \eth - 2\overline{\zeta} \ \overline{\eth} \right). \label{legend25.3}
	\end{align}
	\label{legend25}%
\end{subequations}
Thus, the two first terms in \eqref{legend22} are obtained using 
\eqref{legend21} and \eqref{legend18}, namely
\begin{align}
q_{,\theta\theta}\partial_q + p_{,\theta\theta}\partial_p 
&= \dfrac{1}{2} \left(  \overline{\zeta}\ \eth  + \zeta \ \overline{\eth}   \right).
\label{legend26}
\end{align}
The third term in \eqref{legend22} will be obtained by using Equations \eqref{legend16s} and \eqref{legend25}, namely
\begin{align}
q_{,\theta}^2 \partial_{qq} 
&= \dfrac{\left(\zeta+\overline{\zeta}\right)^2}{16\zeta\overline{\zeta}} \left(\eth^2\  + \overline{\eth}^2 + 2\overline{\eth}\eth -2\zeta \ \eth  -2\overline{\zeta}\ \overline{\eth}  \right).
\label{legend27}
\end{align}
The fourth term in \eqref{legend22} is obtained when \eqref{legend16s}, and \eqref{legend25} are employed, i.e.
\begin{align}
2q_{,\theta}p_{,\theta} \partial_{qp}
&=\dfrac{2 \left( \zeta+\overline{\zeta} \right) \left( \zeta-\overline{\zeta} \right)}{16 \zeta\overline{\zeta}} \left( \overline{\eth}^2 - \eth^2 + 2\zeta \ \eth - 2\overline{\zeta} \ \overline{\eth} \right).
\label{legend28}
\end{align}
After substituting \eqref{legend25} and \eqref{legend16s}, the fifth term in \eqref{legend22} reads
\begin{align}
p_{,\theta}^2\partial_{pp}
&= \dfrac{ \left( \zeta - \overline{\zeta} \right)^2 }{16 \zeta\overline{\zeta} } \left( \eth^2 + \overline{\eth}^2 - 2\overline{\eth}\eth -2\overline{\zeta}\ \overline{\eth} -2\zeta\ \eth   \right).
\label{legend29}
\end{align}
Thus, adding \eqref{legend27} and \eqref{legend29} one obtains
\begin{align*}
q_{,\theta}^2 \partial_{qq}+p_{,\theta}^2\partial_{pp}&=\dfrac{1}{16\zeta\overline{\zeta}}\Big(\left(\left(\zeta+\overline{\zeta}\right)^2+\left(\zeta-\overline{\zeta}\right)^2\right)\left(\eth^2+\overline{\eth}^2-2\zeta\ \eth -2\overline{\zeta}\ \overline{\eth}\right)\nonumber\\
&\hspace{1cm}+2\left(\left(\zeta+\overline{\zeta}\right)^2-\left(\zeta-\overline{\zeta}\right)^2\right)\eth\overline{\eth}\Big).
\end{align*}
Using the last equation and \eqref{legend28} results in
\begin{align}
& q_{,\theta}^2 \partial_{qq}+p_{,\theta}^2\partial_{pp}+2q_{,\theta}p_{,\theta} \partial_{qp} \nonumber\\
=& \dfrac{1}{16\zeta\overline{\zeta}}\Bigg( 4 \zeta^2\left(\overline{\eth}^2  - 2\overline{\zeta}\ \overline{\eth}\right) +4 \overline{\zeta}^2 \left( \eth^2 - 2\zeta \ \eth \right) 
+8  \zeta\overline{\zeta} \eth\overline{\eth}  \Bigg).
\label{legend30}
\end{align}
Then, using \eqref{legend30} and \eqref{legend26}, \eqref{legend22} takes the explicit form
\begin{align}
\partial_{\theta\theta} 
= \dfrac{1}{16\zeta\overline{\zeta}}\Bigg( 4 \zeta^2 \overline{\eth}^2 +4 \overline{\zeta}^2   \eth^2  
+8 \zeta\overline{\zeta}  \eth\overline{\eth}  \Bigg). 
\label{legend31}
\end{align}
Now, the differential operator $\partial_{\phi}$ can be written as
\begin{align}
\partial_\phi&=q_{,\phi}\partial_q+p_{,\phi}\partial_p, 
\label{legend32}
\end{align}
where the coefficients $q_{,\phi}$ and $p_{,\phi}$ are
\begin{align}
q_{,\phi} &=\dfrac{i}{2}(\zeta-\overline{\zeta}),
\label{legend33}
\end{align}
and
\begin{align}
p_{,\phi} &=\dfrac{1}{2}(\overline{\zeta}+\zeta).
\label{legend34}
\end{align}
Then, using the two last relations and \eqref{legend18} one obtains
\begin{align}
\partial_\phi &=i\dfrac{-\overline{\zeta}\eth +  \zeta\eth }{\left(1+\zeta\overline{\zeta}\right)}. 
\label{legend35}
\end{align}
The second order partial derivative $\partial_{\phi\phi}$ can be computed as follows
\begin{align*}
\partial_{\phi\phi} &= q_{,\phi\phi}\partial_q + p_{,\phi\phi}\partial_p + q_{,\phi}\partial_{\phi q}  +p_{,\phi}\partial_{\phi p},
\end{align*}
where
\begin{align*}
q_{,\phi}\partial_\phi\partial_q 
&= q_{,\phi}^2 \partial_{qq} + q_{,\phi}p_{,\phi} \partial_{pq},
\end{align*}
and
\begin{align*}
p_{,\phi}\partial_\phi\partial_p
&=p_{,\phi}q_{,\phi}\partial_{qp}+p_{,\phi}^2\partial_{pp}.
\end{align*}
Then, for this reason
\begin{align}
\partial_{\phi\phi}&= q_{,\phi\phi}\partial_q + p_{,\phi\phi}\partial_p + q_{,\phi}^2 \partial_{qq} + 2q_{,\phi}p_{,\phi} \partial_{pq} + p_{,\phi}^2\partial_{pp}.
\label{legend36}
\end{align}
The factor $q_{,\phi\phi}$ is computed from \eqref{legend33} and with the help of \eqref{legend34}, thus
\begin{align}
q_{,\phi\phi}
&=-\dfrac{1}{2}(\overline{\zeta}+\zeta).
\label{legend37}
\end{align}
The factor $p_{,\phi\phi}$ is calculated from \eqref{legend34}, i.e.,
\begin{align}
p_{,\phi\phi}
&=\dfrac{i}{2}(\zeta-\overline{\zeta}),
\label{legend38}
\end{align}
where we have used \eqref{legend33}. Thus, when \eqref{legend37}, \eqref{legend38} and \eqref{legend18} are substituted into the two first terms of \eqref{legend36} one obtains
\begin{align}
q_{,\phi\phi}\partial_q + p_{,\phi\phi}\partial_p&=- \dfrac{1}{2\left(1+\zeta\overline{\zeta}\right)}\left(\left(\overline{\zeta}+\zeta\right)\left(\eth +\overline{\eth}  \right) + \left(\zeta-\overline{\zeta}\right)\left(\overline{\eth}  - \eth \right)\right)\nonumber\\
&=- \dfrac{ \overline{\zeta}\ \eth + \zeta\ \overline{\eth} }{\left(1+\zeta\overline{\zeta}\right)}. 
\label{legend39}
\end{align} 
The third term in \eqref{legend36} is computed using \eqref{legend25.1} and \eqref{legend33}, i.e.,
\begin{align}
q_{,\phi}^2 \partial_{qq}
&=-\dfrac{\left(\zeta-\overline{\zeta}\right)^2}{4\left(1+\zeta\overline{\zeta}\right)^2}\left(\eth^2\  + \overline{\eth}^2 + 2\overline{\eth}\eth -2\zeta \ \eth  -2\overline{\zeta}\ \overline{\eth}  \right).
\label{legend40}
\end{align}
The fourth term in \eqref{legend36} is obtained from \eqref{legend25.3}, \eqref{legend33} and \eqref{legend34}, namely
\begin{align}
2q_{,\phi}p_{,\phi} \partial_{pq}
&=- \dfrac{2\left(\zeta-\overline{\zeta}\right)\left(\overline{\zeta}+\zeta\right)}{4\left(1+\zeta\overline{\zeta}\right)^2}\left( \overline{\eth}^2 - \eth^2 + 2\zeta \ \eth - 2\overline{\zeta} \ \overline{\eth} \right).
\label{legend41}
\end{align}
The last term in \eqref{legend36} is computed from \eqref{legend25.2} and \eqref{legend34}, resulting in
\begin{align}
p_{,\phi}^2\partial_{pp}&=-\dfrac{\left(\overline{\zeta}+\zeta\right)^2}{4\left(1+\zeta\overline{\zeta}\right)^2}\left( \eth^2 + \overline{\eth}^2 - 2\overline{\eth}\eth -2\overline{\zeta}\ \overline{\eth} -2\zeta\ \eth   \right).
\label{legend42}
\end{align}
The addition of \eqref{legend40} and \eqref{legend42} yields 
\begin{align*}
q_{,\phi}^2 \partial_{qq}+p_{,\phi}^2\partial_{pp}
=& -\dfrac{1}{4\left(1+\zeta\overline{\zeta}\right)^2}\Bigg( \left(\left(\zeta-\overline{\zeta}\right)^2 + \left(\overline{\zeta}+\zeta\right)^2 \right) \left(\eth^2\  + \overline{\eth}^2 -2\zeta \ \eth  -2\overline{\zeta}\ \overline{\eth}  \right)\\
&+2\left(\left(\zeta-\overline{\zeta}\right)^2 - \left(\overline{\zeta}+\zeta\right)^2 \right) \overline{\eth}\eth \Bigg),
\end{align*}
which added to \eqref{legend41} gives
\begin{align}
& q_{,\phi}^2 \partial_{qq}+p_{,\phi}^2\partial_{pp}+2q_{,\phi}p_{,\phi} \partial_{pq} \nonumber\\
=&-\dfrac{\zeta^2 \left(\overline{\eth}^2 -2\overline{\zeta}\ \overline{\eth}  \right) + \overline{\zeta}^2\left( \eth^2 - 2\zeta \ \eth  \right) -2 \zeta\ \overline{\zeta}\   \overline{\eth}\eth}{\left(1+\zeta\overline{\zeta}\right)^2}.
\label{legend43}
\end{align}
Substituting \eqref{legend39} and \eqref{legend43} into \eqref{legend36} one has
\begin{align}
\partial_{\phi\phi}
&=-\dfrac{\overline{\zeta}\ \eth + \zeta\ \overline{\eth}  +  \zeta^2 \left(\overline{\eth}^2 -\overline{\zeta}\ \overline{\eth}  \right) + \overline{\zeta}^2\left( \eth^2 - \zeta \ \eth  \right) -2 \zeta\ \overline{\zeta}\   \overline{\eth}\eth}{\left(1+\zeta\overline{\zeta}\right)^2}.
\label{legend44}
\end{align}
With these results, the explicit form of the Legendrian given in \eqref{legend9} in terms of the $\overline{\eth}$ and $\eth$ operators will be computed. The first term is obtained directly from \eqref{legend10} and \eqref{legend19}, namely
\begin{align}
\dfrac{1}{\tan\theta}\partial_\theta
&=\dfrac{4 \left( \overline{\zeta}\ \eth-\zeta\overline{\zeta}^2\ \eth + \zeta\ \overline{\eth}-\zeta^2\overline{\zeta} \ \overline{\eth}  \right)}{16  \zeta\overline{\zeta} }. 
\label{legend45}
\end{align}
Also, using \eqref{legend11} and \eqref{legend44}, the third term in \eqref{legend9} reads
\begin{align}
\dfrac{1}{\sin^2\theta}\partial_{\phi\phi}
&= - \dfrac{\overline{\zeta}\ \eth + \zeta\ \overline{\eth}  +  \zeta^2 \left(\overline{\eth}^2 -\overline{\zeta}\ \overline{\eth}  \right) + \overline{\zeta}^2\left( \eth^2 - \zeta \ \eth  \right) -2 \zeta\ \overline{\zeta}\   \overline{\eth}\eth }{4\zeta\overline{\zeta}} .
\label{legend46}
\end{align}
Then, substituting \eqref{legend31}, \eqref{legend45} and \eqref{legend46}, one has
\begin{align}
\mathcal L^2 
&=  \eth\overline{\eth}.  
\label{legend47}
\end{align}
This result implies that, the eigenvalue equation \eqref{legend8} can be written as
\begin{equation}
\eth\overline{\eth}Y_{lm}=-l(l+1)Y_{lm}.
\label{legend48}
\end{equation}
Notice that the functional dependence of the spherical harmonics was not written. This was made intentionally because it is valid independently of the coordinate system. We show that through the passage to stereographic coordinates, the expressions of the angular operators $\partial_\theta$, $\partial_\phi$, $\partial_{\theta\theta}$, $\partial_{\theta\phi}$ and $\partial_{\phi\phi}$ in terms of the $\eth$ and $\overline{\eth}$ were obtained.  The spin-weight of the functions in which these operators can be applied was disregarded. Thus, at least for $0$-spin weighted functions an equivalent expression of the Legendrian was found. This relation can be extended to $s$-spin weighted functions and therefore a Legendrian operator for these functions can be constructed. There are at least two ways to do this in a completely consistent manner. One of them is by expressing the operators $\eth$ and $\overline{\eth}$ in spherical coordinates and with them construct the second order operators $\eth^2$, $\overline{\eth}^2$, $\eth\overline{\eth}$ and $\overline{\eth}\eth$, and then compute the eigenvalues of the commutator $[\eth,\overline{\eth}]$. Another way is by expressing these operators in stereographic coordinates and then construct the commutator $[\eth,\overline{\eth}]$. 
\section{The $\eth$ and $\overline{\eth}$ in Spherical Coordinates}
A further generalisation of all the last results can be done, by extending the operators $\eth$ and $\overline{\eth}$ to the case when function with spin-weight different from zero are considered. In order to do so, it is necessary to consider the operators defined in Equations \eqref{eth_spherical_7}, which can be written as
\begin{subequations}
\begin{align}
\eth 
&= \left(\sin\theta\right)^s \left(\partial_\theta + i\csc\theta \partial_\phi  \right) \left(\sin\theta\right)^{-s}, 
\label{eth_spherical_8}
\end{align}
and
\begin{align}
\overline{\eth} 
&= \left(\sin\theta\right)^{-s} \left(\partial_\theta - i\csc\theta \partial_\phi  \right) \left(\sin\theta\right)^{s}.  
\label{eth_spherical_10}
\end{align}
\label{eth_spherical_10s}%
\end{subequations}
It is worth stressing that the operations in \eqref{eth_spherical_10s} are referred to operators, not to scalar functions.
\\From \eqref{eth_spherical_7} one obtains the expressions for $\partial_\theta$ and $\partial_\phi$, namely 
\begin{equation}
\partial_{\theta}=\dfrac{\eth+\overline{\eth}}{2}, \hspace{1cm}
\partial_{\phi}=\dfrac{i\sin\theta}{2}\left(\overline{\eth}-\eth-2s\cot\theta\right), 
\label{eth_spherical_16}
\end{equation}
and the expressions for $\partial_{\theta\theta}$, $\partial_{\theta\phi}$ and $\partial_{\phi\phi}$, namely
\begin{subequations}
\begin{align}
\partial_{\theta\theta}=&\dfrac{{\eth}^2+\left(\overline{\eth},\eth\right)+\overline{\eth}^2}{4}, \label{eth_spherical_30.1}\\
\partial_{\phi\phi} =& -\dfrac{\sin^2\theta}{4} \left({\eth}^2 -\left(\overline{\eth},\eth\right) + \overline{\eth}^2 \right)  - s^2 \cos^2\theta\nonumber\\
& - \sin\theta\cos\theta\left(\left(s+\dfrac{1}{2}\right)\eth -\left(s-\dfrac{1}{2}\right)\overline{\eth}\right), \label{eth_spherical_30.2}\\
\partial_{\theta\phi} =&- \dfrac{i\sin\theta}{4}\left(\eth^2-\overline{\eth}^2\right) -is\cos\theta\dfrac{\eth+\overline{\eth}}{2}  \nonumber\\
& + \dfrac{i\cos\theta}{2}\left(\overline{\eth}-\eth-2s\cot\theta\right) + i\sin\theta\dfrac{s(\cot^2\theta+\csc^2\theta)}{2}, \label{eth_spherical_30.3}
\end{align}
\label{eth_spherical_30}%
\end{subequations}
(see \hyperlink{apendiceB}{Appendix B} for further details of the derivation of these expressions).\\
These operators can be used to transform the field equations projected onto the dyads, in terms of the angular variables $\theta$ and $\phi$ into the eth form, without using the stereographic version of the eth operators. However, most of the characteristic codes use stereographic and gnomonic projections.
\section{Integrals for the Angular Manifold}

In order to compute the inner product of the spin-weighted functions, we will need useful expressions for the integrals involving angular variables, when the $(q,p)$, $(\zeta,\overline{\zeta})$ and the $(\theta,\phi)$ coordinates are used. These integrals are for example of the type
\begin{align}
I=\oiint\limits_{\Omega} d\Omega f(\theta,\phi),
\label{integral1}
\end{align}
where $\Omega$ is the solid angle. In spherical coordinates these quantities are expressed as
\begin{align}
I=\iint\limits_{\Omega} d\phi d\theta \sin\theta f(\theta,\phi).
\label{integral2}
\end{align}

The domain of these integrals can be decomposed into two parts, involving each hemisphere, north and south, in the form
\begin{align}
I&= \iint\limits_{\Omega_N} d\phi d\theta \sin\theta f(\theta,\phi) + \iint_{\Omega_S} d\phi d\theta \sin\theta f(\theta,\phi),\nonumber\\
&= \iint\limits_{\Omega_N} d\phi_N d\theta_N \sin\theta_N f(\theta_N,\phi_N) + \iint_{\Omega_S} d\phi_S d\theta_S \sin\theta_S f(\theta_S,\phi_S),
\label{integral3}
\end{align}
where $\Omega_N$ and $\Omega_S$, label the north and the south regions in which the unitary sphere was decomposed. Both domains share the same boundary that is the equator line. 
\\Now, from the transformation of coordinates \eqref{stereo1}, it is possible to write that
\begin{align}
q=\tan\left(\theta/2\right)\cos{\phi}\hspace{0.5cm}\text{and} \hspace{0.5cm}p=\tan\left(\theta/2\right)\sin{\phi}.
\label{integral4}
\end{align}
Thus, the transformation of coordinates from spherical $(\theta,\phi)$ to stereographic $(q,p)$ can be performed. First, \eqref{integral2} is expressed as
\begin{align}
I=&\iint\limits_{\Omega_N} dq_N dp_N \sin\theta_N J(q_N,p_N) f(q_N,p_N)\nonumber\\
&+\iint\limits_{\Omega_S} dq_S dp_S \sin\theta_S J(q_S,p_S) f(q_S,p_S),
\label{integral5}
\end{align}
where $J(q,p)$ is the Jacobian of the transformation of coordinates\footnote{Here the indices to indicate the hemisphere is suppressed  in the Jacobian, because it has the same form in both.}, which is given by
\begin{equation*}
J(q,p)=\left|\begin{matrix}
\theta_{,q} & \theta_{,p}\\
\phi_{,q} & \phi_{,p}
\end{matrix}\right|.
\end{equation*}
From \eqref{integral4}, the derivatives in the Jacobian read
\begin{align}
\begin{array}{ll}
\theta_{,q}=\dfrac{2q(q^2+p^2)^{1/2}}{1+q^2+p^2}, & \theta_{,p}=\dfrac{2p(q^2+p^2)^{1/2}}{1+q^2+p^2},\\[0.5cm]
\phi_{,q}=-\dfrac{p}{q^2+p^2}, & \phi_{,p}=\dfrac{q}{q^2+p^2},
\end{array}
\label{integral6}
\end{align}
then
\begin{equation}
J(q,p)=\dfrac{2(q^2+p^2)^{-1/2}}{1+q^2+p^2}.
\label{integral7}
\end{equation}
Note that 
\begin{align}
\sin\theta 
&=\dfrac{2(q^2+p^2)^{1/2}}{1+q^2+p^2},
\label{integral8}
\end{align}
thus, the substitution of \eqref{integral7} and \eqref{integral8} into \eqref{integral3} yields 
\begin{align}
I=\iint\limits_{\Omega_N} dq_N dp_N  \dfrac{4f(q_N,p_N)}{\left(1+q_N^2+p_N^2\right)^2}+\iint\limits_{\Omega_S} dq_S dp_S  \dfrac{4f(q_S,p_S)}{\left(1+q_S^2+p_S^2\right)^2}.
\label{integral9}
\end{align}
This last expression is particularly useful when a numerical evaluation of this kind of integrals is performed. From \eqref{integral9}, it is possible to obtain the expressions for the same kind of integrals in terms of the complex stereographic coordinates $(\zeta,\overline{\zeta})$, namely
\begin{align}
I=&\iint\limits_{\Omega_N} d\zeta_N d\overline{\zeta}_N J(\zeta_N,\overline{\zeta}_N)  \dfrac{4f(\zeta_N,\overline{\zeta}_N)}{\left(1+\zeta_N\overline{\zeta}_N\right)^2}\nonumber\\
&+\iint\limits_{\Omega_S} d\zeta_S d\overline{\zeta}_S  J(\zeta_S,\overline{\zeta}_S)\dfrac{4f(\zeta_S,\overline{\zeta}_S)}{\left(1+\zeta_S\overline{\zeta}_S\right)^2},
\label{integral10}
\end{align}
where, the Jacobian of the transformation of coordinates is given by
\begin{equation*}
J(\zeta,\overline{\zeta})=\left|\begin{matrix}
q_{,\zeta} & q_{,\overline{\zeta}}\\
p_{,\zeta} & p_{,\overline{\zeta}}
\end{matrix}\right|.
\end{equation*}
The derivatives in this Jacobian are
\begin{align}
\begin{array}{ll}
q_{,\zeta}=\dfrac{1}{2}, & q_{,\overline{\zeta}}=\dfrac{1}{2},\\[0.5cm]
p_{,\zeta}=-\dfrac{i}{2}, & p_{,\overline{\zeta}}=\dfrac{i}{2}.
\end{array}
\label{integral11}
\end{align}
Thus, the Jacobian of the transformation of the coordinates becomes explicitly
\begin{equation*}
J(\zeta,\overline{\zeta})=\dfrac{i}{2}.
\end{equation*}
Then the integral \eqref{integral10} in terms of $(\zeta,\overline{\zeta})$ is transformed as 
\begin{align}
I=\iint\limits_{\Omega_N} d\zeta_N d\overline{\zeta}_N   \dfrac{2if(\zeta_N,\overline{\zeta}_N)}{\left(1+\zeta_N\overline{\zeta}_N\right)^2}+\iint\limits_{\Omega_S} d\zeta_S d\overline{\zeta}_S  \dfrac{2if(\zeta_S,\overline{\zeta}_S)}{\left(1+\zeta_S\overline{\zeta}_S\right)^2}.
\label{integral12}
\end{align}
The inner product of two functions that depend on the angular variables is defined as
\begin{equation}
\langle f,g\rangle=\oiint\limits_{\Omega} d\Omega\overline{f}g.
\label{integral13}
\end{equation}
Thus, the inner product $\langle \ _0Y_{l'm'},\ _0Y_{lm}\rangle$, where $_0Y_{lm}=Y_{lm}$, can be computed in spherical coordinates as usual, namely
\begin{align}
\langle \ _0Y_{l'm'},\ _0Y_{lm}\rangle&=\int_0^{2\pi} d\phi \int_0^{\pi} d\theta \sin\theta \ _0\overline{Y}_{l'm'}(\theta,\phi) \ _0Y_{lm}(\theta,\phi)\nonumber\\
&=\delta_{ll'}\delta_{mm'}.
\label{integral14}
\end{align}
The explicit form of this inner product in stereographic coordinates $(q,p)$ reads
\begin{align}
\langle \ _0Y_{l'm'},\ _0Y_{lm}\rangle&=\int_{-1}^{1} dq_N \int_{-\sqrt{1-q_N^2}}^{\sqrt{1-q_N^2}} dp_N  \dfrac{4\ _0\overline{Y_N}_{l'm'}(q_N,p_N) \ _0{Y_N}_{lm}(q_N,p_N)}{\left(1+q_N^2+p_N^2\right)^2}\nonumber\\
&+ \int_{-1}^{1} dq_S \int_{-\sqrt{1-q_S^2                                                                                                                   }}^{\sqrt{1-q_S^2}} dp_S  \dfrac{4\ _0\overline{Y_S}_{l'm'}(q_S,p_S) \ _0{Y_S}_{lm}(q_S,p_S)}{\left(1+q_S^2+p_S^2\right)^2}.
\label{integral15}
\end{align}
Now, in order to extend the inner product shown above, to spin-weighted function with spin-weight different from zero, it is important to observe that the $\eth$ and $\overline{\eth}$ operators can be written as
\begin{subequations}
\begin{align}
\eth &=P^{1-s}\partial_{\overline{\zeta}} P^{s},
\label{integral16}
\end{align}
and
\begin{align}
\overline{\eth}
&=P^{s+1}\partial_{{\zeta}}P^{-s},
\label{integral17}
\end{align}
\end{subequations}
where, we have defined the zero spin-weighted function
\begin{equation}
P=1+\zeta\overline{\zeta}.
\label{integral18}
\end{equation}
Noting that
\begin{align}
\eth \overline{\zeta}=P\partial_{\overline{\zeta}}\overline{\zeta}=P,
\label{integral19}
\end{align}
then we have
\begin{align}
\overline{\eth} P=\overline{\eth}\eth\overline{\zeta}=\eth\overline{\eth}\ \overline{\zeta}=\eth P\partial_{\zeta}\overline{\zeta}=0.
\label{integral20}
\end{align}
Also
\begin{align}
\overline{\eth} {\zeta}=P\partial_{{\zeta}} {\zeta}=P,
\label{integral21}
\end{align}
then we obtain
\begin{align}
{\eth} P=\eth\overline{\eth}{\zeta}=\overline{\eth}\eth {\zeta}=\overline{\eth} P\partial_{\overline{\zeta}}{\zeta}=0.
\label{integral22}
\end{align}
Thus, equations \eqref{integral20} and \eqref{integral22} imply that
\begin{equation}
\eth(PA)=P\eth A, \hspace{1cm}\overline{\eth}(PA)=P\overline{\eth}A,
\label{integral23}
\end{equation}
for any spin-weight function $A$.

Then, if  two functions $f$ and $g$ with spin-weight $s$ and $s-1$, respectively, are considered, the inner product of $f$ and $\eth g$ reads
\begin{align*}
\langle f,\eth g\rangle&=\oiint\limits_{\Omega} d\Omega \overline{f}\ \eth g\\
&=\iint\limits_{\Omega} d\zeta d\overline{\zeta}\ \dfrac{2i}{P^2}\overline{f}\ P^{1-s}\partial_{\overline{\zeta}}\left(P^{s} g\right) \\
&=2i\iint\limits_{\Omega} d\zeta d\overline{\zeta}\ \overline{f}\ P^{-(1+s)}\partial_{\overline{\zeta}}\left(P^{s} g\right).
\end{align*}
The last equation can be written as
\begin{align*}
\langle f,\eth g\rangle
=& 2i\iint\limits_{\Omega} d\zeta d\overline{\zeta} \left(\partial_{\overline{\zeta}} \left(\overline{f}\ P^{-(1+s)}P^{s} g\right) - P^{s} g\ \partial_{\overline{\zeta}} \left(\overline{f}\ P^{-(1+s)}\right)\right),
\end{align*}
which results in
\begin{align}
\langle f,\eth g\rangle
=& 2i\left(\iint\limits_{\Omega} d\zeta d\overline{\zeta} \ \partial_{\overline{\zeta}} \left(\overline{f}\ P^{-(1+s)}P^{s} g\right) 
- \iint\limits_{\Omega} d\zeta d\overline{\zeta} P^{s} g\ \partial_{\overline{\zeta}} \left(\overline{f}\ P^{-(1+s)}\right)\right). 
\label{integral24}
\end{align}
The first term in this equation corresponds to
\begin{align*}
2i\iint\limits_{\Omega} d\zeta d\overline{\zeta} \ \partial_{\overline{\zeta}} \left(\overline{f}\ P^{-(1+s)}P^{s} g\right)&=2i\iint\limits_{\Omega} d\zeta d\overline{\zeta} P^{s-1} P^{1-s}\ \partial_{\overline{\zeta}} \left(P^{s}\overline{f}\ P^{-(1+s)} g\right)\nonumber\\
&=2i\iint\limits_{\Omega} d\zeta d\overline{\zeta} P^{s-1} \eth \left(\overline{f}\ P^{-(1+s)} g\right)\nonumber\\
&=2i\iint\limits_{\Omega} d\zeta d\overline{\zeta} P^{s-1} P^{-(1+s)}\eth \left(\overline{f}\  g\right)\nonumber\\
&=2i\iint\limits_{\Omega} d\zeta d\overline{\zeta} P^{-2} \eth \left(\overline{f}\  g\right)\nonumber\\
&=\left\langle 1,\eth(\overline{f} g)\right\rangle.
\end{align*}
Since $f$ has spin-weight $s$ and $g$ has a spin-weight $s-1$, then their product $\overline{f}g$ has spin-weight $s=-1$, consequently $\eth(\overline{f} g)$ is a zero spin-weighted function. Therefore, it can be expanded in spherical harmonics in the form
\begin{equation*}
\eth(\overline{f} g)=\sum_{l,m}a_{lm}\ _0Y_{lm}.
\end{equation*}
Thus,
\begin{align}
\left\langle 1,\eth(\overline{f} g)\right\rangle&=\left\langle 1,\sum_{l,m}a_{lm}\ _0Y_{lm}\right\rangle\nonumber\\
&=\sum_{l,m}a_{lm}\left\langle 1,\ _0Y_{lm}\right\rangle\nonumber\\
&=0.
\label{integral25}
\end{align}
The second term in \eqref{integral24} is given by
\begin{align*}
2i\iint\limits_{\Omega} d\zeta d\overline{\zeta} P^{s} g\ \partial_{\overline{\zeta}} \left(\overline{f}\ P^{-(1+s)}\right)&=2i\iint\limits_{\Omega} d\zeta d\overline{\zeta} \ g P^{-1}P^{s+1}\ \partial_{\overline{\zeta}} \left(P^{-s}\overline{f}\ P^{-1}\right)\\
&=2i \iint\limits_{\Omega}  d\overline{\zeta} d\zeta \ g P^{-1}P^{s+1}\ \partial_{\zeta} \left(P^{-s}\overline{f}\ P^{-1}\right),
\end{align*}
where the integration variables $\zeta$ and $\overline{\zeta}$ have been renamed, thus
\begin{align}
2i\iint\limits_{\Omega} d\zeta d\overline{\zeta} P^{s} g\ \partial_{\overline{\zeta}} \left(\overline{f}\ P^{-(1+s)}\right)&=2i \iint\limits_{\Omega}  d\overline{\zeta} d\zeta \ g P^{-1}\overline{\eth} \left(\overline{f}\ P^{-1}\right) \nonumber\\
&=2i \iint\limits_{\Omega}  d\overline{\zeta} d\zeta \ g P^{-2}\overline{\eth} \left(\overline{f}\right) \nonumber\\
&=-\left\langle\overline{\eth}f,g\right\rangle.
\label{integral26}
\end{align}
Substituting \eqref{integral25} and \eqref{integral26} into \eqref{integral24} one obtains
\begin{equation}
\langle f,\eth g\rangle=-\left\langle\overline{\eth}f,g\right\rangle.
\label{integral27}
\end{equation}
Now, if $f$ and $g$ have spin-weight $s$ and $s+1$, respectively, are considered, then the inner product reads
\begin{align*}
\left\langle f,\overline{\eth}g\right\rangle&=\oiint\limits_{\Omega}d\Omega\overline{f}\ \overline{\eth}g \nonumber\\
&=\iint\limits_{\Omega}d\zeta d\overline{\zeta}\dfrac{2i}{P^2}\overline{f}P^{s+1}\partial_\zeta \left(P^{-s}g\right) \nonumber\\
&=2i\iint\limits_{\Omega}d\zeta d\overline{\zeta}\ \overline{f}P^{s-1}\partial_\zeta \left(P^{-s}g\right).
\end{align*}
This last equation can be written as
\begin{align}
\left\langle f,\overline{\eth}g\right\rangle&=2i\iint\limits_{\Omega}d\zeta d\overline{\zeta} \partial_\zeta \left(\overline{f}P^{s-1} P^{-s}g\right)-2i\iint\limits_{\Omega}d\zeta d\overline{\zeta}  P^{-s}g\partial_\zeta \left(\overline{f}P^{s-1}\right).
\label{integral28}
\end{align}
The first term in \eqref{integral28} is given by
\begin{align*}
2i\iint\limits_{\Omega}d\zeta d\overline{\zeta} \partial_\zeta \left(\overline{f}P^{s-1} P^{-s}g\right)&=2i\iint\limits_{\Omega}d\zeta d\overline{\zeta} P^{-(s+1)} P^{s+1}\partial_\zeta \left( P^{-s} \overline{f}P^{s-1} g\right)\\
&=2i\iint\limits_{\Omega}d\zeta d\overline{\zeta} P^{-(s+1)} \overline{\eth} \left( \overline{f}P^{s-1} g\right),
\end{align*}
i.e.,
\begin{align*}
2i\iint\limits_{\Omega}d\zeta d\overline{\zeta} \partial_\zeta \left(\overline{f}P^{s-1} P^{-s}g\right)&= 2i\iint\limits_{\Omega}d\zeta d\overline{\zeta}  P^{-2} \overline{\eth} \left( \overline{f} g\right)\\
&=\left\langle 1, \overline{\eth} \left( \overline{f} g\right)\right\rangle.
\end{align*}
It is important to observe that, here $\overline{f}g$ has spin-weight $s=1$. Consequently, $\overline{\eth}(\overline{f}g)$ must be a spin-weight zero function. Therefore, it admits a decomposition in the form
\begin{align*}
\overline{\eth} \left( \overline{f} g\right)=\sum_{l,m}a_{0lm}\ _0Y_{lm},
\end{align*}
then
\begin{align}
\left\langle 1, \overline{\eth} \left( \overline{f} g\right)\right\rangle &=\left\langle 1, \sum_{l,m}a_{0lm}\ _0Y_{lm}\right\rangle \nonumber\\
&=\sum_{l,m}a_{0lm}\left\langle 1, \ _0Y_{lm}\right\rangle \nonumber\\
&=0.
\label{integral29}
\end{align}
The second term in \eqref{integral28} is given by
\begin{align*}
2i\iint\limits_{\Omega}d\zeta d\overline{\zeta}  P^{-s}g\partial_\zeta \left(\overline{f}P^{s-1}\right)=2i\iint\limits_{\Omega} d\overline{\zeta} d\zeta P^{-s}g\partial_{\overline{\zeta}} \left(\overline{f}P^{s-1}\right),
\end{align*}
where, the variables $\zeta$ and $\overline{\zeta}$ in the integrals were interchanged. Thus
\begin{align}
2i\iint\limits_{\Omega}d\zeta d\overline{\zeta}  P^{-s}g\partial_\zeta \left(\overline{f}P^{s-1}\right)&=2i\iint\limits_{\Omega} d\overline{\zeta} d\zeta \ g P^{-1} P^{1-s}\partial_{\overline{\zeta}} \left(P^s\overline{f}P^{-1}\right) \nonumber\\
&=2i\iint\limits_{\Omega} d\overline{\zeta} d\zeta g P^{-1} \eth \left(\overline{f}P^{-1}\right)  \nonumber\\
&=2i\iint\limits_{\Omega} d\overline{\zeta} d\zeta g P^{-2} \eth \overline{f} \nonumber\\
&=\left\langle \eth f,g \right\rangle.
\label{integral30}
\end{align}
Thus, substituting \eqref{integral29} and \eqref{integral30} into \eqref{integral28} one obtains
\begin{equation}
\left\langle f,\overline{\eth}g\right\rangle=-\left\langle \eth f,g \right\rangle,
\label{integral31}
\end{equation}
It is worth stressing that \eqref{integral27} and \eqref{integral31} indicate that the $\eth$ operator must be conjugated and the sign interchanged, when the eth operator is passed from one member to the other in the inner product.
\section{Spin-weighted Spherical Harmonics $_sY_{lm}$}
When the Legendrian for zero spin-weighted functions \eqref{legend48} is derived $s$-times, one obtains
\begin{equation}
\eth^s\overline{\eth}\eth\ _0Y_{lm}=-l(l+1)\eth^s\ _0Y_{lm}.
\label{sylm1}
\end{equation}
The left hand side of this equation can be transformed using the commutator \eqref{abase31}, namely
\begin{align*}
\eth^s\overline{\eth}\eth\ _0Y_{lm} &=\eth^{s-1}\left(\eth\overline{\eth}\right)\eth\ _0Y_{lm} \\
&=\eth^{s-1}\left(\overline{\eth}\eth -2\right)\eth\ _0Y_{lm}\\
&=\left(\eth^{s-1}\overline{\eth}\eth^2 -2\eth^{s}\right)\ _0Y_{lm}\\
&=\left(\eth^{s-2}\left(\eth\overline{\eth}\right)\eth^2 -2\eth^{s}\right)\ _0Y_{lm}\\
&=\left(\eth^{s-2}\left(\overline{\eth}\eth-4\right)\eth^2 -2\eth^{s}\right)\ _0Y_{lm}\\
&=\left(\eth^{s-2}\overline{\eth}\eth^3 -(2+4)\eth^{s}\right)\ _0Y_{lm}\\
&\vdots\\
&=\left(\overline{\eth}\eth\eth^s-2\sum_{i=1}^{s}i\ \eth^s\right)\ _0Y_{lm}\\
&=\left(\overline{\eth}\eth\eth^s-s(s+1)\eth^s\right)\ _0Y_{lm};
\end{align*}
thus
\begin{equation*}
\left(\overline{\eth}\eth\eth^s-s(s+1)\eth^s\right)\ _0Y_{lm}=-l(l+1)\eth^s\ _0Y_{lm},
\end{equation*}
or
\begin{equation}
\overline{\eth}\eth\eth^s\ _0Y_{lm}=-\left[l(l+1)-s(s+1)\right]\eth^s\ _0Y_{lm}.
\label{sylm2}
\end{equation}
Then, using \eqref{ethgen27}, it is possible to write
\begin{equation}
\eth^s\ _0Y_{lm}=C_{s}\ _{s}Y_{lm},
\label{sylm3}
\end{equation}
where $C_{s}$ is some unknown complex quantity; this equation defines explicitly the spin-weighted spherical harmonics, consequently
\begin{equation*}
C_s\overline{\eth}\eth\ _sY_{lm}=-C_s\left[l(l+1)-s(s+1)\right]\ _sY_{lm},
\end{equation*}
or
\begin{equation}
\overline{\eth}\eth\ _sY_{lm}=-\left[l(l+1)-s(s+1)\right]\ _sY_{lm}.
\label{sylm4}
\end{equation}
Using again the commutator \eqref{abase31} one obtains
\begin{align}
\eth\overline{\eth}\ _sY_{lm}&=-\left(l(l+1)-s(s+1)+2s\right)\ _sY_{lm} \nonumber\\
&=-\left(l(l+1)-s(s-1)\right)\ _sY_{lm}.
\label{sylm5}
\end{align}
Now, writing the last expression as
\begin{align*}
\eth\overline{\eth}\ _{s+1}Y_{lm}&=-\left(l(l+1)-s(s+1)\right)\ _{s+1}Y_{lm}\nonumber\\
&=\eth A_s\ _{s}Y_{lm},
\end{align*}
one then obtains
\begin{align}
A_s \eth \ _{s}Y_{lm}&=-\left(l(l+1)-s(s+1)\right)\ _{s+1}Y_{lm}.
\label{sylm6}
\end{align}
In order to determine the constant $A_s$, the inner product $\langle A_s \eth \ _{s}Y_{lm},A_s \eth \ _{s}Y_{lm}\rangle$ is computed, namely 
\begin{align}
\langle A_s \eth \ _{s}Y_{lm},A_s \eth \ _{s}Y_{lm}\rangle &=| A_s|^2\langle  \eth \ _{s}Y_{lm}, \eth \ _{s}Y_{lm}\rangle \nonumber\\
&=-| A_s|^2\langle \overline{\eth} \eth \ _{s}Y_{lm}, \ _{s}Y_{lm}\rangle \nonumber \\
&=-| A_s|^2\langle -\left[l(l+1)-s(s+1)\right]\ _sY_{lm}, \ _{s}Y_{lm}\rangle \nonumber\\
&=\left[l(l+1)-s(s+1)\right]| A_s|^2\langle \ _sY_{lm}, \ _{s}Y_{lm}\rangle \nonumber\\
&=\left[l(l+1)-s(s+1)\right]| A_s|^2.
\label{sylm7}
\end{align}
where, Equations \eqref{integral27} and \eqref{sylm4} were used in addition to the fact that these basis are orthonormal, i.e.,
\begin{equation*}
\left\langle\ _sY_{l'm'},\ _sY_{lm}\right\rangle=\delta_{ll'}\delta_{mm'}, \hspace{1cm}\forall s\in \mathbb{Z}.
\end{equation*}
When \eqref{sylm6} is used, the same product gives
\begin{align}
&\langle A_s \eth \ _{s}Y_{lm},A_s \eth \ _{s}Y_{lm}\rangle \nonumber\\
=& \langle -\left[l(l+1)-s(s+1)\right]\ _{s+1}Y_{lm},-\left[l(l+1)-s(s+1)\right]\ _{s+1}Y_{lm}\rangle \nonumber\\
=& \left[l(l+1)-s(s+1)\right]^2 \langle \ _{s+1}Y_{lm},\ _{s+1}Y_{lm}\rangle\nonumber\\
=& \left[l(l+1)-s(s+1)\right]^2.
\label{sylm8}
\end{align}
Then, from \eqref{sylm7} and \eqref{sylm8} one obtains
\begin{align*}
| A_s|^2=l(l+1)-s(s+1),
\end{align*}
or
\begin{align}
| A_s|_{\pm}=\pm\left[l(l+1)-s(s+1)\right]^{1/2}.
\label{sylm9}
\end{align}
Making here the choice $A_s=| A_s|_{-}$ and substituting it into \eqref{sylm6} one has
\begin{align}
\eth \ _{s}Y_{lm}&=\left(l(l+1)-s(s+1)\right)^{1/2}\ _{s+1}Y_{lm}.
\label{sylm10}
\end{align}
Also, from \eqref{sylm5} one obtains
\begin{align*}
\overline{\eth}\eth\ _{s-1}Y_{lm}&=-\left[l(l+1)-s(s-1)\right]\ _{s-1}Y_{lm},\\
&=\overline{\eth}A_s\ _{s}Y_{lm},
\end{align*}
i.e.,
\begin{align}
A_s\overline{\eth}\ _{s}Y_{lm}&=-\left[l(l+1)-s(s-1)\right]\ _{s-1}Y_{lm}.
\label{sylm11}
\end{align}
The inner product $\left\langle A_s\overline{\eth}\ _{s}Y_{lm},A_s\overline{\eth}\ _{s}Y_{lm}\right\rangle$ can be computed by using \eqref{integral31} and \eqref{sylm5}, namely
\begin{align*}
&\left\langle A_s\overline{\eth}\ _{s}Y_{lm},A_s\overline{\eth}\ _{s}Y_{lm}\right\rangle \\
=& |A_s|^2\left\langle \overline{\eth}\ _{s}Y_{lm},\overline{\eth}\ _{s}Y_{lm}\right\rangle\\
=& -|A_s|^2\left\langle \eth\overline{\eth}\ _{s}Y_{lm},\ _{s}Y_{lm}\right\rangle\\
=& -|A_s|^2\left\langle -\left(l(l+1)-s(s-1)\right)\ _sY_{lm},\ _{s}Y_{lm}\right\rangle\\
=& |A_s|^2\left(l(l+1)-s(s-1)\right)\left\langle \ _sY_{lm},\ _{s}Y_{lm}\right\rangle\\
=& |A_s|^2\left(l(l+1)-s(s-1)\right);
\end{align*}
and from the right side of \eqref{sylm11} one has
\begin{align*}
&\left\langle A_s\overline{\eth}\ _{s}Y_{lm},A_s\overline{\eth}\ _{s}Y_{lm}\right\rangle \\
=& \left\langle -\left[l(l+1)-s(s-1)\right]\ _{s-1}Y_{lm},-\left[l(l+1)-s(s-1)\right]\ _{s-1}Y_{lm} \right\rangle\\
=& \left[l(l+1)-s(s-1)\right]^2 \left\langle \ _{s-1}Y_{lm},\ _{s-1}Y_{lm} \right\rangle\\
=& \left[l(l+1)-s(s-1)\right]^2. 
\end{align*}
Equating the two last relations one obtains
\begin{align*}
|A_s|^2=\left[l(l+1)-s(s-1)\right]
\end{align*}
or
\begin{align*}
|A_s|_{\pm}=\pm\left[l(l+1)-s(s-1)\right]^{1/2}.
\end{align*}
Thus, making the choice $A_s=|A_s|_+$ and substituting it into \eqref{sylm11} one obtains
\begin{align}
\overline{\eth}\ _{s}Y_{lm}&=-\left[l(l+1)-s(s-1)\right]^{1/2}\ _{s-1}Y_{lm}.
\label{sylm12}
\end{align}
Now, it is possible to re-write \eqref{sylm10} as 
\begin{align}
\eth \ _{s}Y_{lm} &=\left(l^2-s^2+l-s\right)^{1/2}\ _{s+1}Y_{lm}\nonumber\\
&=\left((l+s)(l-s)+l-s\right)^{1/2}\ _{s+1}Y_{lm} \nonumber\\
&=\left((l+s+1)(l-s)\right)^{1/2}\ _{s+1}Y_{lm},
\label{sylm13}
\end{align}
in which one must observe that $s\le l$.
\\Then, from \eqref{sylm3} and \eqref{sylm13} one has
\begin{align}
&\eth^s\ _0Y_{lm}\nonumber\\
=&\eth^{s-1}\eth\ _0Y_{lm} \nonumber\\
=&\eth^{s-1}\left((l+1)l\right)^{1/2}\ _1Y_{lm} \nonumber\\
=&\eth^{s-2}\left((l+2)(l+1)l(l-1)\right)^{1/2}\ _2Y_{lm} \nonumber\\
=&\eth^{s-3}\left((l+3)(l+2)(l+1)l(l-1)(l-2)\right)^{1/2}\ _3Y_{lm} \nonumber\\
\vdots \nonumber\\
=&\left((l+s)\cdots(l+2)(l+1)l(l-1)(l-2)\cdots(l-(s-1))\right)^{1/2}\ _sY_{lm} \nonumber\\
=&\left(\dfrac{(l+s)!}{(l-s)!}\right)^{1/2}\ _sY_{lm};
\label{sylm14}
\end{align}
note that this relation is true if $0\le s\le l$.
\\Also, it is possible to write \eqref{sylm12} as
\begin{align}
\overline{\eth}\ _{s}Y_{lm} 
&=-\left[l^2-s^2+l+s\right]^{1/2}\ _{s-1}Y_{lm}\nonumber\\
&=-\left[(l-s)(l+s)+l+s\right]^{1/2}\ _{s-1}Y_{lm} \nonumber\\
&=-\left[(l+s)(l-s+1)\right]^{1/2}\ _{s-1}Y_{lm},
\label{sylm15}
\end{align}
in which $s\ge -l$.
\\Then, applying $s$ times the $\overline{\eth}$ operator to \eqref{sylm15} one has
\begin{align*}
\overline{\eth}^s\ _{s}Y_{lm}&=\overline{\eth}^{s-1}\overline{\eth}\ _{s}Y_{lm}\nonumber\\
&=-\left[(l+s)(l-s+1)\right]^{1/2}\overline{\eth}^{s-2}\overline{\eth}\ _{s-1}Y_{lm}\nonumber\\
&=(-1)^2\left[(l+s-1)(l+s)(l-s+1)(l-s+2)\right]^{1/2}\overline{\eth}^{s-2}\ _{s-2}Y_{lm},\nonumber\\
\nonumber 
\end{align*}
thus,
\begin{align}
&\overline{\eth}^s\ _{s}Y_{lm}\nonumber\\
&=(-1)^3\left[(l+s-2)(l+s-1)(l+s)\times\right.\nonumber\\
&\hspace{0.5cm}\left.(l-s+1)(l-s+2)(l-s+3)\right]^{1/2}\overline{\eth}^{s-3}\ _{s-3}Y_{lm}\nonumber\\
&\vdots\nonumber\\
&=(-1)^s\left[(l+1)\cdots(l+s-2)(l+s-1)(l+s)(l-s+1)\cdots l\right]^{1/2}\ _{0}Y_{lm}\nonumber\\
&=(-1)^s\left[\dfrac{(l+s)!}{(l-s)!}\right]^{1/2}\ _{0}Y_{lm}.
\label{sylm16}
\end{align}
From \eqref{sylm14} and \eqref{sylm16} the spin-weighted spherical harmonics $_sY_{lm}$ can be defined by
\begin{equation}
_sY_{lm}=\begin{cases}
\left(\dfrac{(l-s)!}{(l+s)!}\right)^{1/2}\eth^s\ _0Y_{lm} & \text{for} \hspace{0.5cm}0\le s\le l \\
(-1)^s\left(\dfrac{(l+s)!}{(l-s)!}\right)^{1/2}\overline{\eth}^{-s}\ _0Y_{lm} & \text{for} \hspace{0.5cm}-l\le s\le 0
\end{cases}
\label{sylm17},
\end{equation}
in which $\eth^{-1}$ ($\overline{\eth}^{-1}$) is the inverse operator of $\eth$ ($\overline{\eth}$), i.e., 
\begin{equation}
\eth \eth^{-1}\equiv 1, \hspace{1cm}  \hspace{1cm}\overline{\eth} \ \overline{\eth}^{-1}\equiv 1,
\label{sylm18}
\end{equation}
such that
\begin{equation}
\left[\eth,\eth^{-1} \right]\ _s\Psi=0, \hspace{1cm}\left[\overline{\eth},\overline{\eth}^{-1} \right]\ _s\Psi=0,
\label{sylm19}
\end{equation}
for all spin-weighted functions.

Also, as an immediate consequence of \eqref{sylm12} and \eqref{sylm15} one has
\begin{align}
\eth\overline{\eth}\ _sY_{lm}&= \eth\left(-\left[l(l+1)-s(s-1)\right]^{1/2}\ _{s-1}Y_{lm}\right) \nonumber\\
&=-\left[l(l+1)-s(s-1)\right] \ _{s}Y_{lm},
\label{sylm20}
\end{align}
and
\begin{align}
\overline{\eth}\eth\ _sY_{lm}&= \overline{\eth}\left(\left(l(l+1)-s(s+1)\right)^{1/2}\ _{s+1}Y_{lm}\right) \nonumber\\
&=-\left[l(l+1)-s(s+1)\right] \ _{s}Y_{lm},
\label{sylm21}
\end{align}
which show that the spin-weighted spherical harmonics $_sY_{lm}$ are eigenfunctions of the $\eth\overline{\eth}$ and $\overline{\eth}\eth$ operators. It is worth noting that \eqref{sylm21} are the generalisation of \eqref{legend48} when the spin-weight is considered.
\section{Spin-weighted Spherical Harmonics $_sZ_{lm}$}
There exists another base of spherical harmonics in which the functions defined on the surface of the sphere can be expanded, namely $_sZ_{lm}$. They are defined as
\begin{equation}
_sZ_{lm}=\begin{cases}
\dfrac{i}{\sqrt{2}}\left((-1)^m\ _sY_{lm}+\ _sY_{l\ -m}\right) &\text{for} \hspace{0.5cm} m<0 \\
_sY_{lm}&\text{for} \hspace{0.5cm} m=0 \\
\dfrac{1}{\sqrt{2}}\left(_sY_{lm}+(-1)^m\ _sY_{l\ -m}\right) &\text{for} \hspace{0.5cm} m>0.
\end{cases}
\label{spin1}
\end{equation}
Since these spherical harmonics are constructed from linear combinations of $_sY_{lm}$, then they are also eigenfunctions of the $\eth\overline{\eth}$ operator. Also, they are orthonormal \cite{ZGHLW03}.
\\In order to show this, the $_sZ_{lm}$ are written as
\begin{equation}
_sZ_{lm}=A_{lms}\ _sY_{lm}+B_{lms}\ _sY_{l\ -m},\hspace{0.5cm} \text{ for all }\ m,
\end{equation}
therefore
\begin{align*}
	\left\langle \ _sZ_{lm}, \ _sZ_{l'm'}\right\rangle
	=&\left(\overline{A}_{lms}A_{l'm's} + \overline{B}_{lms}B_{l'm's}\right)\delta_{ll'}\delta_{mm'}.
\end{align*}
Evaluating the constants $A_{lms}$ and $B_{lms}$ from \eqref{spin1}, it is possible to write
\begin{align*}
\left\langle \ _sZ_{lm}, \ _sZ_{l'm'}\right\rangle&=\int_{\Omega} d\Omega \ _sZ_{lm} \ _s\overline{Z}_{l'm'} \\
&=\delta_{ll'}\delta_{mm'}.
\label{orto_sZlm}
\end{align*}
Also, they are complete, in exactly the same form as the $_sY_{lm}$, i.e.,
\begin{equation}
\sum_{l=0}^{\infty}\sum_{m=-l}^{l} \ _sZ_{lm}(\theta,\phi) \ _s\overline{Z}_{lm}(\theta',\phi')=\delta(\phi-\phi')\delta(\cos(\theta)-\cos(\theta')).
\label{compl_sZlm}
\end{equation}
This expression is proved in a straightforward way, if it is assumed that any angular function $_s\Psi$ with spin-weight $s$ can be expanded in terms of $_sZ_{lm}$, i.e.,
\begin{equation}
_s\Psi =\sum_{l=0}^{\infty}\sum_{m=-l}^{l} \ _s\Psi_{lm}\ _sZ_{lm}.
\label{exp_sZlm}
\end{equation}
then, the coefficients $_s\Psi_{lm}$ are given by
\begin{equation}
\ _s\Psi_{lm}=\int d\Omega \ _s\overline{Z}_{lm}\ _s\Psi.
\label{compl_sZlm2}
\end{equation}
Substituting \eqref{compl_sZlm2} into \eqref{exp_sZlm} one obtains
\begin{align}
_s\Psi(\theta,\phi) &= \int d\Omega' \ \sum_{l=0}^{\infty}\sum_{m=-l}^{l}  \ _s\overline{Z}_{lm}(\theta',\phi')\ _sZ_{lm}(\theta,\phi) \ _s\Psi(\theta',\phi') \nonumber\\
&= \int d\Omega' \delta(\phi-\phi')\delta(\cos(\theta)-\cos(\theta')) \ _s\Psi(\theta',\phi').
\label{compl_sZlm3}
\end{align}
The $_sZ_{lm}$ spherical harmonics will be important because the Einstein's field equations can be re-expressed  in term of them. The reason to do that, is that the $_sZ_{lm}$ decouple the $m$ mode in the field equations. 
\chapter[THE NON-LINEAR REGIME]{THE INITIAL VALUE PROBLEM AND THE NON-LINEAR REGIME OF THE EINSTEIN'S FIELD EQUATIONS}
This chapter considers the IVP ({\it Initial Value Problem}) in the general relativity context. Essentially, there are three distinct kinds of formulations to evolve a given space-time. The Regge calculus, the  ADM ({\it Arnowitt-Desser-Misner}) or $3+1$ formulations, and the characteristic or null-cone formalisms.  The null cones in these last formalisms can be oriented to the past, to the future or in both directions\footnote{Ingoing, Outgoing and Bi-characteristic null-cone formalisms}.

Here  only the two last formulations are shown, namely, the ADM based and the characteristic formulations. In particular the emphasis lies on the null cone oriented to the future formulation.  In order to do that, this chapter is organised as follows. In the first section the initial value problem is present. Subsequently, some aspects of the ADM formulations are briefly shown. Finally, the principal aspects of the outgoing characteristic formulation are present. 
\section{The Initial Value Problem}
The initial value problem (IVP) consists, essentially, in the evolution of a space-time characterised by a given metric $g_{\mu\nu}$. Here, $g_{\mu\nu}$ and its first derivatives, $g_{\mu\nu,\gamma}$, are specified in an initial three dimensional hypersurface corresponding to $t=t_0$. The evolution of the space-time is then performed using the Einstein's field equations. In addition, in some cases the matter sources are evolved from the conservation laws. The conserved quantities are used to constrain the system of equations, reducing in this manner the degrees of freedom of these physical systems. One example of this is the imposition of specific symmetries, such as axial or reflection symmetries. 

There are several versions of the initial value problem. For example, in the $3+1$ based formulations, which correspond to Hamiltonian formulations of the general relativity, the metric and its derivatives must satisfy certain boundary conditions during the evolution and satisfy some initial conditions in order to start the iteration. Another example is the characteristic initial value problem in which the initial data is specified  on a time-like world tube and on an initial null hypersurface, for which $u=u_0$, where $u$ indicates retarded time. A last example corresponds to the CCM ({\it Cauchy-Characteristic Matching}) formalism in which ADM and Characteristic formulation are used. In this formalism the metric and its derivatives are specified across a world tube which separates the space-time into two distinct regions. The initial conditions are given for the interior of the world tube starting an ADM based evolution, then the boundary conditions generated onto the world tube are used as initial conditions to start a characteristic outgoing evolution which propagate the gravitational radiation to the null infinity.   
\section{Arnowitt-Desser-Misner Formulations (ADM)}
\graphicspath{{./Figuras/ADM/}}
In this section two of the most used ADM based formulations in numerical relativity applications are presented, the ADM formalism and the BSSN ({\it Baumgarte-Shibata-Shapiro-Nakamura}) formulation. The ADM/BSSN equations and their derivations are presented in some detail. In the ADM based formalisms,  the space-time is foliated into space-like hypersurfaces, which are orthogonal to a time-like geodesic, parametrised by an affine parameter $t$.  The BSSN formulation furnishes simulations that result more stable than those based on the original ADM. The constraints and the evolution equations for the metric of the hypersurfaces are given in detail.
\subsection{({\it ADM}) formalism}
\begin{figure}[h!]		
\begin{center}		
	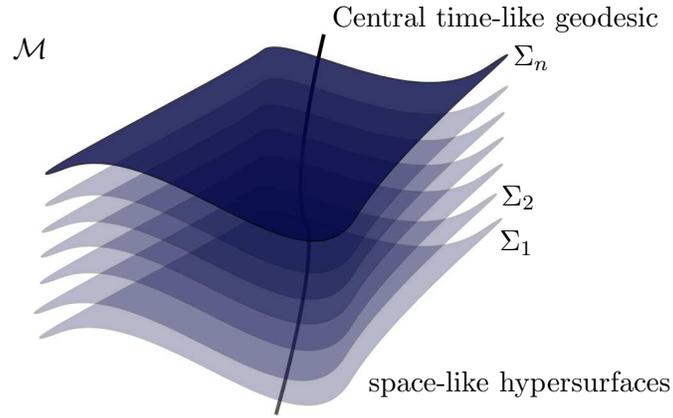
	\caption{Space-time $\mathcal M$ foliated in 3D - hypersurfaces $\Sigma$.}	
	\label{FADM1}	
\end{center}	
\end{figure}

It is supposed that the manifold $\mathcal M$ represents the space-time. $\mathcal M$ is associated with the metric $g_{\mu\nu}$. The space-time is foliated into 3-dimensional space-like hypersurfaces labelled by $\Sigma$, which are orthogonal to the vector $\Omega^\mu$ (at least locally).  $\Omega^\mu$ is defined as the tangent vectors to a central time-like geodesic, in the form
\begin{equation}\label{ADM1}
\Omega_\mu= t_{;\mu}.
\end{equation}
Here, $t$ can be interpreted as a {\it global time}. Also, this time $t$ corresponds to an affine parameter to the arc length described by the central geodesic \cite{A08_book,BS10_Book}. Recall that the intersections between the hypersurfaces $\Sigma$ are forbidden. See Figure \ref{FADM1}
\graphicspath{{./Figuras/ADM/}}
\\The norm $\|\Omega_\mu\|$ is computed from \eqref{ADM1}, namely
\begin{equation}\label{ADM2}
\|\Omega_\mu\|^2= g^{\mu\nu} t_{;\mu} t_{;\nu}.
\end{equation}
From \eqref{ADM2} a scalar function $\alpha$, {\it the lapse function}, is defined such that
\begin{equation}\label{ADM3}
\alpha^ 2=-\frac{1}{\|\Omega_\mu\|^2}.
\end{equation}
\newpage
Thus, $\alpha > 0$ means that $\Omega^\mu$ is a time-like vector. Then at least locally the hypersurfaces $\Sigma$ will be space-like. On other hand, $\alpha < 0$ means $\Omega^\mu$ is space-like. Thus, at least locally, the hypersurfaces $\Sigma$ will be time-like. It measures the lapse between two successive hypersurfaces when measured by an {\it Eulerian observer}\footnote{Namely also {\it Normal observers}, which are moving in normal direction to these hypersurfaces $\Sigma$.}.  

A normalised and irrotational one-form $\omega_\mu=\alpha\Omega_\mu$, is also defined, i.e.,
\begin{equation}\label{ADM4}
\omega_{[\nu}\omega_{\mu;\delta]}=0.
\end{equation}
From the 1-forms $\omega_\mu$ the normal vectors to the hypersurfaces $\Sigma$ can be built as
\begin{equation}\label{ADM7}
n^\nu=-g^{\mu\nu}\omega_\mu,
\end{equation}
where the minus indicates that these vectors are oriented to the future, i.e., they are pointed in the sense in which $t$ increases. Also, the one-forms $\omega_\mu$ and the vectors $n^\nu$ satisfy 
\begin{equation}\label{ADM8}
n^\nu\omega_\nu=-g^{\mu\nu}\omega_\mu\omega_\nu=-1, \hspace{0.5cm} n^\nu n_\nu=1.
\end{equation}

The metric $\gamma_{\mu\nu}$ corresponding to the hypersurfaces $\Sigma$, is  the spacial part of $g_{\mu\nu}$, thus, 
\begin{equation}\label{ADM9}
\gamma_{\mu\nu}=g_{\mu\nu}+n_\mu n_\nu.
\end{equation}
Note that $n^{\mu}\gamma_{\mu\nu}=0$ indicates that $n^{\mu}$ is a normal vector to $\Sigma$. The inverse metric $\gamma^{\mu\nu}$ is given by
\begin{equation}\label{ADM10}
\gamma^{\mu\nu}=g^{\mu\nu}+n^\mu n^\nu.
\end{equation}
From \eqref{ADM9} one  obtains the following projection tensor 
\begin{equation}\label{ADM11}
\gamma^{\mu}_{~\nu}=\delta^{\mu}_{~\nu}+n^\mu n_\nu.
\end{equation}
Then, the tensor that projects in the normal direction to the hypersurfaces is given by 
\begin{equation}\label{ADM12}
N^{\mu}_{~\nu}=-n^\mu n_\nu.
\end{equation}

The covariant derivative compatible\footnote{Compatible means $^3\nabla_\delta \gamma_{\mu\nu}=0$.} with $\gamma_{\mu\nu}$, is obtained from the projection of $\nabla_\mu$ on the hypersurfaces $\Sigma$, namely
\begin{equation}\label{ADM13}
^3\nabla_\nu=-\gamma^\mu_{~\nu} \nabla_\mu.
\end{equation}
These three-dimensional covariant derivatives are expressed in terms of the connection coefficients associated with the hypersurfaces $\Sigma$, i.e.,
\begin{equation}\label{ADM14}
^3\Gamma^\mu_{~\nu\delta}=\frac{1}{2}\gamma^{\mu\theta}(\gamma_{\theta\nu,\delta}+\gamma_{\theta\delta,\nu}-\gamma_{\nu\delta,\theta}).
\end{equation}
On the other hand, the Riemann tensor $^3R^{\gamma}_{~\delta\nu\mu}$ associated to the metric $\gamma_{\mu\nu}$ is defined by
\begin{equation}\label{ADM15}
2 \ ^3\nabla_{[\mu} \ ^3\nabla_{\nu]}v_\delta =\ ^3R^{\gamma}_{~\delta\nu\mu}v_\gamma\hspace{0.5cm} \text{ and }\hspace{0.5cm} ^3R^{\gamma}_{~\delta\nu\mu}n_\gamma=0,
\end{equation}
which are satisfied by any space-like $v_{\gamma}$  and any time-like 1-forms $n_\gamma$. Then, from \eqref{ADM15}, the Riemann tensor $^3R^{\gamma}_{~\delta\nu\mu}$ is defined from the Christoffel symbols $^3\Gamma^\mu_{~\nu\delta}$ as follows
\begin{equation}\label{ADM16}
^3R^{\gamma}_{~\delta\nu\mu}=\ ^3\Gamma^\gamma_{~\delta\mu,\nu} - \ ^3\Gamma^\gamma_{~\nu\mu,\delta} + \ ^3\Gamma^\gamma_{~\sigma\nu} \ ^3\Gamma^\sigma_{~\delta\mu} - \ ^3\Gamma^\gamma_{~\sigma\delta}\ ^3\Gamma^\sigma_{~\nu\mu}.
\end{equation}
The expressions for the Ricci's tensor $^3R_{\mu\nu}=\ ^3R^\gamma_{~\mu\gamma\nu}$ and for the scalar of curvature $^3R=\ ^3R^\mu_{~\mu}$ are obtained from \eqref{ADM16}. 

The 3-dimensional Riemann tensor $^3R^{\gamma}_{~\delta\nu\mu}$ contains only pure spacial information. Then, all quantities derived from it will contain information about the intrinsic curvature of the hypersurfaces $\Sigma$. Thus, it will be necessary to introduce at least one more geometric object to take into account the extrinsic curvature, $K_{\mu\nu}$. This tensor is defined from the projection of the covariant derivatives of the normal vectors onto the hypersurfaces $\Sigma$. Such projections can be decomposed into a symmetric and antisymmetric part, as follows
\begin{eqnarray}\label{ADM17}
\gamma^\beta_{~\delta} \gamma^\alpha_{~\nu}  n_{\alpha;\beta} &=& \gamma^\beta_{~\delta} \gamma^\alpha_{~\nu}  n_{(\alpha;\beta)} + \gamma^\beta_{~\delta} \gamma^\alpha_{~\nu}  n_{[\alpha;\beta]},\nonumber\\
&=&\Theta_{\delta\nu}+\omega_{\delta\nu},
\end{eqnarray}
where $\Theta_{\delta\nu}(\omega_{\delta\nu})$ corresponds to its symmetric (antisymmetric). $\Theta_{\delta\nu}(\omega_{\delta\nu})$ is known as the expansion tensor (rotational 2-form). Note that, given \eqref{ADM4}, $\omega_{\delta\nu}=0$. 
Thus, the extrinsic curvature is defined as
\begin{eqnarray}\label{ADM19}
K_{\mu\nu}&=&-\gamma^\beta_{~\delta} \gamma^\alpha_{~\nu}  n_{\alpha;\beta}, \nonumber\\
&=&-\frac{1}{2}\mathcal L_{\mathbf n} \gamma_{\mu\nu},
\end{eqnarray}
where $\mathcal L_{\mathbf n} \gamma_{\mu\nu}$ is the Lie derivative of $\gamma_{\mu\nu}$ along the vector field $\mathbf n=n^\alpha \mathbf e_\alpha$. Here, $\mathbf e_\alpha$ is any base, which $\mathbf e_\alpha=\partial_\alpha$ when a local coordinate basis is considered.

\begin{figure}[h!]\label{figADM1}
	\begin{center}
		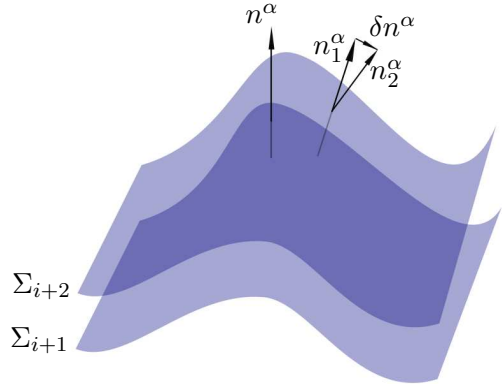
		\caption{Change of the normal vectors to $\Sigma$. The difference $\delta n^a$ only provides information about the change in the direction of the vectors, because they are normalised.}
		\label{FADM2}
	\end{center}
\end{figure}

Note that the extrinsic curvature is symmetric and only spacial and it furnishes information on how much the normal vectors to $\Sigma$ change their directions. Figure \ref{FADM2} shows the change of the normal vectors to the hypersurfaces $\Sigma$. These normal vectors are referred to two distinct and nearly hypersurfaces $\Sigma_{i+1}$ and $\Sigma_{i+2}$. 

The extrinsic curvature $K_{\mu\nu}$ and the metric $g_{\mu\nu}$ give information about the  state of the gravitational field at each instant of time. Consequently, it is possible to do the analogy with the classical mechanics. $K_{\mu\nu}$ is analogue to the velocities, whereas $g_{\mu\nu}$ to the positions in a given set of particles.

The projection of $R_{\alpha\beta\xi\varphi}$ associated with $g_{\mu\nu}$ on $\Sigma$, are related to $K_{\mu\nu}$ and $^3R_{\mu\nu\eta\delta}$, through 
\begin{equation}\label{ADM20}
^3R_{\mu\nu\eta\delta}+K_{\mu\eta}K_{\nu\delta}-K_{\mu\delta}K_{\eta\nu}=\gamma^\alpha_{~\mu} \gamma^\beta_{~\nu} \gamma^\xi_{~\eta} \gamma^\varphi_{~\delta} R_{\alpha\beta\xi\varphi}.
\end{equation}
which is known as the {\it Gauss equation}.

The projection $\gamma^\alpha_{~\mu} \gamma^\beta_{~\eta} \gamma^\xi_{~\nu} n^\varphi R_{\alpha\beta\xi\varphi}$ depends only on the derivatives of $K_{\mu\nu}$. These quantities are functions only of  $\gamma_{\mu\nu}$ and its derivatives, thus
\begin{equation}\label{ADM21}
^3\nabla_\eta K_{\mu\nu}-\ ^3\nabla_\mu K_{\eta\nu}=\gamma^\alpha_{~\mu} \gamma^\beta_{~\eta} \gamma^\xi_{~\nu} n^\varphi R_{\alpha\beta\xi\varphi},
\end{equation}
which in known as {\it Codazzi equation}.

Both \eqref{ADM20} and \eqref{ADM21} lead to the constraint equations, providing the integrability conditions that are propagated along the evolution. The hypersurfaces $\Sigma$ carry the information about $K_{\mu\nu}$ and $\gamma_{\mu\nu}$.

On the other hand, from  the Lie derivative of the extrinsic curvature $K_{\mu\nu}$  along $n^\alpha$, one obtains 
\begin{equation}\label{ADM22}
\mathcal L_{\mathbf n}K_{\mu\nu}=n^\alpha n^\beta\gamma_{~\mu}^\sigma\gamma_{~\nu}^\varphi R_{\alpha\beta\sigma\varphi} -\frac{1}{\alpha}\ ^3\nabla_\mu \ ^3\nabla_\nu \alpha -K_{~\nu}^\sigma K_{\mu\sigma},
\end{equation}
which is known as the {\it Ricci equation}. This equation  expresses the temporal changes in $K_{\mu\nu}$ as a function of $R_{\alpha\beta\sigma\varphi}$, with two of their indices projected in the direction of the time.

Now, contracting the Gauss equation \eqref{ADM20} one obtains \cite{ADM59}
\begin{equation}\label{ADM23}
\gamma^{\alpha\nu} \gamma^{\beta\mu} R_{\alpha\beta\nu\mu}=\ ^3R+K^2-K_{\sigma\varphi}K^{\sigma\varphi},
\end{equation}
where the trace of the extrinsic curvature is $K=\gamma^{\alpha\beta}K_{\alpha\beta}$. From the Einstein's tensor 
\begin{equation}\label{ADM24}
G_{\mu\nu}=R_{\mu\nu}-\frac{1}{2}g_{\mu\nu}R,
\end{equation}
one has
\begin{equation}\label{ADM25}
2n^\mu n^\nu G_{\mu\nu}=\gamma^{\alpha\mu} \gamma^{\beta\nu} R_{\alpha\beta\mu\nu}.
\end{equation}
Therefore \eqref{ADM23} becomes
\begin{equation}\label{ADM26}
2n^\mu n^\nu G_{\mu\nu}=\ ^3R+K^2-K_{\sigma\varphi}K^{\sigma\varphi}.
\end{equation}
If the energy density $\rho$ is defined  as the total energy density as measured by an Eulerian observer, i.e., 
\begin{equation}\label{ADM27}
\rho=n_\mu n_\nu T^{\mu\nu},
\end{equation}
then the projection of the Einstein's field equations \eqref{eq2.24} on the normal vectors to the hypersurfaces $\Sigma$ reads
\begin{equation}\label{ADM28}
^3R+K^2 -K_{\mu\nu}K^{\mu\nu}=16\pi\rho,
\end{equation}
which is a {\it Hamiltonian constrain equation}.

Now, contracting the Codazzi equation one obtains
\begin{equation}\label{ADM29}
^3\nabla_{\varphi} K_\sigma^{~\varphi}  -\ ^3\nabla_{\sigma} K=\gamma^{\alpha}_{~\sigma} \gamma^{\beta\mu}n^\nu R_{\alpha\beta\mu\nu}.
\end{equation}
However, from the Einstein's tensor one has 
\begin{equation}\label{ADM30}
\gamma^{\mu}_{~\sigma} n^{\nu}G_{\mu\nu}=\gamma^{\mu}_{~\sigma} n^{\nu}R_{\mu\nu}.
\end{equation}
Consequently, the Codazzi equation takes the form
\begin{equation}\label{ADM31}
^3\nabla_{\varphi} K_\sigma^{~\varphi}  -\ ^3\nabla_{\sigma} K=8\pi S_\sigma,
\end{equation}
where 
\begin{equation}\label{ADM32}
S_\sigma=-\gamma^\mu_{~\sigma} n^\nu T_{\mu\nu},
\end{equation}
which corresponds to the momentum density as measured by an Eulerian observer. Equation \eqref{ADM31} is usually referred as to {\it the momentum constrain}.

\begin{figure}[h!]
	\begin{center}
		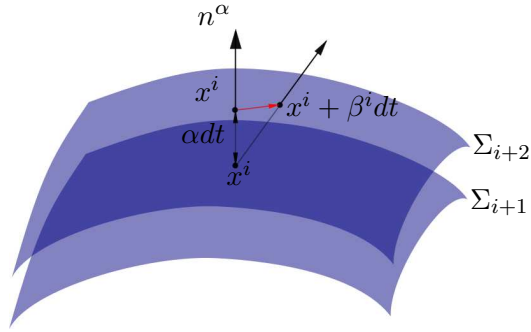
		\caption{Representation of two successive hypersurfaces and the displacement vector $\beta^\mu$ (red).}
		\label{FADM3}
	\end{center}	
\end{figure}

Now, defining a vector $t^\mu$ as follows
\begin{equation}\label{ADM33}
t^\mu=\alpha n^\mu + \beta^\mu,
\end{equation}
where $\beta^\alpha$ is the {\it displacement vector}. This vector indicates the displacement of the Eulerian observers between two successive hypersurfaces (see  Figure \ref{FADM3}). 

Note that the vector $t^\alpha$ is dual to the one-form $\Omega_\alpha$. Thus, from the extrinsic curvature  $K_{\mu\nu}$ one obtains
\begin{equation}\label{ADM34}
\mathcal L_{\mathbf t}\gamma_{\mu\nu}=-2\alpha K_{\mu\nu}+\mathcal L_\beta \gamma_{\mu\nu},
\end{equation}
which is the {\it evolution equation for the metric} $\gamma_{\mu\nu}$ associated with the hypersurfaces $\Sigma$. Taken the Lie derivative of the extrinsic curvature $K_{\mu\nu}$ along $t^a$ one has the following evolution equation 
\begin{equation}\label{ADM35}
\mathcal L_{\mathbf t}K_{\mu\nu} =\alpha\mathcal L_{\mathbf n}K_{\mu\nu}+\mathcal L_{\beta}K_{\mu\nu}.
\end{equation}

However, from the Ricci's equation \eqref{ADM22} and from the Einstein's field equations \eqref{eq2.24} results
\begin{equation}\label{ADM36}
n^\alpha n^\delta\gamma_{~\mu}^\varepsilon\gamma_{~\nu}^\beta R_{\alpha\beta\delta\varepsilon}=\ ^3R_{\mu\nu}+KK_{\mu\nu}-K_{\mu\sigma}K^\sigma_{~\nu}-8\pi \gamma_{~\mu}^\sigma\gamma_{~\nu}^\varphi \left(T_{\sigma\varphi}-\frac{1}{2}g_{\sigma\varphi}T\right),
\end{equation}
where $T$ is the trace of the stress-energy tensor $T=g^{\mu\nu}T_{\mu\nu}$. Defining the spacial part of $T_{\mu\nu}$ and its trace respectively from  
\begin{equation}\label{ADM37}
S_{\mu\nu}=\gamma_{~\mu}^\sigma\gamma_{~\nu}^\varepsilon T_{\sigma\varepsilon} \hspace{0.5cm}\text{and}\hspace{0.5cm}S=S_{~\mu}^\mu,
\end{equation}
and substituting into \eqref{ADM35} one obtains
\begin{eqnarray}\label{ADM38}
\mathcal L_{\mathbf t}K_{\mu\nu}&=&-\ ^3\nabla_\mu \ ^3\nabla_\nu \alpha + \alpha(\ ^3R_{\mu\nu}-2K_{\mu\sigma}K_{~\nu}^\sigma+KK_{\mu\nu})\nonumber\\&& -8\pi\alpha\left(S_{\mu\nu}-\frac{1}{2}\gamma_{\mu\nu}(S-\rho)\right) +\mathcal L_{\beta}K_{\mu\nu}.
\end{eqnarray}
In \eqref{ADM38} all the differential operators as well as the Ricci's tensor are associated to  $\gamma_{\mu\nu}$. The evolution equations given in \eqref{ADM34} and \eqref{ADM38} are coupled and they determine the evolution of $\gamma_{\mu\nu}$ and $K_{\mu\nu}$. These equations together with  the Hamiltonian and momentum constraints  contain the same information present in the Einstein's field equations. 
Furthermore, from these equations it is possible to observe that the differential equations that govern the matter and the space-time dynamics are  differential equations of first order in time. In this sense, they are different from the original field equations, which are of second order. As in any initial value problem, the evolution equations must conserve the constrain equations, therefore, if $\gamma_{\mu\nu}$ and $K_{\mu\nu}$ satisfy the constrain equations, in some hypersurface $\Sigma_i$, then the same constrains must be satisfied along the all temporal evolution, i.e. this conditions must be satisfied for all the hypersurfaces $\Sigma$ in which the space-time is foliated.

At last, specifying the vector $t^\mu=(1,0,0,0)$, and introducing a 3-dimensional basis vectors $e^\mu_{~(i)}$, where $i$ indicates each of three vectors and taking into account that $\Omega_\mu e^\mu_{~(i)}=0$, then it is possible to make the choice that the spatial components of $n_i=0$. Consequently, the displacement vector contains only spacial components, i.e., $\beta^\mu=(0,\beta^i)$, and therefore the normal vectors to the hypersurfaces read $n^\mu=\alpha^{-1}(1,\beta^i)$. Therefore the metric of the space-time can be represented by the matrix
\begin{equation}\label{ADM39}
g_{\mu\nu}=\begin{pmatrix}
-\alpha^2+\beta_k\beta^k & \beta_i \\
\beta_j & \gamma_{ij}
\end{pmatrix},
\end{equation}
or in the form of line element 
\begin{equation}\label{ADM40}
ds^2=-\alpha^2dt^2+\gamma_{ij}(dx^i+\beta^idt)(dx^j+\beta^jdt),
\end{equation}
which is usually known as the {\it line element in the 3+1 form}. 
\subsection{The Baumgarte-Shibata-Shapiro-Nakamura ({\it BSSN}) Equations}
A variant of the ADM formalism is the Baumgarte-Shibata-Shapiro-Nakamura (BSSN) formalism \cite{BS99,SN95}. In this formalism, the metric $\gamma_{ij}$ associated with the hypersurfaces $\Sigma$ is conformal to the metric  $\tilde{\gamma}_{ij}$ and the conformal factor is given by  $e^{i\phi}$, i.e.,
\begin{equation}\label{BSSN1}
	\gamma_{ij}=e^{i\phi}\tilde{\gamma}_{ij},\hspace{1cm}\|\tilde{\gamma}_{ij}\|=1.
\end{equation} 

The fundamental idea is to introduce this conformal factor and evolve both, separately, the conformal factor and the metric. This procedure simplifies the Ricci's tensor and simplifies the numerical codes. In order to obtain the evolution equations, the extrinsic curvature $K_{ij}$ is decomposed into its trace $K$, and the trace-free part, $A_{ij}$, namely 
\begin{equation}\label{BSSN2}
	K_{ij}=A_{ij}+\frac{1}{3}\gamma_{ij}K.
\end{equation}
In addition, $A_{ij}$ is expressed in terms of a trace-free conformal curvature, i.e., 
\begin{equation}\label{BSSN3}
	A_{ij}=e^{i\phi}\tilde{A}_{ij}.
\end{equation}

Contracting the evolution equation for $\gamma_{ij}$ \eqref{ADM35}, one obtains
\begin{equation}\label{BSSN4}
	\partial_t \ln \gamma^{1/2}=\alpha K +\ ^3\nabla_i\beta^i,
\end{equation}
and using \eqref{BSSN1}, results in an evolution equation for $\phi$, namely
\begin{equation}\label{BSSN5}
	\partial_t \phi=-\frac{1}{6}\alpha K + \partial_i\beta^i +\beta^i\partial_i \phi.
\end{equation}

Also, contracting the evolution equation for the extrinsic curvature \eqref{ADM38} one obtains 
\begin{equation}\label{BSSN6}
	\partial_t K=-\ ^3\nabla^2\alpha+\alpha(K_{ij}K^{ij}+4\pi(\rho+S))+\beta^i\ ^3\nabla_iK,
\end{equation}
where 
$$^3\nabla^2 =\gamma^{ij} \ ^3\nabla_i\ ^3\nabla_j,$$ 
such that, substituting \eqref{BSSN2} and using \eqref{BSSN3} one has
\begin{equation}\label{BSSN7}
	\partial_t K=-\ ^3\nabla^2\alpha+\alpha\left(\tilde A_{ij}\tilde A^{ij}+\frac{1}{3}K^2\right)+4\pi\alpha(\rho+S)+\beta^i\partial_i K.
\end{equation}
Subtracting \eqref{BSSN5} from \eqref{ADM35} one obtains the evolution equation for $\tilde \gamma_{ij}$, i.e.,
\begin{equation}\label{BSSN8}
	\partial_t\tilde \gamma_{ij}=-2\alpha \tilde A_{ij}+\beta^k\partial_k\tilde\gamma_{ij}+\tilde \gamma_{kj}\partial_i\beta^k-\frac{2}{3}\tilde \gamma_{ij}\partial_k\beta^k,
\end{equation}
also, subtracting \eqref{BSSN7} from \eqref{ADM38} results in the evolution equation for $\tilde A_{ij}$, namely
\begin{eqnarray}\label{BSSN9}
	\partial_t \tilde A_{ij}&=&e^{4\phi}\left(-(\ ^3\nabla_i \ ^3\nabla_j\alpha)^{\text{TF}}+\alpha(R_{ij}^{\text{TF}}-8\pi S_{ij}^{\text{TF}})\right)+\alpha(K\tilde A_{ij}-2\tilde A_{in}\tilde A^n_{~j})\nonumber\\
	&&+\beta^k\partial_k\tilde A_{ij}+\tilde A_{ik}\partial_j B^k+\tilde A_{kj}\partial_i B^k -\frac{2}{3}\tilde A_{ij}\partial_k\beta^k,
\end{eqnarray}
where the superscript TF indicates trace-free, i.e,
\begin{equation}
	^3R_{ij}^{\text{TF}}=\ ^3R_{ij}-\frac{1}{3}\gamma_{ij}\ ^3R,\hspace{0.5cm} S_{ij}^{\text{TF}}=S_{ij}-\frac{1}{3}\gamma_{ij}\ ^3R,
\end{equation}
and
\begin{equation}
	(\ ^3\nabla_i \ ^3\nabla_j\alpha)^{\text{TF}}=(\ ^3\nabla_i \ ^3\nabla_j\alpha)-\frac{1}{3}\gamma_{ij}(\ ^3\nabla^2\alpha).
\end{equation}

Now, in terms of these variables, the momentum constrain becomes 
\begin{eqnarray}\label{BSSN10}
	\gamma^{ij}\ ^3\tilde\nabla_i\ ^ 3\tilde\nabla_j e^\phi-\frac{1}{8}e^\phi \ ^3\tilde R+\frac{1}{8}e^{5\phi}\tilde A_{ij}\tilde A^{ij}-\frac{1}{12}e^{5\phi}K^2+2\pi e^{5\phi}\rho=0,
\end{eqnarray}
where the operator $^3\tilde\nabla_i=e^{i\phi}\ ^3\nabla_i$, is the hamiltonian constrain, i.e., 
\begin{eqnarray}\label{BSSN11}
	^3\tilde\nabla_j \left(\tilde A^{ji}e^{6\phi}\right)-\frac{2}{3}e^{6\phi} \ ^3\tilde \nabla^i K-8\pi e^{6\phi}S^i=0.
\end{eqnarray}

\section{Outgoing Characteristic Formulation}
\graphicspath{{./Figuras/Characteristic/}}
In this section one of the characteristic formalisms will be described, in which the space-time is foliated into null cones oriented to the future. In order to do so, the Bondi-Sachs metric and the characteristic initial value problem are described, subsequently the non-linear field equations in the characteristic formalism are presented and we finish this section rewriting these equations using the eth formalism previously described.
\subsection{The Bondi-Sachs Metric}
\begin{figure}[h!]
	\begin{center}
		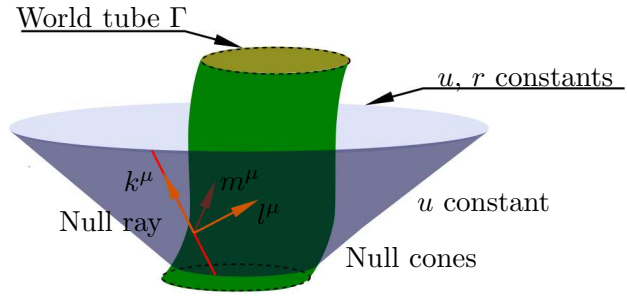
		\caption{Null coordinates construction. Tangent and normal vectors to the null hypersurfaces emanated from a time-like world tube $\Gamma$.}
		\label{FBS1}
	\end{center}
\end{figure}

\citeonline{B_VII,B_VIII} in their remarkable work describe in detail the radiation coordinates construction. Here, these details are reviewed in order to understand the metric and its metric functions. Thus, it is supposed that the manifold $\mathcal{M}$ is doted of a metric tensor such that $g_{\mu\nu}:=g_{\mu\nu}(x^\alpha)$ and have a signature $+2$. We assume a generic scalar function that depends on these unknown and arbitrary coordinates $u:=u(x^\mu)$, such that
\begin{equation}
u_{,\mu}u^{,\mu}=0.
\label{bs1}
\end{equation}
Thus, denoting by $k^\mu=u_{,\nu}g^{\nu\mu}$, one has
\begin{equation}
k_\mu k^\mu=0.
\label{bs2}
\end{equation} 
The hypersurfaces for constant $u$ are null; and its normal vectors $k^\mu$ also satisfy
\begin{equation}
k_{;\mu} k^\mu=0.
\label{bs3}
\end{equation} 
Thus, the lines whose tangent is described by $k^\mu$ are called rays (see Figure \ref{FBS1}). 
\newpage
From \eqref{bs3}, the congruence of rays of null geodesic are also normal to the hypersurfaces for $u$ constant, thus these rays lie on the hypersurfaces and on the normal plane to the null hypersurfaces. The parameter $u$ must be such that the expansion $\xi$ and the shear $\sigma$ of the congruences, formed by these rays (the null cones) satisfy
\begin{equation}
\xi=\dfrac{k^\alpha_{~;\alpha}}{2}\ne 0,\hspace{1cm}|\sigma|^2=\dfrac{k_{\alpha;\beta}k^{\alpha;\beta}}{2}-\rho^2\ne\rho^2.
\end{equation}
It is assumed that $u$ satisfies these conditions for any coordinate system. The parameter $u$ will be selected as the retarded time. The scalar functions $\theta:=\theta(x^\alpha)$, $\phi:=\phi(x^\alpha)$ can be selected such that
\begin{equation}
\theta_{,\alpha}k^\alpha=\phi_{,\alpha}k^\alpha=0,\hspace{0.5cm}\theta_{,\alpha}\theta_{,\beta}\theta_{,\gamma}\theta_{,\delta}g^{\alpha\beta}g^{\gamma\delta}-\left(\theta_{,\alpha}\theta_{,\beta}g^{\alpha\beta}\right)^2=D\ne 0,
\end{equation}
where $D>0$. Thus $\theta$ and $\phi$ are constants along each ray, and therefore, can be identified as optical angles. In addition, it is possible to choose the scalar function $r:=r(x^\alpha)$, such that
\begin{equation}
r^4=\dfrac{1}{D\sin^2\theta},
\end{equation}
in which case $r$ is the luminosity distance, defining hypersurfaces for $u,r$ constants such that its area is exactly $4\pi r^ 2$. Defining $x^\mu=(u,r,\theta,\phi)$ as coordinates with $\mu=1,2,3,4$, and $x^A=(\theta,\phi)$ with $A=3,4$, then the line element that satisfy above conditions reads
\begin{align}
ds^2=&-\left(\frac{Ve^{2\beta}}{r}-r^2h_{AB}U^AU^B\right)du^2-2e^{2\beta}dudr-2r^2h_{AB}U^Bdudx^A\nonumber\\
&+r^2h_{AB}dx^Adx^B,
\label{BS11}
\end{align}
which can be written conveniently as
\begin{align}
ds^2=&-\left(\frac{Ve^{2\beta}}{r}\right)du^2-2e^{2\beta}dudr+r^2h_{AB}\left(U^Adu -dx^A\right)\left(U^Bdu-dx^B\right),
\label{BS22}
\end{align}
where
\begin{align}
2h_{AB}dx^Adx^B&=\left(e^{2\gamma}+e^{2\delta}\right)d\theta^2+4\sin\theta\sinh(\gamma-\delta)d\theta d\phi \nonumber\\
&+ \sin^2\theta\left(e^{-2\gamma}+e^{-2\delta}\right)d\phi^2.
\end{align}
Then, $\|h_{AB}\|=\sin\theta$, that is just the determinant of the unitary sphere, if $\theta$ and $\phi$ can be identified as the usual spherical angles. The line element \eqref{BS22} for $r$ constant, allows us to identify $Ve^{2\beta}/r$ as the square of the lapse function, where $V$ and $\beta$ are related to the Newtonian potential and to the redshift respectively, and $U^\mu$ is the shift displacement between two successive hypersurfaces.
\subsection{Characteristic Initial Value Problem}
\begin{figure}[h!]
	\begin{center}
	\begin{tabular}{cc}
	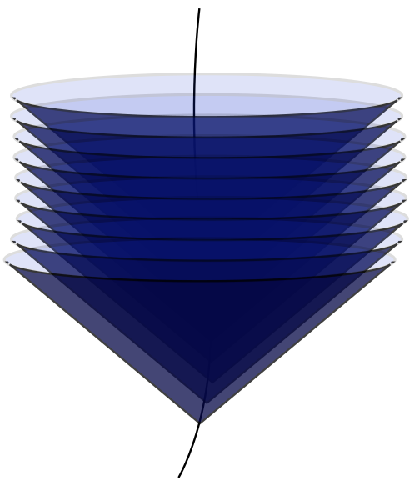 \label{FCHAR1a}&  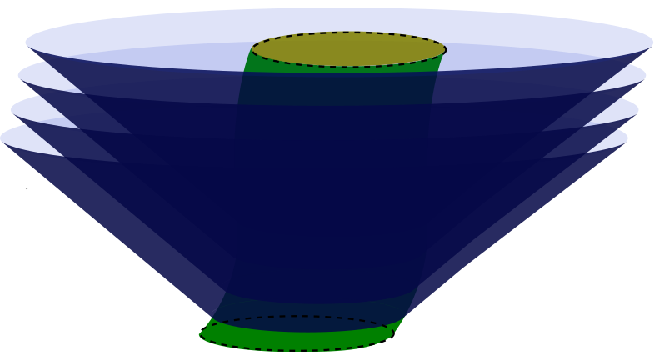 \label{FCHAR1b}\\ 
	(a) & (b)
	\end{tabular}
		\caption{Space-time $\mathcal M$ foliated in 2D - null hypersurfaces $\Sigma$. (a) Null cones emanating from a central time-like geodesic. (b) Null cones as emanating from a central time-like world tube.}
		\label{FCHAR1}
	\end{center}
\end{figure}

As already considered, the initial value problem version in the null cone formalism, is called characteristic initial value problem. In this case, the initial data is specified on a null cone and on the time-like central geodesic, or on a time-like hypersurface (the time-like world tube), which is parametrised through the retarded time $u$, (see Figure \ref{FCHAR1}). In the first version of the null cone formalism (Figure \ref{FCHAR1}a), some evolutions can be carried out in a satisfactory form without caustic formation. However, the second scheme (Figure \ref{FCHAR1}b) is usually implemented, in particular to avoid caustics.

\begin{figure}[h!]
	\begin{center}
		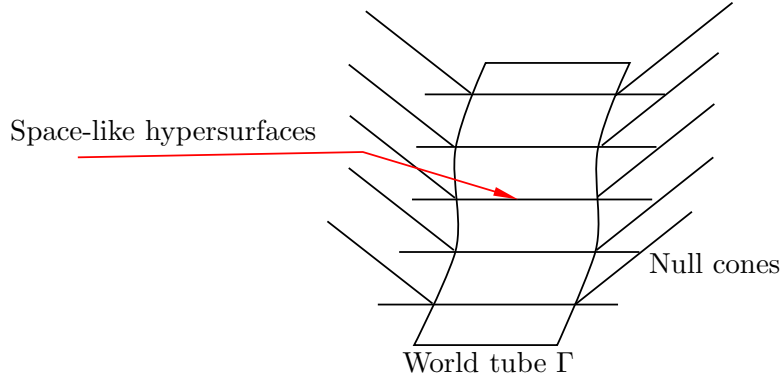
		\vspace{0.5cm}
		\caption{Space-time $\mathcal M$ foliated in 2D - null hypersurfaces $\Sigma$. Section showing the space-like for $t$ constant and characteristic hypersurfaces corresponding to the retarded time $u$ constant.}
		\label{FCHAR2}
	\end{center}
\end{figure}

The common usage for the characteristic formulation is in conjunction with an ADM based formalism, in which the matter is considered inside the world tube $\Gamma$ (see Figure \ref{FCHAR2}). The matter is evolved through a space-like foliation scheme for the space-time. The principal application of such scheme is in binary systems with transfer of momentum and mass. 

The ADM based code determines the initial data needed to perform the characteristic evolution. Specifying it on the common time-like hypersurface $\Gamma$, after that a pure null evolution scheme is used, for example in radial cases the null parallelogram algorithm is applied, or off the spherical symmetry a Crank-Nicolson or a leapfrog algorithms are used. However, in recent works the time evolution is performed using a Runge-Kutta integration scheme (see e.g. the  references \cite{C13,RBP13,HS15}). 
\subsection{The Einstein's Field Equations}
The Einstein's field equations in this formalism can be decomposed into hypersurface, evolution and constraint equations \cite{W12}, namely 
\begin{subequations}
\begin{align}
& R_{22}=0,\hspace{0.5cm}R_{2A}=0,\hspace{0.5cm}h^{AB}R_{AB}=0,\label{CEFE1}\\
& R_{AB}-\frac{1}{2}h_{AB}h^{CD}R_{CD}=0, \label{CEFE2}\\
& R^2_{~A}=0,\hspace{0.5cm}R^2_{~u}=0. \label{CEFE3}
\end{align}
\label{CEFE}%
\end{subequations}

These equations form a  hierarchical system of equations, which can be solved systematically. The first set of equations, \eqref{CEFE1} gives
\begin{subequations}
\begin{align}
\beta_{,r}&=\frac{1}{16}rh^{AC}h^{BD}h_{AB,r}h_{CD,r}, \label{fe1a}\\
\left(r^4e^{2\beta}h_{AB}U^B_{~,r}\right)_{,r}&=2r^4\left(r^{-2}\beta_{,A}\right)_{,r}-r^2h^{BC}h_{AB,r\|C}, \label{fe1b}\\
2V_{,r}&=e^{2\beta}\mathcal{R}-2e^{2\beta}\beta_{\|A}^{~~A}-2e^{2\beta}\beta^{\|A}\beta_{\|A}+r^{-2}\left(r^4U^A\right)_{,r\|A}\nonumber\\
&\hspace{0.5cm}-\frac{r^4e^{-2\beta}}{2}h_{AB}U^A_{~~,r}U^B_{~~,r}, \label{fe1c}
\end{align}
\label{fe1}%
\end{subequations}
for which $u$ is constant, the double vertical lines indicates covariant derivative associated to $h_{AB}$, and $\mathcal{R}$ is the Ricci's scalar associated to $h_{AB}$. The evolution equations \eqref{CEFE2} take the form
\begin{align}
&
\left(r h_{AB,u}\right)_{,r}-\frac{
	\left(rVh_{AB,r}\right)_{,r}}{2r}-\frac{2e^{\beta}e^{\beta}_{~~\|AB}
	}{r}\nonumber\\
&+rh_{AC\|B}U^C_{~,r}
-\frac{r^3e^{-2\beta}h_{AC}h_{BD}U^C_{~,r}U^D_{~,r}}{2}+2U_{B\|A}
\nonumber\\
&+\frac{rh_{AB,r}U^C_{~\|C}
	+rU^Ch_{AB,r\|C}
	}{2}+rh_{AD,r}h^{CD}\left(U_{C\|B}-U_{B\|C}\right)
	\nonumber\\
&=0,
\label{fe2}
\end{align}
in which time derivatives of the $J$ function are involved, and the third set, the constraint equations, must be satisfied for all null cones in which the space-time is foliated, or conversely for all time in the evolution.  
\section{The Einstein's Field Equations in the Quasi-Spherical Approxi-mation}
In this section some results in the quasi-spherical approximation are briefly presented. \citeonline{BGLW96} obtain a decomposition for the field equations using the stereographic dyads $q^A$, separating the linear from the non-linear terms.  When the non-linear terms are disregarded the quasi-linear approximation is obtained. In order to show this, the field equations \eqref{CEFE} are projected as
\begin{subequations}
\begin{align}
& R_{22}=0,\hspace{0.5cm}R_{2A}q^A=0,\hspace{0.5cm}h^{AB}R_{AB}=0,\label{CEFE_QSA_0a}\\
& q^Aq^BR_{AB}=0,\label{CEFE_QSA_0b}\\ 
& R_{11}=0,\hspace{0.5cm}R_{12}=0,\hspace{0.5cm}R_{1A}q^A=0.\label{CEFE_QSA_0c}
\end{align}
\label{CEFE_QSA_0}%
\end{subequations}
It is introduced a quantity to measure the deviation from the sphericity in terms of the connection symbols, considering the higher order terms and therefore, maintaining the non-linear regime without loss of generality.  
\\Thus, the difference between the connexion associated with the unit sphere metric $q_{AB}$ and $h_{AB}$ reads 
\begin{equation}
\Omega^C_{~AB}z_C=\left(\nabla_A-\triangle_A\right)z_B
\label{CEFE_QSA_1}
\end{equation}
which can be written
\begin{equation}
\Omega^C_{~AB}=\dfrac{h^{CD}\left(h_{DB|A}+h_{AD|B}-h_{AB|D}\right)}{2},
\label{CEFE_QSA_2}
\end{equation}
where $f_{|A}=\triangle_A f$. The following quantity is introduced in order to reduce the order of the differential equation \eqref{fe1b} 
\begin{equation}
Q_{A}=r^2e^{-2\beta}h_{AB}U^B_{~,r}.
\label{CEFE_QSA_3}
\end{equation}
Also, the following spin-weighted quantities are introduced,
\begin{equation}
J=\dfrac{h_{AB}q^Aq^B}{2},\hspace{0.5cm}K=\dfrac{h_{AB}q^A\overline{q}^B}{2},\hspace{0.5cm}
Q=Q_Aq^A,\hspace{0.5cm} U=U^Aq_A,\hspace{0.5cm}
\end{equation}
where, the complex scalar $J$, is a $2$-spin-weighted function, and the complex scalar functions $Q$ and $U$ are $1$-spin-weighted functions. The Bondi's gauge $\|h_{AB}\|=\sin\theta$, is translated through these spin-weighted quantities as
\begin{equation}
K^2-J\overline{J}=1.
\end{equation}
where, the overline indicates complex conjugation. Here $J=0$ implies spherical symmetry.
\\Thus \eqref{fe1b} is reduced to the following equations
\begin{subequations}	
\begin{align}
\left(r^2Q_{A}\right)_{,r}&=2r^4\left(r^{-2}\beta_{,A}\right)_{,r}-r^2h^{BC}h_{AB,r\|C},
\label{CEFE_QSA_4a}\\
U^A_{~,r}&=r^{-2}e^{2\beta}h^{AB}Q_{B},
\label{CEFE_QSA_4b}
\end{align}
\label{CEFE_QSA_4}
\end{subequations}
and the field equations \eqref{fe1} adopt the form
\begin{subequations}	
\begin{align}
\beta_{,r}&=N_\beta, \label{CEFE_QSA_5a}\\
\left(r^2Q\right)_{,r}&=-r^2q^Aq^{BC}h_{AB\|C}+2r^4q^A\left(r^{-2}\beta_{,A}\right)_{,r}+N_Q,\label{CEFE_QSA_5b}\\
U_{,r}&=r^2e^{2\beta}Q+N_U, \label{CEFE_QSA_5c}\\
V_{,r}&=\dfrac{e^{2\beta}\mathcal{R}}{2}-e^{2\beta}\beta_{\|A}^{~~A}-e^{2\beta}\beta^{\|A}\beta_{\|A}+\dfrac{r^{-2}\left(r^4U^A\right)_{\|A,r}}{2}+N_w, \label{CEFE_QSA_5d}
\end{align}
whereas the evolution equation \eqref{fe2} becomes
\begin{align}
2\left(rJ\right)_{,ur}-\left(r^{-1}V\left(rJ\right)_{,r}\right)_{,r}=&-r^{-1}\left(r^2\eth U\right)_{,r}+2r^{-1}e^\beta\eth^2 e^\beta\nonumber\\
& -\left(r^{-1}w\right)_{,r}J+N_J,\label{CEFE_QSA_5e}
\end{align}
\label{CEFE_QSA_5}
\end{subequations}
where the non-linear terms are
\begin{subequations}
\begin{align}
N_\beta&=\frac{1}{16}rh^{AC}h^{BD}h_{AB,r}h_{CD,r}, \label{CEFE_QSA_6a}\\
N_Q&=q^A\left(r^2h^{BC}\left(\Omega^D_{~CA}h_{DB,r}+\Omega^D_{~CB}h_{AD,r}\right)-r^2\left(h^{BC}-q^{BC}\right)h_{AB,r|C}\right), \label{CEFE_QSA_6b}\\
N_U&=r^{-2}e^{2\beta}q_A\left(h^{AB}-q^{AB}\right)Q_{B}, \label{CEFE_QSA_6c}\\
N_w&=-e^{\beta}\left(\left(h^{AB}-q^{AB}\right)\left(e^\beta\right)_{|B}\right)_{|A}-\dfrac{r^4e^{-2\beta}h_{AB}U^A_{~,r}U^B_{~,r}}{4}, \label{CEFE_QSA_6d}\\
N_J&=\dfrac{q^Aq^B}{r}\left(-2e^\beta\Omega^C_{~AB}\left(e^\beta\right)_{|C}-h_{AC}\Omega^C_{~DB}\left(r^2U^D\right)_{,r}\nonumber\right.\\
&\hspace{0.4cm}\left.-\left(h_{AC}-q_{AC}\right)\left(r^2U^C\right)_{,r|B}+\dfrac{r^4e^{-2\beta}h_{AC}h_{DB}U^C_{~,r}U^D_{~,r}}{2}\right.\nonumber\\
&\hspace{0.4cm}\left.-\dfrac{r^2h_{AB,r}U^C_{~\|C}}{2}-r^2U^Ch_{AB,r\|C}+2r^2h^{CD}h_{AD,r}U_{[B\|C]}+\dfrac{h_{AB}F}{2}\right), \label{CEFE_QSA_6e}\\
F&=-r^2h^{AB}_{~~~,r}\left(h_{AB,u}-\dfrac{Vh_{AB,r}}{2r}\right)-2e^\beta \left(e^\beta\right)_{\|A}^{~~A}+\left(r^2U^A\right)_{,r\|A}\nonumber\\
&\hspace{0.4cm}-\dfrac{r^4e^{-2\beta}h_{AB}U^A_{~,r}U^B_{~,r}}{2}. \label{CEFE_QSA_6f}
\end{align}
\label{CEFE_QSA_6}%
\end{subequations}
The quasi-spherical approximation is then obtained when $N_\beta=N_Q=N_U=N_w=N_J=0$, which is neither a linear version of the field equations, and nor a spherical version of them. However, this approximation considers slightly deviations from the sphericity.
\section{The Einstein's Field equations Using the {\it Eth} Formalism}
\citeonline{BGLMW97} show that the field equations \eqref{CEFE_QSA_5a}-\eqref{CEFE_QSA_5d} take the following form when the eth formalism is used, 
\begin{subequations}
\begin{align}
\beta_{,r}&=N_\beta,\\
U_{,r}&=r^{-2}e^{2\beta}Q+N_U,\\
\left(r^2Q\right)_{,r}&=-r^2\left(\overline{\eth}J+\eth K\right)_{,r}+2r^4\eth\left(r^{-2}\beta\right)_{,r}+N_Q,\\
w_{,r}&=\dfrac{e^{2\beta}}{2}\mathcal{R}-1-e^{\beta}\eth \overline{\eth}e^{\beta}+\dfrac{r^{-2}}{4}\left(r^4\left(\eth \overline{U}+\overline{\eth} U\right)\right)_{,r}+N_w,
\end{align}
\label{CEFE_ETH_1}
\end{subequations}
where the Ricci's scalar associated to $h_{AB}$ take the form
\begin{equation}
\mathcal{R}=2K-\eth\overline{\eth}K+\dfrac{\overline{\eth}^2J+\eth^2\overline{J}}{2}+\dfrac{\overline{\eth}\overline{J}\eth J -\overline{\eth}J\eth \overline{J} }{4K}.
\label{CEFE_ETH_2}
\end{equation}
The evolution equation \eqref{CEFE_QSA_5e} reads
\begin{align}
 2\left(rJ\right)_{,ur}-\left(r^{-1}(r+W)\left(rJ\right)_{,r}\right)_{,r}=&-r^{-1}\left(r^2\eth U\right)_{,r} + 2r^{-1}e^\beta\eth^2e^{\beta} \nonumber\\
&-\left(r^{-1}w\right)_{,r}J+N_J,
\label{CEFE_ETH_3}
\end{align}
where, the non-linear terms in \eqref{CEFE_QSA_6} become
\begin{subequations}
	\begin{align}
N_\beta&=\dfrac{r\left(J_{,r}\overline{J}_{,r}-K_{,r}^2\right)}{8},\label{CEFE_ETH_4a}\\
N_U&=\dfrac{e^{2\beta}\left(KQ-Q-J\overline{Q}\right)}{r^2},\label{CEFE_ETH_4b}\\
N_Q&=r^2\left((1-K)\left(\eth K_{,r}+\overline{\eth}J_{,r}\right)+\eth\left(\overline{J}J_{,r}\right)+\overline{\eth}\left(JK_{,r}\right)-J_{,r}\overline{\eth}K\right)\nonumber\\
&\hspace{0.4cm}+\dfrac{r^2}{2K^2}\left(\eth \overline{J}\left(J_{,r}-J^2\overline{J}_{,r}\right)+\overline{\eth}J\left(\overline{J}_{,r}-\overline{J}^2J_{,r}\right)\right),\label{CEFE_ETH_4c}\\
N_w&=e^{2\beta}\left((1-K)\left(\eth\overline{\eth}\beta+\eth\beta\overline{\beta}\right)+\dfrac{J\left(\overline{\eth}\beta\right)^2+\overline{J}\left(\eth\beta\right)^2}{2}\right)\nonumber\\
&\hspace{0.4cm}-\dfrac{e^{2\beta}}{2}\left(\eth\beta\left(\overline{\eth}K-\eth\overline{J}\right)+\overline{\eth}\beta\left(\eth K-\overline{\eth}J\right)\right)+\dfrac{e^{2\beta}}{2}\left(J\overline{\eth}^2\beta+\overline{J}\eth^2\beta\right)\nonumber\\
&\hspace{0.4cm}-\dfrac{e^{-2\beta}r^4}{8}\left(2KU_{,r}\overline{U}_{,r}+J\overline{U}^2_{,r}+\overline{J}U^2_{,r}\right),\label{CEFE_ETH_4d}\\
N_J&=\sum_{i=1}^{7}N_{Ji}+\dfrac{J\sum_{n=1}^{4}P_n}{r}\label{CEFE_ETH_4e}.
\end{align}
\label{CEFE_ETH_4}
\end{subequations}
Here, were defined the following terms
\begin{subequations}
\begin{align}
N_{J1}=&-\dfrac{e^{2\beta}}{r}\left(K\left(\eth J \overline{\eth}\beta+2\eth K \eth\beta-\overline{\eth}J\eth\beta\right)+J\left(\overline{\eth}J-2\eth K\right)\overline{\eth}\beta\right.\nonumber\\
&-\left.\overline{J}\eth J \eth\beta\right),\\
N_{J2}=&-\dfrac{\left(\eth J\left(r^2\overline{U}\right)_{,r}+\overline{\eth} J\left(r^2U\right)_{,r}\right)}{2r},\\
N_{J3}=&\dfrac{\left(1-K\right)\eth\left(r^2U\right)_{,r}-J\eth\left(r^2\overline{U}\right)_{,r}}{r},\\
N_{J4}=&\dfrac{r^3e^{-2\beta}\left(K^2U^2_{,r}+2JKU_{,r}\overline{U}_{,r}+J^2\overline{U}^2_{,r}\right)}{2},
\end{align}
\begin{align}
N_{J5}=&-\dfrac{rJ_{,r}\left(\overline{\eth}U+\eth\overline{U}\right)}{2},\\
N_{J6}=&\dfrac{r\left(\overline{U}\eth J + U\overline{\eth}J\right)\left(J\overline{J}_{,r}-\overline{J}J_{,r}\right)}{2} +r\left(JK_{,r}-KJ_{,r}\right)\overline{U}\overline{\eth}J\nonumber\\
&-r\overline{U}\left(\eth J_{,r}-2K\eth H J_{,r}+2J\eth K K_{,r}\right)\nonumber\\
&-rU\left(\overline{\eth}J_{,r}-K\eth \overline{J} J_{,r}+J\eth \overline{J}K_{,r}\right),\\
N_{J7}=&r\left(J_{,r}K-JK_{,r}\right)\left(\overline{U}\left(\overline{\eth}J-\eth K\right)+U\left(\overline{\eth}K-\eth\overline{J}\right)\right.\nonumber\\
&\left.+K\left(\overline{\eth}U-\eth\overline{U}\right)+\left(J\overline{\eth}\overline{U}-\overline{J}\eth U\right)\right),
\end{align}
\label{CEFE_ETH_5}
\end{subequations}
and the $P_n$ terms in \eqref{CEFE_ETH_4e} are defined as
\begin{subequations}
\begin{align}
P_1&=\dfrac{r^2\left(J_{,u}\left(\overline{J}K\right)_{,r}+\overline{J}_{,u}\left(JK\right)_{,r}\right)}{K}-8V\beta_{,r},\\
P_2&=e^{2\beta}\left(-2K\left(\eth\overline{\eth}\beta+\overline{\eth}\beta\eth\beta\right)-\left(\overline{\eth}\beta\eth K+\eth\beta\overline{\eth}K\right)\right.\nonumber\\
&\left.+J\left(\overline{\eth}^2\beta+(\overline{\eth}\beta)^2\right)+\overline{J}\left(\eth^2\beta+(\overline{\eth}\beta)^2\right)+\overline{\eth}J\overline{\eth}\beta+\eth \overline{J}\eth\beta\right),\\
P_3&=\dfrac{\overline{\eth}\left(r^2U\right)_{,r}+\eth\left(r^2\overline{U}\right)_{,r}}{2},\\
P_4&=-\dfrac{r^4e^{-2\beta}\left(2KU_{,r}\overline{U}_{,r}+J\overline{U}^2_{,r}+\overline{J}U^2_{,r}\right)}{4}.
\end{align}
\end{subequations}
Notice that subsequent reductions to a first order equations were made \cite{G01}, improving the performance and the accuracy of the characteristic evolution codes, keeping the problem as a well-possess problem \cite{GF03}.   Also, it is worth mentioning that other approach, for Bondi observers, is obtained by considering the projection of the field equations onto the vectors $m^A$, defined as
\begin{equation}
h_{AB}=m_{(A}\overline{m}_{B)}.
\end{equation}
This kind of approach is used in the analysis of the gravitational radiation near the null infinity \cite{BGLMW97}.

\chapter[Master Equation]{LINEAR REGIME IN THE CHARACTERISTIC FORMULA-\\ TION 
	 AND THE  MASTER  EQUATION SOLUTIONS}
\label{LRME}
The linear regime of the Einstein's field equations leads to different approximations according to how it is made. Depending on the presence of matter, the curvature of the background can be considered in this regime. The perturbations made to the space-time are considered smaller enough to contribute to the curvature, propagating away from the sources. If the curvature is considered then the advanced and retarded Green's functions must be taken into account into the gravitational wave solutions. 

In this section, we show the Einstein's field equations in the outgoing characteristic formalism in the linear regime. These equations result from perturbations to the Minkowski and Schwarzschild's space-times. In order to do this, we shown that, to first order, the Bondi-Sachs metric can be decomposed as a background metric (Minkowski or Schwarzschild) plus a perturbation, which is expressed in terms of the spin-weighted functions $\beta$, $J$, $U$ and $K$ previously defined. After that, the field equations are computed and a decomposition into spin-weighted spherical harmonics is performed, leading to a system of equations for the coefficients used in those multipolar expansions. This system is solved in a completely analytical form, by solving a specific equation obtained for the $J$ metric variable, which is called {\it master equation}. Using their solutions we compute the analytical solutions for the rest of the metric variables for all multipolar orders in terms of Generalised Hypergeometric (Heun) functions for the Minkowski (Schwarzschild) \cite{CA15b}. A simple example is reproduced using this formalism, that is a static spherical thin shell \cite{B05}, whose matter distribution is expressed as a function of the spin-weighted spherical harmonics $_sZ_{lm}$.

Here, we put the Bondi-Sachs metric \eqref{BS11} in terms of the spin-weighted scalars $J,w$ and $\beta$ in stereographic-radiation coordinates, namely
\begin{eqnarray}
\label{bs_explicit}
ds^2&=&-\left(e^{2\beta}\left(1+\frac{w}{r}\right)-r^2(J\bar{U}^2+U^2\bar{J}+2 K U \bar{U})\right)du^2  -2e^{2\beta}dudr \nonumber\\
&&-\frac{2r^2\left((K +\bar{J}) U+(J+K) \bar{U}\right)}{1+|\zeta|^2}dq du\nonumber\\
&&-\frac{2ir^2\left((K -\bar{J}) U+(J-K)	\bar{U}\right)}{1+|\zeta|^2}dpdu+\frac{2 r^2 \left(J+2K+\bar{J}\right)}{(1+|\zeta|^2)^2}dq^2\nonumber\\
&&-\frac{4 i r^2
	\left(J-\bar{J}\right)}{(1+|\zeta|^2)^2}dqdp-\frac{2 r^2 \left(J-2 K+\bar{J}\right)}{(1+|\zeta|^2)^2}dp^2.
\end{eqnarray}

In the weak field limit, i.e., when slight deviations from the Minkowski background are considered i.e., $|g_{\mu\nu}|\ll|\eta_{\mu\nu}|$, and the second order terms are disregarded, the Bondi-Sachs metric is reduced to
\begin{eqnarray}
\label{bs_lin}
ds^2&=&-\left(1-\frac{w}{r}-2\beta\right)du^2 -2(1+2\beta)dudr -2r^2\frac{(U+\overline{U})}{1+|\zeta|^2}dq du
\nonumber \\&& 
-2r^2\frac{i(U-\overline{U})}{1+|\zeta|^2}dp du+2r^2\frac{\left(2+J+\overline{J}\right)}{\left(1+|\zeta|^2\right)^2}dq^2\nonumber\\&&
-4ir^2\frac{(J-\overline{J})}{(1+|\zeta|^2)^2}dqdp-2r^2\frac{\left(-2+J+\overline{J}\right)}{\left(1+|\zeta|^2\right)^2}dp^2,
\end{eqnarray} 
which can be clearly separated as,
\begin{eqnarray}
\label{bs_lin2}
ds^2&=&-du^2-2dudr + \frac{4r^2}{\left(1+|\zeta|^2\right)^2}\left(dq^2+dp^2\right) +\left(\frac{w}{r}+2\beta\right)du^2   \nonumber\\
&&-4\beta dudr -\frac{2r^2}{1+|\zeta|^2}du\left((U+\overline{U})dq -i(U-\overline{U})dp \right)
\nonumber \\&& 
-4ir^2\frac{(J-\overline{J})}{(1+|\zeta|^2)^2}dqdp+\frac{2r^2\left(J+\overline{J}\right)}{\left(1+|\zeta|^2\right)^2}\left(dq^2-dp^2\right),
\label{bs_lin_pert}
\end{eqnarray} 
showing that it corresponds to a Minkowski background plus a perturbation. 
\section{Einstein's Field Equations in the linear }
In the linear regime, the field equations \eqref{CEFE_QSA_0} are reduced to
\begin{subequations}
\begin{align}
& 8 \pi  T_{22}=\frac{4 \beta _{,r}}{r}, \label{field_eq_1}\\
& 8 \pi  T_{2A} q^A=\frac{\overline{\eth }J_{,r}}{2} -\eth \beta _{,r} +\frac{2 \eth \beta}{r} +\frac{\left(r^4 U_{,r}\right)_{,r}}{2r^2}, \label{field_eq_2}\\
& 8 \pi  \left(h^{AB} T_{AB}-r^2 T\right) =-2 \eth \overline{\eth }\beta +\frac{\eth^2\overline{J} + \overline{\eth }^2 J}{2} +\frac{\left(r^4\left(\overline{\eth}U+\eth \overline{U}\right)\right)_{,r}}{2r^2}\nonumber\\
& \hspace{4.0cm}+4 \beta -2 w_{,r}, \label{field_eq_3}\\
& 8 \pi T_{AB}  q^A q^B=-2 \eth^2 \beta + \left(r^2\eth U \right)_{,r} - \left(r^2 J_{,r}\right)_{,r} +2 r\left(rJ\right)_{,ur},\label{field_eq_4}\\
& 8 \pi  \left(\frac{T}{2}+T_{11}\right)=\frac{\eth \overline{\eth }w}{2 r^3} + \frac{\eth \overline {\eth }\beta }{r^2} -\frac{\left(\eth \overline{U} + \overline{\eth} U \right)_{,u}}{2} +\frac{w_{,u}}{r^2} +\frac{w_{,rr}}{2 r}\nonumber\\
&\hspace{3cm}-\frac{2 \beta_{,u}}{r}, \label{field_eq_5}\\
& 8 \pi  \left(\frac{T}{2}+T_{12}\right)=
\frac{\eth \overline{\eth }\beta }{r^2} -\frac{\left(r^2\left(\eth \overline{U} + \overline{\eth   }U\right)\right)_{,r}}{4r^2}
+\frac{w_{,rr}}{2 r}, \label{field_eq_6}
\end{align}
\begin{align}
& 8 \pi  T_{1A} q^A=\frac{\overline{\eth }J_{,u}}{2} -\frac{\eth^2
	\overline{U} }{4} +\frac{\eth \overline{\eth }U}{4} +\frac{1}{2}\left(\frac{\eth w}{r}\right)_{,r} -\eth \beta _{,u} +\frac{\left(r^4U\right)_{,r}}{2r^2}\nonumber\\
& \hspace{2cm}-\frac{r^2 U_{,ur}}{2} +U. \label{field_eq_7}
\end{align}
\label{linear_field_eqs}%
\end{subequations}
which are the field equations corresponding to the perturbed Minkowski or Schwarzschild space-times depending on how the $w$ metric function is defined. This system of equations were originally obtained by Bishop in \cite{B05} and confirmed by us.
\section{Harmonic Decomposition and Boundary Problem}
Now, an expansion of the metric variables in the form of a multipolar series is performed, namely
\begin{equation}
_sf=\sum_{l=0}^\infty \sum_{m=l}^l \Re\left(f_{lm}e^{i|m|\tilde{\phi}}\right)\ \eth^s\ Z_{lm},
\label{expansion}
\end{equation}
where $_sf=\{\beta,w,J,\overline{J},U,\overline{U}\}$, $Z_{lm}=\ _0Z_{lm}$, $\tilde{\phi}$ is a general function of the retarded time, i.e., $\tilde{\phi}:=\tilde{\phi}(u)$, $f_{lm}$ are the spectral components of the function $_sf$, $m\in \mathbb{Z}$, $m\in [-l,l]$ and $l\ge 0$ indicating the multipolar order. In previous works similar expansions were performed, where $\phi=\nu u$ \cite{B05,BBSW09,BPR11,CA15}.

Notice that in \eqref{expansion} the spin-weight of the function $_sf$ is contained in the factor $\eth^s Z_{lm}$. Therefore, substituting \eqref{expansion} into the field equations \eqref{linear_field_eqs} one obtains ordinary differential equations for their spectral components, in which the spin-weighted factors have been eliminated, namely  
\begin{subequations}
	\begin{align}
	&\beta_{lm,r}= 2\pi \int_{\Omega}d\Omega\ \overline{Z}_{lm}\int_{0}^{2\pi}d\tilde\phi \ e^{-i|m|\tilde\phi} r T_{22} 
	\label{field_eq_s1} ,
	\end{align}
	\begin{align}
	& -\frac{(l+2)(l-1)J_{lm,r}}{2} -\beta _{lm,r} +\frac{2 \beta_{lm}}{r} +\frac{\left(r^4 U_{lm,r}\right)_{,r}}{2r^2} 
	\nonumber\\
	&
	= \frac{8 \pi}{\sqrt{l(l+1)}} \int_{\Omega} d\Omega \ \overline{Z}_{lm}\int_{0}^{2\pi} d\tilde\phi \ e^{-i|m|\tilde{\phi}} T_{2A} q^A, 
	\label{field_eq_s2}
	\end{align}
	\begin{align}
	& 2l(l+1) \beta_{lm} +(l-1)l(l+1)(l+2)J_{lm}  +\frac{l(l+1)\left(r^4\left(U_{lm}\right)\right)_{,r}}{r^2}\nonumber\\
	& + 4 \beta_{lm} -2 w_{lm,r} = 8 \pi \int_{\Omega}d\Omega \ \overline{Z}_{lm}\int_{0}^{2\pi}d\tilde\phi\ e^{-i|m|\tilde{\phi}}\left(h^{AB} T_{AB}-r^2 T\right) , \label{field_eq_s3}
	\end{align}
	\begin{align}
	&-2 \beta_{lm} + \left(r^2 U_{lm} \right)_{,r} - \left(r^2 J_{lm,r}\right)_{,r} +2 i|m| r\dot{\tilde{\phi}} \left(rJ_{lm}\right)_{,r} \nonumber\\
	&= \frac{8 \pi}{\sqrt{(l-1)l(l+1)(l+2)}} \int_{\Omega}d\Omega \ \overline{Z}_{lm} \int_{0}^{2\pi}d\tilde{\phi}\ e^{-i|m|\tilde{\phi}}T_{AB}  q^A q^B,
	\label{field_eq_s4}
	\end{align}
	\begin{align}
	& -\frac{l(l+1) w_{lm}}{2 r^3} - \frac{l(l+1)\beta_{lm} }{r^2} +i|m|l(l+1)\dot{\tilde{\phi}} U_{lm}  +\frac{i|m|\dot{\tilde{\phi}} w_{lm}}{r^2}   \nonumber\\
	&+\frac{w_{lm,rr}}{2 r} -\frac{2 i|m|\dot{\tilde{\phi}} \beta_{lm}}{r}+\frac{2\beta_{lm,r}}{r} + \beta_{lm,rr}-2\dot{\tilde{\phi}}\beta_{lm,r}\nonumber\\
	&=8 \pi \int_{\Omega}d\Omega \ \overline{Z}_{lm} \int_{0}^{2\pi}d\tilde{\phi}\ e^{-i|m|\tilde{\phi}} \left(\frac{T}{2}+T_{11}\right), \label{field_eq_s5}\\
	& -\frac{l(l+1)\beta_{lm} }{r^2} +\frac{l(l+1)\left(r^2 U_{lm} \right)_{,r}}{2r^2} +\frac{w_{lm,rr}}{2 r} \nonumber\\
	&=8 \pi \int_{\Omega}d\Omega \ \overline{Z}_{lm} \int_{0}^{2\pi}d\tilde{\phi}\ e^{-i|m|\tilde{\phi}} \left(\frac{T}{2}+T_{12}\right), \label{field_eq_s6}
	\end{align}
	\begin{align}
	& -\frac{i|m|(l+2)(l-1)J_{lm}\dot{\tilde{\phi}}}{2} +\frac{1}{2}\left(\frac{w_{lm}}{r}\right)_{,r} -i|m|\dot{\tilde{\phi}} \beta _{lm} +\frac{\left(r^4U_{lm,r}\right)_{,r}}{2r^2}\nonumber\\
	& -\frac{i|m|r^2\dot{\tilde{\phi}}}{2}U_{lm,r} +U_{lm} =\frac{8 \pi}{\sqrt{l(l+1)}}  \int_{\Omega}d\Omega \ \overline{Z}_{lm} \int_{0}^{2\pi}d\tilde{\phi}\ e^{-i|m|\tilde{\phi}} T_{1A} q^A, \label{field_eq_s7}
	\end{align}
	\label{field_eq_s}%
\end{subequations}
This system of coupled ordinary equations is separable through a simple procedure, as we will show in the next section. Notice that an alternative procedure is presented by M\"adler in \cite{M13}.
\section{The Master Equation}
Here, we sketch the explicit steps to obtain a fourth order equation for $J_{lm}$. This equation is called {\it master equation} and allows one to find the explicit solutions for $U_{lm}$ and $w_{lm}$. 

In order to do that we start making the change of variable $x=r^{-1}$, then, the field equations \eqref{field_eq_s1} - \eqref{field_eq_s4} become 
\begin{subequations}	
	\begin{align}
	&\beta_{lm,x}= -x^2A_{lm},  
	\label{field_eq_x1}\\
	& (l+2)(l-1)xJ_{lm,x} +2x\beta _{lm,x} +4 \beta_{lm} -2U_{lm,x}+xU_{lm,xx} = B_{lm}, 
	\label{field_eq_x2}\\
	& - 2x^3 J_{lm,xx}  -4 i|m|\dot{\tilde{\phi}} x J_{lm,x} +4i|m| \dot{\tilde{\phi}} J_{lm} +4 U_{lm} - 2xU_{lm,x}-4 x\beta_{lm} 
	\nonumber\\
	&
	= 2x D_{lm},
	\label{field_eq_x4}
	\end{align}
\end{subequations}
where the source terms $A_{lm}:=A_{lm}(x)$, $B_{lm}:=B_{lm}(x)$ and $D_{lm}:=D_{lm}(x)$ are explicitly defined \cite{CA15b}, namely
\begin{subequations}
	\begin{align}
	& A_{lm}=2\pi \int_{\Omega}d\Omega\ \overline{Z}_{lm}\int_{0}^{2\pi}d\tilde\phi \ e^{-i|m|\tilde\phi} x T_{22},\\
	& B_{lm}=\frac{16 \pi}{\sqrt{l(l+1)}} \int_{\Omega} d\Omega \ \overline{Z}_{lm}\int_{0}^{2\pi} d\tilde\phi \ e^{-i|m|\tilde{\phi}} x T_{2A} q^A,\\
	& D_{lm}= \frac{8 \pi}{\sqrt{(l-1)l(l+1)(l+2)}} \int_{\Omega}d\Omega \ \overline{Z}_{lm} \int_{0}^{2\pi}d\tilde{\phi}\ e^{-i|m|\tilde{\phi}}T_{AB}  q^A q^B.
	\end{align}
\end{subequations}
In addition, solving \eqref{field_eq_x2} for $4x\beta_{lm}$ 
and substituting it into \eqref{field_eq_x4}, one obtains
\begin{align}
& - 2x^3 J_{lm,xx}  -4 i|m| \dot{\tilde{\phi}}xJ_{lm,x} + x^2(l+2)(l-1)J_{lm,x} +4i|m| \dot{\tilde{\phi}} J_{lm}   \nonumber \\
&  +x^2U_{lm,xx} - 4xU_{lm,x} +4 U_{lm} +2x^2\beta _{lm,x}  = x (2D_{lm}+ B_{lm}).
\label{field_eq_x5}
\end{align}
Thus, the derivative of \eqref{field_eq_x5} with respect to $x$ yields  a third order differential equation for $J_{lm}$, i.e., 
\begin{align}
& - 2x^3 J_{lm,xxx} - 6x^2 J_{lm,xx}  -4 i|m| \dot{\tilde{\phi}}xJ_{lm,xx} + x^2(l+2)(l-1)J_{lm,xx} \nonumber\\
&+ 2x(l+2)(l-1) J_{lm,x}   +x^2U_{lm,xxx}  - 2xU_{lm,xx}   \nonumber \\
&  +4x\beta _{lm,x} +2x^2\beta _{lm,xx} = (2D_{lm}+ B_{lm}) + x (2D_{lm,x}+ B_{lm,x}).
\label{field_eq_x6}
\end{align}
After this, notice that it is possible to obtain $x^2U_{lm,xxx}$ by just deriving \eqref{field_eq_x2} with respect to $x$, namely
\begin{align}
&  x^2U_{lm,xxx} = -x^2(l+2)(l-1)J_{lm,xx} -x(l+2)(l-1)J_{lm,x}  +xU_{lm,xx}  \nonumber\\
&  -6x\beta _{lm,x} -2x^2\beta _{lm,xx} + x B_{lm,x}.
\label{field_eq_x7}
\end{align}
Then, substituting it in \eqref{field_eq_x6} and simplifying one obtains
\begin{align}
& - 2x^3 J_{lm,xxx} - 6x^2 J_{lm,xx}  -4 i|m| \dot{\tilde{\phi}}xJ_{lm,xx} + x(l+2)(l-1) J_{lm,x} \nonumber\\
&  -xU_{lm,xx}  -2x\beta _{lm,x} = 2 x D_{lm,x}+B_{lm}+2 D_{lm}.
\label{field_eq_x8}
\end{align}
Making the derivative of \eqref{field_eq_x8} with respect to $x$, and substituting  $xU_{xxx}$ from \eqref{field_eq_x7}, one  finds a fourth order differential equation for $J_{lm}$, namely
\begin{align}
&  - 2x^4 J_{lm,xxxx} - 12x^3 J_{lm,xxx} - 12x^2 J_{lm,xx}   -4i|m|  \dot{\tilde{\phi}}xJ_{lm,xx} -4i|m|  \dot{\tilde{\phi}} x^2 J_{lm,xxx} \nonumber\\
&+ 2x(l+2)(l-1) J_{lm,x} + 2x^2(l+2)(l-1) J_{lm,xx} +4x\beta _{lm,x} -2x U_{lm,xx}  \nonumber\\
&= 2 x B_{lm,x}+2 x^2 D_{lm,xx}+4 x D_{lm,x}.
\label{field_eq_x10}
\end{align}
Finally, solving \eqref{field_eq_x8} for $U_{lm,xx}$ and substituting into \eqref{field_eq_x10}, a differential equation containing only $J_{lm}$ with source terms is obtained, namely
\begin{align}
&  - 2x^4 J_{lm,xxxx} - 4x^2\left(2x  + i|m|\dot{\tilde{\phi}}\right)J_{lm,xxx} \nonumber\\ &+2x\left(2i|m|  \dot{\tilde{\phi}} + x(l+2)(l-1) \right)J_{lm,xx}  
   = H_{lm}(x),
\label{field_eq_x11}
\end{align}
where 
\begin{align}
H_{lm}(x)=2 x B_{lm,x}+2 x^2 D_{lm,xx}-8 x \beta_{lm,x}-2 B_{lm}-4 D_{lm}
\end{align}
represents the source terms \cite{CA15b}.

In order to reduce the order of this differential equation, one  defines \\  $\tilde{J}_{lm}=J_{lm,xx}$, thus,
\begin{align}
&  - 2x^4 \tilde{J}_{lm,xx} - 4x^2\left(2x  + i|m|\dot{\tilde{\phi}}\right)\tilde{J}_{lm,x} +2x\left(2i|m|  \dot{\tilde{\phi}} + x(l+2)(l-1) \right)\tilde{J}_{lm}  
   = H_{lm}.
\label{field_eq_x12}
\end{align}
For the vacuum, this differential equation turns homogeneous, i.e.,  $H_{lm}=0$, and hence \eqref{field_eq_x12} is reduced to the master equation presented by M\"adler in \cite{M13}, i.e.,
\begin{align}
- x^3 \tilde{J}_{lm,xx} - 2x\left(2x  + i|m|\dot{\tilde{\phi}}\right)\tilde{J}_{lm,x} +\left(2i|m|  \dot{\tilde{\phi}} + x(l+2)(l-1) \right)\tilde{J}_{lm}  
= 0.
\label{field_eq_x13}
\end{align}
Making $l=2$, this master equation reduces to that presented previously in \cite{B05} for the Minkowski's Background i.e.,
\begin{align*}
- x^3 \tilde{J}_{lm,xx} - 2x\left(2x  + i|m|\dot{\tilde{\phi}}\right)\tilde{J}_{lm,x} +2\left(i|m|  \dot{\tilde{\phi}} + 2x \right)\tilde{J}_{lm}  = 0.
\end{align*}
The derivation of the master equation for the Schwarzschild's background follows the same scheme. In this case the master equation is given by
\begin{align}
& J_{lm,xxxx}x^4 (2 M x-1) + J_{lm,xxx} \left(2 x^3 (7 M x-2)-2 i x^2 \dot{\tilde{\phi }}
\left| m\right| \right)\nonumber\\
&+J_{lm,xx} \left(2 i x \dot{\tilde{\phi }} \left| m\right| +(l-1) (l+2) x^2+16 M
x^3\right)=G_{lm}(x),
\label{meqsch1}
\end{align}
where $M$ is the mass of the central static black-hole and $G_{lm}(x)$ represents the source term, which is given by
\begin{equation}
G_{lm}(x)=\frac{H_{lm}(x)}{2}.
\end{equation}
It is important to observe that $M=0$ effectively reduces \eqref{meqsch1} to \eqref{field_eq_x11}. 
\\
Defining $\tilde{J}_{lm}=J_{lm,xx}$, the order of the differential equation \eqref{meqsch1} is reduced \cite{CA15b}, namely 
\begin{align}
& \tilde{J}_{lm,xx}x^4 (2 M x-1) + \tilde{J}_{lm,x} \left(2 x^3 (7 M x-2)-2 i x^2 \dot{\tilde{\phi }}
\left| m\right| \right)\nonumber\\
&+\tilde{J}_{lm} \left(2 i x \dot{\tilde{\phi }} \left| m\right| +(l-1) (l+2) x^2+16 M
x^3\right)=G_{lm}(x).
\label{meqsch2}
\end{align}
\section{Families of Solutions to the Master Equation}
Now, the families of solutions to the master equations \eqref{field_eq_x11} and \eqref{meqsch1} associated with the linear approximation in the Minkowski and the Schwarzschild's space-times are explicitly shown. 

To proceed, consider that $l$ is integer and greater than or equal to zero, i.e., $l\ge 0$, the constants of integration $C_i$ are complexes $C_i\in \mathbb{C}$, $i=1..4$, and arabic lower case letters represent real constants, i.e., $a,b,c,d,e,f,\cdots \in \mathbb{R}$

It is worth stressing that the applicability of the present work has some limitations, since in the context of the characteristic formulation the matter fields must be known {\it a priori} throughout the space-time.

Applications astrophysically relevant for this kind of solutions would be a spherical thick shell obeying some dynamics. This shell can obey an equation of state for some polytropic index. This assumption will destroy the analyticity nature of the master equation and therefore its integration must be numerical. Different polytropic index can lead to different solutions for $J$ and therefore different gravitational patterns. Another possible application would be a star formed by multiple thick layers obeying different equations of state. Also, binaries radiating their eccentricities offers real possibilities of application of the present formalism. In addition, objects gravitating around a Reisner-Norstr\"om black-holes allows one to explore interesting physics. Applications in cosmology are also admitted in this formalism, for example, studying the evolution of  gravitational waves in a de Sitter space-time \cite{B15}. There are other possibilities of applications under a wide spectrum of considerations in $f(R)$ theories. Finally, it is important to note that numerous studies can be made in the linear regime considering the numerical integration of the field equations, for example, the gravitational collapse of a given matter distribution is only one of these possibilities.
\subsection{The Minkowski's Background}
First, let us consider the most simple case corresponding to the non-radiative, $m=0$, Minkowski's master equation without sources \eqref{field_eq_x13}. Assuming the ansatz \\ $J_{lm}=x^k$, we obtain immediately 
\begin{equation*}
(k-l+1) (k+l+2)=0,
\end{equation*}
whose roots lead to the general family of solutions, 
\begin{equation}
\tilde{J}_{l0}(x)= C_1 x^{l-1}+C_2 x^{-(l+2)}.
\label{Msol1}
\end{equation}
Thus, integrating the last equation two times and rearranging the constants one obtains families of solutions to \eqref{field_eq_x11} of four parameters for the vacuum, namely
\begin{equation}
J_{l0}(x)= C_1 x^{l+1}+C_2 x^{-l}+C_3x+C_4.
\label{Msol2}
\end{equation}
When the source term is not null, we find that the non-radiative family of solutions, $m=0$, to the inhomogeneous equation \eqref{field_eq_x12} reads
\begin{align}
\tilde{J}_{l0}(x)=& C_1 x^{l-1} + C_2 x^{-(l+2)}+ x^{-(l+2)} \int_a^x dy\, \frac{H(y) y^{l-1}}{2 l+1} -x^{l-1} \int_b^x dy\, 
\frac{H(y) y^{-(l+2)}}{2 l+1}, 
\label{Msol3}
\end{align}
where $a$ and $b$ are real constants. Therefore, integrating two times with respect to $x$ and rearranging the constants we find the family of solutions to the inhomogeneous master equation \eqref{field_eq_x11}, for $m=0$,
\begin{align}
J_{l0}(x)=& C_1 x^{l+1}+ C_2 x^{-l}+ C_3x +C_4
+ \int_a^x dv\, \int_{b}^{v}dw \, w^{-(l+2)} \int_c^w dy\, \frac{H(y) y^{l-1}}{4 l+2}
\nonumber\\
&  -\int_d^xdv\, \int_{e}^{v} dw\, w^{l-1} \int_f^w dy\, \frac{H(y) y^{-(l+2)}}{4 l+2},
\label{Msol4}
\end{align}
where it is clear that the analyticity of the solutions depends on the existence and analyticity of the integrals. If the source term is disregarded, then \eqref{Msol4} is reduced immediately to \eqref{Msol2}.

Now, we will consider the case for a radiative family of solutions, $m\ne 0, \ |m|\le l$ for $l> 0$, without source term. In this case \eqref{field_eq_x13} becomes a Bessel's type differential equation. \citeonline{M13} previously showed that the general solutions to this master equation can be expressed as a linear combination of the first and second kind spherical Bessel's functions. We find here that the family of solutions to the master equation  \eqref{field_eq_x13} can be expressed in terms only of the first kind Bessel's functions, as
\begin{align}
\tilde{J}_{lm}&=\frac{C_1 2^{\frac{1}{2}-2 l} z^{3/2} e^{\frac{1}{2} i (\pi  l+2 z)} \Gamma \left(\frac{1}{2}-l\right) \left(K
	J_{-l-\frac{1}{2}}+L J_{\frac{1}{2}-l}\right)}{(l-1) l}\nonumber\\
&+\frac{i C_2 2^{2 l+\frac{5}{2}} z^{3/2} e^{i
		z-\frac{i \pi  l}{2}} \Gamma \left(l+\frac{3}{2}\right) \left(K J_{l+\frac{1}{2}}+L
	J_{l-\frac{1}{2}}\right)}{(l+1) (l+2)},
\label{mes2}
\end{align}
where the argument of the first kind Bessel's functions $J_{n}$ are referred to $z$, which is defined as
\begin{equation}
z=\dfrac{|m|\dot{\tilde{\phi}}}{x},
\label{zdef}
\end{equation}
and the coefficients $K$, $L$ and $S$ are given by
\begin{subequations}
	\begin{align}
    K&=-i (l(l-1)+2iz) -2z(l-i z),\\
	L&=-2 z (z-i),\\
	S&=l(l-1)+2iz.
	\end{align}
\end{subequations}
Integrating two times \eqref{mes2} and rearranging the constants we find the family of solutions that satisfies \eqref{field_eq_x11}, i.e.,
\begin{align}
J_{lm}=&-\frac{iC_1 2^{\frac{1}{2}-2 l} \dot{\tilde{\phi}} ^2 \left| m\right| ^2 z^{-1/2}e^{\frac{1}{2} i (\pi  l+2 z)} \Gamma
	\left(\frac{1}{2}-l\right) \left(-2 z J_{\frac{1}{2}-l}+\overline{S}
	J_{-l-\frac{1}{2}}\right)}{l^2 \left(l^2-1\right)}\nonumber\\
&-\frac{C_22^{2 l+\frac{5}{2}} \dot{\tilde{\phi}} ^2 \left| m\right| ^2 z^{-1/2}e^{-\frac{1}{2} i (\pi  l-2 z)} \Gamma
	\left(l+\frac{3}{2}\right) \left(2 z J_{l-\frac{1}{2}}+\overline{S}
	J_{l+\frac{1}{2}}\right)}{l (l+1)^2 (l+2) }\nonumber\\
&+C_3+C_4\frac{\dot{\tilde{\phi}} |m|}{z}.
\end{align}
When matter is considered, we find that the family of solutions to \eqref{field_eq_x13} becomes
\begin{align}
\tilde{J}_{lm}=&\frac{2^{\frac{1}{2}-2 l} z^{3/2} \left(C_1+D_1\right) e^{\frac{i \pi  l}{2}+i z} \Gamma
	\left(\frac{1}{2}-l\right) \left(K J_{-l-\frac{1}{2}}+L J_{\frac{1}{2}-l}\right)}{(l-1) l}\nonumber\\
&+\frac{i 2^{2
		l+\frac{5}{2}} z^{3/2} \left(C_2+D_2\right) e^{i z-\frac{i \pi  l}{2}} \Gamma \left(l+\frac{3}{2}\right)
	\left(K J_{l+\frac{1}{2}}+L J_{l-\frac{1}{2}}\right)}{(l+1) (l+2)},
\label{mes3}
\end{align}
where the coefficients $K$ and $L$ were defined above, and the terms representing the sources are
\begin{subequations}	
\begin{align}
D_1=-&\int_{|m|\dot{\tilde{\phi}}}^{|m|\dot{\tilde{\phi}}/z}d\tilde{z}\, \frac{2^{2 l-\frac{5}{2}} \tilde{z}^{-1/2} e^{-\frac{1}{2} i (\pi  l+2 \tilde{z})} \Gamma \left(l+\frac{1}{2}\right)  \left(K J_{l+\frac{1}{2}}-L
	J_{l-\frac{1}{2}}\right)}{(l+1) (l+2) \dot{\tilde{\phi}} ^2 \left| m\right| ^2}H\left(\frac{\dot{\tilde{\phi}}  \left| m\right| }{\tilde{z}}\right),
\end{align}
and
\begin{align}
D_2=&-i\int_{|m|\dot{\tilde{\phi}}}^{|m|\dot{\tilde{\phi}}/z}d\tilde{z}\, \frac{ 2^{-2 l-\frac{9}{2}}\tilde{z}^{-1/2} e^{\frac{1}{2} i (\pi  l-2 \tilde{z})} \Gamma \left(-l-\frac{1}{2}\right)  \left(K J_{-l-\frac{1}{2}}+L
	J_{\frac{1}{2}-l}\right)}{(l-1) l \dot{\tilde{\phi}} ^2  \left| m\right| ^2} H\left(\frac{\dot{\tilde{\phi}}  \left| m\right| }{\tilde{z}}\right),
\end{align}
\end{subequations}
where the argument of the first kind Bessel's functions $J_n$ is $z$, which is defined just in \eqref{zdef}.
It is worth noting that in this form, it is clear that \eqref{mes3} converges immediately to \eqref{mes2}, when the sources are not considered. 

Integrating \eqref{mes3} two times  we obtain the general family of solutions to the master equation with sources, which reads 
\begin{align}
J_{lm}=&-\frac{iC_1 2^{\frac{1}{2}-2 l} \dot{\tilde{\phi}} ^2 \left| m\right| ^2 z^{-1/2}e^{\frac{1}{2} i (\pi  l+2 z)} \Gamma
	\left(\frac{1}{2}-l\right) \left(-2 z J_{\frac{1}{2}-l}+\overline{S}
	J_{-l-\frac{1}{2}}\right)}{l^2 \left(l^2-1\right)}\nonumber\\
&-\frac{C_22^{2 l+\frac{5}{2}} \dot{\tilde{\phi}} ^2 \left| m\right| ^2 z^{-1/2}e^{-\frac{1}{2} i (\pi  l-2 z)} \Gamma
	\left(l+\frac{3}{2}\right) \left(2 z J_{l-\frac{1}{2}}+\overline{S}
	J_{l+\frac{1}{2}}\right)}{l (l+1)^2 (l+2) }\nonumber\\
&+\int_b^{z}dy\,\int_{a}^{y}d \tilde{z}\, \left(\frac{2^{\frac{1}{2}-2 l} \tilde{z}^{3/2} D_1 e^{\frac{i \pi  l}{2}+i \tilde{z}} \Gamma
	\left(\frac{1}{2}-l\right) \left(K J_{-l-\frac{1}{2}}+L J_{\frac{1}{2}-l}\right)}{(l-1) l}\right.\nonumber\\
&\left.+\frac{i 2^{2
		l+\frac{5}{2}} \tilde{z}^{3/2} D_2 e^{i \tilde{z}-\frac{i \pi  l}{2}} \Gamma \left(l+\frac{3}{2}\right)
	\left(K J_{l+\frac{1}{2}}+L J_{l-\frac{1}{2}}\right)}{(l+1) (l+2)}\right)      \nonumber\\
&+C_3+C_4\frac{\dot{\tilde{\phi}} |m|}{z}.
\label{mes4}
\end{align}
These families of solutions are particularly interesting and useful to explore the dynamics of matter clouds immersed in a Minkowski's background.
\subsection{The Schwarzschild's Background}
Now, we show the non-radiative families of solutions, $m=0$, for the vacuum i.e., $G(x)=0$, for equation \eqref{meqsch2}. The solution is expressed in terms of the hypergeometric functions $_2F_1(a_1,a_2;b;z)$, as
\begin{align}
\tilde{J}_{lm}=& (-2)^{-l-2} C_1  M^{-l-2} x^{-l-2} \, _2F_1(2-l,-l;-2 l;2 M x) \nonumber\\
&+ (-2)^{l-1} C_2  M^{l-1} x^{l-1} \, _2F_1(l+1,l+3;2l+2;2 M x).
\label{sch1}
\end{align}
Integrating two times, we find the family of solutions to \eqref{meqsch1}, namely
\begin{align}
J_{lm}=& \frac{C_1 (-1)^{-l} 2^{-l-2} (M x)^{-l} \, _3F_2(-l-1,2-l,-l;1-l,-2 l;2 M x)}{l (l+1) M^2}\nonumber\\
&+\frac{C_2 (-1)^{l+1}
	2^{l-1} x (M x)^l \, _3F_2(l,l+1,l+3;l+2,2 l+2;2 M x)}{l (l+1) M} + C_3x +C_4,
\label{sch2}
\end{align}
where, $_pF_q(a_1,\cdots a_p ;b_1,\cdots, b_q;z)$ are the generalised hypergeometric functions.

When we consider the source terms, i.e., $H(x)\ne 0$, the non radiative solutions to \eqref{meqsch2} reads
\begin{align}
\tilde{J}_{lm}=&(-1)^{1-l} 2^{-l-2} M^{-l-2} x^{-l-2} \left(A_2 (-1)^{2 l} 2^{2 l+1} M^{2 l+1} x^{2 l+1} \, _2F_1(l+1,l+3;2 l+2;2M x)\right.\nonumber\\
&\left.-A_1 \, _2F_1(2-l,-l;-2 l;2 M x)\right)
 +C_1 (-2)^{-l-2} M^{-l-2} x^{-l-2} \, _2F_1(2-l,-l;-2 l;2 M x)\nonumber\\
&+C_2 (-2)^{l-1} M^{l-1} x^{l-1} \, _2F_1(l+1,l+3;2 l+2;2 M x),
\label{sch3}
\end{align}
where $A_1$, $A_2$ are given by the integrals
\begin{subequations}
\begin{align}
A_1&=-\int_a^xdy\,\frac{(-2)^{l+2} H(y) M^{l+2} y^l \, _2F_1(l+1,l+3;2 (l+1);2 M y)}{B_1+B_2},\\
A_2&=\int_b^x dy\, \frac{(-2)^{1-l} H(y) M^{1-l} y^{-l-1} \, _2F_1(2-l,-l;-2 l;2 M y)}{B_1+B_2},
\end{align}
\end{subequations}
and the functions $B_1$ and $B_2$ are 
\begin{subequations}
\begin{align}
B_1=&(2 M y-1) ((l-2) \, _2F_1(3-l,-l;-2 l;2 M y)\, _2F_1(l+1,l+3;2 (l+1);2 M y),\\
B_2=&\, _2F_1(2-l,-l;-2 l;2 M y) (2 \, _2F_1(l+1,l+3;2 (l+1);2 M y)\nonumber\\
&+(l+1) \,_2F_1(l+2,l+3;2 (l+1);2 M y))).
\end{align}
\end{subequations}

For the radiative ($m\ne 0$) family of solutions to the master equation \eqref{meqsch2} for the vacuum, we find that its most general solution is given by
\begin{align}
\tilde{J}_{lm}=& C_1 L e^{\frac{2\alpha}{xM}} x^{-4}+C_2K \left(2Mx-1\right)^{4\alpha -2}x^{-2-4\alpha} e^{\frac{2\alpha}{xM}},
\label{heunc_sols}
\end{align}
with
\begin{align}
L=H_C\left(-4\alpha ,\beta;\gamma,\delta,\epsilon,\eta\right)\hspace{0.5cm} {\rm and} \hspace{0.5cm}
K=H_C\left(-4\alpha ,-\beta;\gamma,\delta,\epsilon,\eta\right),
\end{align}
where $H_C(\alpha,\beta;\gamma,\delta,\epsilon,\eta)$ are the confluent Heun's functions and their parameters are given by
\begin{subequations}
\begin{align}
\alpha &= i\dot{\tilde{\phi}}m M,\hspace{4.6cm}\beta  = 2-4\alpha,\\
\gamma &= 2, \hspace{5.6cm} \delta = 8\alpha(\alpha-1),\\
\epsilon &=-(l+2)(l-1)-8\alpha(\alpha-1),\hspace{0.8cm}\eta =\frac{2Mx-1}{2Mx}.
\end{align}
\label{heun_par}
\end{subequations}

Finally, we present the analytical family of solutions to \eqref{meqsch2} in the radiative case, $m\ne 0$,  when the source terms are considered, i.e.,
\begin{align}
J_{lm} =&-8M{e^{\frac {2a}{Mx}}} \left( -LMx+{M}^{2}{x}^{2}L+L/4
\right) A_1 x^{-4} \left( 2Mx-1 \right) ^{-2}\nonumber\\
&+2M{e^{\frac {2a}{Mx}}}{x}^{2-4a} \left( 2Mx-1 \right) ^{4a}A_2 K{x}^{-4} \left( 2Mx-1 \right) ^{-2}\nonumber\\
&+C_1 L e^{\frac {2a}{Mx}} x^{-4} + C_2 K e^{\frac {2a}{Mx}} x^{-2-4a}
\left( 2Mx-1 \right)^{-2+4a},
\end{align}
where $A_1$ and $A_2$ are the integrals
\begin{subequations}
\begin{align}
A_1=&\int_a^x d\tilde{x}\, \frac{\tilde{x}^2 H(\tilde{x}) e^{-\frac{2a}{M\tilde{x}}}K}{-4\,
	LKM\tilde{x}+8\,LKaM\tilde{x}-LS+2\,LM\tilde{x}S+KR-2\,KM\tilde{x}R} \label{a1}\\
A_2=&\int_b^x d\tilde{x}\, \frac{4\tilde{x}^{4a}{e^{-\frac {2a}{M\tilde{x}}}}H(\tilde{x})
\left( M\tilde{x}-1/2 \right) ^{2} \left( 2M\tilde{x}-1 \right) ^{-4a}L}{-4
\,LKM\tilde{x}+8\,LKaM\tilde{x}-LS+2\,LM\tilde{x}S+KR-2\,KM\tilde{x}R}\label{a2} ,
\end{align}
\end{subequations}
where $S$ and $R$ are the derivative of the Heun's functions, i.e.,  $S=K'(x)$ and $R=L'(x)$, in which we suppress all indices except one which gives the functional dependence. 

\section{Families of Solutions for $l=2$}
Now, we show that the families of solutions found here are reduced to those previously reported in the literature for $l=2$. Thus, for this particular value of $l$ we obtain that the family of solutions to the master equation for the vacuum, \eqref{field_eq_x13} takes the explicit form
\begin{equation}
\tilde{J}_{lm}=E_1 x+\frac{E_2 e^{\frac{2 i \dot{\tilde{\phi }} \left| m\right| }{x}} \left(6 x^3 \dot{\tilde{\phi }} \left| m\right| -6 i x^2 \dot{\tilde{\phi }}^2 \left| m\right| ^2-4 x
	\dot{\tilde{\phi }}^3 \left| m\right| ^3+2 i \dot{\tilde{\phi }}^4 \left| m\right| ^4+3 i x^4\right)}{4 x^3 \dot{\tilde{\phi }}^5 \left| m\right| ^5}.
\label{fam1}
\end{equation}
Now, substituting $l=2$ in the family of solutions \eqref{mes2}, one obtains
\begin{align}
\tilde{J}_{lm}=&\frac{i C_1 \dot{\tilde{\phi}} ^3 \left| m\right| ^3 e^{\frac{2 i \dot{\tilde{\phi}}  \left| m\right| }{x}}}{6 x^3}-\frac{40 i C_2 \dot{\tilde{\phi}} ^3 \left| m\right| ^3 e^{\frac{2 i \dot{\tilde{\phi}}  \left| m\right|
		}{x}}}{x^3}-\frac{C_1 \dot{\tilde{\phi}} ^2 \left| m\right| ^2 e^{\frac{2 i \dot{\tilde{\phi}}  \left| m\right| }{x}}}{3 x^2} +\frac{80 C_2 \dot{\tilde{\phi}} ^2 \left| m\right| ^2 e^{\frac{2 i \dot{\tilde{\phi}}  \left| m\right|
	}{x}}}{x^2} \nonumber
\\&-\frac{i C_1 \dot{\tilde{\phi}}  \left| m\right|  e^{\frac{2 i \dot{\tilde{\phi}}  \left| m\right| }{x}}}{2 x}+\frac{120 i C_2 \dot{\tilde{\phi}}  \left| m\right|  e^{\frac{2 i \dot{\tilde{\phi}}  \left| m\right|
}{x}}}{x}+\frac{1}{2} C_1 e^{\frac{2 i \dot{\tilde{\phi}}  \left| m\right| }{x}} -120 C_2 e^{\frac{2 i \dot{\tilde{\phi}}  \left| m\right| }{x}}\nonumber
\\&+\frac{i C_1 x e^{\frac{2 i \dot{\tilde{\phi}}  \left| m\right| }{x}}}{4 \dot{\tilde{\phi}} 
\left| m\right| }+\frac{i C_1 x}{4 \dot{\tilde{\phi}}  \left| m\right| }-\frac{60 i C_2 x e^{\frac{2 i \dot{\tilde{\phi}}  \left| m\right| }{x}}}{\dot{\tilde{\phi}}  \left| m\right| }+\frac{60 i  x}{\dot{\tilde{\phi}}  \left|
m\right| }.
\label{fam2}
\end{align} 
Both family of solutions, \eqref{fam1} and \eqref{fam2}, are completely equivalent. Note that, the transformation between the constants, necessary to pass from \eqref{fam1} to \eqref{fam2} is given by
\begin{equation}
E_1= \frac{i \left(C_1+240 C_2\right)}{4 \dot{\tilde{\phi}}  \left| m\right| },\hspace{1cm}
E_2= \frac{1}{3} \left(C_1-240 C_2\right) \dot{\tilde{\phi}} ^4 \left| m\right| ^4.
\end{equation}
Note that for the Schwarzschild case, when no sources are present, the master equation \eqref{meqsch2} for the vacuum and $l=2$ takes the explicit form
\begin{align}
x^2 (2 M x-1) \tilde{J}_{lm,xx}+2 x (7 M x-2) \tilde{J}_{lm,x}+ (16 M x+4)\tilde{J}_{lm}=0,
\end{align}
whose family of solutions is
\begin{equation}
\tilde{J}_{lm}=\frac{C_1}{x^4}-\frac{C_2 \left(16 M^4 x^4+32 M^3 x^3-44 M^2 x^2-4 M x+12 (1-2 M x)^2 \log (1-2 M x)+7\right)}{64
	M^5 x^4 (1-2 M x)^2}.
\end{equation}
Now, specialising the solutions \eqref{sch1} for $l=2$, we find a totally equivalent solution, i.e.,
\begin{equation}
\tilde{J}_{lm}=\frac{D_1}{16 M^4 x^4}+\frac{5 D_2 \left(2 M x \left(2 M^3 x^3+4 M^2 x^2-9 M x+3\right)+3 (1-2 M x)^2 \log (1-2 M
	x)\right)}{8 M^4 x^4 (1-2 M x)^2}.
\end{equation}
Thus, a simple Maclaurin series expansion of both solutions shows that the relationship between the constants is
\begin{align}
D_1= \frac{64 C_1 M^5-7 C_2}{4 M}\hspace{0.5cm}\text{and}\hspace{0.5cm} D_2 = -\frac{C_2}{10 M}.
\end{align}
Finally, given that the known family of solutions for $l=2$ is written in terms of power series around $r=2M$, as shown in \cite{B05}, we expand the radiative family of solutions for the master equation \eqref{meqsch1} around the same point $r=2M$ for $l=2$. Thus, we observe that the Confluent Heun's function $H_C(-4\alpha,\beta;\gamma,\delta,\epsilon,\eta)$ is expressed as a Taylor series for the parameters \eqref{heun_par} around $\eta=0$, namely
\begin{align}
H_C(-4\alpha,\beta;\gamma,\delta,\epsilon,\eta)\simeq& 1+{\frac { \left( (4a+1)^2-5+(l-1)(l+2) \right) \eta}{-3+4\,a}}\nonumber\\
&+\frac{1}{8 (a-1) (4 a-3)}
\left( \left(256 a^4+192 a^3+32 a^2 \left(l^2+l-5\right) \right.\right.\nonumber\\
&+\left.\left. 4 a \left(4 l^2+4 l-39\right)+l^4+2 l^3-17 l^2-18 l+72\right)\eta^2\right),
\label{Heun1}
\end{align}
and for the Confluent Heun's function $H_C(-4\alpha,-\beta;\gamma,\delta,\epsilon,\eta)$, i.e.,
\begin{align}
H_C(-4\alpha,-\beta;\gamma,\delta,\epsilon,\eta)\simeq&1-\frac{\left(4 a+l^2+l\right)\eta}{4 a-1}-\frac{\left(12 a-l^4-2 l^3+l^2+2 l\right)\eta^2 }{8 a (4 a-1)}.
\label{Heun2}
\end{align}
Then, from \eqref{Heun1} and \eqref{Heun2} we obtain that around $r=2M$, \eqref{heunc_sols} at first order for $l=2$ reads
\begin{align}
\tilde{J}_{lm}=& C_1 \left(\frac{16 e^{4 \alpha } (4 \alpha +12) \eta  M^4}{4 \alpha -3}+16 e^{4 \alpha } M^4\right)
-\frac{2^{4 \alpha +2} C_2 e^{4 \alpha } \left(16 \alpha ^2+16 \alpha +2\right) \eta ^{4 \alpha -1}	\left(\frac{1}{M}\right)^{-4 \alpha -2}}{4 \alpha -1}\nonumber
\\
&+\frac{2^{4 \alpha -1} C_2 e^{4 \alpha } \left(256 \alpha^4+576 \alpha ^3+384 \alpha ^2+132 \alpha +24\right) \eta ^{4 \alpha } \left(\frac{1}{M}\right)^{-4 \alpha	-2}}{\alpha  (4 \alpha -1)}\nonumber\\
&-\frac{2^{4 \alpha +1} C_2e^{4 \alpha } \left(256 \alpha ^5+896 \alpha ^4+1056 \alpha^3+636 \alpha ^2+228 \alpha +72\right) \eta ^{4 \alpha +1} \left(\frac{1}{M}\right)^{-4 \alpha -2}}{3 \alpha(4 \alpha -1)}\nonumber\\
&+2^{4 \alpha +2} C_2 e^{4 \alpha } \eta ^{4 \alpha -2} \left(\frac{1}{M}\right)^{-4 \alpha -2},
\end{align}
that are just the family of solutions for the master equation obtained using power series around $r=2M$.

\section{Thin Shell}
In this section we examine a static thin shell in a Minkowski's background, initially studied in \cite{B05}, as an example of application of the solutions of the master equation when the system is restricted to $l=2$ and $\dot{\tilde{\phi}}=0$. This example illustrates the process of solution of the field equations when a static matter distribution such as a spherical thin shell 
is considered. The space-time is divided into two distinct empty regions connected through the jumps imposed into the metric of the space-time and its first derivatives. Here  
boundary conditions at the vertices of the null cones, at the null infinity and on the shell surface are imposed. The master equation is solved for each empty region, which are then connected through the jump conditions on the metric and its derivatives. This procedure fixes the constants of integration, thus the solution to the field equations is found. Physically we are interested in a spherical distribution of matter of radius $r_0$,  centred at the origin of the coordinates for which its density of energy is given by
\begin{equation}
\rho=\rho_0 \delta(r-r_0)\ _0Z_{2m}.
\label{ts_0}
\end{equation} 
Here, the metric variables are restricted to be represented by
\begin{equation}
_sf= \Re\left(f_{0}
\right)\ \eth^s\ Z_{2m},
\label{expansion_ts}
\end{equation}
where $f$ represents any of the $\beta, w, U, J$ functions. Notice that the metric variables do not depend on time, i.e., $_sf_{,u}=0$.

Then, substituting \eqref{expansion_ts} into \eqref{linear_field_eqs}, the system of equations for the vacuum is reduced to
\begin{subequations}
\begin{align}
&\frac{d\beta_0}{dr} =0,\label{ts_2.1}\\
& -4r\frac{dJ_0}{dr}+4r^2\frac{dU_0}{dr}+r^3\frac{d^2U_0}{dr^2}+4\beta_0=0, \label{ts_2.2}\\
& 3r^2\frac{dU_0}{dr}+\frac{dw_0}{dr}-12J_0+12rU_0-8\beta_0 =0,\label{ts_2.3}\\
& -2r\frac{dJ_0}{dr}+r^2\frac{dU_0}{dr}-r^2\frac{d^2J_0}{dr^2}+2rU_0-2\beta_0=0, \label{ts_2.4}\\
& -r^2\frac{d^2w_0}{dr^2}+6w_0 +12r\beta_0 =0, \label{ts_2.5}\\
& 6r^2\frac{dU_0}{dr}+r\frac{d^2w_0}{dr^2}+12rU_0-12\beta_0=0,\label{ts_2.6}\\
&4r^3\frac{dU_0}{dr} + r\frac{dw_0}{dr} + r^4\frac{d^2U_0}{dr^2} + 2r^2U_0 - w_0=0. \label{ts_2.7}
\end{align}
\label{ts_2}%
\end{subequations}
The master equation \eqref{field_eq_x13} for this case, is strongly simplified
\begin{equation}
x^3 \frac{d^2J_2}{dx^2} + 4x^2\frac{dJ_2}{dx} - 4xJ_2 =0,
\label{ts_1}
\end{equation}
where we recall that $x=1/r$. Thus, the family of solutions that satisfy \eqref{ts_1} reads
\begin{equation}
J_2(x)=\tilde{C}_1 x+\frac{\tilde{C}_2}{x^4}.
\label{ts_3}
\end{equation}
Then, integrating \eqref{ts_3} two times one obtains the family of solutions $J_0$, i.e.,
\begin{eqnarray}
J_0(x)&=&\int dx \left(\int dx \ J_2(x)\right), \nonumber\\
&=&\frac{\tilde{C}_1 x^3}{6}+\frac{\tilde{C}_2}{6 x^2}+\tilde{C}_3 x+\tilde{C}_4, 
\label{ts_4}
\end{eqnarray}
or in terms of $r$, it can be written as 
\begin{eqnarray}
J_0(r)&=&C_1+C_2 r^2+\frac{C_3}{r}+\frac{C_4}{r^3},
\label{ts_5}
\end{eqnarray}
where we have done a redefinition of the constants of integration. 

\begin{figure}[h]
\begin{center}
\graphicspath{{./Figuras/Thin_Shell/}}
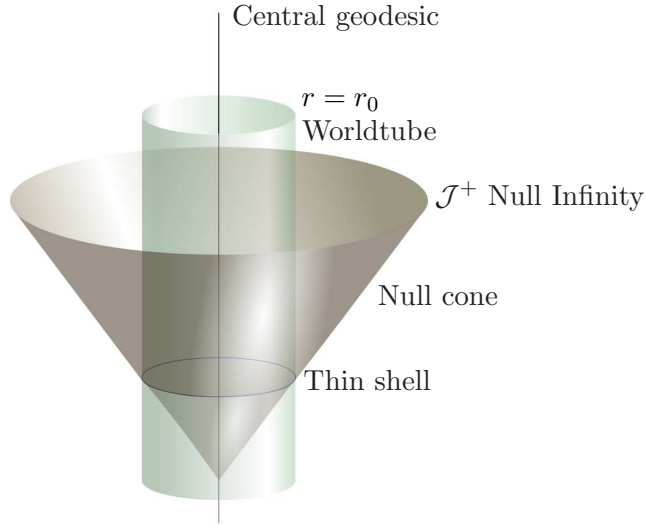
\end{center}
\caption{Sketch of the world tube generated by the thin shell. Here we note the two regions ($r<r_0$ and $r>r_0$) in which the space-time is divided.}
\label{TS1}
\end{figure}

Integrating \eqref{ts_2.1}, and with the family of solutions \eqref{ts_5}, we solve the equations \eqref{ts_2.2} and \eqref{ts_2.3}, thus
\begin{subequations}
\begin{eqnarray}
\beta(r)&=&\beta_0, \label{ts_6.1}\\
U_0(r)&=&-\frac{3 C_4}{r^4}-\frac{C_5}{3 r^3}+\frac{2 C_3}{r^2}+2 C_2 r+C_6+\frac{2 \beta _0}{r},\label{ts_6.2}\\
w_0(r)&=& -6 C_2 r^3-6 C_6 r^2-\frac{6 C_4}{r^2}+12 C_1 r-\frac{C_5}{r}+C_7- 10 r \beta _0. \label{ts_6.3}
\end{eqnarray}
\label{ts_6}
\end{subequations}
When the family of solutions \eqref{ts_5} and \eqref{ts_6} are substituted into  equations \eqref{ts_2.4},\eqref{ts_2.4} and \eqref{ts_2.7} the following constraint conditions are obtained 
\begin{subequations}
\begin{eqnarray}
6 C_6 r^2+\frac{C_5}{r}=0,\label{ts_7.1}\\
 12 C_6 r^2-36 C_1 r+\frac{2 C_5}{r}-3 C_7+24 r \beta _0=0,\label{ts_7.2}\\
 -4 C_6 r^2+\frac{4 C_5}{3 r}-C_7=0,\label{ts_7.3}
\end{eqnarray}
\label{ts_7}%
\end{subequations}
where the constraint given by \eqref{ts_2.6} is satisfied identically. Then, solving $C_5$ in \eqref{ts_7.1} and replacing it in \eqref{ts_7.2} and \eqref{ts_7.3} the constraint equations are reduced to
\begin{subequations}
\begin{eqnarray}
C_5&=&-6 C_6 r^3, \label{ts_8.1}\\
C_7+ 4(3 C_1 -2  \beta _0)r&=&0, \label{ts_8.2}\\
- 12 C_6 r^2 -C_7&=&0. \label{ts_8.3}
\end{eqnarray}
\label{ts_8}
\end{subequations}
Substituting $C_5$ into equations \eqref{ts_6} we obtain
\begin{subequations}
\begin{align}
U_0(r)&=-\frac{3 C_4}{r^4}+3 C_6+\frac{2 C_3}{r^2}+2 C_2 r +\frac{2 \beta _0}{r},\label{ts_9.1} 
\end{align}
\begin{align}
w_0(r)&= -6 C_2 r^3-\frac{6 C_4}{r^2}+12 C_1 r +C_7- 10 r \beta _0. \label{ts_9.2}
\end{align}
\label{ts_9}%
\end{subequations}
Now, since we are considering a spherical and statically thin shell around the origin, then we must consider two separate regions of the space-time formed by the world tube which binds the matter distribution i.e., $r<r_0$ and $r>r_0$. (See Figure \ref{TS1}).
\newpage
We will start with the interior region. In this case the family of solutions can be written as 
\begin{subequations}
\begin{align}
\beta_{0-}(r)=&\beta_{0-}, \label{ts_10.1}\\
J_{0-}(r)=&C_{1-}+ r^2 C_{2-}+\frac{C_{3-}}{r}+\frac{C_{4-}}{r^3}, \label{ts_10.2}\\
U_{0-}(r)=&-\frac{3 C_{4-}}{r^4}+3 C_{6-}+\frac{2 C_{3-}}{r^2}+2 r C_{2-}  +\frac{2 \beta _{0-}}{r}, \label{ts_10.3}\\
w_{0-}(r)=& -6 r^3 C_{2-} -\frac{6 C_{4-}}{r^2}+12 r C_{1-}  +C_{7-}- 10 r \beta _{0-}. \label{ts_10.4}
\end{align}
\label{ts_10}%
\end{subequations}
It is expected that the space-time does not have singularities at the origin of the three space, or in other words at the vertex of the null cones. Then, it is possible to impose convergence of the metric functions given in \eqref{expansion} at this point. To do so, we can expand the metric functions in power series of $r$ around the vertex of the null cones and check if they are convergent at this limit. 

Substituting \eqref{ts_10.4} into \eqref{bs_lin_pert} one obtains
\begin{subequations}
\begin{align}
g_{11-}=& 6 r^2 C_{2-} +\frac{6 C_{4-}}{r^3}-12 C_{1-}  -\frac{C_{7-}}{r}+ 12 \beta _{0-} + 1, \label{ts_11.1}\\
g_{12-}=& -1-2\beta_{0-},\label{ts_11.2} 
\end{align}
\begin{align}
g_{33-}=& \dfrac{2}{(1+\zeta\overline{\zeta})^2}\left[\Re \left( r^2C_{1-}+ r^4 C_{2-}+r C_{3-}+\frac{C_{4-}}{r} \right)\times \right. \nonumber\\
&\left. \left(\eth^2 +\overline{\eth}^2 \right)\ _0Z_{2m} +2r^2\right.{\bigg]},\label{ts_11.3} \\
g_{34-}=&-\dfrac{2i}{(1+\zeta\overline{\zeta})^2}\Re\left(r^2C_{1-}+ r^4 C_{2-}+rC_{3-}+\frac{C_{4-}}{r}\right)\times\nonumber\\
&\left(\eth^2 -\overline{\eth}^2\right)\ _0Z_{2m},\label{ts_11.4} \\
g_{44-}=&-\dfrac{2}{(1+\zeta\overline{\zeta})^2}\left[\Re \left( r^2C_{1-}+ r^4 C_{2-}+r C_{3-}+\frac{C_{4-}}{r} \right) \times \right. \nonumber\\
&\left. \left(\eth^2 +\overline{\eth}^2 \right)\ _0Z_{2m} -2r^2\right.{\bigg]}.\label{ts_11.5}
\end{align}
\label{ts_11}
\end{subequations}
Notice that in this limit, i.e., $r\rightarrow 0$, \eqref{ts_11.1} implies that 
\begin{align}
C_{4-}&=0, \label{ts_12}\\
C_{7-}&=0, \label{ts_13}
\end{align}
thus,
\begin{align}
\lim_{r\rightarrow 0}g_{11-}&=\lim_{r \rightarrow 0}\left(6 r^2 C_{2-} -12 (C_{1-}  + \beta _{0-}) + 1\right),\nonumber\\
&=-12 (C_{1-}  - \beta _{0-}) + 1.
\label{ts_14}
\end{align}
Then, if we expect a flat space-time in the null cone vertices, we must have 
\begin{equation}
C_{1-}  = \beta _{0-}.
\label{ts_15}
\end{equation}
Also, the convergence of $J_{0-}$ is required  at the vertex of the null cones. Thus, from \eqref{ts_10.2} we see that
\begin{equation}
C_{3-}=0,
\label{ts_16}
\end{equation}
and from \eqref{ts_10.3}
\begin{equation}
\beta_{0-}=0.
\label{ts_17}
\end{equation}
Thus, from \eqref{ts_15} one has
\begin{equation}
C_{1-}  = 0.
\label{ts_18}
\end{equation}
It implies that \eqref{ts_8.2} is satisfied identically, whereas from \eqref{ts_8.3} one obtains
\begin{equation}
C_{6-}=0. \label{ts_19}
\end{equation}
Substituting these constants in the families of solutions \eqref{ts_10} we obtain for the interior region that
\begin{subequations}
\begin{align}
\beta_{0-}(r)=&0,          \label{ts_20.1}\\
J_{0-}(r)=&r^2C_{2-},      \label{ts_20.2}\\
U_{0-}(r)=&2 r C_{2-},     \label{ts_20.3}\\
w_{0-}(r)=& -6 r^3 C_{2-}, \label{ts_20.4}
\end{align}
\label{ts_20}%
\end{subequations}
which means that the solution for the interior of the world tube depends only on one parameter. When higher values of $l$ are considered, analogue expressions for the interior solutions are obtained. Thus, the reduction of the degree of freedom for the system at the interior of the world tube is independent on the matter distribution on the shell. 

For the exterior region we have the same set of families of solutions given by \eqref{ts_10}, but replacing the minus sign in the functions and in the constants by a plus sign, i.e., 
\begin{subequations}
\begin{align}
\beta_{0+}(r)=&\beta_{0+}, \label{ts_21.1}\\
J_{0+}(r)=&C_{1+}+ r^2 C_{2+}+\frac{C_{3+}}{r}+\frac{C_{4+}}{r^3}, \label{ts_21.2}\\
U_{0+}(r)=&-\frac{3 C_{4+}}{r^4}+3 C_{6+}+\frac{2 C_{3+}}{r^2}+2 r C_{2+}  +\frac{2 \beta _{0+}}{r}, \label{ts_21.3}\\
w_{0+}(r)=& -6 r^3 C_{2+} -\frac{6 C_{4+}}{r^2}+12 r C_{1+}  +C_{7+}- 10 r \beta _{0+}. \label{ts_21.4}
\end{align}
\label{ts_21}%
\end{subequations}
We expect convergent solutions at the null infinity $\mathcal{J}_+$. At this limit, i.e., when $r\rightarrow \infty$, we see from \eqref{ts_21.2} that
\begin{equation}
 C_{2+}=0.
\label{ts_22}
\end{equation}
Thus,
\begin{align}
\lim_{r\rightarrow\infty}J_{0+}(r)=& \lim_{r\rightarrow\infty}\left( C_{1+}+\frac{C_{3+}}{r}+\frac{C_{4+}}{r^3} \right),\nonumber\\
=&C_{1+}.
\label{ts_23}
\end{align}
We rename this constant as
\begin{equation}
 C_{1+}=J_{0\infty},
\label{ts_24}
\end{equation}
indicating that it is the value of the $J_0(r)$ function at the null infinity.

Using the last results we note that the solutions for $U_{0}$, given by \eqref{ts_21.3}, are simplified to 
\begin{equation}
U_{0+}(r)=-\frac{3 C_{4+}}{r^4}+3 C_{6+}+\frac{2 C_{3+}}{r^2}+\frac{2 \beta _{0+}}{r}.
\label{ts_25}
\end{equation}
Now, it is required that the shift vector at the null infinity be null, thus 
\begin{equation}
\lim_{r\rightarrow\infty}U_{0+}(r)=0,
\label{ts_26}
\end{equation}
then,
\begin{equation}
C_{6+}=0.
\label{ts_27}
\end{equation}
Thus, \eqref{ts_25} takes the form
\begin{equation}
U_{0+}(r)=-\frac{3 C_{4+}}{r^4}+\frac{2 C_{3+}}{r^2}+\frac{2 \beta _{0+}}{r}.
\label{ts_28}
\end{equation}
The constraint \eqref{ts_8.3} fixes the value for $C_{7+}$, namely
\begin{equation}
C_{7+}=0,
\label{ts_29}
\end{equation}
and the conditions \eqref{ts_22} and \eqref{ts_29} simplifies the solution for $w_{0+}$ given by \eqref{ts_21.4}, i.e.,
\begin{equation}
w_{0+}(r)=-\frac{6 C_{4+}}{r^2}+12 r C_{1+} - 10 r \beta_{0+}.
\label{ts_30}
\end{equation}
When \eqref{ts_29} is used on the constraint \eqref{ts_8.2} we find the explicit value for $\beta_{0+}$, namely
\begin{eqnarray}
\beta _{0+}&=& \frac{3}{2} J_{0\infty}.
\label{ts_31}
\end{eqnarray}
With these constants, the families of solutions for the exterior region take the form
\begin{subequations}
\begin{align}
J_{0+}(r)&=J_{0\infty}+\frac{C_{3+}}{r}+\frac{C_{4+}}{r^3}, \label{ts_32.1}\\
U_{0+}(r)&=-\frac{3 C_{4+}}{r^4}+\frac{2 C_{3+}}{r^2}+\frac{3 J_{0\infty}}{r}, \label{ts_32.2}\\
w_{0+}(r)&=-\frac{6 C_{4+}}{r^2}- 3 r J_{0\infty}. \label{ts_32.3}
\end{align}
\label{ts_32}%
\end{subequations}
It is worth noting that the family of solutions for the exterior region depends only on two constants. For values of $l>2$, the same situation is repeated, i.e., for each $l$ greater than two the exterior solutions will depend only on two constants. 

Now, in order to fix the constants of integration $C_{2-}$, $C_{3+}$ and $C_{4+}$, we impose the jump conditions across the world tube generated by the thin shell, i.e., at $r=r_0$. These conditions are
\begin{equation}
\beta_{0}\big|_{r_{0-}}^{r_{0+}}=2\pi r_0\rho_0, \hspace{0.5cm} J_0\big|_{r_{0-}}^{r_{0+}}=0,\hspace{0.5cm}U_0\big|_{r_{0-}}^{r_{0+}}=0,\hspace{0.5cm} w_0\big|_{r_{0-}}^{r_{0+}}=-2r_0\beta_{0+},\hspace{0.5cm}
\label{ts_33}%
\end{equation}
where $\beta_{0+}=\beta_0(r_{0+})$, and
\begin{align}
\frac{dU_0}{dr}\Bigg|_{r_{0-}}^{r_{0+}} = \dfrac{2\beta_{0+}}{r_{0}^2}, 
\hspace{1cm}\frac{d J_0}{dr}\Big|_{r_{0-}}^{r_{0+}}  =0.
\label{ts_34}%
\end{align}

From \eqref{ts_33}, \eqref{ts_20.1} and \eqref{ts_21.1}  the function $\beta(r)$ results in
\begin{equation}
\beta(r)=\beta_{0+}\Theta(r-r_0),
\label{beta_fun}
\end{equation}
where $\Theta(r)$ is the Heaviside's function, namely
\begin{equation}
\Theta(r)=\begin{cases}
	0 &  \text{for} \ \ r\le 0\\
	1 & \text{for} \ \ r>0
\end{cases}.
\end{equation}
Evaluating the continuity conditions \eqref{ts_33} for $J_0$ and $w_0$ one has  
\begin{align}
r_0^2C_{2-}&=J_{0\infty} +\frac{C_{3+}}{r_0}+\frac{C_{4+}}{r_0^3}, \label{continuidade7}\\
2r_0^2C_{2-}&=-\frac{C_{3+}}{r_0}-3\frac{C_{4+}}{r_0^3}. \label{continuidade8}
\end{align}
Adding them, we obtain 
\begin{equation}
3r_0^2C_{2-}=J_{0\infty} -2\frac{C_{4+}}{r_0^3}.
\label{continuidade9}
\end{equation}
Evaluating the continuity conditions for $U_0$ \eqref{ts_33} and for $dU_0/dr$ \eqref{ts_34} one obtains
\begin{align}
& 2 r_0^2 C_{2-}=3 J_{0\infty}+\frac{2 C_{3+}}{r_0} -\frac{3 C_{4+}}{r_0^3}, \label{continuidade10}\\
& -\beta_{0+}-2\frac{C_{3+}}{r_0} +6\frac{C_{4+}}{r_0^3} - r_0^2C_{2-} =\beta_{0+}. \label{continuidade11}
\end{align}
Thus, from \eqref{continuidade8} and \eqref{continuidade10} $C_{3+}$ is determined, resulting in
\begin{equation}
C_{3+}=- r_0 J_{0\infty},
\label{continuidade12}
\end{equation}
which substituting in \eqref{continuidade11} one finds the following condition 
\begin{equation}
6\frac{C_{4+}}{r_0^3} - r_0^2C_{2-} =J_{0\infty}.
\label{continuidade13}
\end{equation} 
Solving \eqref{continuidade9} and \eqref{continuidade13} we find
\begin{align}
C_{4+}&=\frac{1}{5}r_0^3J_{0\infty},\\
C_{2-}&=\frac{1}{5r_0^2}J_{0\infty}.
\end{align}
Thus, we determine the three constants $C_{2-},\ C_{3+}$ and $C_{4+}$ for the shell and therefore we determine completely the solution of the system. 

Thus, the solution of the field equations reads
\begin{subequations}
\begin{align}
\beta(r)&=\frac{3}{2} J_{0\infty}\Theta(r-r_0),\\
J(r)&=\frac{r^2}{5r_0^2}J_{0\infty}\left(1-\Theta(r-r_0)\right)+J_{0\infty}\left(1-\frac{r_0 }{r}+\frac{r_0^3}{5r^3}\right)\Theta(r-r_0),\\
U(r)&=\frac{2r}{5r_0^2}J_{0\infty}\left(1-\Theta(r-r_0)\right)+J_{0\infty}\left(-\frac{3 r_0^3}{5r^4}-\frac{2 r_0 }{r^2}+\frac{3 }{r}\right)\Theta(r-r_0),\\
w(r)&=-\frac{6r^3}{5r_0^2}J_{0\infty}\left(1-\Theta(r-r_0)\right)-J_{0\infty}\left(\frac{6 r_0^3}{5r^2}+ 3 r \right)\Theta(r-r_0).
\end{align}
\label{TS2}
\end{subequations}
It is important to note that, from \eqref{ts_31} and \eqref{ts_33} one obtains
\begin{equation}
J_{0\infty}=\dfrac{4}{3}\pi r_0 \rho_0,
\end{equation}
which relates the value of the $J_0$ function at the null infinity with the density and the radius of the shell.

\begin{figure}[ht!]
\begin{center}
\begin{tabular}{@{\hspace{-1.3cm}}c@{\hspace{-2.5cm}}c}
\includegraphics[height=5.0cm]{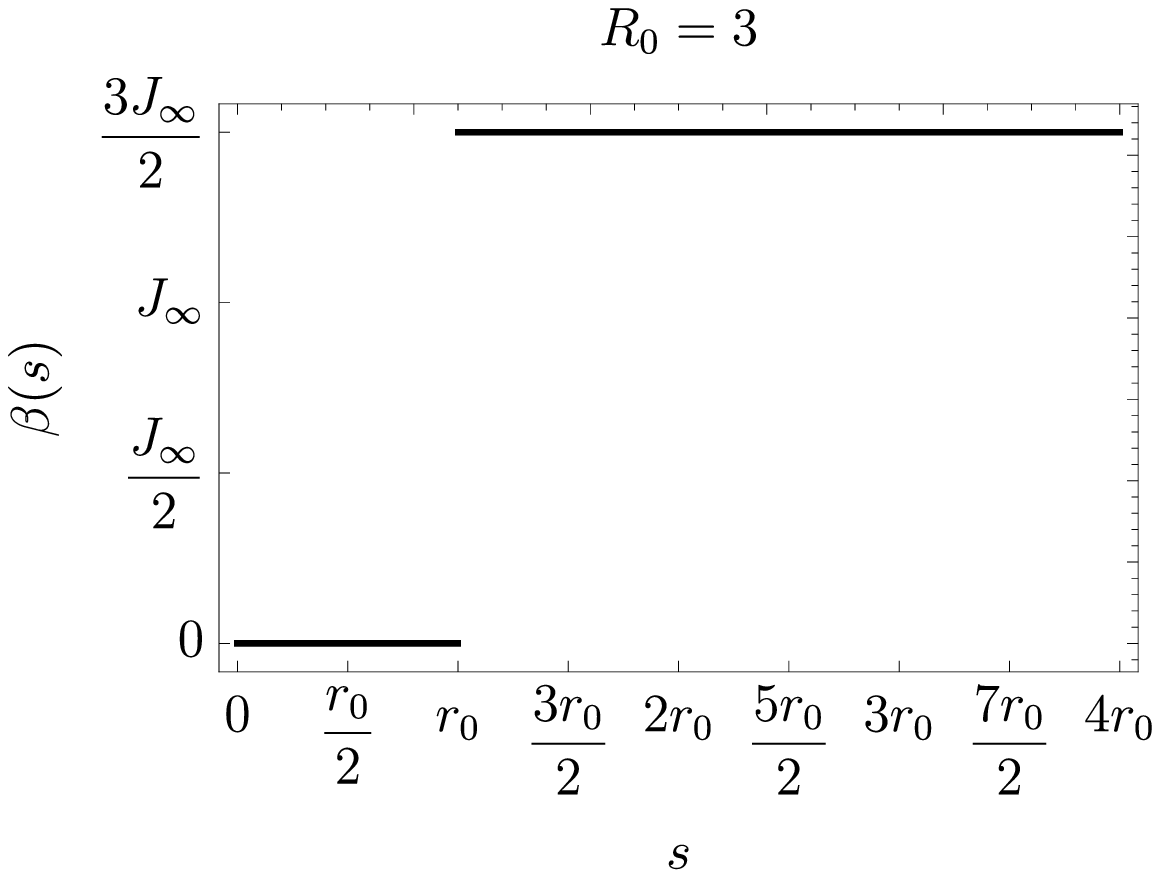} & \includegraphics[height=5.0cm]{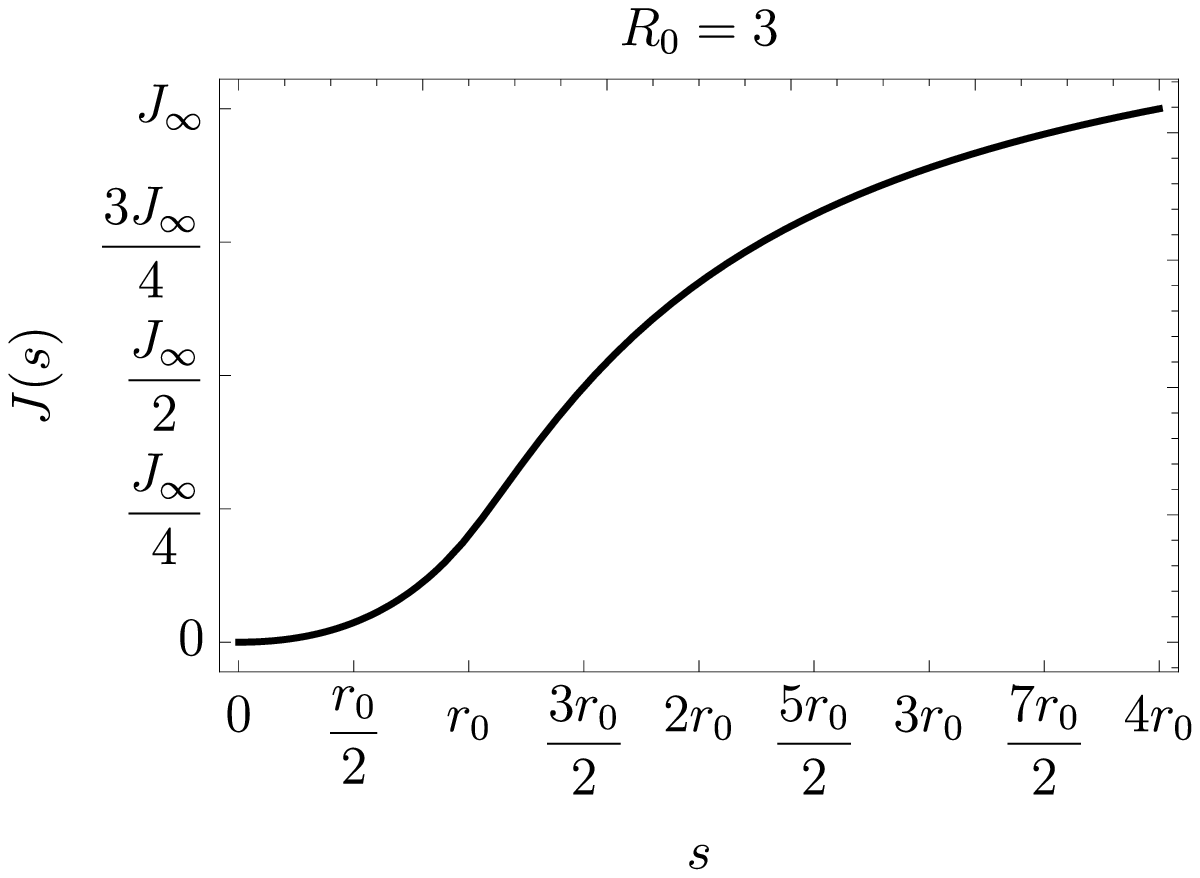} \\ 
(a) & (b) \\ 
\includegraphics[height=5.0cm]{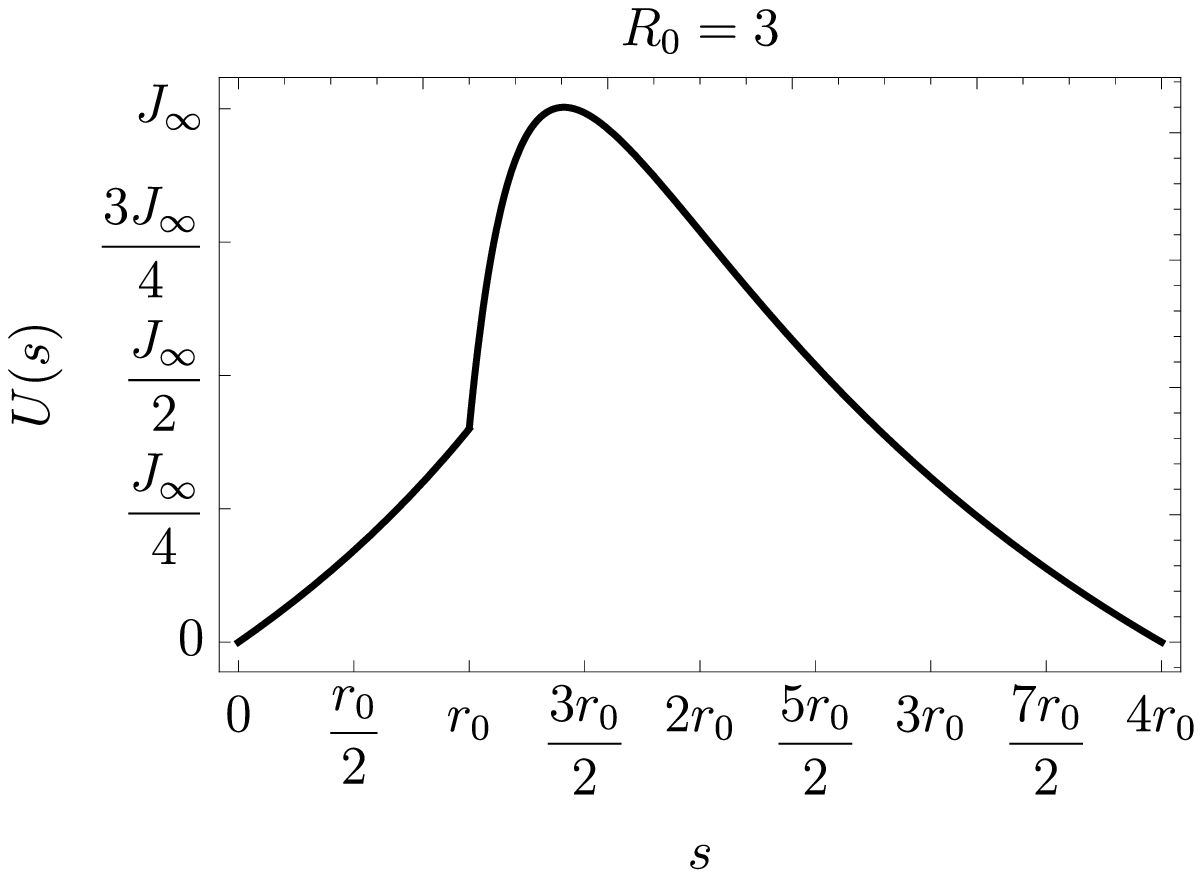} & \includegraphics[height=5.0cm]{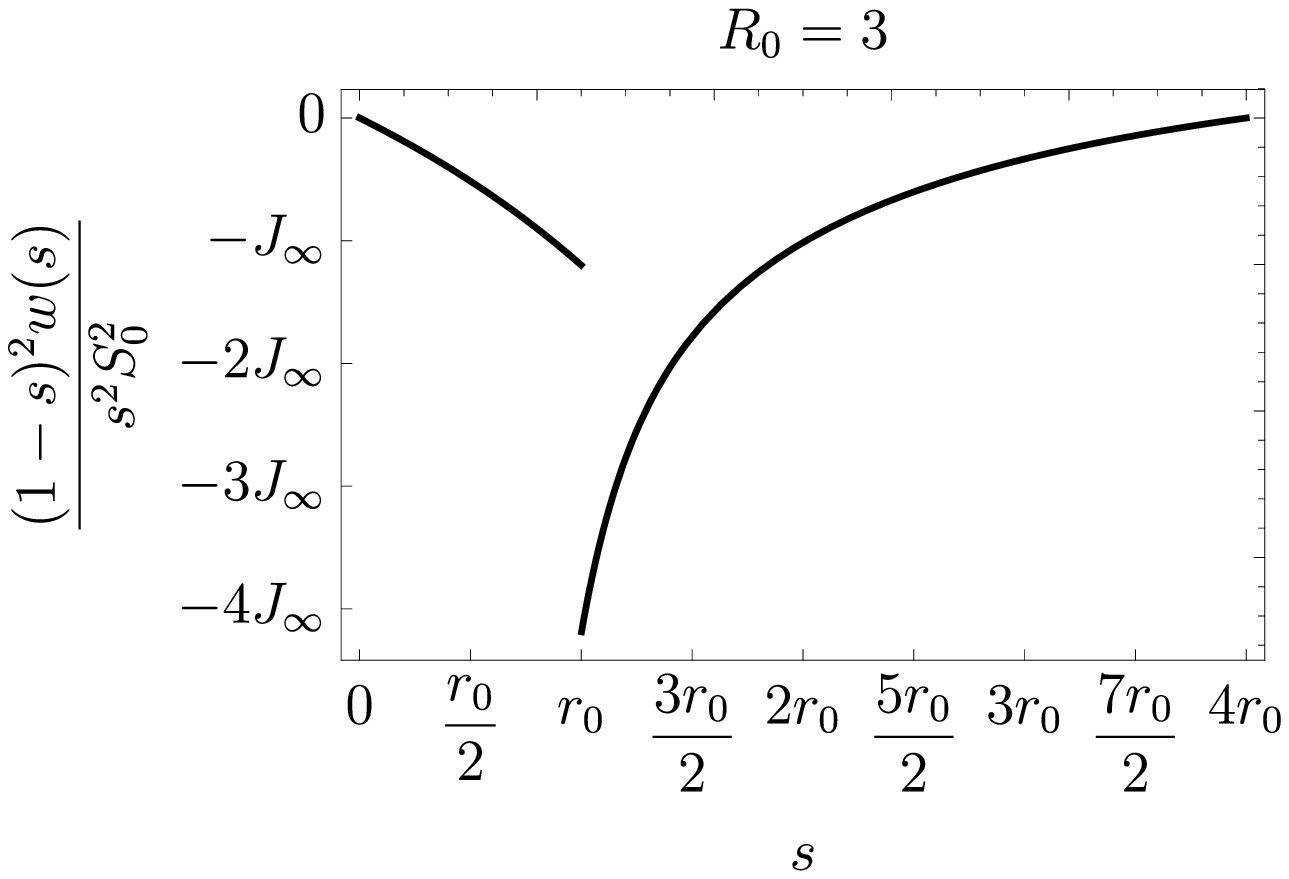} \\ 
(c) & (d)
\end{tabular}
\end{center}
\caption{Metric variables as a function of the compactified coordinate $s$ for a thin shell of $r=r_0$, centred at the origin. (a) $\beta_0:=\beta_0(s)$, (b) $J_0:=J_0(s)$, (c) $U_0:=U_0(s)$, (d) $w_0:=w_0(s)$}
\label{FTS}
\end{figure}

We plot the solutions \eqref{TS2} in Figure \ref{FTS}, in terms of a compactified coordinate $s$, which we define as
\begin{equation}
s=\dfrac{r}{r+R_0}
\label{comp}
\end{equation}
where $R_0$ is called {\it a compactification parameter}.  The transformation \eqref{comp} maps the luminosity distance, $0\le r<\infty$, into a finite interval $0\le s < 1$. 
Note that, if $r+R_0=0$, then $s$ would have singular points. Thus, considering that $r\ge 0$, the condition $R_0>0$ guaranties that the transformation \eqref{comp} will not have singular points and therefore it will be invertible.

\chapter{APPLICATIONS}
\label{APP}
Here, we study two novel applications of the solutions to the master equation. These applications are related to point particle binary systems. The first generalises a previous study \cite{B05}, now considering point particle binary systems of different masses in circular orbits \cite{CA15}; and the second considers binaries with elliptical orbits \cite{CA16b}. In both applications, the gravitational radiation patterns are obtained from the Bondi's News functions. Here, we generalise the boundary conditions \cite{B05,BPR11,K121,K13} imposed across the world tubes generated by the orbits of the binaries. The problem of the jump conditions imposed on the metric and its derivatives across a given time-like or space-like hypersurface, separating two regions of the space-time is not new \cite{T57,I58,I66,T80,BV81,G94,G96}.
\section{Point Particle Binary System with Different Masses}
Here, a study found in literature, in which the authors \cite{BPR11} considered particles with equal masses is generalised. It is worth stressing that one of our aims is to study the well-known problem of a system of two point particles with different masses orbiting each other in circular orbits. In the end, we show that the Peters and Mathews result for the power radiated in gravitational waves \cite{PM63} 
 can be obtained by using the characteristic formulation and the News function. 
 
In our study the particles are held together by their mutual gravitational interaction. The particles are far enough from each other such that at first order, the interaction between them can be considered essentially Newtonian. This assumption is valid if one considers the weak field approximation, in which the Bondi-Sachs metric in stereographic null coordinates is reduced to \eqref{bs_lin_pert}.

Note that writing $g_{11}\simeq -1+2\Phi$, then $\Phi=\beta+w/(2r)$ represents the Newtonian potential, as usual in this kind of approximation.  

\begin{figure}[h!]
	\begin{center}
		\includegraphics[height=6cm]{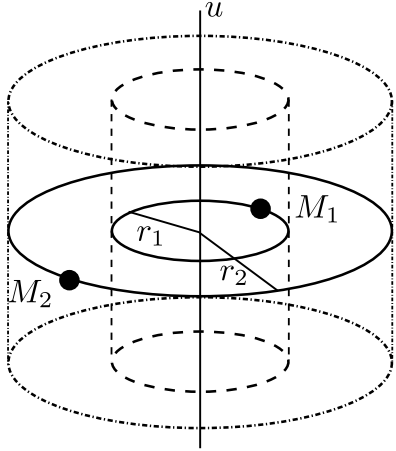}
	\end{center}
	\caption{Binary system with the world tubes of each orbit extended along the direction of the retarded time, separating the space-time into three regions.}
	\label{fig1}
\end{figure}

We consider that these two particles are in a Minkowski's background, in exactly the same way Peters and Mathews did in their paper of 1963 \cite{PM63} and Bishop et. al. did in \cite{BPR11}. Such a system allows one to explore in full detail the boundary conditions across the hypersurfaces generated by their orbits (see Figure \ref{fig1}). 
\graphicspath{{./Figuras/BinarySystem/}}
\\The density that describes the binary system is given by
\begin{equation}
\rho=\frac{\delta(\theta-\pi/2)}{r^2}\left(M_1\delta(r-r_1)\delta(\phi-\nu u) + M_2\delta(r-r_2)\delta(\phi-\nu u-\pi)\right),
\label{dens}
\end{equation}
where, $r_i\ (M_i)$ are the orbital radius (mass) of each particle and $r_1 < r_2$. 

The orbit of each mass generates world tubes, which are extended along the retarded time, allowing the separation of the space-time into three empty regions: inside, between and outside the matter distribution. 

In order to solve the field equations \eqref{field_eq_1}-\eqref{field_eq_7} for the vacuum, the metric variables are expanded as in \eqref{expansion}, taking $\tilde{\phi}=\nu u$.   
Thus, the substitution of equations \eqref{expansion} into \eqref{linear_field_eqs} provides the system of ordinary differential equations \eqref{field_eq_s} for the coefficients in the above expansions. The families of solutions, for $l=2$, satisfying this system of equations for the vacuum read
\begin{subequations}
\begin{align}
\beta_{2m}(r)=&D_{1\beta 2m},
\label{gen_sol_1_l2}\\
J_{2m}(r)=&\frac{2 i D_{1\beta 2m}}{\nu  r \left| m\right| }-\frac{D_{1J2m} (\nu  r \left| m\right| -1) (\nu  r \left| m\right| +1)}{6 r^3} \nonumber \\
& -\frac{i D_{2J2m} e^{2 i \nu  r \left| m\right| } (\nu  r \left| m\right| +i)^2}{8 \nu ^5 r^3 \left| m\right| ^5} +\frac{D_{3J2m} (\nu  r \left| m\right| -3 i)}{\nu  r \left| m\right| }, \label{gen_sol_2_l2}
\end{align}
\begin{align}
U_{2m}(r)=&\frac{2 D_{1\beta 2m} (\nu  r \left| m\right| +2 i)}{\nu  r^2 \left| m\right|
} -\frac{D_{1J2m} \left(2 \nu ^2 r^2 \left| m\right|^2 + 4 i \nu  r \left| m\right| +3\right)}{6
r^4} \nonumber\\
& -\frac{D_{2J2m} e^{2 i \nu  r \left| m\right| } (2 \nu  r \left| m\right| +3 i)}{8 \nu ^5 r^4 \left| m\right|^5} -\frac{i D_{3J2m} \left(\nu ^2 r^2 \left| m\right| ^2+6\right)}{\nu  r^2 \left| m\right| } , 
\label{gen_sol_3_l2} \\
w_{2m}(r)=& -10 r D_{1\beta 2m} +6 r D_{3J2m} (2+i \nu  r \left| m\right| ) -\frac{3 i D_{2J2m} e^{2 i \nu  r \left| m\right| }}{4 \nu ^5 r^2 \left|m\right| ^5} \nonumber\\
&  -\frac{i D_{1J2m} ((1+i) \nu  r \left| m\right| -i)	(1+(1+i) \nu  r \left| m\right| )}{r^2}, \label{gen_sol_4_l2}
\end{align}
\label{gen_sol_l2}%
\end{subequations}
where the constants of integration are represented by $D_{nFlm}$; here $n$ numbers the constant and $F$ corresponds to the metric function whose integration generates it. This set of families of solutions depends only on four constants, namely, $D_{1\beta 2m}$, $D_{3J2m}$, $D_{1J2m}$ and $D_{2J2m}$. This is so because the families of solutions for the coefficients $\beta_{2m}$, $J_{2m}$, $U_{2m}$ and $w_{2m}$ resulting from \eqref{field_eq_1}-\eqref{field_eq_4} are constrained by using \eqref{field_eq_5}-\eqref{field_eq_7}.  This fact is independent of $l$, and thus the set of families of solutions for any $l$ will have four degrees of freedom. 

A unique solution for the whole space-time cannot be determined by only imposing regularity of the metric variables at the null cone vertices and at the null infinity.
Therefore, additional boundary conditions must be imposed. In particular, this can be done by imposing boundary conditions on other hypersurfaces, such as in the case of the thin shells studied by \citeonline{B05}, in which the additional conditions are imposed across the world tubes generated by the shell itself. Once the above constants are determined, one readily obtains the metric functions $\beta$, $J$, $U$, and $w$ for the whole space-time.

As divergent solutions are not expected at the vertices of the null cones, regularity at these points must be imposed for the metric. In order to do so, an expansion of the metric variables around $r=0$ in power series of $r$ is made  and the divergent terms are disregarded. This procedure establishes relationships between the coefficients, leading to a family of solutions for the interior that depends only on one parameter to be determined, where in particular $\beta_{lm-}(r)=0$. One obtains, for example, for $l=2$
\begin{subequations}
\begin{align}
\beta_{2m-}(r)= & 0, \label{par_sol_int_1_l2}\\
J_{2m-}(r)= & \frac{D_{2J2m-}}{24 \nu^5 r^3 \left| m\right|^5} \left(2 \nu ^3 r^3 \left| m\right|^3 - 3 i \nu^2 r^2 \left| m\right|^2 e^{2 i \nu  r \left| m\right| } -3 i \nu ^2 r^2 \left| m\right|^2   \right.\nonumber\\
&\left. + 6 \nu  r \left| m\right|  e^{2 i \nu  r \left| m\right| } +3 i e^{2 i \nu  r \left| m\right| }-3 i\right),\label{par_sol_int_2_l2}
\end{align}
\begin{align}
U_{2m-}(r)= & -\frac{i D_{2J2m-}}{24 \nu ^5 r^4 \left| m\right| ^5}\left(2 \nu ^4 r^4 \left| m\right|^4+6 \nu ^2 r^2 \left| m\right| ^2 - 6 i \nu  r \left| m\right|  e^{2 i \nu  r \left| m\right| }  \right. \nonumber \\ 
& \left.  -12 i \nu  r \left| m\right| +9 e^{2 i \nu  r \left| m\right| }-9\right), \label{par_sol_int_3_l2}\\
w_{2m-}(r)= & \frac{D_{2J2m-}}{4 \nu^5 r^2 \left| m\right|^5} \left(2 i \nu ^4 r^4 \left| m\right|^4 + 4 \nu ^3 r^3 \left| m\right| ^3 -6 i \nu ^2 r^2 \left| m\right| ^2 -6 \nu  r \left| m\right| \right. \nonumber \\
&\left. -3 i e^{2 i \nu  r \left| m\right| }+3 i\right).  \label{par_sol_int_4_l2}
\end{align}
\end{subequations}
For the intermediate region, the same structure of the general solutions is maintained, for the case of $l=2$ given by \eqref{gen_sol_1_l2}-\eqref{gen_sol_4_l2}. That is so because there is no reason to discard any particular term, or to establish any relationship between the constants as occurs for the interior region. Since  regularity is required at the null infinity, the coefficient of the exponential factor $\left(\exp(2i\nu r |m|)\right)$ must be null in the exterior solutions. This means that all constants of the form $D_{2Jlm+}$, with $l=2,3,\cdots$, must be zero. Therefore, the number of degrees of freedom for the exterior family of solutions is reduced in one parameter. Thus, a family of solutions for the field equations \eqref{field_eq_1}-\eqref{field_eq_7}, with eight parameters to be determined, for describing the whole space-time is obtained. Now, in order to fix these eight constants, it is necessary to impose additional boundary conditions in particular across the time-like world tubes generated by their orbits.

These boundary conditions across the world tubes, i.e. when $r=r_i$, $i=1,2$, come from imposing discontinuities on the metric coefficients, i.e.,
\begin{eqnarray}
&&\left[g_{11}\right]_{r_i}=0, \hspace{0.2cm}\left[g_{12}\right]_{r_i}=\left.\Delta g_{12}\right|_{r_i},\hspace{0.2cm}\left[g_{1A}\right]_{r_i}=0, \hspace{0.2cm} \left[g_{22}\right]_{r_i}=0,\nonumber\\
&& \left[g_{2A}\right]_{r_i}=0,\hspace{0.2cm}\left[g_{3\mu}\right]_{r_i}=0,\hspace{0.2cm}\left[g_{4\mu}\right]_{r_i}=0,
\label{bound_cond_1}
\end{eqnarray}
and on their first derivatives,
\begin{eqnarray}
&&\left[g_{\mu\nu}'\right]_{r_i}=\Delta g_{\mu\nu}', \hspace{0.2cm}\mu,\nu=1,\cdots 4,
\label{bound_cond_2}
\end{eqnarray}
where the brackets mean $[f(r)]_{r_i}=\left.f(r)\right|_{r_i+\epsilon}-\left.f(r)\right|_{r_i-\epsilon}$. 
From the linearised Bondi-Sachs metric \eqref{bs_lin}, and from the two sets of jump conditions \eqref{bound_cond_1} and \eqref{bound_cond_2}, the coefficients $\beta_{lm}, \ J_{lm},\ U_{lm}$ and $w_{lm}$ are restricted to satisfy
\begin{eqnarray}
& \left[w_{lm}(r_j)\right]=\Delta w_{jlm},& \hspace{0.2cm} \left[\beta_{lm}(r_j)\right]=\Delta \beta_{jlm}, \nonumber\\
& \left[J_{lm}(r_j)\right]=0, & \hspace{0.2cm} \left[U_{lm}(r_j)\right]=0,
\label{bound_cond_3}
\end{eqnarray}
and for their first derivatives
\begin{eqnarray}
& \left[w'_{lm}(r_j)\right]=\Delta w'_{jlm},& \hspace{0.2cm} \left[\beta_{lm}'(r_j)\right]=\Delta \beta_{jlm}', \nonumber\\
& \left[J'_{lm}(r_j)\right]=\Delta J'_{jlm}, & \hspace{0.2cm} \left[U'_{lm}(r_j)\right]=\Delta U'_{jlm},
\label{bound_cond_4}
\end{eqnarray}
where $j=1,2$, and $\Delta w_{jlm}$, $\Delta \beta_{jlm}$, $\Delta w_{jlm}'$, $\Delta \beta_{jlm}'$, $\Delta J'_{jlm}$ and $\Delta U'_{jlm}$ are functions to be determined. 

Solving equations \eqref{bound_cond_3} and \eqref{bound_cond_4}, simultaneously for both world tubes, the boundary conditions are explicitly obtained. We find that 
\begin{subequations}	
\begin{equation}
\Delta \beta_{jlm}=b_{jlm},\hspace{0.5cm}\Delta w_{jlm}=-2r_jb_{jlm},
\label{bound_cond_5}
\end{equation}
where $b_{jlm}$ are constants. Note that, this last fact implies that $\Delta \beta'_{jlm}=0$. 
We obtain that the jumps for the first derivative of the $J_{lm}$ and $U_{lm}$ functions are given by
\begin{eqnarray}
&&\Delta J'_{jlm}=\frac{8\nu ^2 r_j  b_{jlm} \left| m\right|^2}{(l-1)l(l+1)(l+2)},\label{bound_cond_6}\\
&&\Delta U'_{jlm}=2b_{ilm}\left(\frac{1}{r_i^2}-\frac{4i\nu |m|}{l(l+1)r_i}\right).\label{bound_cond_7}
\end{eqnarray}
\label{bound_condss}%
\end{subequations}
Thus, the boundary conditions \eqref{bound_cond_6} and \eqref{bound_cond_7} fix all parameters of the families of solutions, providing the specific solutions for the coefficients $\beta_{lm}, \ J_{lm},\ U_{lm}$ and $w_{lm}$. Therefore, these coefficients can be written as 
\begin{align}
f_{lm}(r)=& f_{1lm}(r)\left(1-\Theta(r-r_1)\right)+f_{2lm}(r)\left(\Theta(r-r_1)-\Theta(r-r_2)\right)\nonumber\\
&+f_{3lm}(r)\Theta(r-r_2),
\label{Sols_Coeff}
\end{align}
where $f_{lm}$ represents $\beta_{lm}$, $J_{lm}$, $U_{lm}$ and $w_{lm}$, with the first subscript on the right hand side terms indicating the interior (1), the middle (2) and the exterior (3) solutions. 

These solutions depend explicitly on two specific parameters, namely $b_{jlm}$, with $j=1,2$, which are related to the density of matter. The specific form of these relationships is obtained by just integrating the first field equation \eqref{field_eq_1} across each world tube. As a result one obtains 
\begin{equation}
b_{jlm}=2\pi r_j \rho_{jlm} \left(1+v_j^2\right),
\label{betaf}
\end{equation} 
where, $\rho_{jlm}$ are given by
\begin{equation}
\rho_{jlm}=\frac{1}{\pi}\int_S d(\nu u) \int_\Omega d\Omega\int_{I_j} dr \ _0\overline{Z}_{lm} e^{-i|m|\nu u}\rho,
\label{denscomps}
\end{equation}
in which $S = \ [0,2\pi)$, $v_j$ is the physical velocity of the particle $j$ in the space, and $I_j$ is an interval $\epsilon$ around $r_j$ that is given by $I_j=(r_j-\epsilon/2,r_j+\epsilon/2)$, with $\epsilon>0$.

Before proceeding, it is worth noticing that the above procedure is a generalisation of Section 3 of the paper by \citeonline{BPR11}, in which the binary components have equal masses. In particular, the boundary conditions are also generalised since in the present case there exist two independent world tubes. Another interesting aspect has to do with the fact that our solution is fully analytical.

\begin{figure}[h!]
\vspace{-0.5cm}
\begin{center}
	\begin{tabular}{c@{\hspace{0.7cm}}c}
		\includegraphics[height=4.7cm]{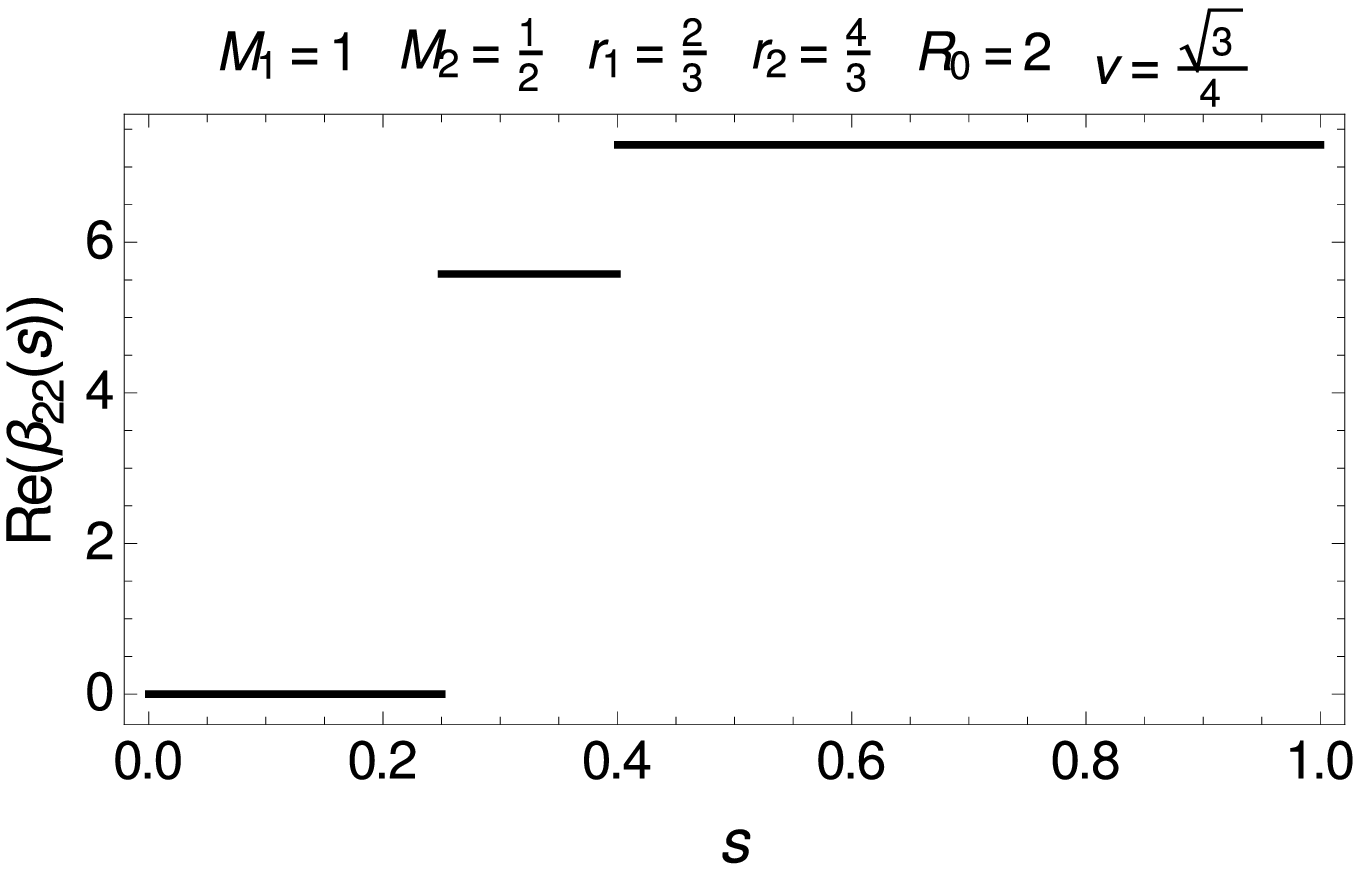} & \includegraphics[height=4.7cm]{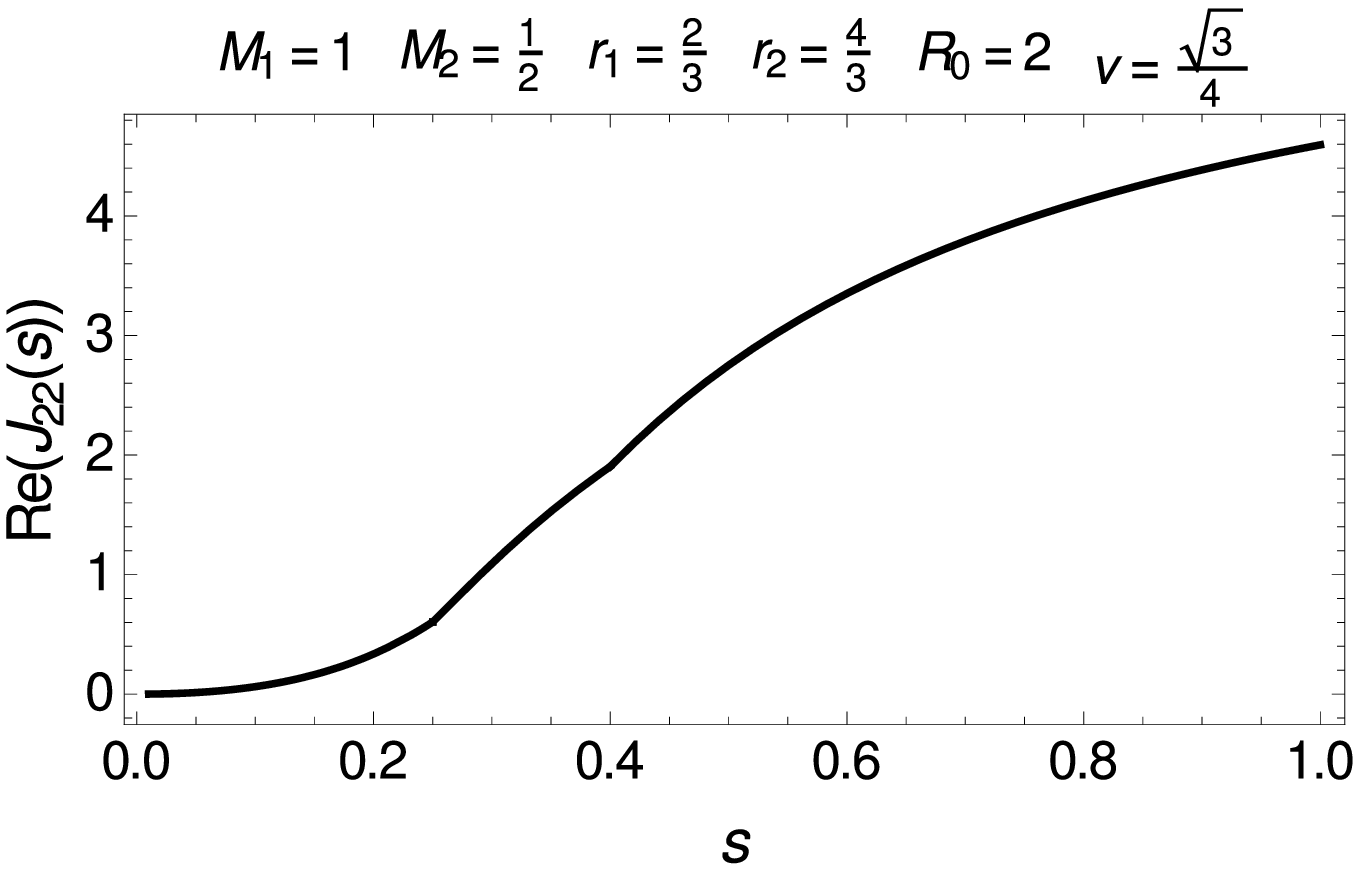}\\[-0.2cm]
		(a) & (b)\\
		\includegraphics[height=4.7cm]{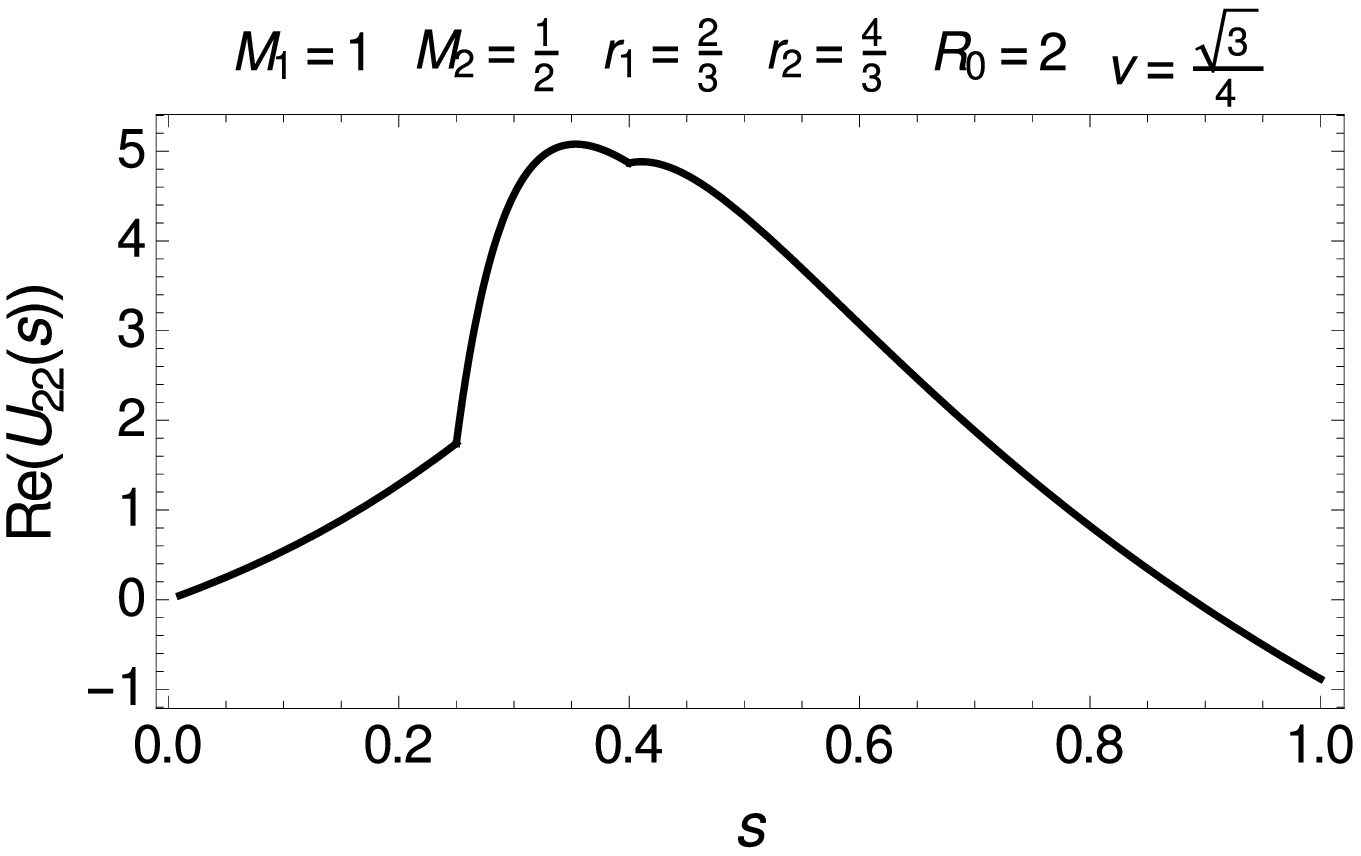}    & \includegraphics[height=4.7cm]{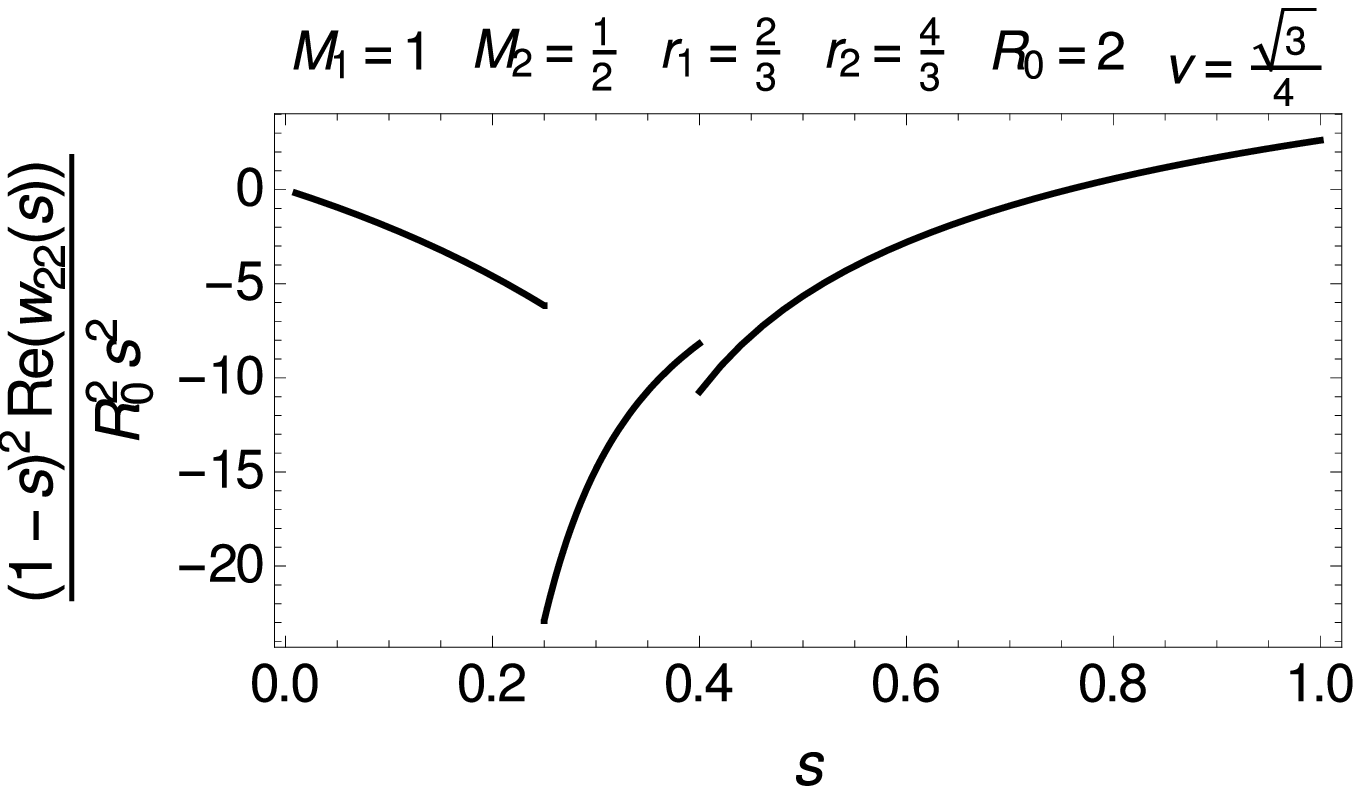}\\[-0.2cm]
		(c) & (d)
	\end{tabular}
\end{center}
	\caption{Real part of some components of the metric functions ( $l=m=2$ ) versus the compactified coordinate $s$ (see the text) for a binary system with $M_1=1/2$, $M_2=1$. The angular velocity is computed by means of Kepler's third law. Here $r_1$ and $r_2$ are referred to the center of mass of the system.}
	\label{metric_functions}
\end{figure}

Figure \ref{metric_functions} shows some of the coefficients of the expansion of the metric variables in terms of the compactified coordinate $s$ (defined just below) for $l=m=2$.

In order to include the null infinity, which is reached when $r$ tends to infinity, a radial compactified coordinate $s$ is defined as follows 
\begin{equation*}
s=\frac{r}{r+R_0} ,
\end{equation*}
where $R_0$ is a compactification parameter. Thus, $0\le s\le 1$, where $s=0$ and $s=1$ corresponds to the null cone vertices and the null infinity, respectively. 
\\Here $M_1=1/2$, $M_2=1$, $R_0=2$, and the radius of each orbit is referred to the centre of mass of the system, namely
\begin{equation}
r_j=\frac{\mu}{M_j}d_0,  \hspace{0.2cm}j=1,2,
\label{rdef}
\end{equation}
where $\mu$ is the reduced mass of the system and $d_0$ is the distance between the masses. The frequency of 
rotation $\nu$ is computed by means of Kepler's third law, i.e., 
\begin{equation}
\nu=\sqrt{\frac{M_1+M_2}{d_0^3}}.
\label{freqnu}
\end{equation} 
It is worth noting that the jumps in $\beta_{lm}$ and $w_{lm}$ functions are present at exactly $r_1$ and $r_2$, whereas for $J_{lm}$ and $U_{lm}$ only their first derivatives present discontinuities, in agreement with the boundary conditions \eqref{bound_cond_6} and \eqref{bound_cond_7}.
 
\begin{figure}[h!]
\begin{center}
	\begin{tabular}{c@{\hspace{0.5cm}}c}
		\includegraphics[height=6.0cm]{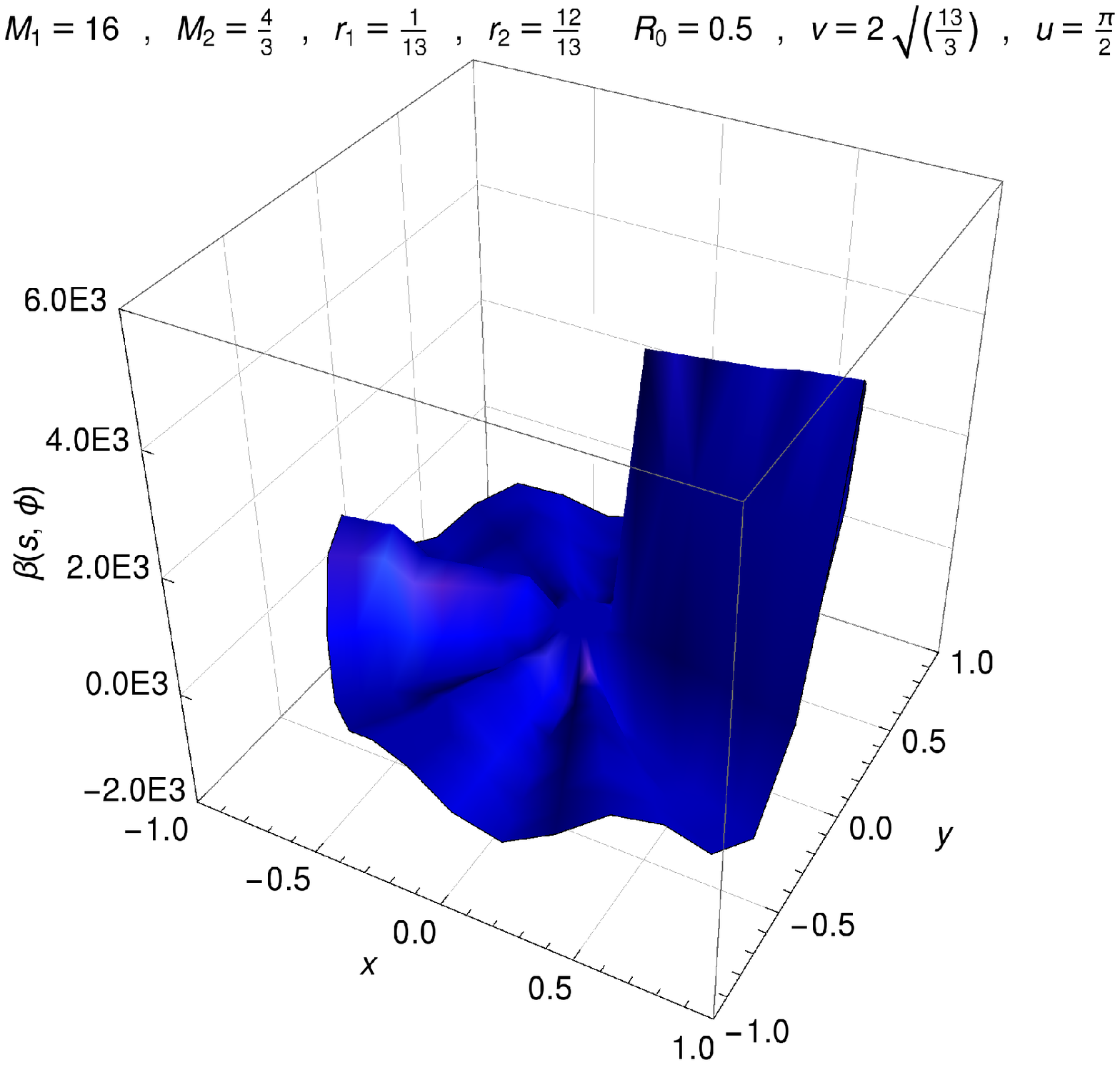} & \includegraphics[height=6.0cm]{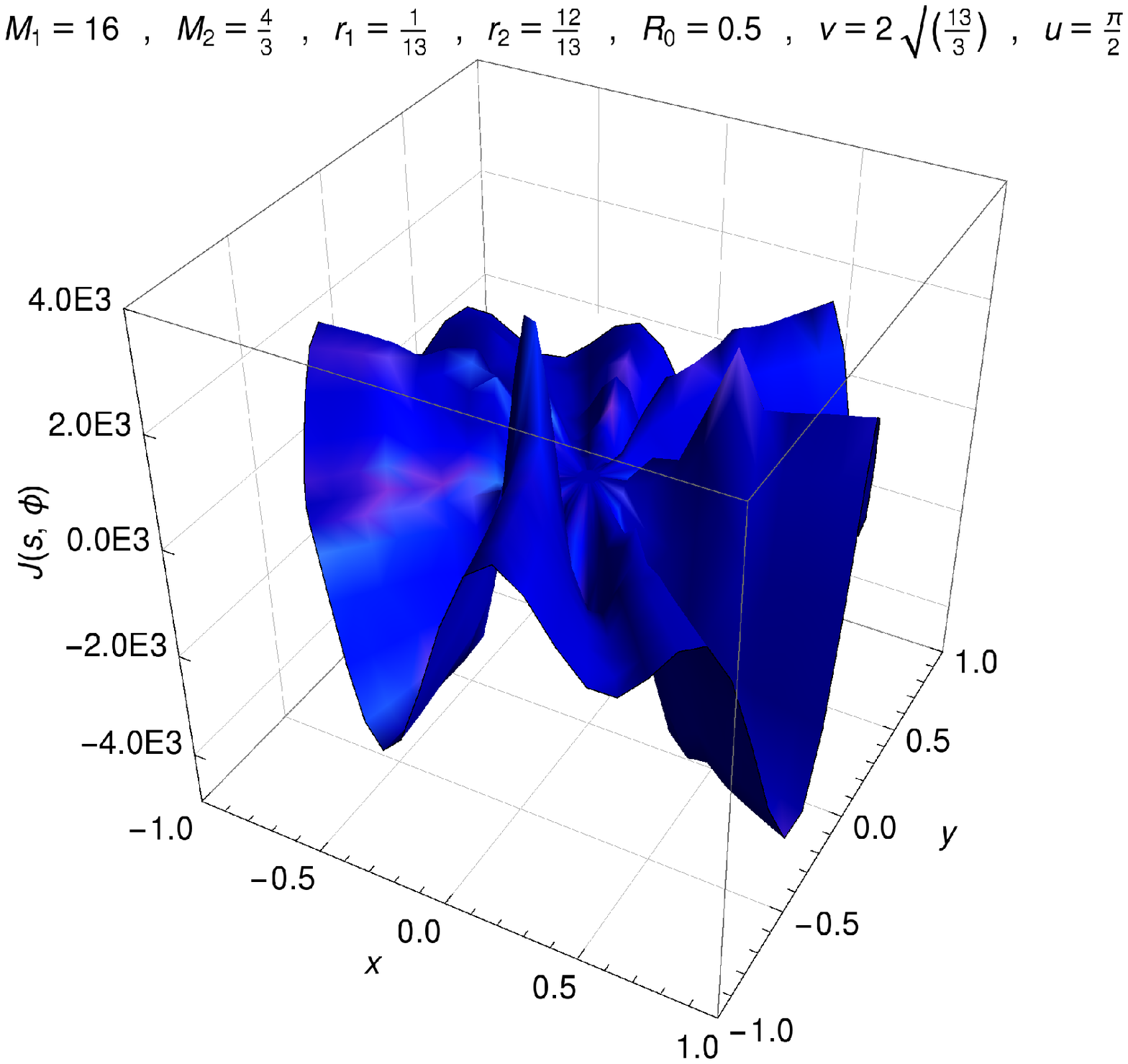}\\
		(a) & (b) \\
		\includegraphics[height=6.0cm]{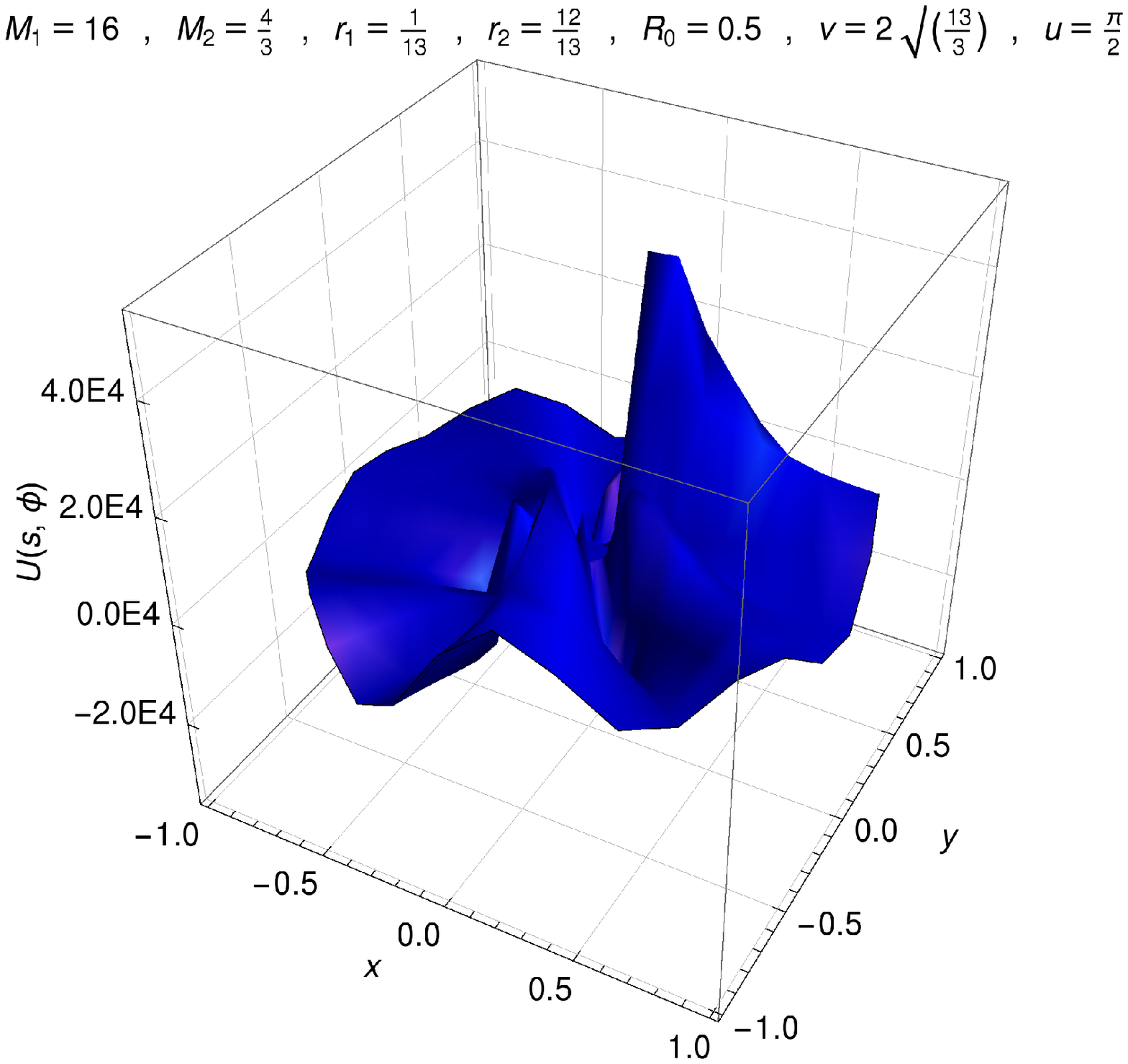}    & \includegraphics[height=6.0cm]{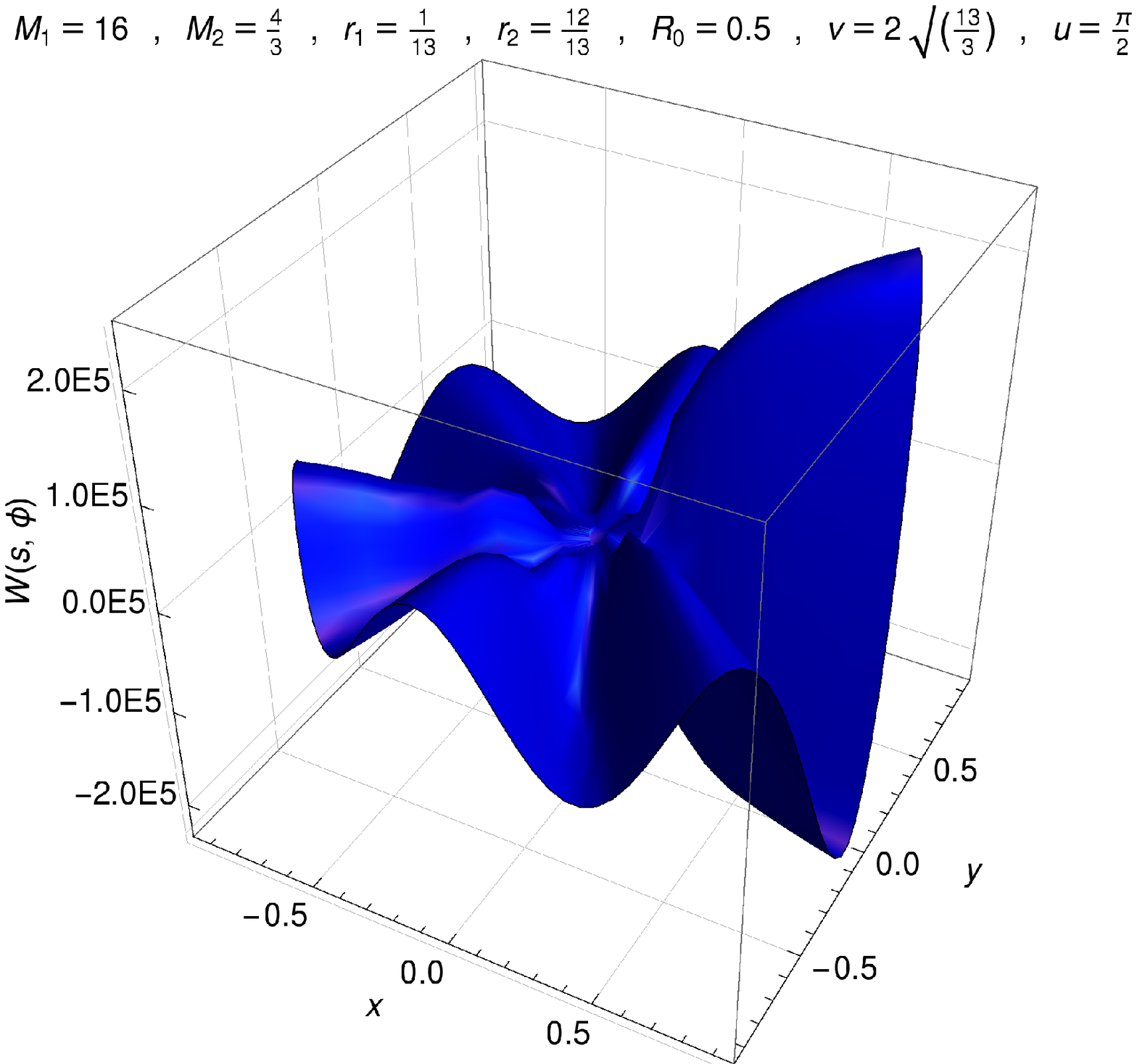}\\
		(c) & (d) 
	\end{tabular}
\end{center}
	\caption{Snapshots of the metric variables as seen from the equatorial plane ($\theta=\pi/2$), as a function of $s$ and $\phi$ for $u=\pi/2$. Here $M_1=16$, $M_2=4/3$, $r_1=1/13$, $r_2=12/13$, $R_0=1/2$ and $\nu=2\sqrt{13/3}$. (a) $\beta(s,\phi)$, (b) $J(s,\phi)$, (c) $U(s,\phi)$ and (d) $W(s,\phi)=w(s,\phi)(1-s^2)/(s^2 R_0^2)$.}
	\label{metric_functions1}
\end{figure}

To illustrate the behaviour of $\beta$, $J$, $U$ and $w$ we present them in Figure \ref{metric_functions1} as a function of $s$ and $\phi$ for a particular value of the retarded time $u$. These functions are constructed by using  Equations \eqref{expansion}
, and the solutions for the coefficients for each $l$ and $m$. In this case, we use $l\le 8$.

As expected, the metric functions $\beta$ and $w$ and the first derivatives of $J$ and $U$ show jumps at $(r,\theta,\phi)=(r_1,\pi/2,\nu u)$ and $(r,\theta,\phi)=(r_2,\pi/2,\nu u-\pi)$, which are just the positions of the masses, in agreement with the boundary conditions initially imposed. 

Note that since the first field equation for the vacuum $\beta_{,r}=0$ implies that $\beta_{lm}$ are constants along $r$, as sketched in Figures \ref{metric_functions}, and that $\beta$ is a gauge term for the gravitational potential. Then,  $\Phi$ can be redefined as $\Phi=w/(2r)$. These facts make the choice of the angular velocity $\nu$ as obeying Kepler's third law, completely consistent and natural.
\newpage
\subsection{Gravitational Radiation from the Binary System}
Now, we proceed with the calculation of the power lost by the binary system via gravitational wave emission. We show that the approach presented here is robust because we can obtain the well-known result 	obtained by \citeonline{PM63} for the power emitted by binary systems 
in circular orbits, now using the News function.

Following \citeonline{B05}, the Bondi's News function in the weak field approximation is given by
\begin{equation}
\mathcal{N}=\lim\limits_{r\rightarrow\infty}\left(-\frac{r^2J_{,ur}}{2}+\frac{\eth^2 \omega}{2} +\eth^2\beta\right).
\label{BNews}
\end{equation}
Substituting here the metric variables given in \eqref{expansion}
, one obtains the News function for $l\ge 2$ and $-l\le m\le l$, namely
\begin{equation}
\mathcal{N}=\lim\limits_{r\rightarrow\infty}\sum_{l,m}\Re\left(\left(-\frac{i|m|\nu r^2\left(J_{lm}\right)_{,r}}{2}+\beta_{lm}+\frac{l(l+1)J_{lm}}{4}\right)e^{i|m|\nu u}\right)\ \eth^2\ _0Z_{lm}.
\end{equation}
Now, substituting the coefficients of the metric variables for the exterior region, one obtains
\begin{eqnarray}
\mathcal{N}=&& -\frac{i \nu ^3 \ _2S_{21} }{\sqrt{6}}-4 i \sqrt{\frac{2}{3}} \nu^3 \ _2S_{22}  -\frac{i \nu ^4 \ _2S_{31} }{\sqrt{30} (\nu -3 i)}-\frac{8 i \sqrt{\frac{2}{15}} \nu ^4 \ _2S_{32} }{2 \nu -3 i}  \nonumber
\\&& -\frac{9 i \sqrt{\frac{3}{10}} \nu ^4 \ _2S_{33} }{\nu -i} - \frac{i \nu ^5 \ _2S_{41} }{3 \sqrt{10} \left(\nu^2-7 i \nu -14\right)}-\frac{8 i \sqrt{\frac{2}{5}} \nu ^5 \ _2S_{42} }{3 \left(2 \nu ^2-7i \nu -7\right)}   \nonumber
\\&&-\frac{81 i \nu ^5 \ _2S_{43} }{\sqrt{10} \left(9 \nu ^2-21 i \nu -14\right)} -\frac{256 i \sqrt{\frac{2}{5}} \nu ^5 \ _2S_{44} }{3 \left(8 \nu ^2-14 i \nu -7\right)}    \nonumber
\\&&+\frac{\nu ^6 \ _2S_{51} }{\sqrt{210} \left(i \nu ^3+12 \nu ^2-54 i \nu-90\right)} +\frac{16 \sqrt{\frac{2}{105}} \nu ^6 \ _2S_{52} }{4 i \nu ^3+24 \nu ^2-54 i\nu -45}  \nonumber\\
&&+\frac{27 \sqrt{\frac{3}{70}} \nu ^6 \ _2S_{53} }{3 i \nu ^3+12 \nu ^2-18 i \nu-10}+\frac{1024 \sqrt{\frac{2}{105}} \nu ^6 \ _2S_{54} }{32 i \nu ^3+96 \nu ^2-108 i \nu-45}\nonumber\\
&&+\frac{625 \sqrt{\frac{5}{42}} \nu ^6 \ _2S_{55} }{25 i \nu ^3+60 \nu ^2-54 i \nu -18} + \cdots, 
\label{News_all}
\end{eqnarray}
where we define the spin 2 quantity $_2S_{lm}$ as
\begin{equation}
_2S_{lm}= \frac{\left(\Re(D_{1Jlm+}e^{i|m|\nu u}) \ \eth^2\ _0Z_{lm}+ \Re(D_{1Jl-m+}e^{i|m|\nu u}) \ \eth^2\ _0Z_{l\ -m}\right)}{\sqrt{(l-1)l(l+1)(l+2)}}.
\end{equation}
Since the binary system is confined to a plane, then a natural choice to simplify the problem of expressing the News function, without loss of generality, is to put the masses to move on the equatorial plane $\theta=\pi/2$. This means symmetry of reflection for the density of matter and, consequently, for the space-time. Thus, this choice restricts the components of the density, obtained from  \eqref{denscomps}, to have the following form
\begin{equation}
\rho_{lm}=
\begin{cases}
	\tilde{\rho}_{lm}\dfrac{M_2  r_1 ^2 \delta \left(r-r_2 \right)+M_1  r_2^2 \delta \left(r-r_1
		\right)}{r_1^2 r_2^2} &  \text{if}\   l,m \ \text{even} \\[0.5cm]
	\tilde{\rho}_{lm} \dfrac{M_1  r_2^2 \delta \left(r-r_1 \right)-M_2  r_1^2 \delta \left(r-r_2
		\right)}{r_1^2 r_2^2} & \text{if}\ l,m \ \text{odd},
\end{cases}
\label{rholm}
\end{equation}
where $\tilde{\rho}_{lm}$ are numerical constants. Therefore, for binaries of different masses, 
the News function \eqref{News_all} is simplified to
\begin{eqnarray}
\mathcal{N}=&& -4 i \sqrt{\frac{2}{3}} \nu^3 \ _2S_{22} -\frac{i \nu ^4 \ _2S_{31}}{\sqrt{30} (\nu -3 i)} -\frac{9 i \sqrt{\frac{3}{10}} \nu ^4 \ _2S_{33}}{\nu -i} -\frac{8 i \sqrt{\frac{2}{5}} \nu ^5 \ _2S_{42}}{3 \left(2 \nu ^2-7 i \nu -7\right)} \nonumber
\\&&  -\frac{256 i \sqrt{\frac{2}{5}} \nu ^5 \ _2S_{44}}{3 \left(8 \nu ^2-14 i \nu -7\right)} +\frac{\nu ^6 \ _2S_{51} }{\sqrt{210} \left(i \nu ^3+12 \nu ^2-54 i \nu
	-90\right)}  \nonumber
\\&& +\frac{27 \sqrt{\frac{3}{70}} \nu ^6 \ _2S_{53}}{3 i \nu ^3+12 \nu ^2-18 i \nu
	-10} +\frac{625 \sqrt{\frac{5}{42}} \nu ^6 \ _2S_{55} }{25 i \nu ^3+60 \nu ^2-54 i \nu -18} + \cdots 
\label{News_restricted}
\end{eqnarray}

When the explicit solutions are used, the News functions for the binary system take the form
\begin{eqnarray}
\mathcal{N}&=& 8\sqrt{\frac{2 \pi }{5}} \ _2L_{22} \left(\mathcal{M}_{21}+\mathcal{M}_{22}\right) \nu ^3 +\frac{1}{3} i \sqrt{\frac{\pi }{35}} \ _2L_{31} \left(\mathcal{M}_{31}-\mathcal{M}_{32}\right) \nu ^4 \nonumber\\
&-& 9 i \sqrt{\frac{3 \pi }{7}} \ _2L_{33} \left(\mathcal{M}_{31}-\mathcal{M}_{32}\right) \nu ^4 +\frac{8}{63}\sqrt{2 \pi } \ _2L_{42} \left(\mathcal{M}_{41}+\mathcal{M}_{42}\right) \nu ^5 \nonumber\\
&-& \frac{128}{9} \sqrt{\frac{2 \pi }{7}} \ _2L_{44} \left(\mathcal{M}_{41}+\mathcal{M}_{42}\right) \nu ^5
\frac{1}{180} i \sqrt{\frac{\pi }{154}} \ _2L_{51} \left(\mathcal{M}_{51} - \mathcal{M}_{52}\right) \nu ^6 \nonumber\\
&-&\frac{27}{40} i \sqrt{\frac{3 \pi }{11}} \ _2L_{53}  \left(\mathcal{M}_{51}-\mathcal{M}_{52}\right) \nu ^6
+\frac{625}{24} i \sqrt{\frac{5 \pi }{33}} \ _2L_{55} \left(\mathcal{M}_{51}-\mathcal{M}_{52}\right) \nu ^6\nonumber\\
&+&\cdots,
\label{News_binary}
\end{eqnarray}
where,
\begin{equation}
\mathcal{M}_{lj}=M_j r_j^l(v_j^2+1),
\end{equation}
and $_2L_{lm}$ are defined as
\begin{equation}
_2L_{lm}=\left(\ _2Z_{l\ -m}\Re(e^{ i |m|\nu  u}) -\Re(i e^{ i |m|\nu  u}) \ _2Z_{lm}\right).
\end{equation}
Note that, as consequence of \eqref{rholm}, for $M_1=M_2=M_0$ the terms with $l$ odd disappear from the News function \eqref{News_binary}. Thus, as expected, one obtains immediately
\begin{eqnarray}
\mathcal{N}&=&  16 \sqrt{\frac{2 \pi }{5}} \nu ^3 M_0  r_0^2 \left(V_0^2+1\right) \ _2L_{22} +\frac{16}{63} \sqrt{2 \pi } \nu ^5 M_0  r_0 ^4 \left(V_0^2+1\right) \ _2L_{42} \nonumber \\
&&-\frac{256}{9} \sqrt{\frac{2 \pi }{7}} \nu ^5 M_0  r_0 ^4 \left(V_0^2+1\right) \ _2L_{44} +\frac{32 \sqrt{\frac{2 \pi }{13}} }{1485} \nu^7 M_0  r_0 ^6 \left(V_0^2+1\right) \ _2L_{62} \nonumber\\
&&-\frac{8192}{495} \sqrt{\frac{\pi }{195}} \nu ^7 M_0  r_0^6 \left(V_0^2+1\right) \ _2L_{64}
+\frac{2592}{5} \sqrt{\frac{2 \pi }{715}} \nu ^7 M_0  r_0 ^6\left(V_0^2+1\right) \ _2L_{66}\nonumber\\
&&+\cdots.
\end{eqnarray}
where $V_0$ is the physical velocity of the masses, which is obviously tangent to the circular orbit.

The energy lost by the system $dE/du$ is related to the News function \cite{B05}, via
\begin{equation}
\frac{dE}{du}=\frac{1}{4\pi}\int_{\Omega} d\Omega \ \mathcal{N}\overline{\mathcal{N}},
\end{equation} 
which results  for $M_1\ne M_2$ in
\begin{align}
\frac{dE}{du}=&\frac{32}{5} \nu ^6 \left(\mathcal{M}_{21}+\mathcal{M}_{22}\right)^2 +\frac{2734}{315} \nu ^8 \left(\mathcal{M}_{31}-\mathcal{M}_{32}\right)^2 \nonumber \\
&+\frac{57376}{3969} \nu ^{10} \left(\mathcal{M}_{41}+\mathcal{M}_{42}\right)^2 +\frac{4010276}{155925}\nu ^{12} \left(\mathcal{M}_{51}-\mathcal{M}_{52}\right)^2\nonumber\\
&+\cdots.
\label{power_circ}
\end{align}
Notice that the first term on the right side of the above equation is nothing but
the power lost obtained by \citeonline{PM63} for circular orbits and the other terms stand for the octupole, hexadecapole, etc contributions.
\section{Eccentric Point Particle Binary System}
Here the eccentricity in the binary systems in the characteristic formulation is introduced, generalising the study of the previous section. From the density of energy and from an angular velocity that is variable on time, we deduce boundary conditions at the orbits, generalising those boundary conditions found for circular orbits. Also, we found the expression for the power emitted by the binary in gravitational radiation from the characteristic formulation, in agreement with the Peter and Mathews expression \cite{PM63}. In order to do that, we consider in the News, those terms related to the angular velocity, disregarded in the circular case \cite{CA16b}. 

\begin{figure}
	\centering
	\begin{tabular}{cc}
		\includegraphics[height=4cm]{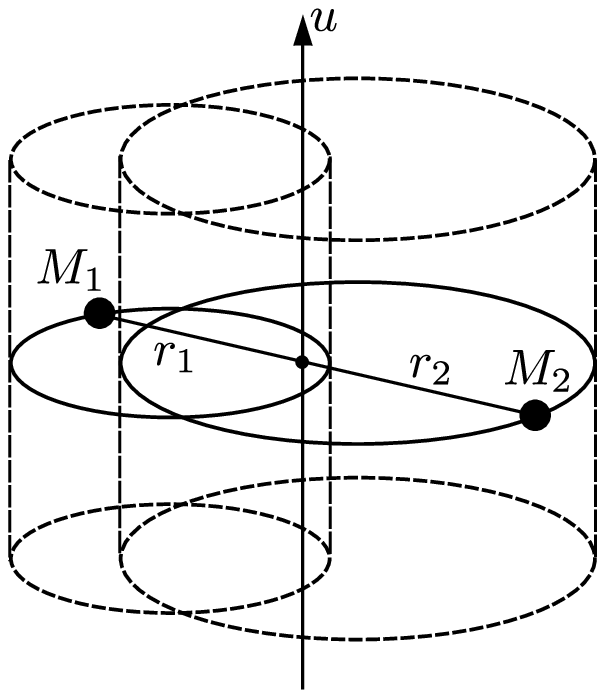}&\includegraphics[height=4cm]{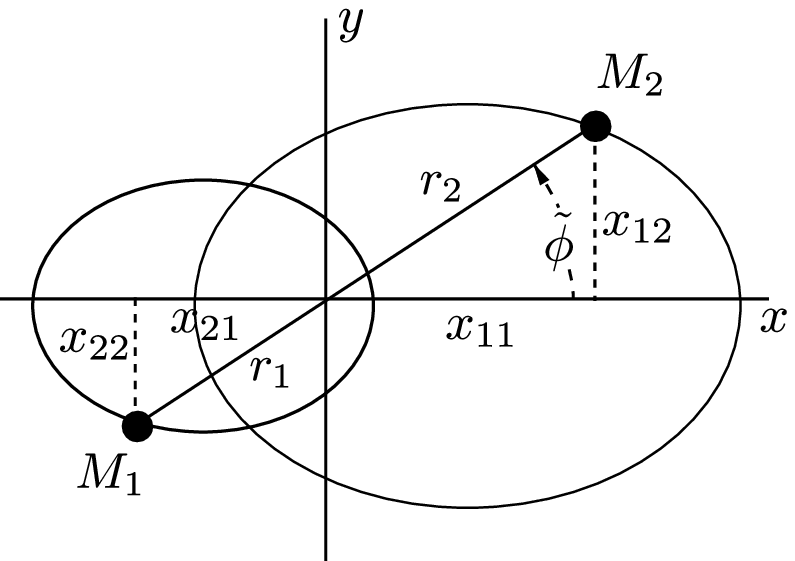}\\
		(a) & (b)
	\end{tabular}
	\caption{(a) Eccentric binary system with the world tubes of their orbits extended along the central time-like geodesic. (b) Top view of the point particle binary system, where the angular position $\tilde{\phi}$ is indicated.}
	\label{figure1}
\end{figure}

In this case, the density that describes the point particle binary is given by
\begin{equation}
\rho=\frac{\delta(\theta-\pi/2)}{r^2}\left(M_1\delta(r-r_1)\delta(\phi-\tilde\phi)+ M_2\delta(r-r_2)\delta(\phi-\tilde\phi-\pi)\right),
\label{dens}
\end{equation}
where, $r_i\ (M_i)$ are the orbital radius (mass) of each particle, $r_1 < r_2$ and \ $\tilde\phi:=\tilde\phi(u)$ is the angular position as indicated in Figure \ref{figure1}.

The instantaneous radius of the particle's orbits reads 
\begin{equation}
r_j=\frac{\mu d}{M_j}, \hspace{0.5cm} \mu=\frac{M_1M_2}{M_1+M_2}, \hspace{0.5cm} j=1,2,
\label{bin1}
\end{equation}
where the separation between the masses $d$, is given by
\begin{equation}
d=\frac{a(1-\epsilon^2)}{1+\epsilon \cos\tilde{\phi}},
\label{bin2}
\end{equation}
in which $\epsilon$ represents the eccentricity, and $a$ is the semi-major axis which becomes the radius of the orbits when the eccentricity is zero. 
For Keplerian orbits, the angular velocity reads
\begin{equation}
\dot{\tilde{\phi}}=\frac{\sqrt{a(1-\epsilon^2)(M_1+M_2)}}{d^2},
\label{bin3}
\end{equation}
which depends on time. Note that \eqref{bin1}-\eqref{bin3} are the same expressions given in \eqref{PM_ang_vel_and_rad}.

Using the expansion \eqref{expansion} of the metric variables, substituting them into the field equations and assuming the same boundary conditions presented in \eqref{bound_cond_3} and \eqref{bound_cond_4}, one obtains that the boundary conditions \eqref{bound_condss} can be easily extended for a general function $\tilde{\phi}:=\tilde{\phi}(u)$ and a radial function $r_j:=r_j(u)$, namely
\begin{subequations}
	\begin{align}
	\Delta \beta_{jlm}&=b_{jlm},\label{bound_cond_5}\\
	\Delta w_{jlm}&=-2r_jb_{jlm},\label{bound_cond_6} \\
		\Delta J'_{jlm}&=\frac{8\dot{\tilde{\phi}} ^2 r_j  b_{jlm} \left|  m \right|^2} {(l-1)l(l+1)(l+2)}, \label{bound_cond_7} 
	\end{align}
	\begin{align}
	\Delta U'_{jlm}&=2b_{ilm}\left(\frac{1}{r_i^2}-\frac{4i\dot{\tilde{\phi}} |m|}{l(l+1)r_i}\right), \label{bound_cond_8}
	\end{align}	
	\label{bound_conds}
\end{subequations}
\noindent where $b_{jlm}$ are constants, which implies that $\Delta \beta'_{jlm}=0$. Also, the constants $D_{nFlm}$ depend on two parameters, namely $b_{1lm}$ and $b_{2lm}$. As an example, we show $D_{1J2m+}$ for $|m|\ne 0$, i.e.
\begin{align}
D_{1J2m+}=&\frac{i r_1^2 b_{12m} e^{-2 i r_1 \dot{\tilde{\phi }} \left| m\right|
	}}{\dot{\tilde{\phi }} \left| m\right| }-\frac{i r_1^2 b_{12m}}{\dot{\tilde{\phi }} \left| m\right| }+\frac{2 r_1  b_{12m} e^{-2 i r_1\dot{\tilde{\phi }} \left| m\right| }}{\dot{\tilde{\phi }}^2 \left| m\right| ^2} +\frac{2 r_1 b_{\text{12m}}}{\dot{\tilde{\phi }}^2 \left| m\right|^2} \nonumber\\
	&-\frac{3 i b_{12m} e^{-2 i r_1  \dot{\tilde{\phi }} \left| m\right| }}{\dot{\tilde{\phi }}^3
		\left| m\right| ^3} - \frac{3 b_{12m} e^{-2 i r_1  \dot{\tilde{\phi }} \left| m\right|
		}}{r_1 \dot{\tilde{\phi }}^4 \left| m\right| ^4} -\frac{3 b_{12m}}{r_1 \dot{\tilde{\phi }}^4 \left| m\right| ^4}  -\frac{3i b_{12m}}{r_1^2 \dot{\tilde{\phi }}^5 \left| m\right| ^5}\nonumber\\
		& +\frac{3 i b_{12m} e^{-2 i r_1 \dot{\tilde{\phi }} \left| m\right| }}{r_1^2 \dot{\tilde{\phi }}^5 \left| m\right| ^5} +\frac{3 i
			b_{12m}}{\dot{\tilde{\phi }}^3 \left| m\right| ^3} +\frac{i r_2^2 b_{22m} e^{-2 i r_2  \dot{\tilde{\phi }} \left| m\right| }}{\dot{\tilde{\phi }}
			\left| m\right| } -\frac{i r_2^2 b_{22m}}{\dot{\tilde{\phi }} \left| m\right| } \nonumber\\
		& +\frac{2 r_2  b_{\text{22m}} e^{-2 i r_2  \dot{\tilde{\phi }} \left| m\right|
			}}{\dot{\tilde{\phi }}^2 \left| m\right| ^2}+\frac{2 r_2  b_{22m}}{\dot{\tilde{\phi
			}}^2 \left| m\right| ^2}-\frac{3 i b_{22m} e^{-2 i r_2  \dot{\tilde{\phi }} \left|
			m\right| }}{\dot{\tilde{\phi }}^3 \left| m\right| ^3} -\frac{3b_{22m}}{r_2 \dot{\tilde{\phi }}^4 \left| m\right| ^4}\nonumber\\
	& -\frac{3 b_{22m} e^{-2 i r_2 \dot{\tilde{\phi }} \left| m\right| }}{r_2 \dot{\tilde{\phi }}^4 \left| m\right| ^4}+\frac{3 i
		b_{22m} e^{-2 i r_2  \dot{\tilde{\phi }} \left| m\right| }}{r_2^2 \dot{\tilde{\phi }}^5 \left| m\right| ^5}-\frac{3 i b_{22m}}{r_2^2\dot{\tilde{\phi }}^5 \left| m\right| ^5}+\frac{3 i b_{22m}}{\dot{\tilde{\phi }}^3\left| m\right| ^3}.
	\end{align}
	The parameters $b_{jlm}$, $j=1,2$ are  determined directly from \eqref{bound_cond_5} and \eqref{field_eq_s1}. In particular for the binary system, 
	\begin{equation}
	b_{jlm}=2M_j\int_0^{2\pi} d\tilde{\phi}\ \frac{e^{-i|m|\tilde\phi }\overline{Z}_{lm}(\pi/2,\tilde{\phi}+\pi \delta_{2j})}{r_j^2}.
	\label{bes}
	\end{equation}
	where it is important to note that the spin-weighted spherical harmonics $Z_{lm}$ become real on the equatorial plane $\theta=\pi/2$, but in general these functions are complex.
	
Specifically, the non-null $b_{jlm}$, for the firsts $l$ and $m$, are given in Table \ref{table1}.
\begin{table}[h!]
	\caption{First non-null values for the constants $b_{jlm}$.}
	\begin{center}
	\begin{tabular}{|c|c|c|c|c|}\hline
	$l$ &  2 & 2&   3 &  3
	\\\hline
	$m$ & -2 & 0 &  -3 & -1  
	\\\hline 
	& & & &  \\[0.05cm]
	$\dfrac{a \mu (\epsilon^2-1)b_{jlm}}{M_j^2}$ 
	& $\dfrac{i\sqrt{15 \pi}}{2}$ & $\sqrt{5\pi}$  & $-\dfrac{i}{2}\sqrt{\dfrac{35 \pi}{2}}$ & $\dfrac{i}{2}\sqrt{\dfrac{21 \pi}{2}}$ 
	\\[0.05cm]
	& & & &  \\\hline
	\end{tabular}
	\end{center}
	\label{table1}
	\end{table}
	
Here, these coefficients are written only for $m<0$, because the others can be obtained, recalling that
\begin{equation}
_s\overline{Z}_{lm}=\left(-1\right)^{s+m}\ _{-s}Z_{l(-m)}.
\end{equation}
Thus, for $m\ne0$, one has
\begin{equation}
b_{jlm}=ib_{jl(-m)}\hspace{0.5cm}j=1,2.
\end{equation}
\subsection{Gravitational Radiation Emitted by the Binary}
The power emitted in gravitational waves is computed from the Bondi's News function \eqref{BNews}. In terms of the coefficients $_sf_{lm}$, this function reads
\begin{align}
\mathcal{N}=&\sum_{l,m}\lim\limits_{r\rightarrow\infty}\Re\left(\left(-\frac{ir^2\dot{\tilde{\phi}}|m| J_{lm,r}}{2}-\frac{r^2\dot{\tilde{\phi}} J_{lm,\tilde{\phi} r} }{2}\right.\right.\nonumber\\
&\hspace{2.5cm}+\left.\left.\frac{l(l+1)J_{lm}}{4}+\beta_{lm}\right)e^{i|m|\dot{\tilde{\phi}}}\right)\eth^2Z_{lm},
\label{news2}
\end{align}
where the sum indicates that the News is constructed from the contribution of several multipole terms. Here it is important to note that the coefficients $J_{lm}$ depend directly on the source angular position, represented by $\tilde{\phi}$. For this reason the retarded-time derivative $J_{,ur}$ is re-expressed using the chain rule.    
 
When the solutions to the field equations, for $r>r_2$ are substituted in \eqref{news2}, for $l=2$, one finds,
\begin{align}
\mathcal{N}=& 2i\sqrt{\frac{2}{3}} \dot{\tilde{\phi}} \left(\Re(e^{2i\tilde{\phi}}D_{2J22+} )\ _2Z_{2\ 2}+\Re(e^{2i\tilde{\phi}}D_{2J2-2+})\ _2Z_{2\ -2}\right)\nonumber\\
&+\frac{1}{2}\sqrt{\frac{3}{2}}\dot{\tilde{\phi}}\left(\Re(e^{2i\tilde{\phi}}D_{2J22+}')\ _2Z_{2\ 2}+\Re(e^{2i\tilde{\phi}}D_{2J2-2+}')\ _2Z_{2\ -2}\right),
\label{Nf}
\end{align}
where the prime indicates derivation with respect to $\tilde{\phi}$. It is worth noting that the $D_{2Jlm+}$ depends on $\dot{\tilde{\phi}}$, just as indicated in \eqref{field_eq_x13}. Given that $\dot{\tilde{\phi}}:=\dot{\tilde{\phi}}(\tilde{\phi})$, then they are functions of the retarded angular position. Likewise, it is important to note that the absence of terms for $|m|=1$ in the News expression is because $b_{j21}=b_{j2-1}=0$ as indicated in Table \ref{table1}. In addition, despite $b_{jl0}\ne 0$ for $l=2,3,\cdots$, the terms for $m=0$ do not enter in the News, which indicates that they are non-radiative terms.

In the limit of low velocities, $r_1\dot{\tilde{\phi}}\ll c$, $r_2\dot{\tilde{\phi}} \ll c$ and for $l=2$, we find that the power lost by gravitational radiation emission reads
\begin{align}
\frac{dE}{du}=&\frac{32M_1^2M_2^2\left(M_1+M_2\right)\left(1+\epsilon\cos\tilde{\phi}\right)^6}{5a^5(1-\epsilon^2)^5}\nonumber\\
&+\frac{8 M_1^2M_2^2\left(M_1+M_2\right)\epsilon^2\sin^2\tilde{\phi}(1+\epsilon\cos\tilde{\phi})^4}{15 a^5(1-\epsilon^2)^5},
\label{power}
\end{align}
which is nothing but the Peters and Mathews expression for the energy lost by binary systems directly computed from the quadrupole radiation formulae \eqref{PM_pot_ecc} (see \cite{PM63}). 

The agreement between our results and those by Peters and Mathews is in fact expected, since the system under study is the same. On the other hand, this agreement shows that characteristic formalism in the linear regime has been properly applied in the present text. Recall that Winicour in the 1980s decade showed that the Bondi's News function in the Quasi-Newtonian regime \cite{IWW85,W87} is just 
\begin{equation}
\mathcal{N}=\dddot{Q},
\end{equation}
with \begin{equation}
Q=q^A\overline{q}^BQ_{AB}.
\end{equation} 
Likewise, it is important to note that the first term in the power expression \eqref{power} represents approximately $97\%$ of the power emitted by the source. Thus, for $\epsilon<0.5$ a reasonable approximation is just given by the first term of \eqref{power}.

It is worth noting that, for the case of circular orbits, the two first terms in the News \eqref{Nf}, lead directly to
\begin{align}
\frac{dE}{du}=&\frac{32M_1^2M_2^2\left(M_1+M_2\right)}{5a^5}.
\end{align} 
which corresponds in fact to the first term of \eqref{power_circ}.
\chapter{CONCLUSIONS, FINAL REMARKS AND PERSPECTIVES}

In this work we report new solutions to the master equation when a flat background is considered, generalising the results obtained by M\"adler \cite{M13} with the inclusion of source terms. Likewise, we re-express the family of solutions for the vacuum using only Bessel's functions of the first kind \cite{CA15b}. 

We also report for the first time in the literature the exact solutions to the master equation in terms of the Hypergeometric (Heun's function) for the non-radiative (radiative) modes with and without source terms when a Schwarzschild background is considered. Considering the solutions for $l=2$ we also show the equivalence between our solution and those reported in the literature \cite{CA15b}. Thus, this work extends the results shown by \citeonline{B05}, who already found the solutions to the field equations in the space-time exterior to a static and spherically symmetric black-hole. He treats the case for $l=2$, but only by expanding the metric variables in power series around the coordinate singularity $r=2M$, and in an asymptotic expansion near the null infinity. However, his solutions depend on the order of the expansion and in this sense they are approximations.  

It is worth stressing that the importance of these analytical results is in the fact that they can be useful in the construction of semi-analytical models for matter distributions in the linear regime, like thin and thick shells or stars composed of layers obeying some equation of state.   However, as already mentioned, it is important to bear in mind that the matter fields must be known {\it a priori} throughout the space-time.

Another important aspect is that the solutions when matter is present are valid only when the light ray bending is negligible and consequently the linear regime is valid. Out of this consideration, caustics could be formed and, consequently, the radiation coordinates becomes meaningless and, in this case, the space-time could not be represented by this kind of metric.

We generalised a previous work by \citeonline{B05}, in particular that concerned with binary systems composed of two components of equal masses in a Minkowski's background \cite{CA15}. Here we considered the case in which the components of the binary systems have different masses, although still in circular orbits.

We showed that, instead of two regions, as in the case of binaries with equal components, the space-time needs now to be separated in three regions, namely, interior, between and outside the two world tubes. As a result, the matched conditions need to be applied now for two hypersurfaces generated by the circular orbits of the two (different) masses. 

In that event, it has been necessary to generalise the boundary conditions satisfied by the coefficients in the spin-weighted spherical harmonics expansion, on the two hypersurfaces generated by the circular orbits of these two (different) masses. Also, the procedure developed here allows one to perform calculations for arbitrary values of the $l$ and $m$ modes.

It is worth stressing, that one of the most interesting aspects of this  study has do with the development of a procedure that can be applied in problems in which multi layers are present. 

We also calculate the energy lost by the emission of gravitational waves by means of the Bondi's News function. Again, we do that for arbitrary multipoles, in other words, for different values of the $l$ and $m$ modes. The interesting point here is that for different masses the emission of gravitational radiation occurs for all values (multipoles) of $l\ge 2$; for the particular case of binary systems with equal components, the multipole terms for odd values of $l$ vanish.

We also study for the first time in the literature a binary system composed of point particles of unequal masses in eccentric orbits in the linear regime of the characteristic formulation of general relativity \cite{CA16b}. This work generalises previous studies \cite{B05} (\cite{CA15}) in which a system of equal (different) masses in circular orbits is considered. Also, it was considered that in general the angular velocity is a temporal function, which allows the inclusion of the terms responsible for the the contributions of the eccentricity in the power emitted by the system.

We show that the boundary conditions on the time-like world tubes \eqref{bound_conds} can be extended beyond circular orbits. Concerning the power lost by the emission of gravitational waves, it is directly obtained from the Bondi's News function.

Since the contribution of the several multipole terms ($l>2$) to the power is smaller than the contribution given by mode for $l=2$, the terms for $l>2$ are disregarded in the power expression \eqref{power}. In addition, the second term in \eqref{power} is smaller than the first one. For example, for eccentricities $\epsilon \lesssim 0.5$ the first term contributes with almost $97\%$ of the power emitted in gravitational waves \cite{CA16b}. 

It is worth noting that our results are completely consistent, because we obtain the same result for the power derived by Peters and Mathews using a different approach. Recall that the News function in the Quasi-Newtonian limit corresponds to the third derivative with respect to the retarded time of the quadrupole moment contracted with the tangent vectors $q^A$, i.e., $\mathcal{N}=\dddot{Q}$, where $Q=q^A\overline{q}^BQ_{AB}$ \cite{W87}.

Finally, the present study constitutes a powerful tool to construct extraction schemes in the characteristic formalism to obtain the gravitational radiation produced by binary systems during the inspiralling phase. This can be done in regions that are far enough from the sources where the space-time can be essentially considered flat. 

This work contributes to extend analytical previous results obtained in \cite{B05,M13}. This new extensions can be applicable to relevant astrophysical sources as thick shells in which the dynamics obeys  particular equations of state. Also, it is possible to generalise this results to a star formed by concentric thick layers. With the introduction of the eccentricity and an angular velocity depending on the position, it is possible to generalise the form of such layers to spheroidal layers in order to include this into the gravitational signature of such kind of objects. In addition, from the linear version of the field equations it is possible to integrate them numerically and reproduce the quasi-normal modes for Schwarzschild and Reissner-Nordstr\"om solutions. There are also possible extensions of this work in cosmology, in $f(R)$ theories, using radiation coordinates and the eth formalism. Finally, it is worth mentioning that from the linear version of the field equations in the characteristic formalism and in order to avoid the numerical angular treatment, it is possible to study the gravitational collapse of a matter distribution by using the multipolar expansions present here.   
\bibliography{bib/references}
\hypertarget{estilo:apendice1}{} 
\renewcommand{\thechapter}{}%
\chapter[Appendix A]{Appendix A - Explicit Form for the $\eth$ and $\overline\eth$ Operators in Stereographic Coordinates}
\renewcommand{\thechapter}{A}%
\label{apendiceA}	
Considering that the covariant derivative $\Psi^{\tilde{B}_{1m}}_{\hspace{0.65cm}\tilde{A}_{1n}|D}$ associated to $q_{AB}$ is
\begin{align}
\Psi^{\tilde{B}_{1m}}_{\hspace{0.65cm}\tilde{A}_{1n}|D}=&\Psi^{\tilde{B}_{1m}}_{\hspace{0.65cm}\tilde{A}_{1n},D}+\Omega^{B_{1}}_{~~~DC}\Psi^{C \tilde{B}_{2m}}_{\hspace{0.75cm}\tilde{A}_{1n}}+\cdots
+\Omega^{B_{m}}_{~~~DC}\Psi^{\tilde{B}_{1(m-1)}C}_{\hspace{1.3cm}\tilde{A}_{1n}}\nonumber\\ &-\Omega^C_{~~A_1D}\Psi^{\tilde{B}_{1m}}_{\hspace{0.6cm}C\tilde{A}_{2m}}
-\cdots -\Omega^C_{~~A_mD}\Psi^{\tilde{B}_{1m}}_{\hspace{0.6cm}\tilde{A}_{1(m-1)}C},
\label{ethgen3_v2}
\end{align}
then, when substituted into \eqref{ladder4} yields
\begin{align}
\eth \ _s\Psi=& q^D\tilde{\Lambda}_{\tilde{B}_{1m}}\tilde{\Lambda}^{\tilde{A}_{1n}} \Psi^{\tilde{B}_{1m}}_{\hspace{0.6cm}\tilde{A}_{1n},D} 
+ q^D \tilde{\Lambda}_{\tilde{B}_{1m}} \tilde{\Lambda}^{\tilde{A}_{1n}} \Omega^{B_{1}}_{~~~DC}\Psi^{C \tilde{B}_{2m}}_{\hspace{0.8cm}\tilde{A}_{1n}} +\cdots\nonumber\\
&+q^D\tilde{\Lambda}_{\tilde{B}_{1m}}\tilde{\Lambda}^{\tilde{A}_{1n}}\Omega^{B_{m}}_{~~~DC}\Psi^{\tilde{B}_{1(m-1)C}}_{\hspace{1.3cm}\tilde{A}_{1n}} - q^D\tilde{\Lambda}_{\tilde{B}_{1m}} \tilde{\Lambda}^{\tilde{A}_{1n}}\Omega^C_{~~A_1D} \Psi^{\tilde{B}_{1m}}_{\hspace{0.6cm}C\tilde{A}_{2n}} \nonumber
\\&-\cdots-q^D\tilde{\Lambda}_{\tilde{B}_{1m}}\tilde{\Lambda}^{\tilde{A}_{1n}}\Omega^C_{~~A_mD}\Psi^{\tilde{B}_{1m}}_{\hspace{0.6cm}\tilde{A}_{1(n-1)}C}.
\label{ethgen4_v2}
\end{align}
Notice that the first term of the last equation can be written as
\begin{align}
q^D\tilde{\Lambda}_{\tilde{B}_{1m}}\tilde{\Lambda}^{\tilde{A}_{1n}}\Psi^{\tilde{B}_{1m}}_{\hspace{0.6cm}\tilde{A}_{1n,D}}
=& q^D\left(_s\Psi\right)_{,D} 
- q^D\Lambda_{B_1,D}\tilde{\Lambda}_{\tilde{B}_{2m}} \tilde{\Lambda}^{\tilde{A}_{1n}} \Psi^{\tilde{B}_{1m}}_{\hspace{0.6cm}\tilde{A}_{1n}}-\cdots\nonumber\\
&-q^D\Lambda_{B_{m},D}\tilde{\Lambda}_{\tilde{B}_{1(m-1)}} \tilde{\Lambda}^{\tilde{A}_{1n}} \Psi^{\tilde{B}_{1m}}_{\hspace{0.6cm}\tilde{A}_{1n}}-\cdots \nonumber\\
&-q^D\Lambda^{A_{1}}_{~~~,D} \tilde{\Lambda}_{\tilde{B}_{1m}} \tilde{\Lambda}^{\tilde{A}_{2n}} \Psi^{\tilde{B}_{1m}}_{\hspace{0.6cm}\tilde{A}_{1n}}\nonumber\\
&-\cdots-q^D \Lambda^{A_n}_{~~~,D} \tilde{\Lambda}_{\tilde{B}_{1m}} \tilde{\Lambda}^{\tilde{A}_{1(n-1)}} \Psi^{\tilde{B}_{1m}}_{\hspace{0.6cm}\tilde{A}_{1n}}.
\label{ethgen5_v2}
\end{align}
Thus, substituting \eqref{ethgen5_v2} into \eqref{ethgen4_v2}, reorganising the sums and changing the name of some indices  one obtains
\begin{align}
\eth \ _s\Psi=& q^D\left(_s\Psi\right)_{,D} - q^D\left(\Lambda_{B_1,D}-\Lambda_{C}\Omega^{C}_{~~~B_1 D}\right) \tilde{\Lambda}_{\tilde{B}_{2m}} \tilde{\Lambda}^{\tilde{A}_{1n}}\Psi^{\tilde{B}_{1m}}_{\hspace{0.6cm}\tilde{A}_{1n}}  -\cdots\nonumber\\
&-q^D\left(\Lambda_{B_m,D}-\Lambda_{C}\Omega^{C}_{~~~B_{m}D}\right) \tilde{\Lambda}_{\tilde{B}_{1(m-1)}} \tilde{\Lambda}^{\tilde{A}_{1n}}\Psi^{\tilde{B}_{1m}}_{\hspace{0.6cm}\tilde{A}_{1n}} \nonumber\\
&- q^D\left(\Lambda^{A_1}_{~~~,D}+\Lambda^{C}\Omega^{A_1}_{~~~CD}\right)  \tilde{\Lambda}_{\tilde{B}_{1m}} \tilde{\Lambda}^{\tilde{A}_{2n}}\Psi^{\tilde{B}_{1m}}_{\hspace{0.6cm}\tilde{A}_{1n}}-\cdots \nonumber\\
&-q^D\left(\Lambda^{A_n}_{~~~,D}+\Lambda^{C}\Omega^{A_n}_{~~CD}\right) \tilde{\Lambda}_{\tilde{B}_{1m}} \tilde{\Lambda}^{\tilde{A}_{1(n-1)}}\Psi^{\tilde{B}_{1m}}_{\hspace{0.6cm}\tilde{A}_{1n}}.
\label{ethgen7_v2}
\end{align}
Recognising the covariant derivatives for the $\Lambda$ symbols in the brackets, one obtains
\begin{align}
\eth \ _s\Psi=& q^D\left(_s\Psi\right)_{,D} - q^D \Lambda_{B_1|D} \tilde{\Lambda}_{\tilde{B}_{2m}} \tilde{\Lambda}^{\tilde{A}_{1n}}\Psi^{\tilde{B}_{1m}}_{\hspace{0.6cm}\tilde{A}_{1n}}  -\cdots
\nonumber\\&
-q^D\Lambda_{B_m|D} \tilde{\Lambda}_{\tilde{B}_{1(m-1)}} \tilde{\Lambda}^{\tilde{A}_{1n}}\Psi^{\tilde{B}_{1m}}_{\hspace{0.6cm}\tilde{A}_{1n}} 
- q^D\Lambda^{A_1}_{~~~|D} \tilde{\Lambda}_{\tilde{B}_{1m}} \tilde{\Lambda}^{\tilde{A}_{2n}}\Psi^{\tilde{B}_{1m}}_{\hspace{0.6cm}\tilde{A}_{1n}}-\cdots \nonumber\\
&-q^D\Lambda^{A_n}_{~~~|D} \tilde{\Lambda}_{\tilde{B}_{1m}} \tilde{\Lambda}^{\tilde{A}_{1(n-1)}}\Psi^{\tilde{B}_{1m}}_{\hspace{0.6cm}\tilde{A}_{1n}}.
\label{ethgen8_v2}
\end{align}
Now, it is important to observe that 
\begin{align}
q^D\Lambda^{A_k}_{~~~|D}&=\dfrac{q^{A_k}q^D\overline{q}^{C}\Lambda_{C|D}+\overline{q}^{A_k}q^D{q}^{C}\Lambda_{C|D}}{2},
\label{ethgen9}
\end{align}
since
\begin{align}
q^Aq^{C}q_{C|A}&=\dfrac{q^Aq^{C}q_{C|A}+q^Aq^{C}q_{C|A}}{2}\nonumber\\
&=\dfrac{q^A\left(q^{C}q_{C|A}+q_{C}q^{C}_{~~|A}\right)}{2}\nonumber\\
&=0.
\label{ethgen10}
\end{align}
Then, \eqref{ethgen9} reads
\begin{align}
q^D\Lambda^{A_k}_{~~~|D}&=\begin{cases}
\dfrac{q^{A_k}q^D\overline{q}^{C}q_{C|D}}{2} & \text{for} \hspace{0.5cm}\Lambda^{A_k}=q^{A_k}\\[.4cm]
-\dfrac{\overline{q}^{A_k}q^D\overline{q}^{C}q_{C|D}}{2} & \text{for} \hspace{0.5cm} \Lambda^{A_k}=\overline{q}^{A_k}
\end{cases},
\label{ethgen13}
\end{align}
Also, since in stereographic coordinates
\begin{align}
q^A\overline{q}^{C}q_{C|A}&=q^A\overline{q}^{C}q_{C,A}-q^A\overline{q}^{C}q_D\Omega^D_{~~CA}\nonumber\\
&=-2\zeta,
\label{ethgen11}
\end{align}
where
\begin{equation}
q^A\overline{q}^{C}q_D\Omega^D_{~~CA}=0,
\label{ethgen12}
\end{equation}
then, \eqref{ethgen13} is simplified to
\begin{align}
q^D\Lambda^{A_k}_{~~~|D}&=\begin{cases}
q^{A_k}\zeta & \text{for} \hspace{0.5cm}\Lambda^{A_k}=q^{A_k}\\[.4cm]
-\overline{q}^{A_k}\zeta & \text{for} \hspace{0.5cm} \Lambda^{A_k}=\overline{q}^{A_k}
\end{cases}.
\label{ethgen14}
\end{align}
Thus, from \eqref{ethgen14}, lowering the index with the metric $q_{AB}$, one obtains that
\begin{align}
q^D\Lambda^{}_{A_k|D}&=\begin{cases}
q_{A_k}\zeta & \text{for} \hspace{0.5cm}\Lambda_{A_k}=q_{A_k}\\[.4cm]
-\overline{q}_{A_k}\zeta & \text{for} \hspace{0.5cm} \Lambda_{A_k}=\overline{q}_{A_k}
\end{cases}.
\label{ethgen15}
\end{align}
Writing \eqref{ethgen8_v2} in the form, 
\begin{align}
\eth \ _s\Psi=& q^D\left(_s\Psi\right)_{,D} - \left(q^D \Lambda_{B_1|D} \tilde{\Lambda}_{\tilde{B}_{2m}} +\cdots + q^D\Lambda_{B_m|D} \tilde{\Lambda}_{\tilde{B}_{1(m-1)}} \right)\tilde{\Lambda}^{\tilde{A}_{1n}} \Psi^{\tilde{B}_{1m}}_{\hspace{0.6cm}\tilde{A}_{1n}}  
\nonumber\\&
- \left(q^D\Lambda^{A_1}_{~~~|D}  \tilde{\Lambda}^{\tilde{A}_{2n}}+\cdots 
+q^D\Lambda^{A_n}_{~~~|D}  \tilde{\Lambda}^{\tilde{A}_{1(n-1)}}\right) \tilde{\Lambda}_{\tilde{B}_{1m}} \Psi^{\tilde{B}_{1m}}_{\hspace{0.6cm}\tilde{A}_{1n}},
\label{ethgen16}
\end{align}
one observes that the first bracket corresponds to
\begin{align*}
& q^D \Lambda_{B_1|D}\tilde{\Lambda}_{\tilde{B}_{2m}}  +\cdots+q^D\Lambda_{B_m|D}\tilde{\Lambda}_{\tilde{B}_{1(m-1)}}\\
=& q^D \Lambda_{B_1|D}\tilde{\Lambda}_{\tilde{B}_{2x}} \tilde{\Lambda}_{\tilde{B}_{(x+1)m}}+\cdots+q^D\Lambda_{B_x|D} \tilde{\Lambda}_{\tilde{B}_{1(x-1)}} \tilde{\Lambda}_{\tilde{B}_{(x+1)m}}  \nonumber\\
&+q^D\Lambda_{B_{x+1}|D} \tilde{\Lambda}_{\tilde{B}_{1x}} \tilde{\Lambda}_{\tilde{B}_{(x+2)m}} +\cdots+q^D\Lambda_{B_m|D} \tilde{\Lambda}_{\tilde{B}_{1x}} \tilde{\Lambda}_{\tilde{B}_{(x+1)(m-1)}}\\
&=\zeta\Bigg(\underbrace{- \tilde{q}_{\tilde{B}_{1x}} \tilde{\overline{q}}_{\tilde{B}_{(x+1)m}}-\cdots- \tilde{q}_{\tilde{B}_{1x}} \tilde{\overline{q}}_{\tilde{B}_{(x+1)m}}}_{x \ \text{terms}}\\
&\hspace{1cm} +\underbrace{ \tilde{q}_{\tilde{B}_{1x}} \tilde{\overline{q}}_{\tilde{B}_{(x+1)m}} 
	+\cdots+ \tilde{q}_{\tilde{B}_{1x}} \tilde{\overline{q}}_{\tilde{B}_{(x+1)m}}}_{m-x \ \text{terms}}\Bigg), 
\end{align*}
i.e.,
\begin{align}
 q^D \Lambda_{B_1|D} \tilde{\Lambda}_{\tilde{B}_{2m}}  +\cdots+q^D\Lambda_{B_m|D} \tilde{\Lambda}_{\tilde{B}_{1(m-1)}}
=\zeta(m-2x) \tilde{\Lambda}_{\tilde{B}_{1m}};
\label{ethgen17}
\end{align}
whereas the second bracket is
\begin{align*}
& q^D\Lambda^{A_1}_{~~~|D} \tilde{\Lambda}^{\tilde{A}_{2n}}+\cdots + q^D\Lambda^{A_n}_{~~~|D}\tilde{\Lambda}^{\tilde{A}_{1(n-1)}}\nonumber\\
&= -q^{A_1}\zeta \tilde{q}^{\tilde{A}_{2y}} \tilde{\overline{q}}^{\tilde{A}_{(y+1)n}}
\Psi^{\tilde{B}_{1m}}_{\hspace{0.6cm}\tilde{A}_{1n}}
-\cdots-q^{A_y}\zeta \tilde{q}^{\tilde{A}_{1(y-1)}} \tilde{\overline{q}}^{\tilde{A}_{(y+1)n}}
\nonumber\\
&\hspace{0.5cm}+\overline{q}^{A_{y+1}}\zeta \tilde{q}^{\tilde{A}_{1y}} \tilde{\overline{q}}^{\tilde{A}_{(y+2)n}}\Psi^{\tilde{B}_{1m}}_{\hspace{0.6cm}\tilde{A}_{1n}} +\cdots +\overline{q}^{A_{n}}\zeta \tilde{q}^{\tilde{A}_{1y}} \tilde{\overline{q}}^{\tilde{A}_{(y+1)(n-1)}}\nonumber\\
&=\zeta\Bigg(\underbrace{-\tilde{q}^{\tilde{A}_{1y}} \tilde{\overline{q}}^{\tilde{A}_{(y+1)n}} -\cdots- \tilde{q}^{\tilde{A}_{1y}} \tilde{\overline{q}}^{\tilde{A}_{(y+1)n}}}_{y \ \text{terms}}\nonumber\\
&\hspace{1cm}+ \underbrace{\tilde{q}^{\tilde{A}_{1y}} \tilde{\overline{q}}^{\tilde{A}_{(y+1)n}} +\cdots + \tilde{q}^{\tilde{A}_{1y}} \tilde{\overline{q}}^{\tilde{A}_{(y+1)n}}}_{n-y\ \text{terms}}\Bigg),
\end{align*}
or
\begin{align}
q^D\Lambda^{A_1}_{~~~|D} \tilde{\Lambda}^{\tilde{A}_{2n}}+\cdots + q^D\Lambda^{A_n}_{~~~|D}\tilde{\Lambda}^{\tilde{A}_{1(n-1)}}
=\zeta(n-2y)\tilde{\Lambda}^{\tilde{A}_{1n}}.
\label{ethgen18}
\end{align}
Thus, substituting \eqref{ethgen17} and \eqref{ethgen18} into \eqref{ethgen16} one obtains a simple expression in stereographic coordinates
\begin{align}
\eth \ _s\Psi=& q^D\left(_s\Psi\right)_{,D} - \zeta(m-2x+n-2y) \tilde{\Lambda}_{\tilde{B}_{1m}} \tilde{\Lambda}^{\tilde{A}_{1n}} \Psi^{\tilde{B}_{1m}}_{\hspace{0.6cm}\tilde{A}_{1n}}\nonumber\\
=& q^D\left(_s\Psi\right)_{,D} + s\, \zeta \ _s\Psi.
\label{ethgen19a}
\end{align}
From the definition for the {\it eth bar} operator,
\begin{equation}
\overline{\eth} \ _s\Psi=\overline{q}^D \tilde{\Lambda}_{\tilde{B}_{1m}} \tilde{\Lambda}^{\tilde{A}_{1n}}  \Psi^{\tilde{B}_{1m}}_{\hspace{0.6cm}\tilde{A}_{1n}|D},
\label{ethgen18_v2a}
\end{equation} 
and using \eqref{ethgen3_v2} one obtains that
\begin{align}
\overline{\eth} \ _s\Psi&=\overline{q}^D\left(_s\Psi\right)_{,D} - \overline{q}^D \Lambda_{B_1|D} \tilde{\Lambda}_{\tilde{B}_{2m}} \tilde{\Lambda}^{\tilde{A}_{1n}}\Psi^{\tilde{B}_{1m}}_{\hspace{0.6cm}\tilde{A}_{1n}}  -\cdots\nonumber\\
&-\overline{q}^D\Lambda_{B_m|D} \tilde{\Lambda}_{\tilde{B}_{1(m-1)}} \tilde{\Lambda}^{\tilde{A}_{1n}}\Psi^{\tilde{B}_{1m}}_{\hspace{0.6cm}\tilde{A}_{1n}} 
- \overline{q}^D\Lambda^{A_1}_{~~~|D} \tilde{\Lambda}_{\tilde{B}_{1m}} \tilde{\Lambda}^{\tilde{A}_{2n}} \Psi^{\tilde{B}_{1m}}_{\hspace{0.6cm}\tilde{A}_{1n}} -\cdots \nonumber\\
&-\overline{q}^D\Lambda^{A_n}_{~~~|D} \tilde{\Lambda}_{\tilde{B}_{1m}} \tilde{\Lambda}^{\tilde{A}_{1(n-1)}}\Psi^{\tilde{B}_{1m}}_{\hspace{0.6cm}\tilde{A}_{1n}},
\label{ethgen19_v2}
\end{align}
where
\begin{align}
\overline{q}^D\Lambda^{A_k}_{~~~|D}&=\dfrac{q^{A_k}\overline{q}^D\overline{q}^{C}\Lambda_{C|D}+\overline{q}^{A_k}\overline{q}^D{q}^{C}\Lambda_{C|D}}{2},
\label{ethgen20}
\end{align}
and
\begin{align}
\overline{q}^D\Lambda_{B_k|D}&=\dfrac{q_{B_k}\overline{q}^D\overline{q}^{C}\Lambda_{C|D}+\overline{q}^D\overline{q}_{B_k}{q}^{C}\Lambda_{C|D}}{2}.
\label{ethgen21}
\end{align}
Thus, the two last equations result in
\begin{subequations}
\begin{align}
\overline{q}^D\Lambda^{A_k}_{~~~|D}&=\begin{cases}
q^{A_k}\overline{\zeta} & \text{if} \hspace{0.5cm}\Lambda^{A_k}=q^{A_k}\\ 
-\overline{q}^{A_k}\overline{\zeta} & \text{if} \hspace{0.5cm} \Lambda^{A_k}=\overline{q}^{A_k}
\end{cases},
\label{ethgen22}
\end{align}
and
\begin{align}
\overline{q}^D\Lambda_{B_k|D}&=\begin{cases}
q_{B_k}\overline{\zeta} & \text{for} \hspace{0.5cm} \Lambda_{B_k}=q_{B_k}\\ 
-\overline{q}_{B_k}\overline{\zeta}  & \text{for} \hspace{0.5cm} \Lambda_{B_k}=\overline{q}_{B_k}
\end{cases}.
\label{ethgen23}
\end{align}
\label{ethgen23s}%
\end{subequations}
Then, the expression associated with the $\overline{\eth}$ operator acting on the $_s\Psi$ becomes 
\begin{align}
\overline{\eth} \ _s\Psi&=\overline{q}^D\left(_s\Psi\right)_{,D} -\overline{\zeta}(2x-m)\ _s\Psi -\overline{\zeta}(2y-n)\ _s\Psi\nonumber\\
&=\overline{q}^D\ _s\Psi_{,D} -s\overline{\zeta} \ _s\Psi.
\label{ethgen24}
\end{align} 
\hypertarget{estilo:apendice2}{} 
\renewcommand{\thechapter}{}%
\chapter[Appendix B]{Appendix B - Angular Operators $\partial_{\theta\theta}$, $\partial_{\theta\phi}$ and $\partial_{\phi\phi}$ in terms of $\eth$ and $\overline\eth$}
\label{apendiceB}
\renewcommand{\thechapter}{B}
From
\begin{equation}
\partial_{\theta}=\dfrac{\eth+\overline{\eth}}{2}, \hspace{1cm}
\partial_{\phi}=\dfrac{i\sin\theta}{2}\left(\overline{\eth}-\eth-2s\cot\theta\right). 
\label{eth_spherical_16}
\end{equation}
and from Equations \eqref{eth_spherical_7}, it is possible to obtain the expressions for $\eth^2$, $\overline{\eth}^2$, $\overline{\eth}\eth$ and $\eth\overline{\eth}$, which lead to the expressions for $\partial_{\theta\theta}$, $\partial_{\theta\phi}$ and $\partial_{\phi\phi}$, namely
\begin{align}
\eth^2 
=&\left(\partial_\theta + i\csc\theta\partial_\phi - (s+1)\cot\theta\right)\partial_\theta + i\left(\partial_\theta + i\csc\theta\partial_\phi \right.\nonumber\\
&\left. - (s+1)\cot\theta\right)  \csc\theta\partial_\phi - s \left(\partial_\theta + i\csc\theta\partial_\phi - (s+1)\cot\theta\right)  \cot\theta.
\label{eth_spherical_17}
\end{align}
The first term in \eqref{eth_spherical_17} is 
\begin{align}
\left(\partial_\theta + i\csc\theta\partial_\phi - (s+1)\cot\theta\right)\partial_\theta &= \partial_{\theta\theta} + i\csc\theta\partial_{\phi\theta} - (s+1)\cot\theta\partial_\theta,
\label{eth_spherical_18}
\end{align}
the second term in \eqref{eth_spherical_17} is given by
\begin{align}
&i\left(\partial_\theta + i\csc\theta\partial_\phi - (s+1)\cot\theta\right) \csc\theta\partial_\phi \nonumber\\
=& i\big(-\csc\theta\cot\theta\partial_\phi + \csc\theta \partial_{\theta\phi} + i\csc^2\theta\partial_{\phi\phi} 
- (s+1)\cot\theta  \csc\theta\partial_\phi \big),
\label{eth_spherical_19}
\end{align}
finally the third term in \eqref{eth_spherical_17} reads
\begin{align}
&- s\left(\partial_\theta + i\csc\theta\partial_\phi - (s+1)\cot\theta\right)  \cot\theta\nonumber\\
=& - s\left(-\csc^2\theta + \cot\theta \partial_\theta  + i\csc\theta\cot\theta\partial_\phi   - (s+1)\cot^2\theta  \right).
\label{eth_spherical_20}
\end{align}
Thus, the substitution of \eqref{eth_spherical_18}-\eqref{eth_spherical_20} into \eqref{eth_spherical_17} leads to
\begin{align}
\eth^2
=& \partial_{\theta\theta} - \csc^2\theta\partial_{\phi\phi} + 2i\csc\theta\partial_{\phi\theta} - (2s+1)\cot\theta\partial_\theta  \nonumber\\
&- 2i (s+1)\cot\theta  \csc\theta\partial_\phi + s\left((s+1)\cot^2\theta + \csc^2\theta   \right).
\label{eth_spherical_21}
\end{align}
From \eqref{eth_spherical_7} we construct $\overline{\eth}^2$ as follows
\begin{align}
\overline{\eth}^2 =&\partial_\theta\left[\partial_\theta - i\csc\theta\partial_\phi + s\cot\theta\right] - i\csc\theta\partial_\phi\left[\partial_\theta - i\csc\theta\partial_\phi + s\cot\theta\right]\nonumber\\
& + (s-1)\cot\theta\left[\partial_\theta - i\csc\theta\partial_\phi + s\cot\theta\right].
\label{eth_spherical_22}
\end{align}
The first term in \eqref{eth_spherical_22} corresponds to 
\begin{align}
\partial_\theta\left[\partial_\theta - i\csc\theta\partial_\phi + s\cot\theta\right]
=&\partial_{\theta\theta} - i\csc\theta\partial_{\theta\phi} + i\csc\theta\cot\theta\partial_\phi  \nonumber\\
&- s\csc^2\theta + s\cot\theta\partial_\theta,
\label{eth_spherical_23}
\end{align}
the second term in \eqref{eth_spherical_22} is given by
\begin{align}
&- i\csc\theta\partial_\phi\left[\partial_\theta - i\csc\theta\partial_\phi + s\cot\theta\right]\nonumber\\
=& - i\left[\csc\theta\partial_{\phi\theta} - i\csc^2\theta\partial_{\phi\phi} + s\csc\theta\cot\theta\partial_\phi\right];
\label{eth_spherical_24}
\end{align}
finally the third term in \eqref{eth_spherical_22} is given by
\begin{align}
&(s-1)\cot\theta\left[\partial_\theta - i\csc\theta\partial_\phi + s\cot\theta\right]\nonumber\\
=&(s-1)\left[\cot\theta\partial_\theta - i\cot\theta\csc\theta\partial_\phi + s\cot^2\theta\right].
\label{eth_spherical_25}
\end{align}
Thus, substituting \eqref{eth_spherical_23}-\eqref{eth_spherical_25} into \eqref{eth_spherical_22}, one obtains
\begin{align}
\overline{\eth}^2&=\partial_{\theta\theta}-\csc^2\theta\partial_{\phi\phi}-2i\csc\theta\partial_{\theta\phi} + (2s-1)\cot\theta\partial_\theta  \nonumber\\
& - 2i(s-1)\cot\theta\csc\theta\partial_\phi + s\left((s-1)\cot^2\theta-\csc^2\theta\right).
\label{eth_spherical_26}
\end{align}
Also, from \eqref{eth_spherical_7} one obtains
\begin{align}
\eth\overline{\eth}
=&\partial_{\theta\theta}+\csc^2\theta\partial_{\phi\phi}+\cot\theta\partial_\theta+2is\csc\theta\cot\theta\partial_\phi-s\left(s\cot^2\theta+1\right),
\label{eth_spherical_27}
\end{align}
and
\begin{align}
\overline{\eth}\eth 
=&\partial_{\theta\theta}+\csc^2\theta\partial_{\phi\phi} + \cot\theta\partial_\theta + 2is\cot\theta\csc\theta\partial_\phi  - s \left( s \cot^2\theta - 1\right).
\label{eth_spherical_28}
\end{align}
In order to check the above expressions, the commutator $\left[\overline{\eth},\eth\right]$ is computed, resulting in the well-known result 
\begin{equation*}
\left[\overline{\eth},\eth\right]= 2s,
\end{equation*}
and its anti-commutator reads
\begin{equation}
\left(\overline{\eth},\eth\right)= 2\left(\partial_{\theta\theta}+\csc^2\theta\partial_{\phi\phi} + \cot\theta\partial_\theta + 2is\cot\theta\csc\theta\partial_\phi  -  s^2 \cot^2\theta\right). 
\label{eth_spherical_29}
\end{equation}
Thus, from \eqref{eth_spherical_21}, \eqref{eth_spherical_26}, \eqref{eth_spherical_27}, \eqref{eth_spherical_28} the explicit form of the second order angular operators in terms of the spin-weighted  operators are given by
\begin{subequations}
\begin{align}
\partial_{\theta\theta}=&\dfrac{{\eth}^2+\left(\overline{\eth},\eth\right)+\overline{\eth}^2}{4}, \label{eth_spherical_30.1}\\
\partial_{\phi\phi} =& -\dfrac{\sin^2\theta}{4} \left({\eth}^2 -\left(\overline{\eth},\eth\right) + \overline{\eth}^2 \right)  - s^2 \cos^2\theta\nonumber\\
& - \sin\theta\cos\theta\left(\left(s+\dfrac{1}{2}\right)\eth -\left(s-\dfrac{1}{2}\right)\overline{\eth}\right), \label{eth_spherical_30.2}\\
\partial_{\theta\phi} =&- \dfrac{i\sin\theta}{4}\left(\eth^2-\overline{\eth}^2\right) -is\cos\theta\dfrac{\eth+\overline{\eth}}{2}  \nonumber\\
& + \dfrac{i\cos\theta}{2}\left(\overline{\eth}-\eth-2s\cot\theta\right) + i\sin\theta\dfrac{s(\cot^2\theta+\csc^2\theta)}{2}. \label{eth_spherical_30.3}
\end{align}
\label{eth_spherical_30}%
\end{subequations}	
\end{document}